\documentclass{ejm}
\usepackage{amsmath}
\usepackage{afterpage}
\usepackage[dvips]{graphicx}
\usepackage[normal]{subfigure}
\usepackage{amssymb}
\usepackage{subfig}
\def\ni{\noindent}
\def\proof{{\ni \bf \underline{Proof:} }}

\newcommand{\bsub}{\begin{subequations}}
\newcommand{\esub}{\end{subequations}$\!$}

\newcommand{\ac}{ \mathcal{A}}

\newcommand{\R}{{\mathbb{R}}}
\newcommand{\eps}{{\displaystyle \varepsilon}}
\newcommand{\lam}{{\lambda}}
\newcommand{\alp}{{\alpha}}

\newcommand{\p}{\prime}

\newcommand{\xb}{\mathbf{x}}
\newcommand{\xib}{\mathbf{x}_0}

\newcommand{\eb}{\mathbf{e}}

\newcommand{\newsection}[1]{{\setcounter{equation}{0}}\section{#1}}
\renewcommand{\theequation}{\arabic{section}.\arabic{equation}}

\newcommand{\stacksum}{\sum_{\stackrel{i=1}{i\neq j}}^{\scriptscriptstyle k}}

\newtheorem{example}{Example}[section]

\newcommand{\bex}{\begin{example}\rm}
\newcommand{\eex}{\end{example}}

\renewcommand{\theequation}{\arabic{section}.\arabic{equation}}
\renewcommand{\thesection}{\arabic{section}}

\graphicspath{{./figures/}}

\title[Localized Spot Patterns in the Two-Dimensional Gray-Scott
Model] {The Stability and Dynamics of Localized Spot Patterns in the
Two-Dimensional Gray-Scott Model}

\author[W. Chen, M. J. Ward]{%
  W. \ns C\ls H\ls E\ls N, \ns    M.\ns J. \ns W\ls A\ls R\ls D}
\affiliation{Wan Chen; Department of Mathematics,  University of British 
 Columbia, Vancouver, British Columbia, V6T 1Z2, Canada, (Currenntly at
 OCCAM, Oxford University, Oxford, U.K.)

 Michael J. Ward; Department of Mathematics,  University of British Columbia, 
 Vancouver, British Columbia, V6T 1Z2, Canada (corresponding author)}

\date{10 September 2010}
\pubyear{2010}
\volume{000}
\pagerange{\pageref{firstpage}--\pageref{lastpage}}

\begin{document}

\label{firstpage}
\maketitle

\baselineskip=12pt

\begin{abstract}
The dynamics and stability of multi-spot patterns to the Gray-Scott
(GS) reaction-diffusion model in a two-dimensional domain is studied
in the singularly perturbed limit of small diffusivity $\eps$ of one
of the two solution components. A hybrid asymptotic-numerical approach
based on combining the method of matched asymptotic expansions with
the detailed numerical study of certain eigenvalue problems is used to
predict the dynamical behavior and instability mechanisms of
multi-spot quasi-equilibrium patterns for the GS model in the limit
$\eps\to 0$.  For $\varepsilon\to 0$, a quasi-equilibrium $k$-spot
pattern is constructed by representing each localized spot as a
logarithmic singularity of unknown strength $S_j$ for $j=1,\ldots,k$
at unknown spot locations ${\mathbf x}_j\in \Omega$ for $j=1,\ldots,
k$. A formal asymptotic analysis is then used to derive a differential
algebraic ODE system for the collective coordinates $S_j$ and
${\mathbf x}_j$ for $j=1,\ldots,k$, which characterizes the slow
dynamics of a spot pattern. Instabilities of the multi-spot pattern
due to the three distinct mechanisms of spot self-replication, spot
oscillation, and spot annihilation, are studied by first deriving
certain associated eigenvalue problems by using singular perturbation
techniques. From a numerical computation of the spectrum of these
eigenvalue problems, phase diagrams representing in the GS parameter
space corresponding to the onset of spot instabilities are obtained
for various simple spatial configurations of multi-spot patterns. In
addition, it is shown that there is a wide parameter range where a
spot instability can be triggered only as a result of the intrinsic
slow motion of the collection of spots.  The construction of the
quasi-equilibrium multi-spot patterns and the numerical study of the
spectrum of the eigenvalue problems relies on certain detailed
properties of the reduced-wave Green's function. The hybrid
asymptotic-numerical results for spot dynamics and spot instabilities
are validated from full numerical results computed from the GS model
for various spatial configurations of spots.

\end{abstract}


\baselineskip=16pt 

\setcounter{equation}{0}
\setcounter{section}{0}
\section{Introduction} \label{section:intro}

Spatially localized spot patterns have been observed in a wide variety
of experimental settings including, the ferrocyanide-iodate-sulphite
reaction (cf.~\cite{lmps}), the chloride-dioxide-malonic acid reaction
(cf.~\cite{dbdd}), and certain electronic gas-discharge systems
(cf.~\cite{as1}, \cite{as2}).  Furthermore, numerical simulations of
certain simple reaction-diffusion systems, such as the two-component
Gray-Scott model (cf.~\cite{patternGS_Pearson:1993},
\cite{pattern_Muratov:2001}) and a three-component gas-discharge model
(cf.~\cite{schnek}), have shown the occurrence of complex
spatio-temporal localized spot patterns exhibiting a wide range of
different instabilities including, spot oscillation, spot
annihilation, and spot self-replication behavior.  A survey of
experimental and theoretical studies, through reaction-diffusion (RD)
modeling, of localized spot patterns in various physical or chemical
contexts is given in \cite{vanag}.  These experimental and numerical
studies have provided considerable impetus for developing a
theoretical understanding of the dynamics, stability, and spot
self-replication behavior associated with localized solutions to
singularly perturbed RD systems. A brief survey of some new directions
and open problems for the theoretical study of localized patterns in
various applications is given in \cite{kn_survey}. More generally, a
wide range of topics in the analysis of far-from-equilibrium patterns
modeled by PDE systems are discussed in \cite{book_Nishiura}.

In this context, the goal of this paper is to provide a detailed case
study of the dynamics, stability, and self-replication behavior, of
localized multi-spot patterns for the Gray-Scott (GS) RD system in a
two-dimensional domain. The GS model is formulated in dimensionless
form as (cf.~\cite{autosoliton_Muratov:2000})
\begin{equation}
v_t = \eps^2\, \Delta v - v + A u v^2 \,, \quad
\tau u_t = D\, \Delta u +(1-u) - u v^2\,, \quad {\mathbf x} \in \Omega\in
 \R^2 \,; \qquad \partial_n u =\partial_n v = 0\,, \quad \mathbf{x} \in 
  \partial \Omega. \label{1:GS_2D}
\end{equation}
Here $\eps>0$, $D>0$, $\tau>1$, and $A$ are constants. The parameter
$A$ is called the feed-rate parameter. For various ranges of these
parameters, \eqref{1:GS_2D} has a rich solution structure consisting
of oscillating spots, spot annihilation behavior, spot
self-replication, and stripe and labyrinthian patterns. For the
specific choice $D=2\eps^2$, the complexity and diversity of these
patterns were first studied numerically in
\cite{patternGS_Pearson:1993} in the unit square. Further numerical
studies that clearly exhibit the distinguishing phenomena of spot
self-replication in a two-dimensional domain include
\cite{pattern_Muratov:2001} and \cite{replication_Reynolds:1997}.

For the study of localized patterns in the GS model \eqref{1:GS_2D},
there are two distinguished limits for the diffusion coefficient $D$
in \eqref{1:GS_2D}; the weak interaction regime with $D={\mathcal
  O}(\eps^2)$, where the original numerical simulations of the GS
model were performed (cf.~\cite{patternGS_Pearson:1993}), and the
semi-strong interaction regime $D={\mathcal O}(1)$, where many
analytical studies have been focused. In the weak interaction regime,
there is only an exponentially weak coupling between any two spots in
the multi-spot pattern. This weak coupling arises from the exponential
decay of a local spot profile. In contrast, in the semi-strong
interaction regime, and for a certain range of $A$, the spots are more
strongly coupled through the long-range effect of the slowly varying $u$
component in (\ref{1:GS_2D}). In this way, for $D={\mathcal O}(1)$ the
dynamics of each individual spot in a multi-spot pattern is rather
strongly influenced by the locations of the other spots in the
pattern, as well as by the geometry of the confining domain.

In a one-dimensional spatial domain, there has been much work over the
past decade in analyzing the stability, dynamics, and self-replication
of spike patterns for the GS model \eqref{1:GS_2D}. For the weak
interaction regime where $D={\mathcal O}(\eps^2)$ and $A={\mathcal
O}(1)$, the mechanism for spike self-replication put forth in
\cite{replication_Nishiura:1999} (see also
\cite{dynamics_Ueyama:1999}) was based on the occurrence of a
nearly-coinciding hierarchical saddle-node global bifurcation
structure for the global bifurcation branches of multi-spike
solutions. This mechanism was also shown to occur for the related
Gierer-Meinhardt (GM) system (cf.~\cite{book_Nishiura}). In this
one-dimensional context, it was shown in \cite{enu} that typically
only the spikes at the edges of a multi-spike pattern can undergo
splitting.  The possibility of spatial-temporal chaotic behavior of
spike patterns due to repeated annihilation and self-replication
events was explored in \cite{chaos_Nishiura:2001} from a global
bifurcation viewpoint. The study of solution behavior in this weak
interaction regime relies heavily on the use of numerically computed
global bifurcation diagrams, since it appears to be analytically
intractable to study the local problem near each spike. In contrast,
for the semi-strong interaction regime $D={\mathcal O}(1)$ there are
many analytical studies of spike behavior for \eqref{1:GS_2D} for
different ranges of the parameter $A$. For the range
${\mathcal O}(\eps^{1/2}) \leq A \ll {\mathcal O}(1)$, oscillatory
instabilities of the spike profile, characterized in terms of
threshold values of the time-constant $\tau$ in \eqref{1:GS_2D}, have
been analyzed in \cite{dgk2}, \cite{matchasymp_Doelman:1998},
\cite{stability_Muratov:2002}, \cite{low_KWW:2005}, and
\cite{osci1D_Chen:2008}. Competition or annihilation instabilities of
the spike profile, characterized by threshold values of the
diffusivity $D$, have been analyzed in \cite{low_KWW:2005} for the
range $A={\mathcal O}(\eps^{1/2})$.  In addition,
self-replication instabilities of spike patterns have been shown to
occur only in the regime $A={\mathcal O}(1)$, and they have been
well-studied in \cite{replication_Reynolds:1997},
\cite{dynamics_Reynolds:1994}, \cite{matchasymp_Doelman:1998},
\cite{autosoliton_Muratov:2000}, \cite{pulsesplit_KWW:2005}, and
\cite{DKP}. Weak translation, or drift, instabilities of spike
patterns have been analyzed in \cite{translation_KWW:2006} and
\cite{pulsesplit_KWW:2005}. Finally, there have been several studies
of the dynamical behavior of spike patterns for the one-dimensional GS
model including, two-spike dynamics for the infinite-line problem
(cf.~\cite{2pulse1_Doelman:2000}, \cite{2pulse2_Doelman:2000}) and in
a bounded domain (cf.~\cite{2spike_SWR:2005}), and multi-spike
patterns on a bounded domain (cf.~\cite{osci1D_Chen:2008}). Related
studies on the stability and dynamics of spike solutions for the GM
model in a one-dimensional domain are given in \cite{iww}, \cite{iw},
\cite{ww_2003}, \cite{dkpr}, and \cite{pd} (see also the references
therein). For the semi-strong regime $D={\mathcal O}(1)$,
one key feature of the GS model \eqref{1:GS_2D} in one spatial
dimension is that the parameter regime $A={\mathcal
O}(\eps^{1/2})$ where spike competition instabilities occur is
well-separated in parameter space from the range $A={\mathcal O}(1)$
where spike self-replication occurs. As we discuss below, this feature
with the one-dimensional GS model is in distinct contrast to the
two-dimensional GS model \eqref{1:GS_2D} where several distinct spot
instability mechanisms occur in nearby parameter regimes for $A$.

Although the stability properties of spike patterns for the
one-dimensional case is rather well-understood, there are only a few
studies of the stability of multi-spot patterns for singularly
perturbed two-component RD systems in two dimensional domains. In
particular, for the GS model (\ref{1:GS_2D}) on the infinite plane
$\Omega = \mathbb{R}^2$, the existence and the stability, with respect
to locally radially symmetric perturbations, of a one-spot solution to
\eqref{1:GS_2D} was studied in \cite{2D_Wei:2001} for the range $A
={\mathcal O}(\eps (-\ln \eps)^{1/2})$ with either $D={\mathcal O}(1)$
or $D={\mathcal O}(\nu^{-1})$, where $\nu\equiv {-1/\ln\eps}$ . This
rigorous study was based on first deriving, and then analyzing, a
certain nonlocal eigenvalue problem (NLEP). For the same range of $A$,
in \cite{2Dmulti_Wei:2003} the one-spot NLEP stability analysis of
\cite{2D_Wei:2001} was extended to treat the case of multi-spot
patterns on a bounded domain. A further extension of this
theory to study certain asymmetric multi-spot patterns was made in
\cite{2Dasym_Wei:2003}. The $k$-spot NLEP stability analysis of
\cite{2Dmulti_Wei:2003} for the regime $A = {\mathcal O}(\eps (-\ln
\eps)^{1/2})$ was based on retaining only the leading-order term in
powers of $\nu \equiv - {1/\ln\eps}$ in the construction of the spot
profile, and it pertains to the class of locally radially
symmetric perturbations near each spot. This theory characterizes
competition and oscillatory profile instabilities for the parameter
range $A={\mathcal O}(\eps (-\ln \eps)^{1/2})$ with either
$D={\mathcal O}(1)$ or $D={\mathcal O}(\nu^{-1})$. In this
leading-order theory, the stability thresholds depend only on the
number of spots, and not on their spatial locations in the multi-spot
pattern.  Spot self-replication instabilities were not accounted for in
these NLEP studies, as this instability is triggered by a locally
non-radially symmetric perturbation near each spot for the nearby
parameter regime $A={\mathcal O}(\eps (-\ln \eps))$
(cf.~\cite{stability_Muratov:2002}). A survey of NLEP stability theory
as applied to other two-component singularly perturbed RD systems,
such as the GM and Schnakenburg models, is given in \cite{survey_Wei:2008}.

With regards to the dynamics of spots, there are only a few analytical
results characterizing spot dynamics for singularly perturbed RD
systems in two-dimensional domains. These include,
\cite{dynamics_Chen:2001}, \cite{reducewaveGM_KW:2003} and
\cite{pinningGM:2002} for a one-spot solution of the GM model,
\cite{pulse_Ei:2002}, \cite{spot_Ei:2006} and \cite{metastable_Ei:2002}
for exponentially weakly interacting metastable spots in various
contexts, \cite{Schnaken_KWW:2008} for the Schnakenburg model, and
\cite{schnek} for a three-component gas-discharge RD model. We are not
aware of any previous study of the dynamics of spots for the GS model
(\ref{1:GS_2D}) in a two-dimensional domain.

We emphasize that the previous NLEP stability studies for the GS model
(\ref{1:GS_2D}) (cf.~\cite{2D_Wei:2001}, \cite{2Dasym_Wei:2003},
\cite{2Dmulti_Wei:2003}) are based on a leading-order theory in powers
of $\nu={-1/\ln\eps}$ for the parameter range $A= {\mathcal O}(\eps
(-\ln \eps)^{1/2})$ with either $D={\mathcal O}(1)$ or $D={\mathcal
O}(\nu^{-1})$. Therefore, since $\nu$ is not very small unless $\eps$
is extremely small, it is desirable to obtain a stability theory for
multi-spot solutions that accounts for all terms in powers of
$\nu$. However, more importantly, since the scaling regime $A =
{\mathcal O}(-\eps \ln \eps)$ where a spot-replication instability can
occur for \eqref{1:GS_2D} (cf.~\cite{stability_Muratov:2002}) is
logarithmically close to the low feed-rate regime $A = {\mathcal
O}(\eps (-\ln \eps)^ {1/2})$ studied in \cite{2Dmulti_Wei:2003} and
\cite{2D_Wei:2001}, where only competition or oscillatory profile
instabilities can occur, it is highly desirable to develop an
asymptotic theory that incorporates these two slightly different
scaling regimes into a single parameter regime where all three types
of spot instability can be studied simultaneously. The leading-order
theory in \cite{2Dmulti_Wei:2003} is not sufficiently accurate to
study the three types of spot profile instability (competition,
oscillatory, and self-replication) within a single parameter regime.

For the simpler Schnakenburg RD model, where competition and
oscillatory instabilities do not occur when $D={\mathcal O}(1)$, a
hybrid asymptotic-numerical method was developed in
\cite{Schnaken_KWW:2008} to study the dynamics and self-replication
instabilities of a collection of spots for this specific RD model in
an arbitrary two-dimensional domain. The theory, which accounts for
all terms in powers of $\nu={-1/\ln\eps}$, was illustrated explicitly
for the square and the disk, and the results from this theory were
favorably compared with full numerical computations of the RD system.

The main goal of this paper is to extend the theoretical framework of
\cite{Schnaken_KWW:2008} for the Schnakenburg model to study
the dynamics and three types of instabilities associated with a
collection of spots for the GS model \eqref{1:GS_2D} in the
semi-strong parameter regime $D={\mathcal O}(1)$ with $A={\mathcal
O}(-\eps\ln\eps)$. In our theory we account for all terms in
powers of $\nu={-1/\ln\eps}$.  In contrast to the Schnakenburg model
of \cite{Schnaken_KWW:2008}, we emphasize that there are three
distinct instability mechanisms for a collection of spots to the GS
model \eqref{1:GS_2D} in this scaling regime for $A$ and $D$ that must
be considered.  

We now give an outline of the paper.  In \S \ref{sec:quasi} and \S
\ref{sec:dyn} the method of matched asymptotic expansions is used to
construct a quasi-equilibrium $k$-spot pattern for (\ref{1:GS_2D})
that evolves slowly over a long ${\mathcal O}(\eps^{-2})$ time scale.
For this pattern, the spatial profile for $v$ concentrates at a set of
points $\mathbf{x}_j\in \Omega$ for $j=1,\ldots,k$ that drift with an
asymptotically small ${\mathcal O}(\eps^2)$ speed. Within an
${\mathcal O}(\eps)$ neighborhood of each spot centered at
$\mathbf{x}_j$, and at any instant in $t$, the local spot profiles for
$u$ and $v$ are radially symmetric to within ${\mathcal O}(\eps)$
terms and satisfy a coupled system of BVP, referred to as the {\em
core problem}, on the (stretched) infinite plane. In the outer region,
each spot at a given instant in time is represented as a Coulomb
singularity for $u$ of strength $S_j$. The {\em collective
coordinates} characterizing the slow dynamics of this
quasi-equilibrium $k$-spot pattern are the locations
$\mathbf{x}_1,\ldots,\mathbf{x}_k$ of the spots and their
corresponding source strengths $S_1,\ldots,S_k$, which measure the
far-field logarithmic growth of the (inner) core solution for $u$ near
each spot. By asymptotically matching the inner and outer solutions
for $u$, we derive a differential algebraic system (DAE) of ODE's for
the slow time evolution of these collective coordinates. At any
instant in time, the quasi-equilibrium solution is characterized as in
Principal Result 2.1, where the source strengths $S_j$ for
$j=1,\ldots,k$ are shown to satisfy a coupled nonlinear algebraic
system that depends on the instantaneous spot locations $\mathbf{x}_j$
for $j=1,\ldots,k$, together with certain properties of the
reduced-wave Green's function $G(\mathbf{x};\mathbf{x}_j)$ and its
regular part $R_{jj}$ defined by \bsub \label{3:Green}
\begin{gather}
\label{3:Green_1} \Delta G - \frac{1}{D} G = -
\delta(\mathbf{x}-\mathbf{x}_j)\,, \quad \mathbf{x}\in \Omega \,; \qquad 
\partial_n G = 0 \,, \quad \mathbf{x}\in \partial\Omega \,, \\
 G(\mathbf{x}; \mathbf{x}_j) \sim - \frac{1}{2\pi} 
 \ln \left|{\mathbf{x}-\mathbf{x}_j} \right| + R_{j,j} + o(1) \,, \quad
 \mbox{as} \quad \mathbf{x}\to \mathbf{x}_j \,. \label{3:gloc}
\end{gather}
\esub In Principal Result 3.1 of \S \ref{sec:dyn} the dynamical
behavior of the collection of spots is characterized in terms of the
source strengths and certain gradients of the reduced-wave Green's
function. The overall DAE ODE system for ${\mathbf x}_j$ and $S_j$,
for $j=1,\ldots,k$, incorporates the interaction between the spots and
the geometry of the domain, as mediated by the reduced-wave Green's
function and its regular part, and it also accounts for all
logarithmic correction terms in powers of $\nu={-1/\ln\eps}$ in the
asymptotic expansion of the solution. In this DAE ODE system there are
two nonlinear functions of $S_j$, defined in terms of the solution to
the core problem, that must be computed numerically.

In \S \ref{sec:eig_nrad} we study spot self-replication instabilities
by first deriving a local eigenvalue problem near the $j^{\mbox{th}}$
spot that characterizes any instability due to a non-radially
symmetric local deformation of the spot profile.  We emphasize that in
this stability analysis the local eigenvalue problems near each spot
are not coupled together, except in the sense that the source
strengths $S_1,\ldots,S_k$ must be determined from a globally coupled
nonlinear algebraic system.  The spectrum of the local two-component
linear eigenvalue problem near the $j^{\mbox{th}}$ spot is studied
numerically, and we show that there is a critical value
$\Sigma_2\approx 4.31$ of the source strength $S_j$ for which there is
a peanut-splitting linear instability for any $S_j>\Sigma_2$. These
results for spot-splitting parallel those for the Schnakenburg model,
as given in \cite{Schnaken_KWW:2008}. As a new result, we derive and
then numerically study a certain time-dependent elliptic-parabolic
core problem near the $j^{\mbox{th}}$ spot. Our computations from this
time-dependent core problem strongly suggest that the localized
peanut-splitting linear instability of the quasi-equilibrium spot
profile is in fact subcritical, and robustly triggers a nonlinear spot
self-replication event for the $j^{\mbox{th}}$ spot whenever
$S_j>\Sigma_2$.  In summary, our hybrid asymptotic-numerical theory
predicts that if $S_J>\Sigma_2\approx 4.31$ for some $J\in
\lbrace{1,\ldots,k\rbrace}$ then the $J^{\mbox{th}}$ spot will undergo
a nonlinear spot self-replication event. Alternatively, the spots are
all stable to self-replication whenever $S_j<\Sigma_2$ for
$j=1\ldots,k$.

In \S \ref{sec:eig_rad} we use the method of matched asymptotic
expansions to formulate a novel global eigenvalue problem associated
with competition or oscillatory instabilities in the spot amplitudes
for a $k$-spot quasi-equilibrium solution to \eqref{1:GS_2D} for the
parameter range $A={\mathcal O}(-\eps\ln\eps)$ with $D={\mathcal
O}(1)$.  This global eigenvalue problem, as formulated in Principal
Result 4.1, is associated with a locally radially symmetric
perturbation near each spot, and it accounts for all terms in powers
of $\nu$. It differs from the eigenvalue problem characterizing spot
self-replication in that now the local eigenfunction for the
perturbation of the $u$ component in \eqref{1:GS_2D} has a logarithmic
growth away from the center of each spot. This logarithmic growth
leads to a global eigenvalue problem that effectively couples together
the local problems near each spot. A key component in the formulation
of this global eigenvalue problem is a certain eigenvalue-dependent 
Green's matrix, with entries determined in terms of
properties the Green's function $G_{\lam}(\mathbf{x};\mathbf{x}_j)$
satisfying $\Delta G - D^{-1}(1+\tau\lam) G=-\delta(x-x_j)$ for
$\mathbf{x}\in \Omega$, with $\partial_n G=0$ for $\mathbf{x}\in
\partial\Omega$. This globally coupled eigenvalue problem can be
viewed, essentially, as an extended NLEP theory that accounts for all
terms in powers of $\nu$. In \S \ref{sec:12inf}--\S \ref{sec:sym} we
show that it determines thresholds for competition and oscillatory
instabilities very accurately. However, in contrast to the
leading-order-in-$\nu$ NLEP stability studies (cf.~\cite{2D_Wei:2001},
\cite{2Dmulti_Wei:2003}) for $A={\mathcal
O}\left(\eps[-\ln\eps]^{1/2}\right)$, our globally coupled eigenvalue
problem is not readily amenable to rigorous analysis. In Appendix B we
briefly review the NLEP theory of \cite{2D_Wei:2001} and
\cite{2Dmulti_Wei:2003}, and we show how our globally coupled
eigenvalue problem can be reduced to leading order in $\nu$ to the
NLEP problems of \cite{2D_Wei:2001} and \cite{2Dmulti_Wei:2003}
when $A={\mathcal O}\left(\eps[-\ln\eps]^{1/2}\right)$ and
$D={\mathcal O}(\nu^{-1})$.

In our stability analysis of \S \ref{sec:eig} we linearize the GS
model \eqref{1:GS_2D} around a quasi-equilibrium solution where the
spots are assumed to be at fixed locations
$\mathbf{x}_1,\ldots,\mathbf{x}_k$, independent of time.  However,
since the spots locations undergo a slow drift with speed ${\mathcal
O}(\eps^2)$, the source strengths $S_j$ for $j=1,\ldots,k$ also vary
slowly in time on a time-scale $t={\mathcal O}(\eps^{-2})$. As a
result of this slow drift, there can be triggered, or dynamically
induced, instabilities of a quasi-equilibrium spot pattern that is
initially stable at time $t=0$. To illustrate this, suppose that the
pattern is initially stable to spot self-replication at $t=0$ in the
sense that $S_j < \Sigma_2$ at $t=0$ for $j=1,\ldots,k$. Then, it is
possible, that as the $J^{\mbox{th}}$ spot drifts toward its
equilibrium location in the domain, that $S_J > \Sigma_2$ after a
sufficiently long time of order $t={\mathcal O}(\eps^{-2})$. This will
trigger a nonlinear spot self-replication event for the
$J^{\mbox{th}}$ spot. In a similar way, we show that dynamically-triggered
oscillatory and competition instabilities can also occur for
a multi-spot pattern. This dynamical bifurcation phenomena is similar
to that for other ODE and PDE slow passage problems (cf.~\cite{BK},
\cite{me}) that have triggered instabilities generated by a slowly
varying {\em external} bifurcation, or control, parameter. The key
difference here, is that the dynamically-triggered instabilities for
the GS model (\ref{1:GS_2D}) occur as a result of the {\em intrinsic} motion
of the collection of spots, and is not due to the tuning of an
external control parameter.

In our numerical computations of competition and oscillatory
instability thresholds from our globally coupled eigenvalue problem of
\S \ref{sec:eig_rad}, we will for simplicity only consider $k$-spot
quasi-equilibrium spot configurations
$\mathbf{x}_1,\ldots,\mathbf{x}_k$ for which a certain Green's matrix
is circulant symmetric.  For instance, this circulant matrix structure
occurs when $k$ spots are equally spaced on a circular ring that is
concentric within a circular disk, and it also occurs for other spot
patterns with sufficient spatial symmetry in other domains. Examples
of such patterns are given in \S \ref{sec:12inf} and \ref{sec:sym}
below.  Under this condition, we show in \S \ref{sec:quasi_circ} that
the source strengths $S_j$ for $j=1,\ldots,k$ have a common value.  In
addition, by calculating the spectrum of the circulant symmetric
Green's matrix, we show in Principal Result 4.3 of \S
\ref{sec:eig_circ} that the globally coupled eigenvalue problem
simplifies to $k$ separate transcendental equations for the eigenvalue
parameter. In Appendix C we outline the numerical methods that we use
to compute the instability thresholds from the globally coupled
eigenvalue problem under the circulant Green's matrix assumption.

Spot patterns that give rise to this special circulant matrix
structure are the direct counterpart of {\em equally-spaced} $k$-spike
patterns with spikes of a common amplitude, treated in almost all of
the previous NLEP stability studies of the GS and related RD models on
a one-dimensional domain (cf.~\cite{dgk2},
\cite{matchasymp_Doelman:1998}, \cite{low_KWW:2005},
\cite{2pulse1_Doelman:2000}, \cite{2pulse2_Doelman:2000}, \cite{iww},
\cite{ww_2003}, \cite{dkpr}, \cite{pd}). In one spatial dimension, the
only NLEP stability studies of an {\em arbitrarily-spaced} slowly
evolving $k$-spike quasi-equilibrium solution are the
asymptotic-numerical study of dynamic competition instabilities for
the Gierer-Meinhardt model with $\tau=0$ in \cite{iw}, and the study
of oscillatory instabilities in \cite{osci1D_Chen:2008} for the
one-dimensional GS model for the range ${\mathcal O}(\eps^{1/2})\ll A
\ll {\mathcal O}(1)$. For this range of $A$ it was shown in
\cite{osci1D_Chen:2008} that the $k$ separate NLEP problems can be
reduced, via a scaling law, to only one single NLEP problem. To date,
there has been no NLEP stability study of a slowly evolving {\em
arbitrarily-spaced} $k$-spike quasi-equilibrium spike pattern in a
one-dimensional domain that takes into account both competition and
oscillatory instabilities. As a result, in our two-dimensional
setting, it is a natural first step to study the global eigenvalue
problem, which governs competition and oscillatory instabilities,
under the circulant matrix condition, which allows for spots of a common
source strength.

In \S \ref{sec:12inf} the asymptotic theory of \S \ref{sec:quasi}-- \S
\ref{sec:eig} is illustrated for the case of both one and two-spot
quasi-equilibrium solutions to the GS model (\ref{1:GS_2D}) on the
infinite plane. Phase diagrams characterizing the GS parameter ranges
for the different types of instabilities are derived for these simple
spot patterns. In particular, for two spots that are
sufficiently far apart, we show that spot self-replication
instabilities will occur when $A$ exceeds some threshold. In contrast,
a competition instability will occur if the two spots are too closely
spaced. For a very large, but finite, domain the full numerical
simulations in \S \ref{sec:2inf_large} are used to validate the
stability results from the asymptotic theory.

In \S \ref{sec:sym} the asymptotic theory of \S \ref{sec:quasi}-- \S
\ref{sec:eig} is implemented and compared with full numerical results
computed from (\ref{1:GS_2D}) for various special multi-spot
patterns on the unit disk and square for which a certain Green's
matrix has a circulant matrix structure. For these domains, the
explicit formulae for the reduced-wave Green's function and its
regular part, as derived in Appendix A, are used to numerically
implement the asymptotic theory. The overall hybrid
asymptotic-numerical approach provides phase diagrams in parameter
space characterizing both the stability thresholds and the possibility
of dynamically-triggered instabilities.  One key theoretical advantage
of considering the unit disk is that the reduced-wave Green's function
can be well-approximated for $D\gg 1$ by the Neumann Green's function,
which has a simple explicit formula in the unit disk. By using this
explicit formula, the hybrid asymptotic-numerical framework of \S
\ref{sec:quasi}-- \S \ref{sec:eig} can be studied, to a large extent,
analytically for the case of $k$ equally-spaced spots on a ring that
is concentric within the unit disk.

In \S \ref{sec:asy} we compare our theoretical predictions for spot
dynamics and spot self-replication instabilities with full numerical
results computed from (\ref{1:GS_2D}) for a few simple ``asymmetric''
spot patterns for which the associated Green's matrix is not
circulant.  We show that the dynamics in Principal Result 3.1
accurately determines spot dynamics before a self-replication event,
and with a re-calibration of the initial spot locations, it accurately
predict spot dynamics after a spot-splitting event. We emphasize that
since the local eigenvalue problems near each spot are decoupled for
the case of locally non-radially symmetric perturbations, the onset of
spot self-replication behavior only depends on the source strength of
an individual spot. Therefore, given any initial spatial configuration
of spots at $t=0$, we need only solve the nonlinear algebraic system
for $S_j$, $j=1, \cdots, k$ at $t=0$ to predict that the
$j^{\mbox{th}}$ spot undergoes splitting starting at $t=0$ when
$S_j>\Sigma_2\approx 4.31$.  A special asymmetric pattern that we
consider in some detail in \S \ref{sec:asy} is a $k$-spot pattern
consisting of $k-1$ equally-spaced spots on a ring concentric within
the unit disk, with an additional spot at the center of the unit disk.
For $D\gg 1$, we use the simple explicit formula for the Neumann
Green's function to explicitly predict the occurrence of
dynamically-triggered spot self-replication instabilities for this
special pattern.

Although the hybrid asymptotic-numerical framework developed herein to
study the stability and dynamics of multi-spot quasi-equilibrium
patterns for the GS model (\ref{1:GS_2D}) is related to that initiated 
for the Schnakenburg model in \cite{Schnaken_KWW:2008}, there are some key
differences in the analysis and in the results obtained.  The primary
difference between these two models is that the GS model
(\ref{1:GS_2D}) admits three types of instability mechanisms, whereas
only spot self-replication instabilities can occur for the
Schnakenburg model of \cite{Schnaken_KWW:2008} when $D={\mathcal
  O}(1)$.  In addition, in contrast to our study in \S \ref{sec:12inf}
of one- and two-spot patterns to the GS model on the infinite plane,
the Schnakenburg model of \cite{Schnaken_KWW:2008} is ill-posed in
$\mathbb{R}^2$. Finally, our results for the GS model (\ref{1:GS_2D})
show that there is a wide parameter range and many simple spot
configurations for which we can theoretically predict the occurrence
of dynamically-triggered instabilities due to either competition,
oscillation, or splitting. These dynamically-triggered bifurcation
events can occur even within the very simple context of a multi-spot
pattern with a common source strength. For these special patterns,
such instabilities cannot occur for the Schnakenburg model.

Finally, although this paper focuses only on the study of spot patterns,
we remark that the GS model (\ref{1:GS_2D}) supports patterns
of increasing complexity as the feed-rate parameter $A$ increases. In
particular, in the range ${\mathcal O}(\eps^{1/2}) \ll A \ll {\mathcal
  O}(1)$, the GS model \eqref{1:GS_2D} with $D={\mathcal O}(1)$ on a
two-dimensional domain does not admit spots, but instead allows for
solutions for which $v$ concentrates on a higher dimensional set such
as on a one-dimensional stripe or a one-dimensional ring inside a
two-dimensional domain. A stability analysis of a planar stripe inside
a square domain or a concentric ring inside a disk was given in
\cite{mk} and in \cite{kww}.

\setcounter{equation}{0}
\setcounter{section}{1}
\section{$K$-Spot Quasi-Equilibrium Solutions} \label{sec:quasi}

We first construct a $k$-spot quasi-equilibrium solution to
\eqref{1:GS_2D} by using the method of matched asymptotic expansions.
We denote the center of the $j^{\mbox{th}}$ spot by 
$\mathbf{x}_j = (x_j, y_j)\in \Omega$ for $j=1,\ldots,k$. We assume that the
spots are well-separated in the sense that $|\mathbf{x}_i -
\mathbf{x}_j|={\mathcal O}(1)$ for $i\neq j$, and
$\mbox{dist}(\mathbf{x}_j,\partial\Omega)={\mathcal O}(1)$ for
$j=1,\ldots,k$. In an ${\mathcal O}(\eps)$ neighborhood near the
$j^{\mbox{th}}$ spot, we get $v={\mathcal O}(\eps^{-1})$ and $u={\mathcal
O}(\eps)$. Thus, we introduce the local variables $U_j$, $V_j$, and
$\mathbf{y}$, defined by
\begin{equation}
 u = \frac{\eps}{A \sqrt{D}} U_j \,, \qquad v = \frac{\sqrt{D}}{\eps}
V_j \,, \qquad \mathbf{y} = \eps^{-1} (\mathbf{x}-\mathbf{x}_j) \,. 
\label{3:2dinnvar}
\end{equation}
In terms of these local variables, \eqref{1:GS_2D} transforms to
\begin{equation}
\label{3:2Dcore} \Delta_{\mathbf{y}} V_j - V_j + U_jV_j^2 = 0 \,, \qquad
\Delta_{\mathbf{y}} U_j - U_jV_j^2 +\frac{\eps A}{\sqrt{D}} - 
\frac{\eps^2}{D} U_j = 0 \,,\qquad \mathbf{y} \;\in\; \mathbb{R}^2 \,.
\end{equation}
We look for a radially symmetric solution to \eqref{3:2Dcore} of the
form $U_j=U_j(\rho)$ and $V_j=V_j(\rho)$, where $\rho \equiv
|y|$. Then, to leading order in $\eps$, $U_j$ and $V_j$ are the solutions
to the radially symmetric problem
\bsub \label{3:2Dcore_sol}
\begin{gather}
U_{j}^{\p\p} + \frac{1}{\rho} U_{j}^{\p} - U_jV_j^2 = 0\,, \qquad
V_{j}^{\p\p} + \frac{1}{\rho} V_{j}^{\p}- V_j + U_jV_j^2 = 0 \,,
 \qquad   0<\rho<\infty \,, \label{3:2Dcore_rad} \\
 V_j^{\p}(0)=0 \,, \qquad U_j^{\p}(0)=0 \,; \qquad 
 V_j(\rho) \to 0 \,, \qquad U_j(\rho) \sim S_j \ln \rho + \chi
(S_j) + o(1) \,, \quad \mbox{as} \quad \rho \to \infty \,. \label{3:bdc2} 
\end{gather}
\esub This leading-order coupled inner problem is referred to as the
\emph{core problem}, and is the same as that derived in
\cite{Schnaken_KWW:2008} for the Schnakenburg model.  We refer to
$S_j$ as the \emph{source strength} of the $j^{\mbox{th}}$ spot. From
the divergence theorem, it follows from the $U_j$ equation in
\eqref{3:2Dcore_sol} that $S_j = \int^{\infty}_0 U_j V_j^2 \rho\,
d\rho > 0$. In the far-field behavior \eqref{3:bdc2} for $U_j$, the
constant $\chi$ is a nonlinear function of the source strength $S_j$,
which must be computed numerically from the solution to
\eqref{3:2Dcore_sol}.

\begin{figure}[htbp]
\begin{center}
\subfigure[$\chi$ vs.~$S_j$] { \label{fig:core:a}
  \includegraphics[width=2.2in, height=1.8in]{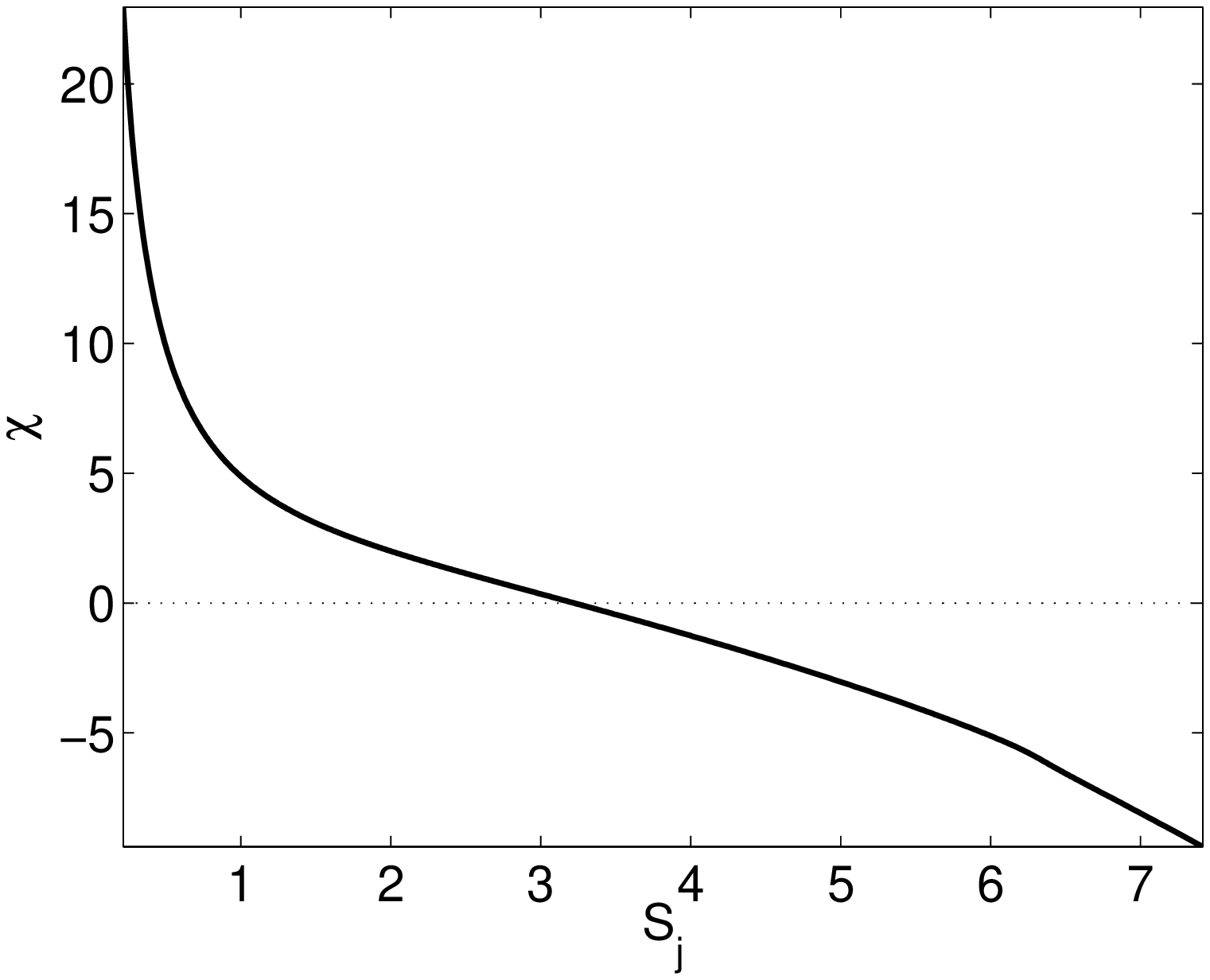}}
\subfigure[$V_j(0)$ vs.~$S_j$] { \label{fig:core:b}
    \includegraphics[width=2.2in, height=1.8in]{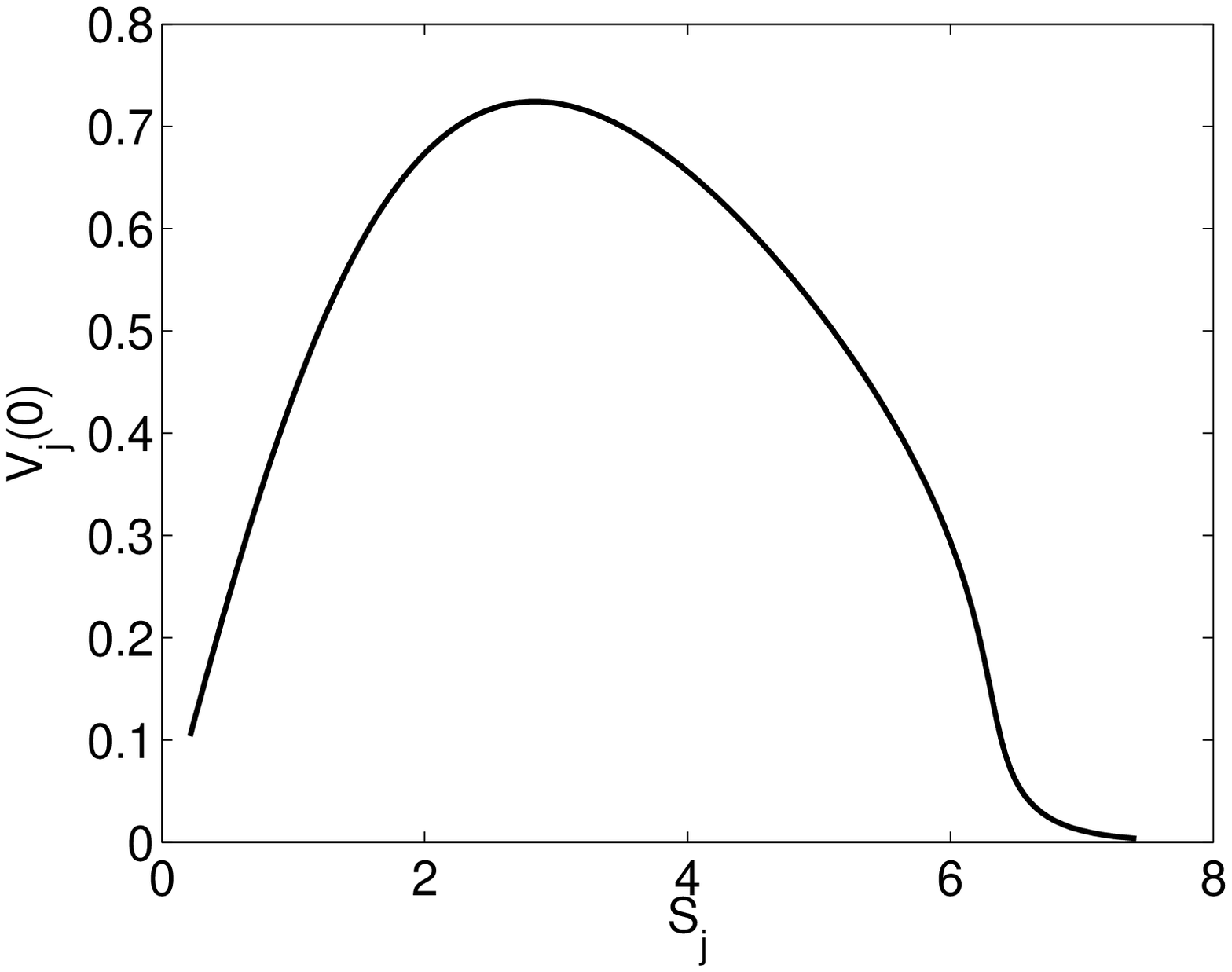}}
\subfigure[$V_j$ vs.~$\rho$] {\label{fig:core:d}
    \includegraphics[width=2.2in, height=1.8in]{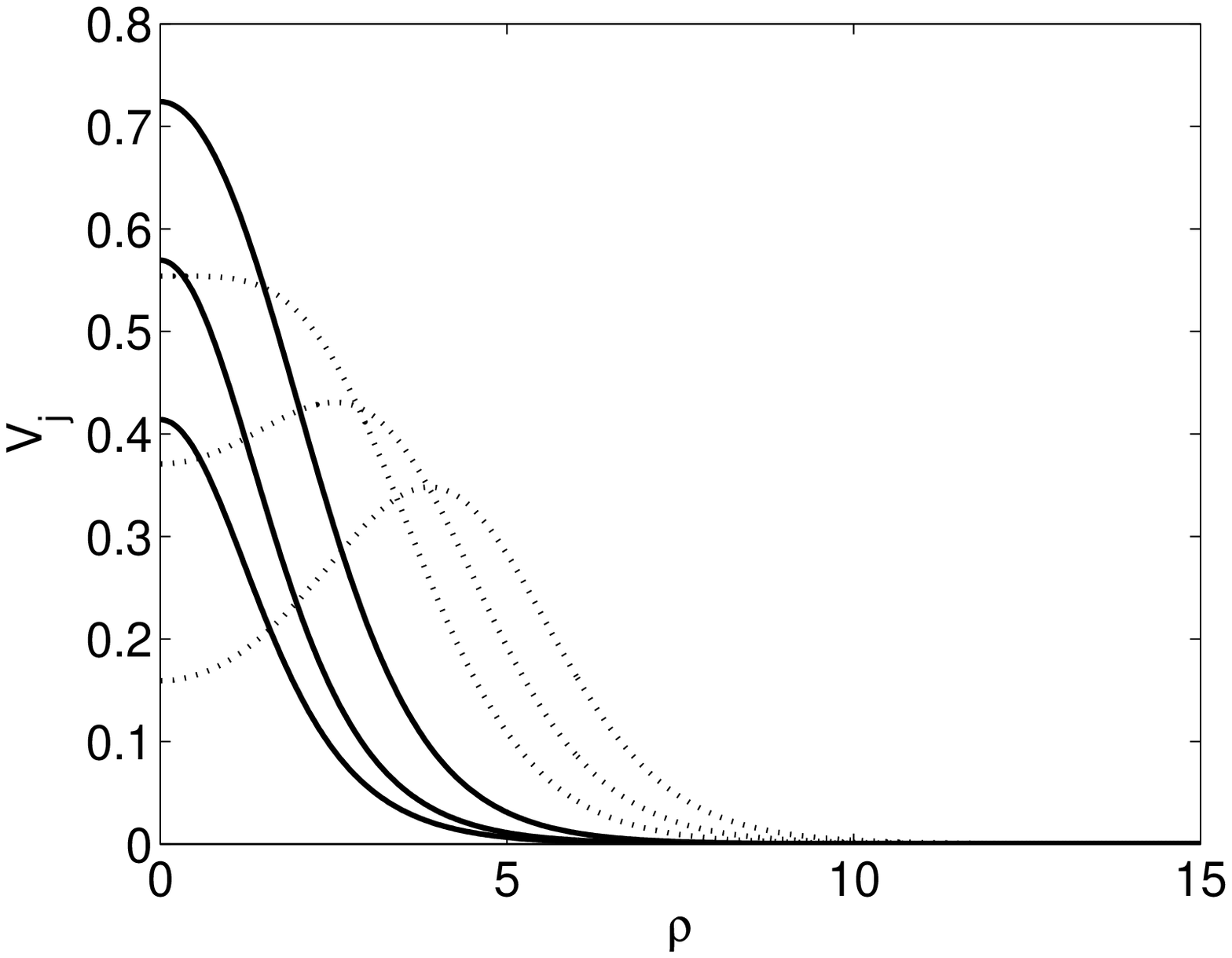} }
\caption[Numerical solution of the core problem \eqref{3:2Dcore_sol}]
{\em Numerical results for the core problem \eqref{3:2Dcore_sol} (a)
The function $\chi$ vs.~$S_j$; (b) $V_j(0)$ vs.~$S_j$; (c) The spot
profile $V_j(\rho)$ for $S_j=0.94,\, 1.45,\, 2.79$ (solid curves where
$V_{j}(0)$ is the maximum of $V_j(\rho)$), and the volcano profile
$V_j(\rho)$ for $S_j= 4.79,\, 5.73, \,6.29$ (the dotted curves
correspond to cases where the maximum of $V_j(\rho)$ occurs for
$\rho>0$). }
\label{fig:core}
\end{center}
\end{figure}

The solution to \eqref{3:2Dcore_sol} is calculated numerically for a
range of values of $S_j>0$ by using the BVP solver COLSYS
(cf.~\cite{colsys_Ascher:1979}). In Fig.~\ref{fig:core}, we plot
$\chi(S_j)$, $V_j(0)$ versus $S_j$, and $V_j(\rho)$
for a few different values of $S_j$. For $S_j > S_v \approx 4.78$, the
profile $V_j(\rho)$ has a volcano shape, whereby the maximum of
$V_j$ occurs at some $\rho > 0$. These computations give numerical
evidence to support the conjecture that there is a unique solution to
\eqref{3:2Dcore_sol} for each $S_j>0$.  

Since $v$ is localized near each $\mathbf{x}_j$ for $j=1,\ldots,k$,
and is exponentially small in the outer region away from the spot
centers, the effect of the nonlinear term $uv^2$ in the outer region
can be calculated in the sense of distributions as
$uv^2 \sim \eps^2 \sum_{j=1}^k \left( \int_{\mathbb{R}^2}
 \sqrt{D} \left(A \eps\right)^{-1} U_j V_j^2 \, dy\right) \,
 \delta(\mathbf{x}-\mathbf{x}_j) \; \sim 2 \pi \eps
\sqrt{D} A^{-1} \sum_{j=1}^k  S_j \,\delta(\mathbf{x}-\mathbf{x}_j)$.
Therefore, in the quasi-steady limit, the outer problem for $u$ from
\eqref{1:GS_2D} is
\bsub \label{3:uout_i}
\begin{gather}
D \Delta u  +(1-u) = \frac{2 \pi \sqrt{D} \,\eps}{A} \sum_{j=1}^k
S_j \,\delta(\mathbf{x}-\mathbf{x}_j) \,, \quad \mathbf{x}\in \Omega \,;
  \qquad \partial_n u = 0 \,, \quad \mathbf{x}\in \partial\Omega\,, \\
  u \sim  \frac{\eps}{A \sqrt{D}} \Big( S_j \ln
|\mathbf{x}-\mathbf{x}_j| - S_j \ln \eps + \chi (S_j)
\Big)\,,\quad\mbox{as}\;\;\mathbf{x} \to \mathbf{x}_j \,,\quad j=1,\ldots,k \,.
  \label{3:uout_ising}
\end{gather}
\esub 
The singularity condition \eqref{3:uout_ising} for $u$ as
$\mathbf{x}\to\mathbf{x}_j$ was derived by matching the outer solution
for $u$ to the far-field behavior \eqref{3:bdc2} of the core solution,
and by recalling $u = {\eps U_j/(A \sqrt{D})}$ from
\eqref{3:2dinnvar}. The problem \eqref{3:uout_i} suggests that we
introduce new variables $\ac={\mathcal O}(1)$ and $\nu\ll 1$ defined
by
\begin{equation}
       \nu = {-1/\ln\eps} \,, \qquad \ac = \nu A \sqrt{D} /\eps = 
  {A \sqrt{D}/\left[-\eps\ln\eps\right]} \,.  \label{3:pval}
\end{equation}
In terms of these new variables, \eqref{3:uout_i} transforms to
\bsub \label{3:uout}
\begin{gather}
 \Delta u  +\frac{(1-u)}{D} = \frac{2 \pi \nu}{\ac} \sum_{j=1}^k
S_j \,\delta(\mathbf{x}-\mathbf{x}_j) \,, \quad \mathbf{x}\in \Omega \,; 
  \qquad \partial_n u = 0 \,, \quad \mathbf{x}\in \partial\Omega\,, \\
  u \sim  \frac{1}{\ac} \left[ S_j \nu \ln|\mathbf{x}-\mathbf{x}_j| + S_j 
  + \nu \chi (S_j) \right]\,,\quad\mbox{as}\;\;\mathbf{x} \to \mathbf{x}_j \,, 
  \quad j=1,\ldots,k \,.  \label{3:uout_sing}
\end{gather}
\esub 
We emphasize that the singularity behavior in
\eqref{3:uout_sing} specifies both the strength of the logarithmic
singularity for $u$ and the regular, or non-singular, part of this
behavior.  This pre-specification of the regular part of this
singularity behavior at each $\mathbf{x}_j$ will yield a nonlinear
algebraic system for the source strengths $S_1,\ldots,S_k$.

The solution to (\ref{3:uout}) is represented as $u = 1 -
\sum_{i=1}^k 2 \pi \nu \ac^{-1} S_i G(\mathbf{x};\mathbf{x}_i)$, where
$G(\mathbf{x};\mathbf{x}_i)$ is the reduced-wave Green's function
defined by (\ref{3:Green}). By expanding $u$ as
$\mathbf{x}\to\mathbf{x}_j$, and then equating the resulting
expression with the required singularity behavior in
\eqref{3:uout_sing}, we obtain the following nonlinear algebraic
system for $S_1,\ldots,S_k$:
\begin{equation}
 \ac = S_j (1+ 2\pi \nu R_{j,j}) + \nu \chi(S_j) +
2 \pi \nu \stacksum S_i G(\mathbf{x}_j;
\mathbf{x}_i) \,, \quad j=1,\ldots,k \,. \label{3:ASsmallD}
\end{equation}
Given the GS parameters $A$, $\eps$ and $D$, we first
calculate $\ac$ and $\nu$ from \eqref{3:pval}, and then solve
\eqref{3:ASsmallD} numerically for the source strengths
$S_1,\ldots,S_k$. With $S_j$ known, the quasi-equilibrium solution
in each inner region is determined from \eqref{3:2Dcore_sol}. As a
remark, since $\ac={\mathcal O}(1)$, then $A={\mathcal
O}(-\eps\ln\eps)$ from \eqref{3:pval}. Therefore, the error made
in approximating \eqref{3:2Dcore} by \eqref{3:2Dcore_sol} in the inner
region is of the order ${\mathcal O}(-\eps^2 \ln\eps)$. We summarize
our result as follows:

\vspace*{0.2cm}
\noindent {\bf \underline{Principal Result 2.1}:} {\em 
For $\eps \to 0$ assume that $A={\mathcal O}(-\eps\ln\eps)$, and define
$\nu$ and $\ac$ by  $\nu = - 1/ \ln \eps$ and $\ac = \nu A
\sqrt{D}/\eps $, where $\ac={\mathcal O}(1)$. Then, the solution $v$
and the outer solution for $u$, corresponding to a $k$-spot 
quasi-equilibrium solution of the GS model \eqref{1:GS_2D}, are given 
asymptotically by}
\bsub \label{3:quasi}
\begin{equation}
 u(\mathbf{x}) \sim 1 - \frac{2\pi\nu}{\ac} \sum_{j=1}^k S_j
G(\mathbf{x}; \mathbf{x}_j)\,, \qquad v(\mathbf{x}) \sim
\frac{\sqrt{D}}{\,\eps} \sum_{j=1}^k V_j\left(
\eps^{-1}|\mathbf{x}-\mathbf{x}_j|\right) \,.
\end{equation}
{\em Moreover, the inner solution for $u$, defined in an ${\mathcal O}(\eps)$
neighborhood of the $j^{\mbox{th}}$ spot, is}
\begin{equation}
   u(\mathbf{x}) \sim \frac{\nu}{\ac} U_{j}\left( 
 \eps^{-1}|\mathbf{x}-\mathbf{x}_j|\right) \,.
\end{equation}
\esub
 {\em Here $\mathbf{x}_1, \ldots, \mathbf{x}_k$ is the spatial
configuration of the centers of the spots, and $G(\mathbf{x};
\mathbf{x}_j)$ is the reduced-wave Green's function satisfying
\eqref{3:Green}.  In \eqref{3:quasi}, each spot profile $V_j(\rho)$
and $U_j(\rho)$ for $j=1,\ldots,k$ satisfies the coupled BVP system
\eqref{3:2Dcore_sol}, where the source strength $S_j$ in \eqref{3:bdc2}
is to be calculated from the nonlinear algebraic system
\eqref{3:ASsmallD}.}

We emphasize that the nonlinear algebraic system (\ref{3:ASsmallD})
determines the source strengths $S_j$ for $j=1,\ldots,k$ to within an
error smaller than any power of $\nu={-1/\ln\eps}$. As such, our
construction of the quasi-equilibrium pattern is accurate to all
orders in $\nu$. Similar techniques for summing logarithmic expansions
in the context of {\em linear} elliptic PDE's or eigenvalue problems
in two-dimensional perforated domains containing small holes have been
developed in a variety of contexts (cf.~\cite{sumlog_ward},
\cite{trap_KTW:2005}, \cite{coombs}, \cite{pillay}). Our construction
here of a quasi-equilibrium solution for the GS model (\ref{1:GS_2D})
extends this previous methodology for treating logarithmic expansions
to an RD system for which the local problem is nonlinear with a
logarithmic far-field behavior.  The nonlinear algebraic system
\eqref{3:ASsmallD} for the source strengths is the mechanism through
which the spots interact and sense the presence of the
domain $\Omega$.  This global coupling mechanism, which is not of
nearest-neighbor type as in the case of the exponentially weak spot
interactions studied in \cite{spot_Ei:2006} and
\cite{metastable_Ei:2002}, is rather significant since
$\nu=-{1/\ln\eps}$ is not very small unless $\eps$ is extremely small.

The quasi-equilibrium solution in Principal Result 2.1 exists only
when the spatial configuration of spots and the GS parameters are such
that the nonlinear algebraic system (\ref{3:ASsmallD}) for
$S_1,\ldots,S_k$ has a solution. Determining precise conditions for
the solvability of this system is a difficult issue. Therefore, in \S
\ref{sec:12inf}--\ref{sec:asy} we will primarily consider spot
patterns where the source strengths have a common value, for which
(\ref{3:ASsmallD}) reduces to a scalar nonlinear equation.

Next, we derive analytical approximations for the solution to
\eqref{3:ASsmallD} by first re-writing \eqref{3:ASsmallD} in matrix
form. To do so, we define the Green's matrix $\mathcal{G}$, the vector
of source strengths $\mathbf{s}$, the vector
$\mathbf{\chi}(\mathbf{s})$, and the identity vector $\mathbf{e}$ by
\begin{equation}
 {\cal G} \equiv
\left ( 
\begin{array}{ccccccc}
 R_{1,1}  & G_{1,2} & \cdots & G_{1,k} \\
 G_{2,1}  & \ddots  & \ddots  & \vdots \\
 \vdots   &\ddots   & \ddots & G_{k-1,k} \\ 
 G_{k,1}  &\cdots   & G_{k,k-1} & R_{k,k} 
\end{array}
\right ) \,, \qquad \mathbf{s} \equiv
\left (
\begin{array}{c}
S_1 \\ \vdots \\ S_k
\end{array}
\right ) \,, \qquad
\eb \equiv
\left (
\begin{array}{c}
1 \\ \vdots \\ 1
\end{array}
\right ) \,, \qquad
\mathbf{\chi}(\mathbf{s}) \equiv
\left (
\begin{array}{c}
\chi(S_1) \\ \vdots \\ \chi(S_k)
\end{array}
\right ) \,. \label{3:Gmatrix}
\end{equation}
Here $G_{i,j} \equiv G(\mathbf{x}_i; \mathbf{x}_j)$, and $G_{i,j} =
G_{j,i}$ by reciprocity, so that $\mathcal{G}$ is a symmetric matrix.
Then, \eqref{3:ASsmallD} becomes
\begin{equation}
\label{3:ASmatrix} \ac \,\mathbf{e} = \mathbf{s} + 2 \pi
\nu \mathcal{G}\, \mathbf{s} + \nu \mathbf{\chi}(\mathbf{s}) \,.
\end{equation}

We will consider \eqref{3:ASmatrix} for two ranges of $D$;
$D={\mathcal O}(1)$ and $D={\mathcal O}(\nu^{-1})$. For $D={\mathcal
O}(1)$, we can obtain a two-term approximation for the source
strengths in terms of $\nu\ll 1$ by expanding $\mathbf{s} =
\mathbf{s}_0 + \nu \mathbf{s}_1 + \cdots$. This readily yields
\begin{equation}
  \mathbf{s} = \ac \mathbf{e} -\nu \left[ 2 \pi \ac \, \mathcal{G}\,
\mathbf{e} + \chi(\ac) \mathbf{e} \right] + {\mathcal O}(\nu^2)
\,. \label{3:s2t}
\end{equation}
Therefore, for $\nu\ll 1$, the leading-order approximation for
$\mathbf{s}$ is the same for all of the spots. However, the ${\mathcal
O}(\nu)$ correction term depends on the spot locations and the domain
geometry.

Next, we consider \eqref{3:ASmatrix} for the distinguished limit where
$D ={D_0/\nu}\gg 1$ with $D_0 = {\mathcal O}(1)$. Since the
reduced-wave Green's function $G$, satisfying \eqref{3:Green}, depends
on $D$ we first must approximate it for $D$ large. Assuming that
$\Omega$ is a bounded domain, we expand $G$ and its regular part $R$
for $D\gg 1$ as
 \[ G \sim D G_{-1} + G_0 + \frac{1}{D} G_1 + \cdots\,, 
\quad R \sim D R_{-1} + R_0 + \frac{1}{D} R_1 + \cdots \,.\]
Substituting this expansion into \eqref{3:Green}, and collecting
powers of $D$, we obtain that $G_{-1}$ is constant and that 
 \[ \Delta G_0 = G_{-1} - \delta(\mathbf{x}-\mathbf{x}_j)\,, \quad
 \mathbf{x} \in \Omega\,; \qquad \partial_n G_0 = 0 \,, \quad
 \mathbf{x} \in \partial \Omega \,.\] The divergence theorem then
 shows that $G_{-1}= |\Omega|^{-1}$, where $|\Omega|$ is the area of
 $\Omega$.  In addition, the divergence theorem imposed on the $G_1$
 problem enforces that $\int_\Omega G_0\, d\mathbf{x}=0$, which makes
 $G_0$ unique. Next, from the singularity condition \eqref{3:gloc} for
 $G$ we obtain for $\mathbf{x}\to \mathbf{x}_j$ that $D G_{-1} +
 G_0(\mathbf{x}; \mathbf{x}_j) + \cdots \sim - (2\pi)^{-1} \ln
 |\mathbf{x}-\mathbf{x}_j| + D R_{-1} + R_0(\mathbf{x}; \mathbf{x}_j)
 + \cdots$. Since $G_{-1}=|\Omega|^{-1}$, we conclude that $R_{-1}
 =G_{-1} = 1/|\Omega|$. In this way, we obtain the following two-term
 expansion
\begin{equation}
 R_{j,j} \sim \tilde{R}_{j,j} \equiv \frac{D}{|\Omega|} + R^{(N)}_{j,j} +
\cdots \,, \qquad G(\mathbf{x};\mathbf{x}_j) \sim \tilde{G}(\mathbf{x};
 \mathbf{x}_j) \equiv \frac{D}{|\Omega|} + 
  G^{(N)} (\mathbf{x};\mathbf{x}_j)+\cdots \,, \qquad \mbox{for}
 \,\,\, D\gg 1 \,. \label{3:G2GN}
\end{equation} 
Here $G^{(N)}(\mathbf{x};\mathbf{x}_j)$ is the Neumann Green's function 
with regular part $R_{j,j}^{(N)}$, determined from the unique solution to
\bsub \label{3:gneum}
\begin{gather}
   \Delta G^{(N)} = \frac{1}{|\Omega|} - \delta(\mathbf{x}-\mathbf{x}_j)\,, 
  \quad \mathbf{x} \in \Omega \,; \qquad
  \partial_n G^{(N)} = 0 \,, \quad \mathbf{x} \in \partial \Omega \,;
  \qquad \int_{\Omega} G^{(N)} \, d\mathbf{x} = 0 \,, \\
 G^{(N)}(\mathbf{x}; \mathbf{x}_j)  \sim - \frac{1}{2\pi} \ln
|\mathbf{x}-\mathbf{x}_j| + R_{j,j}^{(N)} + o(1)  \,, \quad
  \mbox{as} \quad \mathbf{x} \to \mathbf{x}_j \,.
\end{gather}
\esub

\begin{figure}[htbp]
\centering \subfigure[$R_{1,1}$ and $\tilde{R}_{1,1}$ vs.~$D$]
{ \label{fig:Green_squ_D:a}
\includegraphics[width=3.0in, height=1.8in]{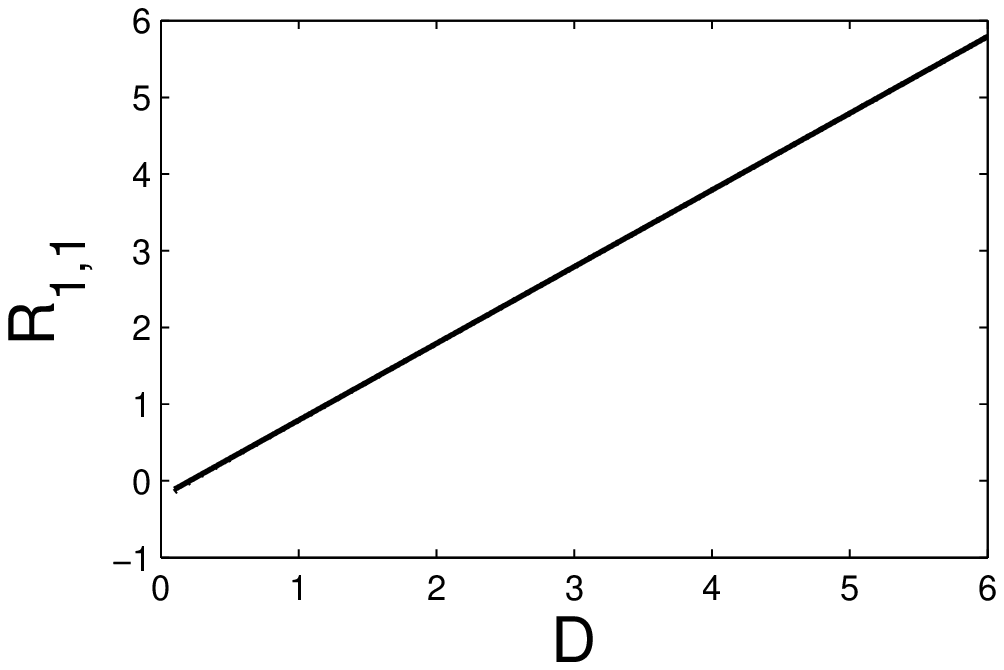}}
\subfigure[$R_{1,1}$ and $\tilde{R}_{1,1}$ vs.~$D$] { \label{fig:Green_squ_D:b}
\includegraphics[width=3.0in, height=1.8in]{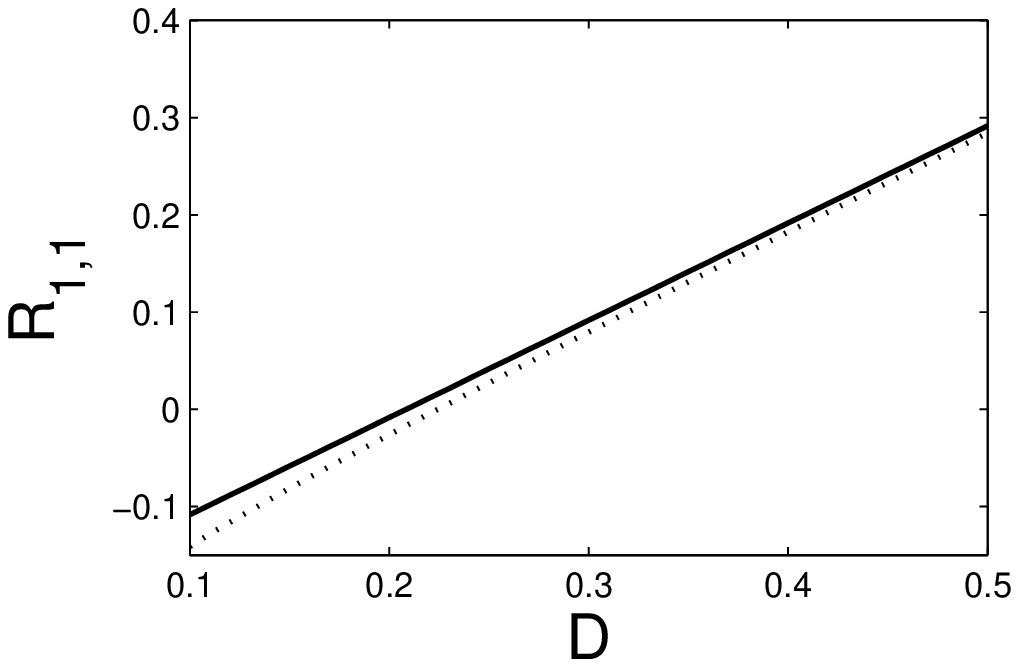}}
\caption{{\em Consider a single spot centered at the midpoint
 $\mathbf{x}_1 = (0.5, 0.5)$ of the unit square. We plot $R_{1,1}$ vs.~$D$ 
(solid curve) and its two-term large $D$ approximation
$\tilde{R}_{1,1}$ given in \eqref{3:square_r} (dotted curve). (a) $D \in [0.1,
6]$; (b) $D \in [0.1, 0.5]$. }} \label{fig:gcomp_square}
\end{figure}

\begin{figure}[htbp]
\centering \subfigure[$R_{1,1}$ and $\tilde{R}_{1,1}$ vs.~$D$]
{ \label{fig:Green_cir_D:a}
\includegraphics[width=3.0in, height=1.8in]{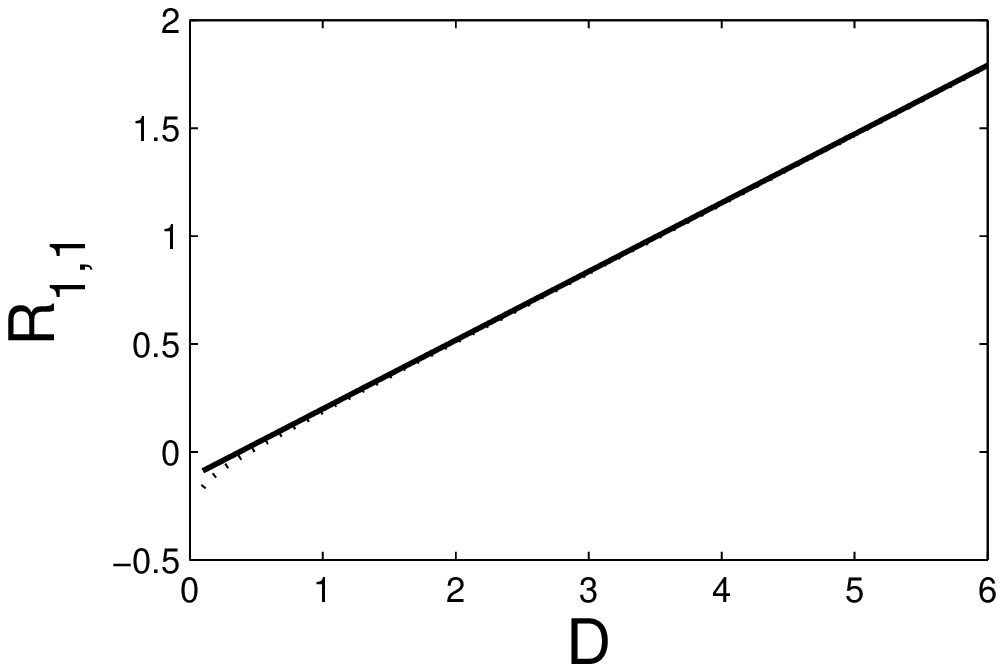}}
\subfigure[$R_{1,1}$ and $\tilde{R}_{1,1}$ vs.~$D$] { \label{fig:Green_cir_D:b}
\includegraphics[width=3.0in, height=1.8in]{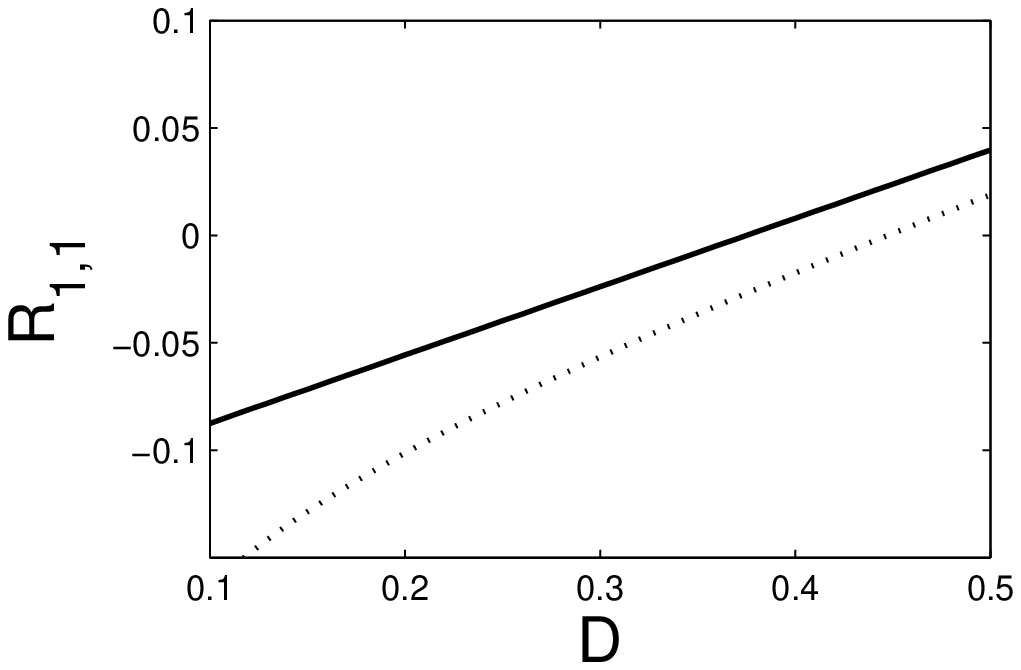}}
\caption{{\em Consider a single spot located at the center 
 $\mathbf{x}_1 = (0,0)$ of the unit disk. We plot $R_{1,1}$ vs.~ $D$ 
given in (\ref{3:disk_r}) (dotted curve) and its two-term large $D$ 
approximation $\tilde{R}_{1.1}$ given in (\ref{3:disk_r}) (solid curve).
 (a) $D \in [0.1,6]$; (b) $D \in [0.1, 0.5]$.}} \label{fig:gcomp_disk}
\end{figure}

For the disk and the square, in Appendix A we analytically
calculate both $G(\mathbf{x};\mathbf{x}_j)$ and its regular part
$R_{j,j}$, as well as the Neumann Green's function
$G^{(N)}(\mathbf{x};\mathbf{x}_j)$ and its regular part
$R^{(N)}_{j,j}$. For the case of a one-spot solution centered at the
midpoint $\mathbf{x}_{1}$ of the unit square $[0,1]\times [0,1]$, we
use some of the explicit formulae from Appendix A to compare the
two-term approximation $\tilde{R}_{1,1}$, given in \eqref{3:G2GN},
with the reduced-wave regular part $R_{1,1}$, as computed from
(\ref{3:squaregreen}).  These results are shown in
Fig.~\ref{fig:gcomp_square}. From \eqref{A:RN00}, the two-term
approximation \eqref{3:G2GN} for large $D$ is
\begin{equation}
  \tilde{R}_{1,1}= D -\frac{1}{\pi}\sum_{n=1}^{\infty} 
  \ln\left( 1- q^n\right) + \frac{1}{12} - \frac{1}{2\pi}\ln(2\pi) \,, 
   \qquad q\equiv e^{-2\pi}\,. \label{3:square_r}
\end{equation}
A similar comparison is made in Fig.~\ref{fig:gcomp_disk} for the case
of a single spot located at the center $\mathbf{x}_1=(0,0)$ of the
unit disk. For this radially symmetric case $R_{1,1}$ can be found
explicitly, and its two-term approximation for large $D$ is obtained
from (\ref{3:G2GN}) and (\ref{gr:neum_disk_r}). In this way, we get
\begin{equation}
  R_{1,1}=\frac{1}{2\pi} \left[ \frac{1}{2}\ln{D} + \ln{2} - \gamma_e 
  + \frac{  K_{1}\left( D^{-1/2}\right)}{I_{1}\left( D^{-1/2}\right)} 
  \right] \,, \qquad \tilde{R}_{1.1}= \frac{D}{\pi} - \frac{3}{8\pi}\,.
    \label{3:disk_r}
\end{equation}
Here $\gamma_e\approx 0.5772$ is Euler's constant, while $I_{1}(r)$
and $K_{1}(r)$ are the modified Bessel functions of order one.  From
Fig.~\ref{fig:gcomp_square} and Fig.~\ref{fig:gcomp_disk} we observe
that for both domains the two-term approximation \eqref{3:G2GN} involving the
regular part of the Neumann Green's provides a decent approximation of
$R_{1,1}$ even for only moderately large values of $D$. 

Next, we use the large $D$ asymptotics to find an approximate
solution to the nonlinear algebraic system \eqref{3:ASmatrix} in the
limit $D={D_0/\nu}\gg 1$, where $D_0={\mathcal O}(1)$ and $\nu\ll
1$. Upon substituting \eqref{3:G2GN} into \eqref{3:ASmatrix} we obtain
\[\ac \mathbf{e} = \mathbf{s} + \frac{2\pi D_0}{|\Omega|} \mathbf{e\,e}^T 
  \mathbf{s} + 2 \pi \nu \mathcal{G}^{(N)} \mathbf{s} + \nu
  \mathbf{\chi}(\mathbf{s}) \,,\] 
where $\mathcal{G}^{(N)}$ is the Green's matrix associated with the
Neumann Green's function, i.e.
$\mathcal{G}^{(N)}_{i,j}=G^{(N)}(\mathbf{x}_i;\mathbf{x}_j)$ for
$i\neq j$, and $\mathcal{G}^{(N)}_{j,j}\equiv R^{(N)}_{j,j}$. By expanding
$\mathbf{s}$ as $\mathbf{s}= \mathbf{s}_0 + \nu \mathbf{s}_1 + \cdots$
for $\nu\ll 1$, we then obtain that $\mathbf{s}_0$ and $\mathbf{s}_1$
satisfy
\begin{equation}
  \left(I +  \frac{2\pi D_0}{\,|\Omega|}
\mathbf{e}\mathbf{e}^T \right) \mathbf{s}_0 =\ac \mathbf{e}\,, \qquad
 \left(\mathcal{I} +  \frac{2\pi D_0}{\,|\Omega|} \mathbf{e}\mathbf{e}^T
  \right) \mathbf{s}_1 = - 2 \pi \mathcal{G}^{(N)} \mathbf{s}_0 -
  \mathbf{\chi}(\mathbf{s}_0) \,,  \label{3:slarged}
\end{equation}
where $I$ is the $k\times k$ identity matrix.  Since
$\mathbf{e}^T \mathbf{e} = k$, the leading-order approximation
$\mathbf{s}_0$ shows that the source strengths have an asymptotically
common value $S_c$ given by
\begin{equation}
  \mathbf{s}_0 = S_c  \mathbf{e} \,, \qquad S_c  \equiv 
 \frac{\ac}{1+ \mu k} \,, \qquad \mu \equiv \frac{ 2 \pi D_0}
  { |\Omega|} \,.    \label{3:def_mu_sc}
\end{equation}
The next order approximation $\mathbf{s}_1$ from \eqref{3:slarged}
yields
\begin{equation}
\mathbf{s}_1 = - \left(I +  \mu
\mathbf{e}\mathbf{e}^T \right)^{-1} \Big( 2 \pi S_c
\mathcal{G}^{(N)} + \chi(S_c) \Big) \mathbf{e} \,. \label{3:s1dlarge}
\end{equation}
Since the matrix $I + \mu \mathbf{e e}^T$ is a rank-one perturbation
of the identity, its inverse is readily calculated from the
Shermann-Woodbury-Morrison formula as $(I + \mu\mathbf{e}\mathbf{e}^T
)^{-1} = I - {\mu \mathbf{e}\mathbf{e}^T/( 1 + \mu k)}$, which
determines $\mathbf{s}_1$ from \eqref{3:s1dlarge}. In this way, 
for $D={D_0/\nu}$ and $\nu\ll 1$ we obtain the two-term expansion for $s$ 
given by
\begin{equation}
\mathbf{s} = S_c \mathbf{e} - \left(\frac{\chi(S_c)}{1 + \mu k}
\mathbf{e} + \frac{2 \pi \ac}{1 + \mu k}\Big(\mathcal{G}^{(N)} -
\frac{\mu\, F}{1 + \mu k} I\,\Big)\,\mathbf{e} \right)\nu +
  {\mathcal O}(\nu^2) \,. \label{3:2term_dlarge}
\end{equation}
Here $S_c$ and $\mu$ are defined in (\ref{3:def_mu_sc}), while the
scalar function $F(\mathbf{x}_1, \ldots, \mathbf{x}_k)$ is defined by
\begin{equation}
F(\mathbf{x}_1, \ldots, \mathbf{x}_k) =
\mathbf{e}^T \mathcal{G}^{(N)} \, \mathbf{e} = \sum_{i=1}^{k}
 \sum_{j=1}^{k} \mathcal{G}^{(N)}_{i,j} \,.
\label{3:FsumG} 
\end{equation}
In contrast to the leading-order approximation in (\ref{3:s2t}) when
$D={\mathcal O}(1)$, the leading-order approximation $S_c$ in
(\ref{3:2term_dlarge}) depends on the number of spots and the area of
the domain, with $S_c$ increasing as the area $|\Omega|$ increases.

We now illustrate our asymptotic theory for the construction of
quasi-equilibria for the case of a one-spot solution centered at the
midpoint of either the unit square or disk. Many additional examples of the
theory are given in \S \ref{sec:12inf}--\ref{sec:sym}.

\begin{figure}[htbp]
\centering
\subfigure[${\mathcal A}$ vs.~$S_1$]
{\label{fig:fold:a} 
\includegraphics[width=3.0in,height=2.0in]{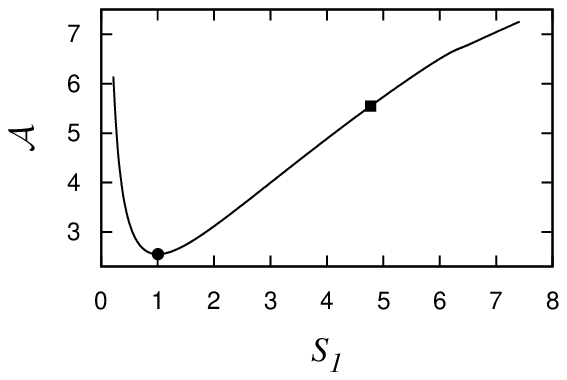}}
\subfigure[$u({\mathbf{x}_1})$ vs.~${\mathcal A}$ ]
{\label{fig:fold:b} 
\includegraphics[ width=3.3in,height=2.0in]{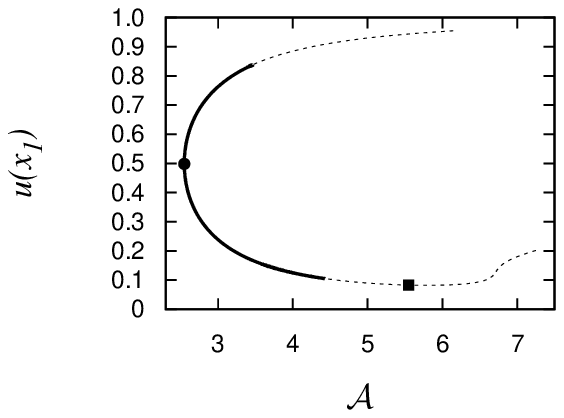}}
\caption{{\em Let $\Omega = \{\mathbf{x}| |\mathbf{x}| \leq 1\}$,
$\varepsilon=0.02$, $D=1$, and ${\mathbf x}_1={\mathbf 0}$. (a)
$\ac$ vs.~$S_1$; the square marks the volcano threshold $S_v \approx
4.78$, and the circle marks the fold point $\mathcal{A}_f \approx
2.55$ at which $S_f \approx 1.01$. (b) $u(\mathbf{x}_1)$
vs.~$\mathcal{A}$; the square marks $S_v \approx 4.78$, and the circle
marks the fold point $S_f \approx 1.01$. The upper branch is for $S_1
< S_f$, and the lower branch is for $S_1>S_f$.}}
\label{fig:fold}
\end{figure}

For our first example, we consider a one-spot solution with a spot
located at the center $\mathbf{x}_1={\mathbf 0}$ of the unit disk with
$\eps = 0.02$. We fix $D=1$, so that from \eqref{3:disk_r} the regular
part of the Green's function for a spot at ${\mathbf x}_1$ is $R_{1,1}  \approx
0.1890$.  Then, \eqref{3:ASsmallD} reduces to the following scalar nonlinear
algebraic equation for the source strength $S_1$ in terms of $\ac$:
\begin{equation}
\label{3:AS1spot} \ac =S_1\left(1+ 2\pi\nu R_{1,1}\right)+\nu \chi(S_1)\,.
\end{equation}
In Fig.~\ref{fig:fold:a} we plot $\ac$ versus $S_1$, showing the
existence of a fold point at $\ac_f \approx 2.55$ corresponding to
$S_f \approx 1.01$. Thus, $\ac \geq \ac_f$ is required for the
existence of a one-spot quasi-equilibrium solution located at the
center of the unit disk. In Fig.~\ref{fig:fold:b} the asymptotic
result for $u(\mathbf{x}_1) = {\nu \,U_1(0) / \ac}$ vs.~$\ac$ is shown
by the dotted curve, with a fold point at $u_f(\mathbf{x}_1) \approx
0.50$. The fold point is marked by a circle in both figures, and the
critical value $\mathcal{A}_v \approx 5.55$, $u_v(\mathbf{x}_1)
\approx 0.083$ for a volcano-type solution corresponding to $S_v \approx
4.78$ is marked by a square.  For $\ac>\ac_f$, $u(\mathbf{x}_1)$ has
two solution branches. The upper branch corresponds to $S_1<S_f$,
while the lower branch is for the range $S_1>S_f$.

To validate the asymptotic result for solution multiplicity, we solve
the steady-state GS model (\ref{1:GS_2D}) in the unit disk by using
the Matlab BVP solver {\em BVP4C}. By varying $u(0)$, we then compute
the corresponding value of $\ac$. The resulting full numerical result
for $u(0)$ vs.~$\ac$ is shown by the heavy solid curve in
Fig.\ref{fig:fold:b}, which essentially overlaps the asymptotic
result. This shows that when $\eps=0.02$, the asymptotic result for
the bifurcation diagram, based on retaining all terms in powers of
$\nu$, agrees very closely with the full numerical result.

\begin{figure}[htbp]
\centering \subfigure[$\ac$ vs.~$S_1$] { \label{fig:AS:a}
\includegraphics[width=3.0in,height=2.0in]{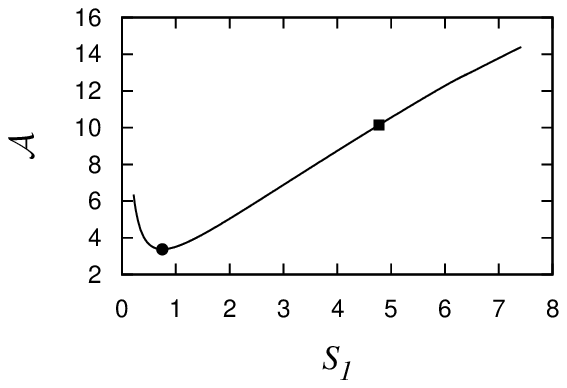}}
\subfigure[$u(\mathbf{x}_1)$ vs.~$\ac$]{ \label{fig:AS:b}
\includegraphics[width=3.3in,height=2.0in]{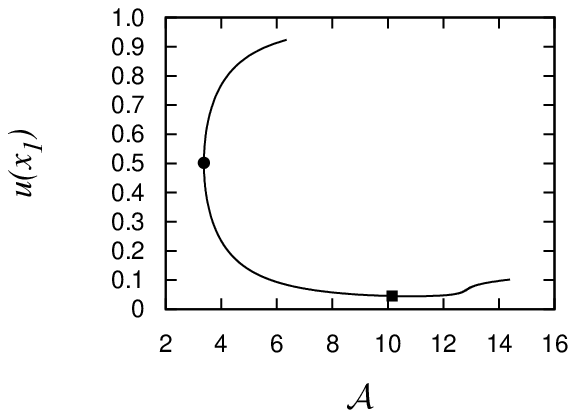} }
\caption[]{{\em Let $\Omega = [0, 1] \times [0, 1]$, $\eps = 0.02$,
$D=1$, and $\mathbf{x}_1 = (0.5, 0.5)$. (a) $\ac$ vs.~$S_1$; the
square marks the volcano threshold $S_v \approx 4.78$, and the circle
marks the fold point $\ac_f \approx 3.3756$ at which $S_f \approx
0.7499$. (b) $u(\mathbf{x}_1)$ vs.~$\ac$; the square marks $S_v
\approx 4.78$, and the circle marks the fold point $S_f \approx
0.7499$. The upper branch is for $S_1 < S_f$, and the lower branch is
for $S_1>S_f$.}}\label{fig:AS}
\end{figure}

For our second example, we consider a one-spot solution centered at
the midpoint $\mathbf{x}_1=(0.5,0.5)$ of the unit square $\Omega = [0,
1] \times [0, 1]$. We fix $\eps = 0.02$ and $D=1$.  Then, by using
\eqref{3:squaregreen} for the reduced-wave Green's function, as given
in Appendix A, we obtain that $R_{1,1} \approx 0.7876$. We remark that
even if we use the two-term large $D$ asymptotics (\ref{3:square_r})
for the unit square, then we get the rather good estimate
$R_{1,1}\approx 0.7914$. In Fig.~\ref{fig:AS} we use (\ref{3:AS1spot})
to plot $\ac$ versus $S_1$ and $u(\mathbf{x}_1) = {\nu \,U_1(0) /
\ac}$ versus $\ac$. For $\ac>\ac_f\approx 3.376$, $u(\mathbf{x}_1)$
versus $\ac$ has two solution branches, with the upper branch
corresponds to $S_1<S_f\approx 0.750$ and the lower branch for
$S_1>S_f$.  In contrast to the previous example for the unit disk,
where the full GS model (\ref{1:GS_2D}) is reduced to a coupled set of BVP
for ODE's, we cannot readily verify this asymptotic result for
solution multiplicity in a square from a full numerical solution of
(\ref{1:GS_2D}).

\subsection{Symmetric Spot Patterns and a Circulant Matrix} 
\label{sec:quasi_circ}

A special case for multi-spot patterns, which features prominently in
\S \ref{sec:eig}--\ref{sec:asy} below, is when the spatial
configuration $\mathbf{x}_1,\ldots,\mathbf{x}_k$ of spots within
$\Omega$ is sufficiently symmetric so that the Green's matrix
$\mathcal{G}$ is a circulant matrix.

When $\mathcal{G}$ is circulant, then it has the eigenpair
$\mathcal{G}\mathbf{e}=\theta \mathbf{e}$, where
$\theta=k^{-1}\sum_{i=1}^{k}\sum_{j=1}^{k} \left({\mathcal
G}\right)_{i,j}$.  For this special case, \eqref{3:ASmatrix} has a
solution for which the spots have a common source strength $S_j=S_c$
for $j=1,\ldots,k$, where $S_c$ is the solution to the single nonlinear
algebraic equation
\begin{equation}
   \ac = S_c + 2\pi \nu \theta S_c + \nu \chi(S_c) \,. \label{3:circscalar}
\end{equation}
For $\nu\ll 1$, a two-term approximation for this common source 
strength $S_c$ is
\begin{equation}
  S_c = \ac  -\nu \left[ 2 \pi \theta \ac  + \chi(\ac) \right] + 
 {\mathcal O}(\nu^2) \,. \label{3:circscalar_p}
\end{equation}
For the distinguished limit $D={\mathcal O}(\nu^{-1})$, and with
${\mathcal G}$ a circulant matrix, we calculate from
(\ref{3:2term_dlarge}) that a two-term asymptotic approximation for
the source strengths is given in terms of
$F(\mathbf{x}_1,\ldots,\mathbf{x}_k)$, as defined in (\ref{3:FsumG}),
by
\begin{equation}
\label{3:SlargeDk} \mathbf{s} = S_c \mathbf{e} - \nu
\left(\frac{\chi(S_c)}{1 + \mu k} + \frac{2 \pi \ac \theta^{(N)} }{
  (1 + \mu k)^2} \right) \mathbf{e} + {\mathcal O}(\nu^2) \,, \qquad
  S_c  \equiv \frac{\ac}{1 + \mu k} \,, \qquad \mu \equiv
  \frac{2 \pi D_0}{ |\Omega|} \,,
\end{equation}
Here $\theta^{(N)}={F/k}$ is the eigenvalue for the eigenvector
$\mathbf{e}$ of the circulant Neumann Green's matrix
$\mathcal{G}^{(N)}$.

Owing to the fact that the nonlinear algebraic system
\eqref{3:ASmatrix} can be reduced to the scalar nonlinear problem
(\ref{3:circscalar}) when the Green's matrix is circulant, in the
majority of our numerical experiments for multi-spot patterns in \S
\ref{sec:12inf}--\S \ref{sec:asy} below we will consider $k$-spot
quasi-equilibrium patterns that lead to this special matrix structure.

\setcounter{section}{2}
\section{The Slow Dynamics of a Collection of Spots} \label{sec:dyn}

In this section we derive the slow dynamics for the spot locations
corresponding to a $k$-spot quasi-equilibrium solution of the GS model
\eqref{1:GS_2D}. At each fixed time $t$, the spatial profile of the
spot pattern is characterized as in Principal Result 2.1.  In the
inner region near the $j^{\mbox{th}}$ spot, we introduce $\mathbf{y} =
\eps^{-1}\left[\mathbf{x}-\mathbf{x}_j(\xi)\right]$, where $\xi\equiv
\eps^2 t$ is the slow time variable, and we expand the inner solution
as
\begin{equation}
 u = \frac{\eps}{A \sqrt{D}} \left( U_{0j}(\rho) + 
\eps U_{1j}(\mathbf{y}) + \ldots \right) \,, \qquad v = \frac{\sqrt{D}}{\eps}
 \left(V_{0j}(\rho) + \eps V_{1j}(\mathbf{y}) + \ldots \right) \,.
  \label{4:uexp}
\end{equation}
Here the subscript $0$ in $U_{0j}, V_{0j}$ denotes the order of the
expansion, while $j$ denotes the $j^{\mbox{th}}$ inner region. In the
analysis below we omit the subscript $j$ if there is no confusion in
the notation. The leading-order terms $U_{0j}$ and $V_{0j}$ are
solutions of the core problem \eqref{3:2Dcore_sol}.  Define
$\mathbf{w}_j\equiv (V_{1j},U_{1j})^T$, where $T$ denotes
transpose. At next order, we get from \eqref{4:uexp} and 
\eqref{1:GS_2D} that $\mathbf{w}_j$ satisfies 
\begin{equation}
\label{3:dyn2_all} \Delta_{\mathbf{y}} \mathbf{w}_j + \mathcal{M}_j
\mathbf{w}_j = \mathbf{g}_j \,, \qquad \mathbf{y} \in \mathbb{R}^2 \,;
\qquad \mathcal{M}_j \equiv \left( \begin{array}{cc}  -1+2 U_0 V_0  &V_0^2 \\
 - 2 U_0 V_0 &-V_0^2 \end{array} \right)\,, \qquad \mathbf{g}_j \equiv 
\left( \begin{array}{c} - V^{\p}_0 \,\mathbf{x}^{\p}_j
 \cdot \mathbf{e}_{\theta} \\ 0 \end{array} \right) \,.
\end{equation}
In \eqref{3:dyn2_all}, $\cdot$ denotes dot product, and $\mathbf{e}_{\theta}
\equiv (\cos \theta, \sin \theta)^T$, where $\theta$ is the polar angle
for the vector $(\mathbf{x} - \mathbf{x}_j)$.

To determine the dynamics of the spots we must calculate the gradient
terms in the local expansion of the outer solution for $u$, as given
in \eqref{3:quasi}, in the limit $\mathbf{x}\to\mathbf{x}_j$. To do
so, we must calculate a further term in the local behavior as $\mathbf{x}
\to \mathbf{x}_j$ of the reduced-wave Green's function satisfying
(\ref{3:Green}). In terms of the inner variable $\mathbf{y}$, we get
\[ G(\mathbf{x}; \mathbf{x}_j) \sim - \frac{1}{2\pi} 
\ln \left|\mathbf{y}\right| + R_{j,j} + \eps 
\nabla R(\mathbf{x}_j; \mathbf{x}_j)  \cdot \mathbf{y} + \ldots \,,
 \] 
where we have defined $\nabla R(\mathbf{x}_j; \mathbf{x}_j) \equiv
\nabla_{\mathbf{x}} R(\mathbf{x}; \mathbf{x}_j)
\Big\vert_{\mathbf{x}=\mathbf{x}_j}$.  By comparing the higher-order
terms in the matching condition between the inner expansion
\eqref{4:uexp} for $u$ and the outer expansion for $u$, we obtain the
required far-field behavior as $|\mathbf{y}|\to\infty$ of the inner
solution $U_{1j}$. In this way, we obtain that $\mathbf{w}_j$
satisfies \eqref{3:dyn2_all} with far-field behavior
\begin{equation}
\label{3:dynbc} \mathbf{w}_j  \to  \left( \begin{array}{c}  0 \\
- \mathbf{f}_j \cdot \mathbf{y} \end{array} \right) \,, \quad \mbox{as}
\quad \mathbf{y} \to \infty \,, \qquad 
 \mathbf{f}_j =
\left( \begin{array}{c} f_{j1} \\ f_{j2}
\end{array} \right) \equiv  2 \pi \left(S_j \nabla
R(\mathbf{x}_j;\mathbf{x}_j) + \stacksum S_i \nabla 
 G(\mathbf{x}_j;\mathbf{x}_i) \right) \,. 
\end{equation}

To determine the dynamics of the $j^{\mbox{th}}$ spot we must
formulate the solvability condition for \eqref{3:dyn2_all} and
\eqref{3:dynbc}. We define
$\hat{P}_j^*(\rho) = (\hat{\phi}^*_j(\rho), \hat{\psi}^*_j(\rho))^T$
to be the radially symmetric solution of the adjoint
problem
\begin{equation}
\label{3:adjoint} \Delta_{\rho} \hat{P}_j^* + \mathcal{M}_j^T
\hat{P}_j^* = 0 \,, \qquad 0<\rho <\infty \,,
\end{equation}
subject to the far-field condition that $\hat{P}_j^* \to \,(0,
1/\rho)^T$ as $\rho \to \infty$, where $\Delta_{\rho} \equiv
\partial_{\rho \rho} + \rho^{-1} \partial_{\rho } - \rho^{-2}$. We
look for solutions $P^{c}_j$ and $P^{s}_j$ to the homogeneous adjoint
problem $\Delta_{\mathbf{y}} P_j^{*} + \mathcal{M}_j^T P_j^{*} = 0 $
for $\mathbf{y}\in \R^2$ in the form $P_{j}^{*}=P^{c}_j\equiv
\hat{P}_j^* \cos \theta$ and $P_{j}^{*}= P^{s}_j \equiv \hat{P}_j^*
\sin \theta$, where $\hat{P}_j^{*}(\rho)$ is the radially symmetric
solution of \eqref{3:adjoint}.

In terms of the adjoint solution $P^c_j$, the solvability condition
for \eqref{3:dyn2_all}, subject to \eqref{3:dynbc}, is that
\begin{equation}
\label{3:solv} \lim_{\sigma \to \infty} \int_{B_{\sigma}} P_j^c
\cdot \mathbf{g}_j \, d \mathbf{y} = \lim_{\sigma \to \infty}
\int_{\partial B_{\sigma}} \Big[ P_j^c \cdot \partial_{\rho}
\mathbf{w}_j - \mathbf{w}_j \cdot \partial_{\rho} P_j^c \Big]
\Big|_{\rho = \sigma} d \mathbf{y} \,.
\end{equation}
Here $B_{\sigma}$ is a ball of radius $\sigma$, i.e~
$|\mathbf{y}|=\sigma$.  Upon using the far-field condition
\eqref{3:dynbc}, and writing $\mathbf{x}_j=(x_{j1},x_{j2})^T$ in
component form, we reduce \eqref{3:solv} to
\begin{equation}
 x_{j1}^{\p} \int^{2 \pi}_0 = \int_{0}^{\infty} \hat{\phi}_j^* V^{\p}_0
\cos^2 \theta \, \rho \,d\rho\, d\theta \, -  x_{j2}^{\p} \int^{2 \pi}_0
\int_{0}^{\infty}
\hat{\phi}_j^* V^{\p}_0 \cos \theta \sin \theta \, \rho \,d\rho\, d\theta 
 = \lim_{\sigma \to \infty} \int^{2 \pi}_0 \left( \frac{2 \cos\theta}{\,
  \sigma} \mathbf{f}_j \cdot \mathbf{e}_{\theta} \right)\, 
 \sigma \, d\theta \,.
\end{equation}
Therefore, since $\int^{2 \pi}_0 \cos \theta \sin \theta \, d\theta =
0$, we obtain ${d x_{j1}/d \xi} = {2 f_{j1}/\left(\int_{0}^{\infty}
\hat{\phi}_j^{\star} V^{\p}_0\rho \, d\rho\right)}$.  Similarly, the
solvability condition for \eqref{3:dyn2_all}, subject to
\eqref{3:dynbc}, with respect to the homogeneous adjoint solution
$P^s_j$, yields ${ d x_{j2}/d \xi} = {2 f_{j2}/\left(
\int_{0}^{\infty} \hat{\phi}_j^* V^{\p}_0\rho \, d\rho\right) }$.
Upon recalling the definition of $(f_{j1},f_{j2})^T$ in
\eqref{3:dynbc}, we can summarize our result for the slow dynamics as
follows:

\vspace*{0.2cm}\noindent {\bf \underline{Principal Result 3.1}:}\;
{\em Consider the GS model \eqref{1:GS_2D} with $\eps\ll 1$,
$A={\mathcal O}(-\eps\ln\eps)$, and $\tau\ll {\mathcal O}(\eps^{-2})$.
Then, provided that each spot is stable to any profile instability,
the slow dynamics of a collection $\mathbf{x}_1,\ldots,\mathbf{x}_k$
of spots satisfies the differential-algebraic (DAE) system}
\begin{equation}
\label{3:dyn} \frac{\,d \mathbf{x}_j}{d t} \sim -2\pi\eps^{2}
\gamma(S_j) \left(S_j \nabla R(\mathbf{x}_j;\mathbf{x}_j) + 
\stacksum S_i \nabla  G(\mathbf{x}_j;\mathbf{x}_i) \right) \,, 
 \quad j=1,\ldots, k \,; \qquad \gamma(S_j) \equiv \frac{-2}
 {\int_{0}^{\infty} \hat{\phi}_j^* V^{\p}_0\rho \,d\rho} \,.
\end{equation}
{\em In \eqref{3:dyn}, the source strengths $S_j$, for $j=1,\ldots,k$,
 are determined in terms of the instantaneous spot locations and the
 parameters $\ac$ and $\nu$ of \eqref{3:pval} by the nonlinear
 algebraic system \eqref{3:ASsmallD}.  In the definition of
 $\gamma(S_j)$, $V_0$ satisfies the core problem \eqref{3:2Dcore_sol},
 while $\hat{\phi}_j^*$ is the first component of the solution to the
 radially symmetric adjoint problem \eqref{3:adjoint}. Finally, the
 equilibrium spot locations $\mathbf{x}_{je}$ and spot strengths
 $S_{je}$, for $j=1,\ldots,k$, satisfy}
\begin{equation}
\label{3:finaleq1} S_{je} \nabla R(\mathbf{x}_{je};\mathbf{x}_{je})
 + \stacksum S_{ie} \nabla G(\mathbf{x}_{je};\mathbf{x}_{ie}) = 0 \,, 
 \quad j = 1,\ldots,k \,,
\end{equation}
{\em subject to the nonlinear algebraic system \eqref{3:ASsmallD}, which
relates the source strengths to the spot locations.}

\vspace*{0.2cm} The ODE system \eqref{3:dyn} coupled to the nonlinear
algebraic system \eqref{3:ASsmallD} constitutes a DAE system for the
time-dependent spot locations $\mathbf{x}_j$ and source strengths
$S_j$ for $j=1,\ldots,k$. These collective coordinates evolve slowly
over a long time-scale of order $t={\mathcal O}(\eps^{-2})$, and
characterizes the slow evolution of the quasi-equilibrium
pattern. From a numerical computation of $\hat{\phi}_j^{\star}$, the
function $\gamma(S_j)$ in \eqref{3:dyn} was previously computed
numerically in Fig.~3 of \cite{Schnaken_KWW:2008}, where it was shown
that $\gamma(S_j)>0$ when $S_j>0$. This plot is reproduced below in
Fig.~\ref{fig:AS_inf:f}.  In \S \ref{sec:asy} we will compare the
dynamics \eqref{3:dyn} with corresponding full numerical results for
different spot patterns in the unit square.

We emphasize that the DAE system in Principal Result 3.1 for the slow
spot evolution is only valid if each spot is stable to any spot
profile instability that occurs on a fast ${\mathcal O}(1)$
time-scale. One such spot profile instability is the peanut-splitting
instability, studied below in \S \ref{sec:eig_nrad}, that is triggered
whenever $S_J>\Sigma_2\approx 4.31$ for some $J \in
\lbrace{1,\ldots,k\rbrace}$. The other profile instabilities are
locally radially symmetric instabilities and, roughly speaking,
consist of a temporal oscillation of the spot amplitude if $\tau$ is
sufficiently large, or a spot over-crowding competition instability,
which is triggered when either the spots are too closely spaced or,
equivalently, when $D$ is too large. A new global eigenvalue problem
characterizing these latter two types of instabilities is formulated
in \S \ref{sec:eig_rad}.

For the equilibrium problem \eqref{3:finaleq1}, it is analytically
intractable to determine all possible equilibrium solution branches
for $k$-spot patterns in an arbitrary two-dimensional domain as the
parameters $\ac$ and $D$ are varied. However, some partial analytical
results are obtained in \S \ref{sec:sym} for the special case of $k$
spots equally spaced on a circular ring that lies within, and is
concentric with, a circular disk domain. For this special case, the
Green's matrix in \eqref{3:ASmatrix} is circulant, and the equilibrium
problem is reduced to determining the equilibrium ring radius for the
pattern.

For the related Schnakenburg model, it was shown in \S 2.4 of
\cite{Schnaken_KWW:2008} that near the spot self-replication
threshold, i.e. for $S_j$ near $\Sigma_2$, the direction at which the
spot splits is always perpendicular to the direction of the motion of
the spot. This result was derived in \cite{Schnaken_KWW:2008} from a
center-manifold type calculation involving the four dimensional
eigenspace associated with the two independent translation modes and the two
independent directions of splitting. Since this calculation
in \cite{Schnaken_KWW:2008} involves only the inner region near an
individual spot, it also applies directly to the GS model
\eqref{1:GS_2D}.  This qualitative result is stated as follows:

\vspace*{0.2cm}\noindent {\bf \underline{Principal Result 3.2}:}\;
{\em Consider the GS model \eqref{1:GS_2D} with $\eps\ll 1$,
$A={\mathcal O}(-\eps\ln\eps)$, and $\tau\ll {\mathcal O}(\eps^{-2})$.
Suppose that $S_J> \Sigma_2$, with $S_J-\Sigma_2\to 0^{+}$ for some unique
index $J$ in the set $j=1,\ldots,k$. Then, the direction of splitting
of the $J^{\mbox{th}}$ spot is perpendicular to the direction of its
motion.}

\setcounter{equation}{0}
\setcounter{section}{3}
\section{Fast Instabilities of the Quasi-Equilibrium Spot Pattern}
\label{sec:eig}

In this section we study the stability of a $k$-spot quasi-equilibrium
pattern to either competition, oscillatory, or self-replication,
instabilities that can occur on a fast ${\mathcal O}(1)$ time-scale
relative to the slow motion, of speed ${\mathcal O}(\eps^2)$, of the
spot locations. The stability analysis below with regards to spot
self-replication is similar to that done in \cite{Schnaken_KWW:2008}
for the Schnakenburg model. The formulation of a globally coupled
eigenvalue problem governing competition and oscillatory instabilities
is a new result.

Let $u_e$ and $v_e$ denote the quasi-equilibrium solution of Principal
Result 2.1. We introduce the perturbation
\[u(\mathbf{x},t) = u_e + e^{\lambda\,t} \eta(\mathbf{x})\,, 
\quad v(\mathbf{x},t) = v_e + e^{\lambda\,t} \phi(\mathbf{x})\,,\] 
for a fixed spatial configuration $\mathbf{x}_1,\ldots,\mathbf{x}_k$ of
spots. Then, from \eqref{1:GS_2D}, we obtain the eigenvalue problem
\begin{equation}
 \eps^2 \Delta \phi - (1 +\lambda)\phi + 2 A u_e v_e \phi + A v_e^2 \eta =
  0 \,, \qquad  D \Delta \eta - (1+\tau \lambda) \eta - 2 u_e v_e \phi - 
 v_e^2 \eta = 0 \,. \label{3:eigen}
\end{equation}
In the inner region near the $j^{\mbox{th}}$ spot, we recall from
\eqref{3:2dinnvar} that $u_e \sim \frac{\eps}{A \sqrt{D}} U_j$ and
$v_e \sim \frac{\sqrt{D}}{\eps} V_j$, where $U_j, V_j$ is the radially
symmetric solution of the core problem \eqref{3:2Dcore_sol}. Next, we
define
\begin{equation}
 \mathbf{y} = \eps^{-1}(\mathbf{x}-\mathbf{x}_j) \,, \qquad
 \eta = \frac{\eps}{A \sqrt{D}}N_j \,, \qquad \phi =\frac{\sqrt{D}}{\eps} 
 \Phi_j \,, \label{3:inn_stab_var}
\end{equation}
so that \eqref{3:eigen} transforms to
\begin{equation}
 \Delta_{\mathbf{y}}\Phi_j - (1 + \lambda)\Phi_j  + 2 U_j V_j \Phi_j + 
 V_j^2 N_j = 0 \,, \qquad
 \Delta_{\mathbf{y}} N_j  -  V_j^2 N_j  - 2 U_j V_j \Phi_j = \frac{\eps^2}{D}
  \left(1+\tau \lam\right) N_j  \,. \label{3:innereigen}
\end{equation}
Then, assuming that $D={\mathcal O}(1)$ and $\tau \ll {\mathcal
O}(\eps^{-2})$, we can neglect the right-hand side of the
equation for $N_j$ in \eqref{3:innereigen}.

Next, we look for angular perturbations of the form $\Phi_j = e^{i m
\theta} \hat{\Phi}_j (\rho)\,,\,\, N_j = e^{i m \theta} \hat{N}_j
(\rho)$, where $m\geq 0$ is a non-negative integer,
$\theta=\mbox{arg}(\mathbf{y})$, and $\rho = |\mathbf{y}|$. Then, from
\eqref{3:innereigen}, $\hat{N}_j(\rho)$ and $\hat{\Phi}_j(\rho)$
satisfy \bsub
\label{3:eigenmode}
\begin{gather}
 \hat{\Phi}_{j}^{\p\p} + \frac{1}{\rho} \hat{\Phi}_{j}^{\p} - 
\frac{m^2}{\rho^2} \hat{\Phi}_j - (1 + \lambda)\hat{\Phi}_j  + 2 U_j V_j
 \hat{\Phi}_j + V_j^2 \hat{N}_j = 0 \,, \quad 0<\rho<\infty \,, 
  \label{3:eigenmode:a}\\
 \hat{N}_{j}^{\p\p} + \frac{1}{\rho} \hat{N}_{j}^{\p} - 
\frac{m^2}{\rho^2} \hat{N}_j   -  V_j^2 \hat{N}_j  - 2 U_jV_j \hat{\Phi}_j 
 = 0 \,, \quad 0<\rho<\infty \,, \label{3:eigenmode:b}
\end{gather}
with boundary conditions 
\begin{equation}
\label{3:eigenbdc} \hat{\Phi}^{\p}_j(0)=0,\;\;\; \hat{N}^{\p}_j(0) =0,\quad
\hat{\Phi}_{j}(\rho) \to 0, \;\; \mbox{as} \;\;\;\rho \to \infty \,.
\end{equation}
\esub Since the far-field behavior of $\hat{N}_j$ is different for
$m=0$ and $m\geq 2$, we will consider these two different cases
separately below. For $m=1$, which corresponds to translation
invariance, it follows trivially that $\lam=0$ is an eigenvalue of the
local eigenvalue problem. For this translation eigenvalue, a
higher-order analysis would show that $\lam={\mathcal O}(\eps^2)$ when
$\eps \to 0$. Since any weak instability of this type should be
reflected by the properties of the Hessian of the DAE system of
Principal Result 3.1 for the slow spot dynamics, the mode $m=1$ is not
considered here.

\subsection{Non-Radially Symmetric Local Perturbations: Spot Self-Replication 
Instabilities}\label{sec:eig_nrad}

An instability of \eqref{3:eigenmode} for the mode $m =2$ is
associated with the initiation of a peanut-splitting instability.
Instabilities for the higher modes $m\ge 3$ suggest the possibility of
the initiation of more spatially intricate spot self-replication
events. Thus, the eigenvalue problem \eqref{3:eigenmode} with angular
modes $m \geq 2$ initiate angular deformations of the spot
profile. For this range of $m$, the linear operator for $N_j$ in
\eqref{3:eigenmode:b} allows for algebraic decay of $N_j$ as $\rho\to
\infty$ owing to the ${m^2 \hat{N}_j/\rho^2}$ term. As such, for
$m\geq 2$ we impose the far-field boundary condition that $\hat{N}_j
\to 0$ as $\rho \to \infty$.

For $m\ge 2$, the eigenvalue problem \eqref{3:eigenmode} is coupled to
the core problem \eqref{3:2Dcore_sol} for $U_j$ and $V_j$, and can
only be solved numerically.  To do so, we first solve the BVP
\eqref{3:2Dcore_sol} numerically by using COLSYS
(cf.~\cite{colsys_Ascher:1979}). Then, we discretize
\eqref{3:eigenmode} by a centered difference scheme to obtain a matrix
eigenvalue problem. By using the linear algebra package LAPACK
\cite{lapack_Anderson:1999} to compute the spectrum of this matrix
eigenvalue problem, we estimate the eigenvalue $\lambda_0$ of
\eqref{3:eigenmode} with the largest real part as a function of the
source strength $S_j$ for different angular modes $m\geq 2$. The
instability threshold occurs when $\mbox{Re}(\lam_0)=0$. We find
numerically that $\lam_0$ is real when $S_j$ is large enough. In the
left subfigure of Fig.~\ref{fig:eigm2}, we plot $\mbox{Re}(\lambda_0)$
as a function of the source strength $S_j$ for $m = 2, 3, 4$. Our
computational results show that the instability threshold for the
modes $m\ge 2$ occurs at $S_j=\Sigma_m$, where $\Sigma_2 \approx
4.31$, $\Sigma_3 \approx 5.44$, and $\Sigma_4 \approx 6.14$.  In the
right subfigure of Fig.~\ref{fig:eigm2}, we plot the eigenfunction
$(\hat{\Phi}_j, \hat{N}_j)$ corresponding to $\lambda_0 = 0$ with
$m=2$ at $S_j =\Sigma_2$. 

We emphasize that since $\hat{N}_j\to 0$ as $\rho\to\infty$, the initiation
of a spot self-replication instability is determined through a {\em local}
stability analysis near the $j^{\mbox{th}}$ spot. We summarize the result
in the following statement:

\begin{figure}[htbp]
\centering \subfigure[$\mbox{Re}(\lambda_0)$ vs.~$S_j$] { \label{fig:eigm2:a}
\includegraphics[width=3.0in,height=1.8in]{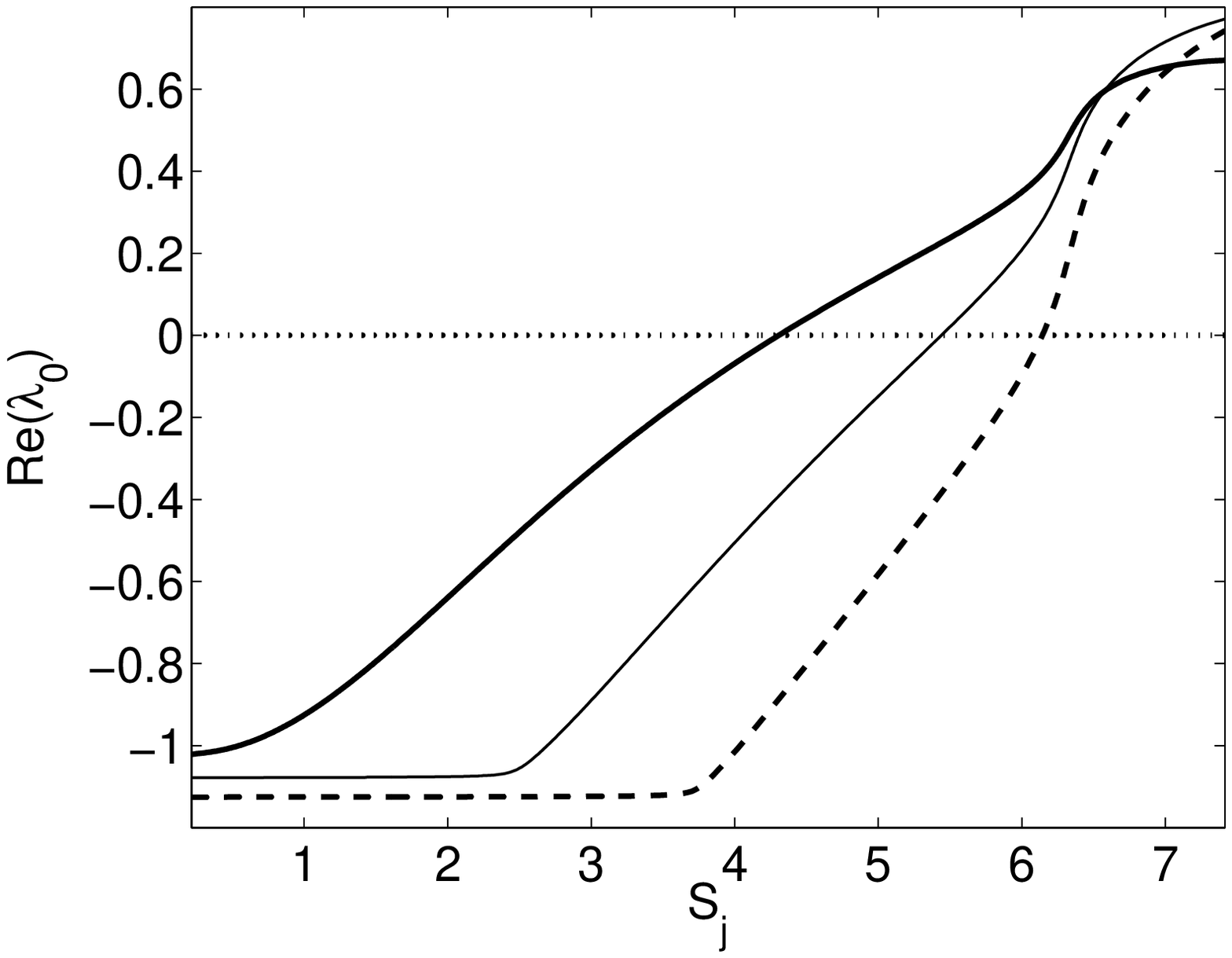}}
\subfigure[$(\hat{\Phi}_j, \hat{N}_j)$ vs.~$\rho$]  {\label{fig:eigm2:b}
\includegraphics[width=3.0in,height=1.8in]{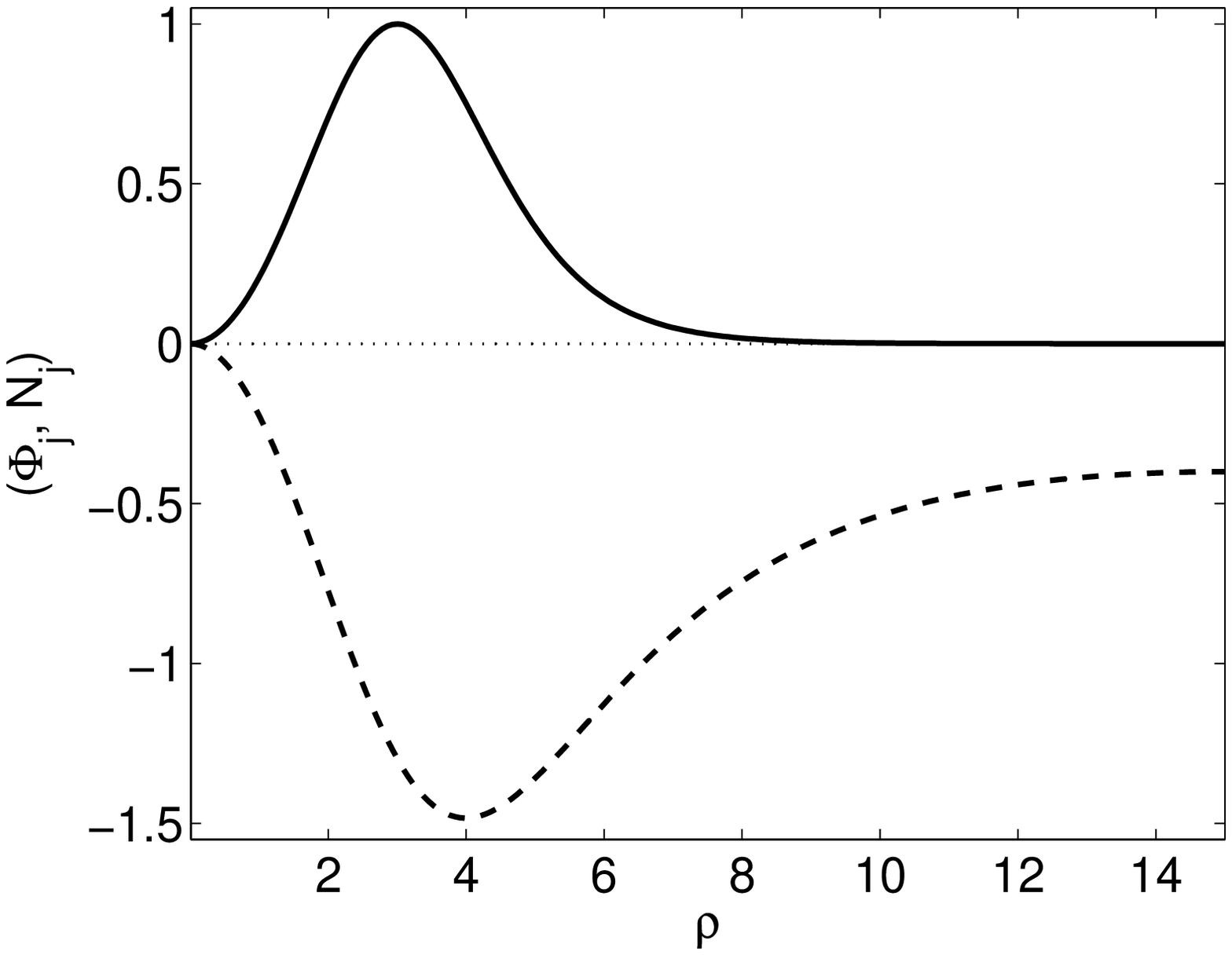}}
\caption[Eigenvalue of \eqref{3:eigenmode} with modes $m \geq 2$]{{\em
Numerical results for the principal eigenvalue $\lam_0$ of \eqref{3:eigenmode}
with mode $m \geq 2$. (a) $\mbox{Re}(\lambda_0)$ vs.~$S_j$; heavy
solid curve is for $m=2$ with $\Sigma_2 = 4.31$, the solid curve is for
$m=3$ with $\Sigma_3 = 5.44$, and the dashed curve is for $m=4$ with
$\Sigma_4 = 6.14$.  (b) For $m=2$, the eigenfunctions
$(\hat{\Phi}_j(\rho), \hat{N}_j(\rho))$ near $\lambda_0 = 0$ with $S_j
= \Sigma_2 \approx 4.31$ are shown. The solid curve is
$\hat{\Phi}_j(\rho)$, and the dashed curve is
$\hat{N}_j(\rho)$. In this subfigure the maximum value of $\hat{\Phi}_j$ 
has been scaled to unity.}}\label{fig:eigm2}
\end{figure}

\vspace*{0.2cm}\noindent {\bf \underline{Principal Result 4.1}:} {\em
Consider the GS model \eqref{1:GS_2D} with $\eps\ll 1$, $A={\mathcal
O}(-\eps\ln\eps)$, and $\tau\ll {\mathcal O}(\eps^{-2})$.  We define
$\nu$ and $\ac$ as in \eqref{3:pval}. In terms of $\ac$, $D$, and
$\nu$, we calculate $S_1,\ldots,S_k$ for a $k$-spot quasi-equilibrium
pattern from the nonlinear algebraic system \eqref{3:ASsmallD}. Then,
if $S_j < \Sigma_2 \approx 4.31$, the $j^{\mbox{th}}$ spot is linearly
stable to a spot deformation instability for modes $m\geq
2$. Alternatively, for $S_j > \Sigma_2$, it is linearly unstable to
the peanut-splitting mode $m=2$.}

We now show numerically that the peanut-splitting linear instability
leads to a nonlinear spot self-replication event. This suggests that
the bifurcation as $S_j$ increases above $\Sigma_2$ is subcritical.
To show this, we formulate a time-dependent inner, or core, problem
near a single spot, defined in terms of the local inner variables
\begin{equation}
 u = \frac{\eps}{A \sqrt{D}} \, U\left( \mathbf{y},t\right) \,, \qquad
 v = \frac{ \sqrt{D}}{\eps} V\left( \mathbf{y},t\right) \,, \qquad 
  \mathbf{y}=\eps^{-1}\left( \mathbf{x} - \mathbf{x}_j \right) \,.
 \label{2Dreduced_var}
\end{equation}
Then, from \eqref{1:GS_2D}, we obtain to leading order that $U$ and $V$
satisfy the time-dependent parabolic-elliptic problem
\begin{equation}
   V_t = \Delta_{\mathbf{y}} V - V + U V^2 \,, \qquad 
 \Delta_{\mathbf{y}} U - U V^2 = 0\,,
\quad \mathbf{y} \in \mathbb{R}^2 \,; \qquad
 V \to 0 \,, \quad U \to S \ln |\mathbf{y}| \,, \quad \mbox{as} \,\,\, 
 |\mathbf{y}| \to \infty \,.  \label{2Dreduced}
\end{equation}

\begin{figure}[htpb]
\begin{center}
\subfigure[$t=0$]
{\includegraphics[width=2.5cm,clip]{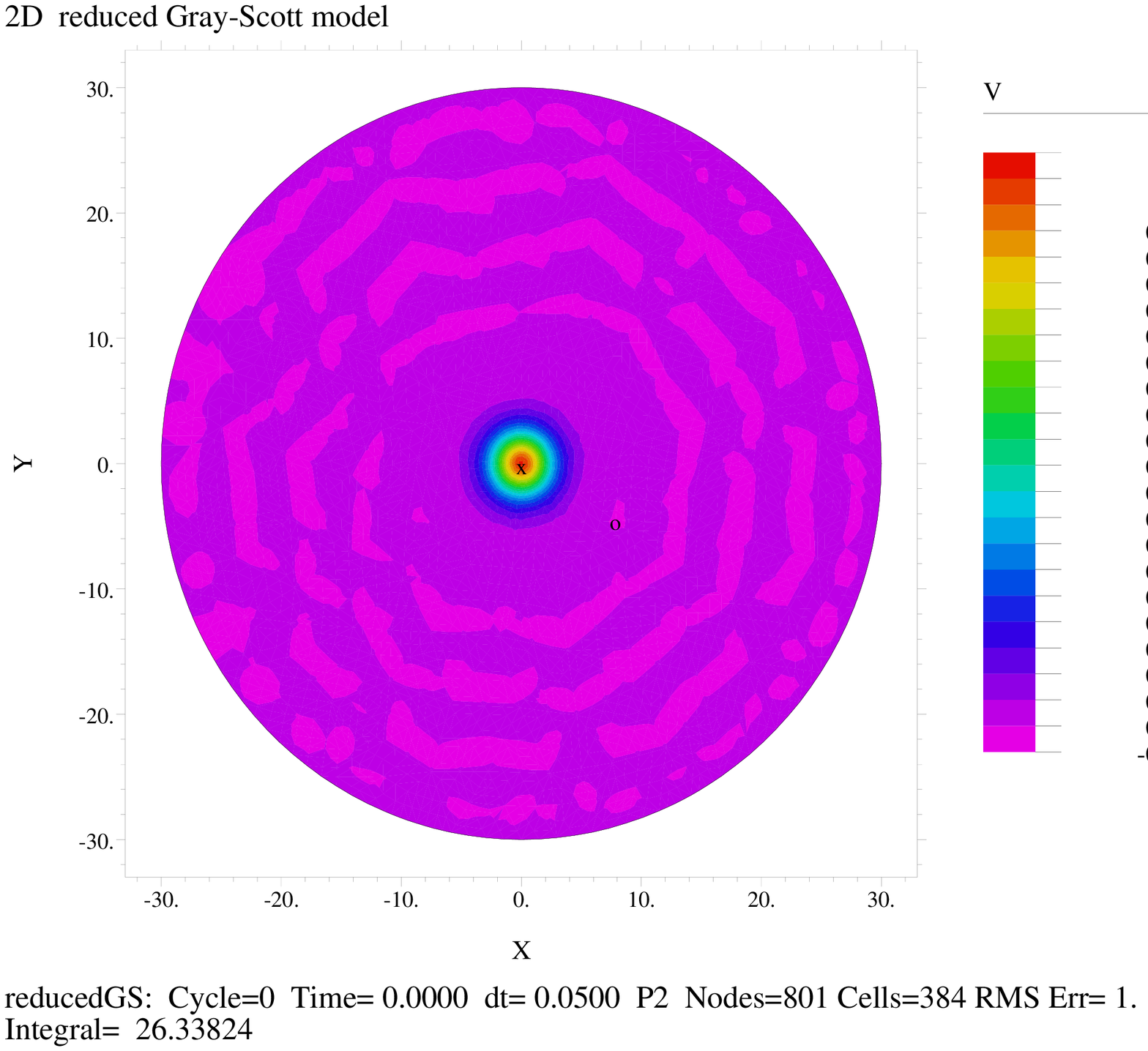}}
\subfigure[$t=100$]
{\includegraphics[width=2.5cm,clip]{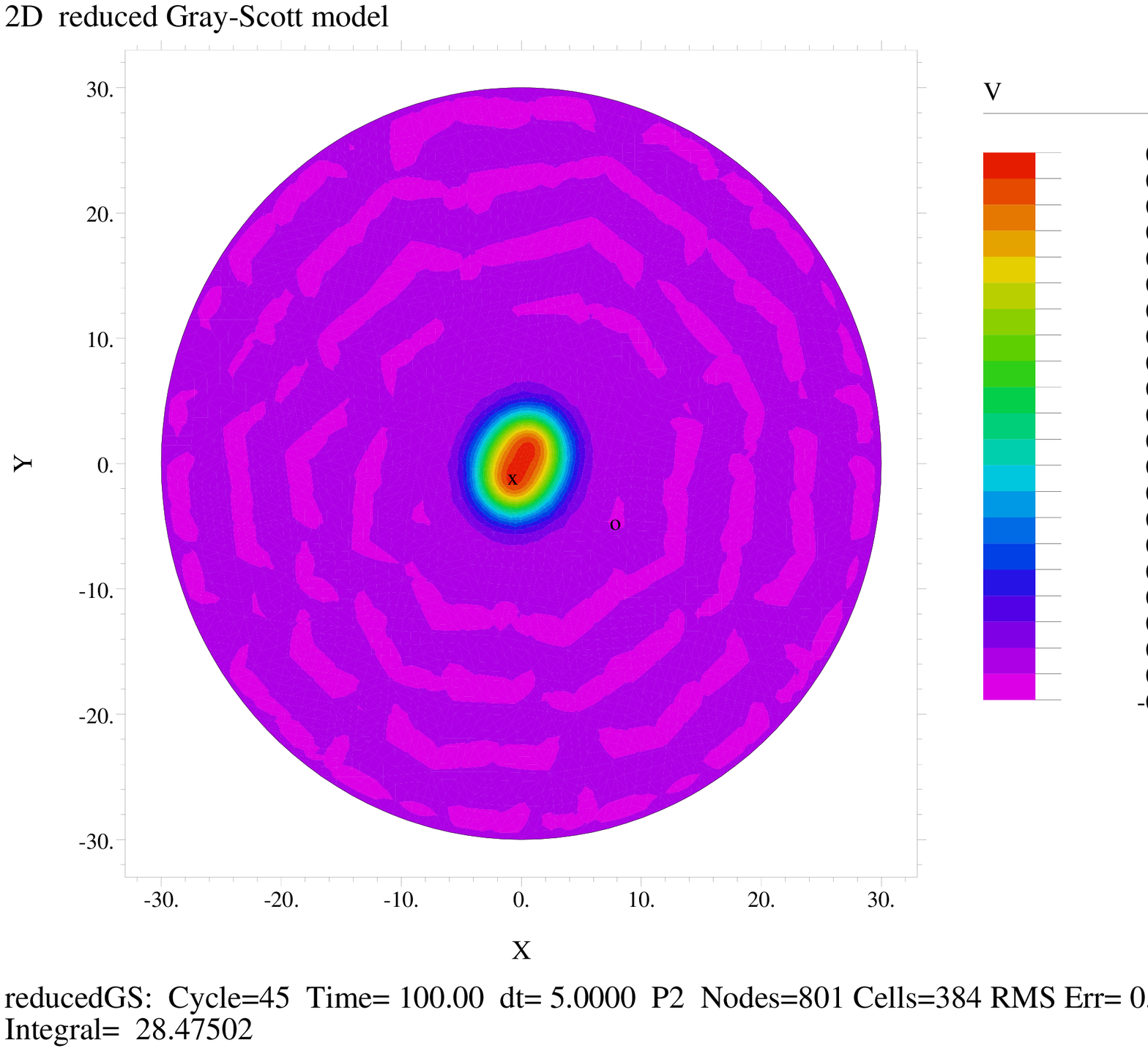}}
\subfigure[$t=130$]
{\includegraphics[width=2.5cm,clip]{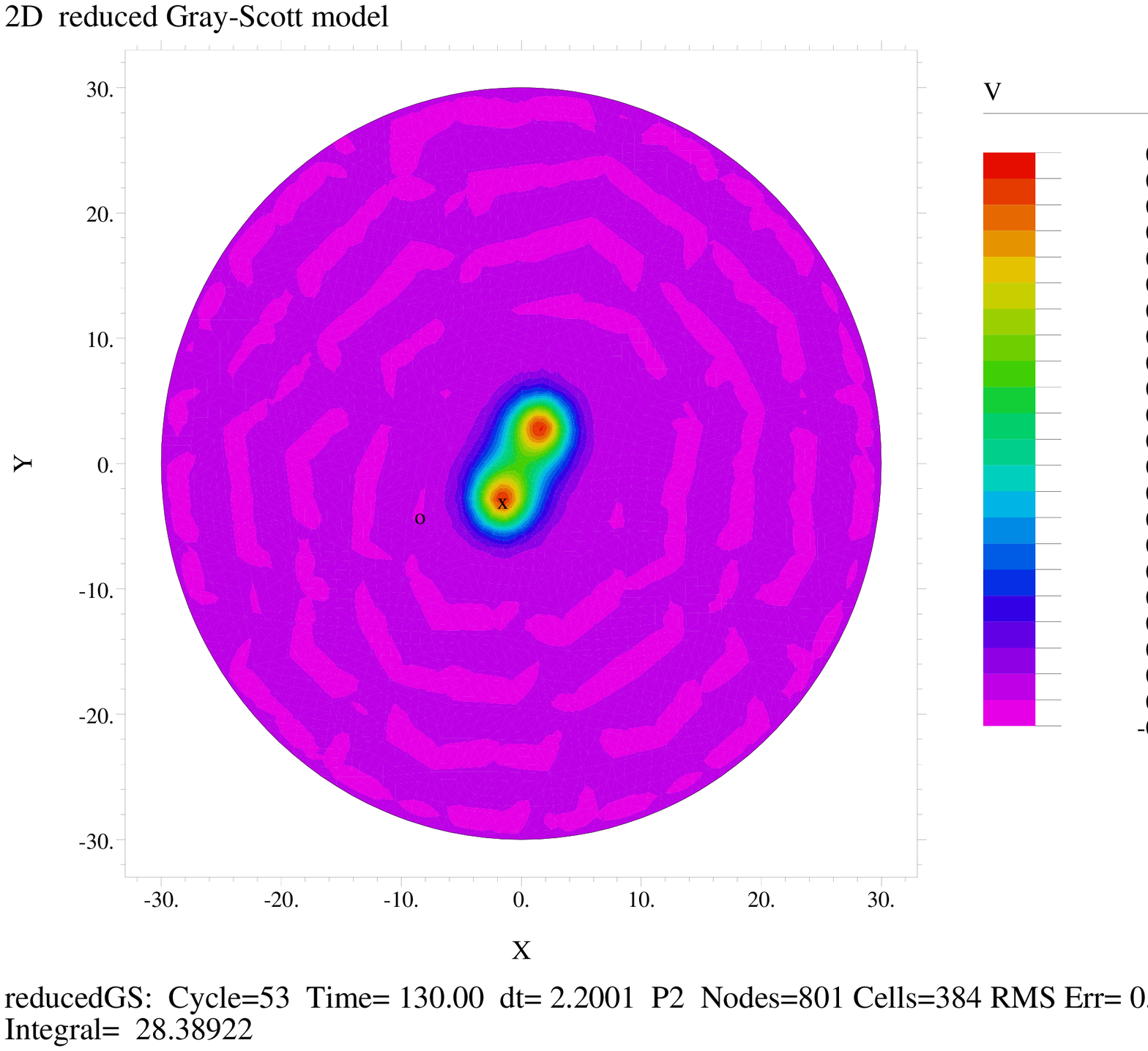}}
\subfigure[$t=140$]
{\includegraphics[width=2.5cm,clip]{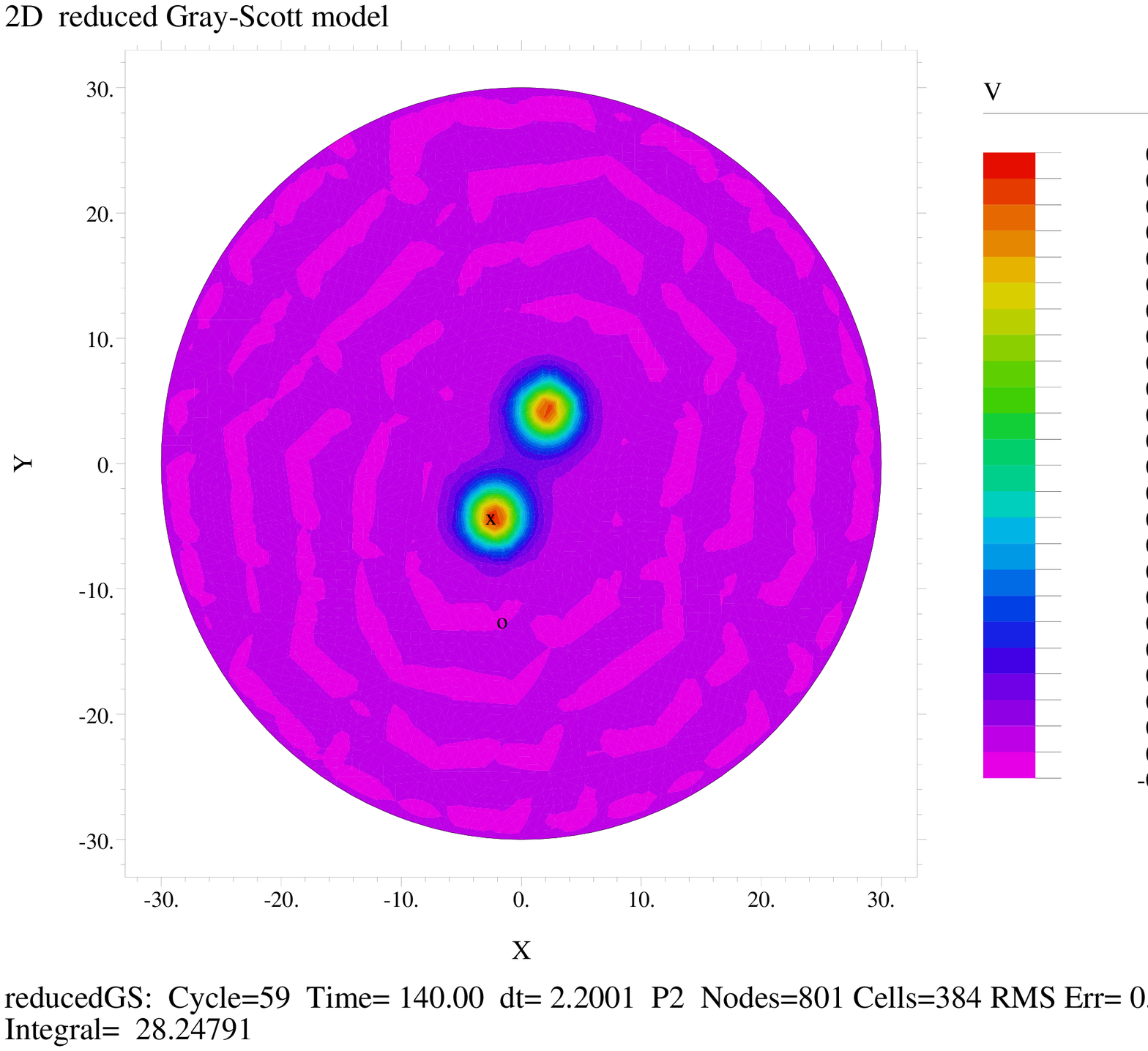}}
\subfigure[$t=170$]
{\includegraphics[width=2.5cm,clip]{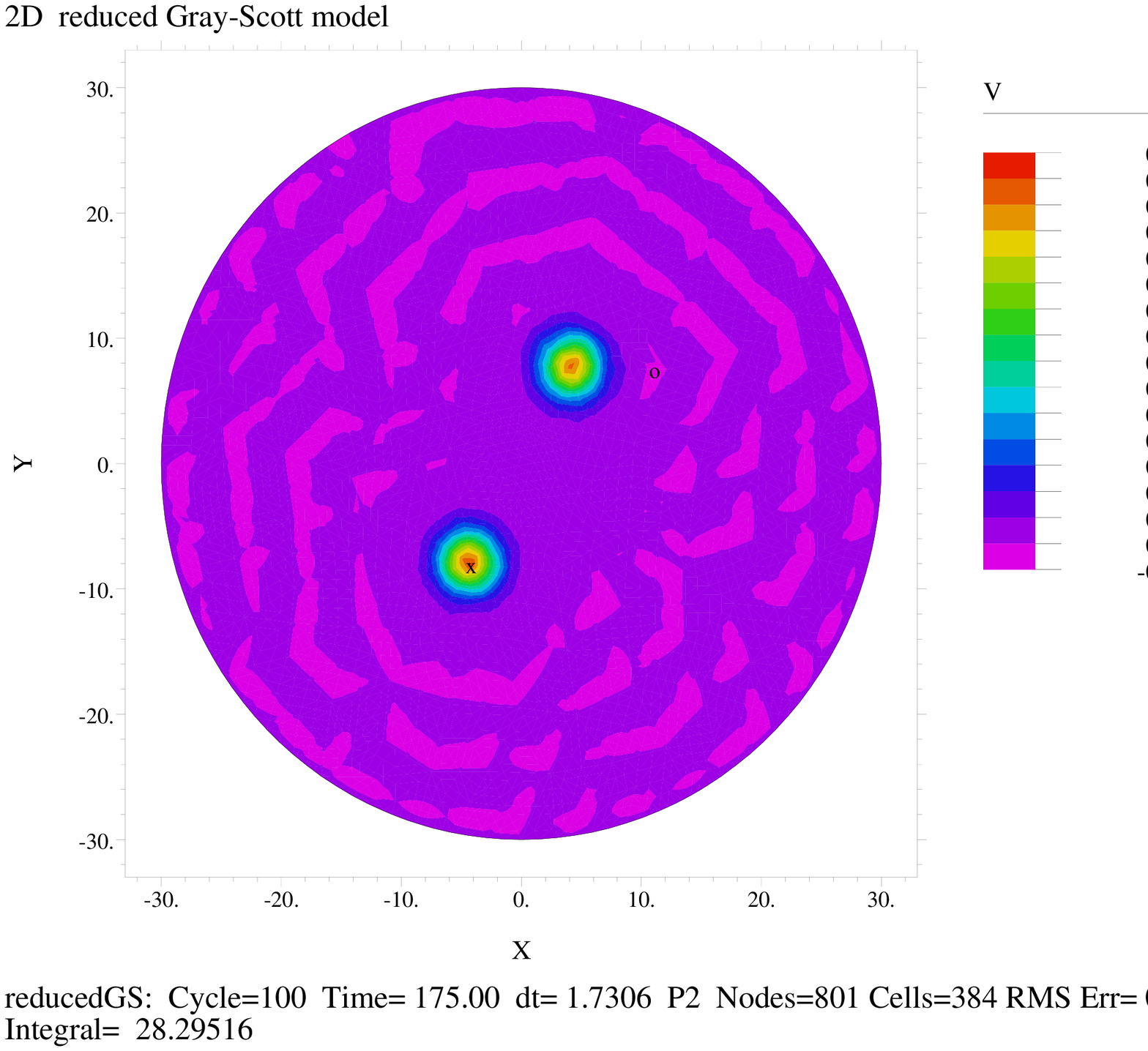}}
\subfigure[$t=300$]
{\includegraphics[width=2.5cm,clip]{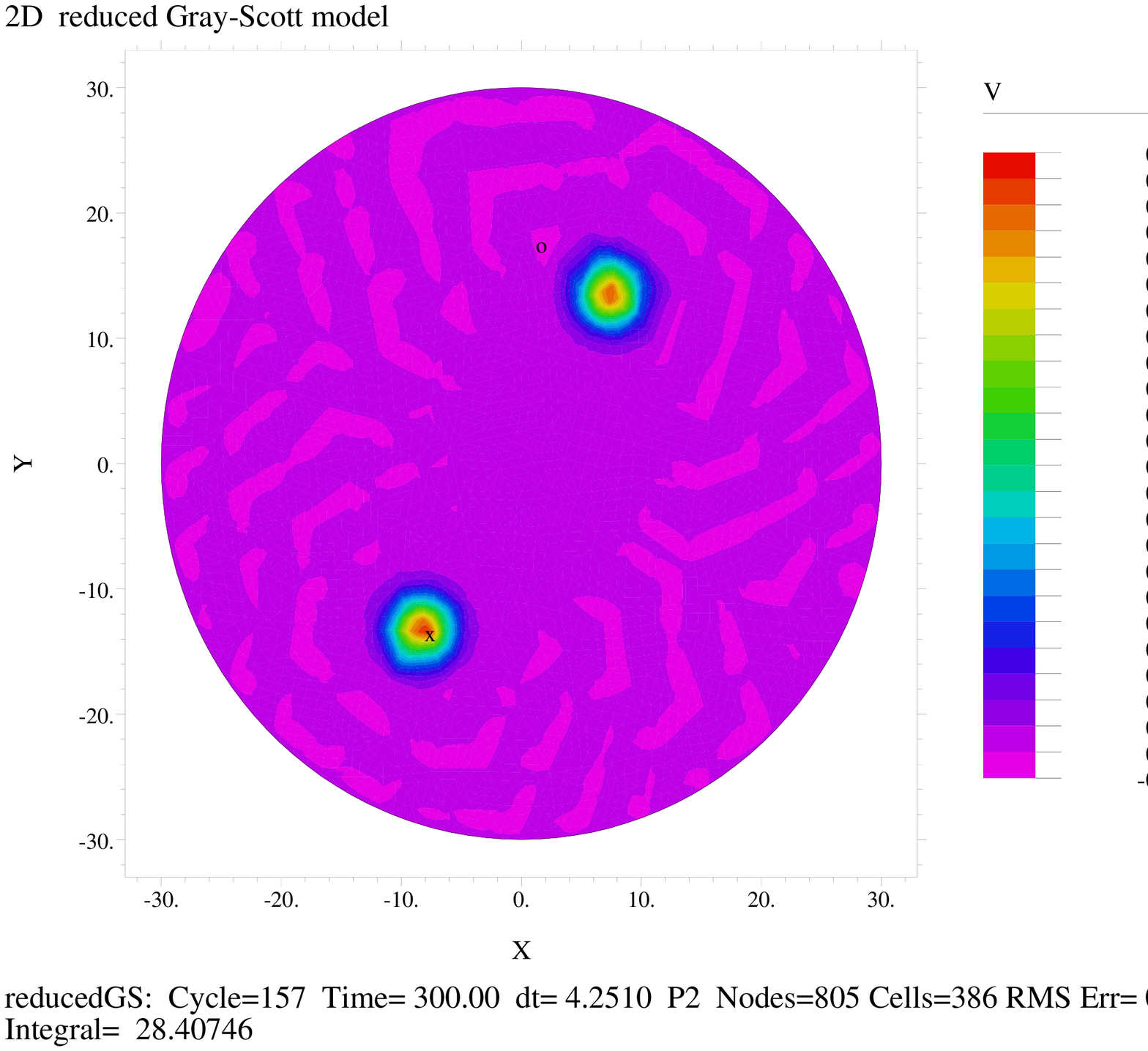}}
\caption{{\em In a circular domain with radius $R_m=30$, we compute
numerical solutions to \eqref{2Dreduced} by \emph{FlexPDE}
(cf.~\cite{flexpde}) using the initial condition in
\eqref{2Dreduced_IC}. The solution $V$ with $S=4.5$ is plotted at
$t=0, 10, 100, 130, 140, 170, 300$.}}
\label{fig:reducedGS_S4d5}
\end{center}
\end{figure}

From our eigenvalue computations, based on \eqref{3:eigenmode}, the
radially symmetric equilibrium solution to \eqref{2Dreduced} for $U$
and $V$ exhibits a peanut-splitting linear instability when $S >
\Sigma_2 \approx 4.31$.  To determine whether this linear instability
leads to a nonlinear spot self-replication event when $S>\Sigma_2$, we
use \emph{FlexPDE} (cf.~\cite{flexpde}) to compute solutions to
\eqref{2Dreduced} in a large disk of radius $|\mathbf{y}|=R_{m}\equiv
30$, and with initial data
\begin{equation}
V(\mathbf{y},0) = \frac{3}{2} \mbox{sech}^2(|\mathbf{y}|/2)\,, \quad 
 U(\mathbf{y},0) = 
 1 - \frac{\,\cosh\left( R_{m} -|\mathbf{y}| \right)}{\cosh R_{m}} \,.
  \label{2Dreduced_IC}
\end{equation}
The asymptotic boundary condition $\partial_{|\mathbf{y}|} U =
{S/|\mathbf{y}|}$ is imposed at $|\mathbf{y}|=R_{m}=30$. We set
$S=4.5>\Sigma_2$, and in Fig.~\ref{fig:reducedGS_S4d5} we plot the
solution $V$ for $t < 300$ showing a nonlinear spot self-replication
event. Owing to the rotational symmetry of this problem, the direction
of spot-splitting observed in Fig.~\ref{fig:reducedGS_S4d5} is likely
due to small numerical errors or grid effects. In contrast, if we
choose $S=4.1$ and the same initial condition, then there is no spot
self-replication for \eqref{2Dreduced} (not shown).  Therefore, this
numerical evidence supports the conjecture that the peanut-splitting
instability associated with the $m=2$ mode initiates a nonlinear spot
self-replication event when $S>\Sigma_2$.

Next, we numerically investigate spot-splitting for values of $S_j$
that well-exceed the threshold $\Sigma_2\approx 4.31$ for the
peanut-splitting instability. From Fig.~\ref{fig:eigm2:a}, the
threshold values of $S_j$ for higher splittings are $\Sigma_3 \approx
5.44$ and $\Sigma_4 \approx 6.14$, corresponding to $m=3$ and $m=4$,
respectively. From Fig.~\ref{fig:eigm2:a} we observe that the growth
rates associated with these further unstable modes are comparable to
that for the mode $m=2$ when $S_j\approx 6.5$.

\begin{figure}[htbp]
\centering
\subfigure{\includegraphics[width=6in]{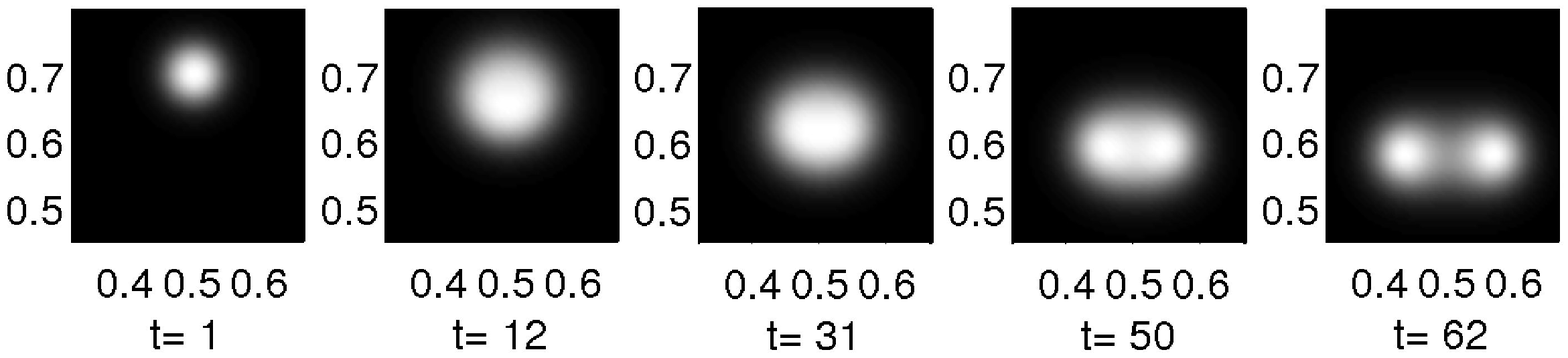}}
\subfigure{\includegraphics[width=6in]{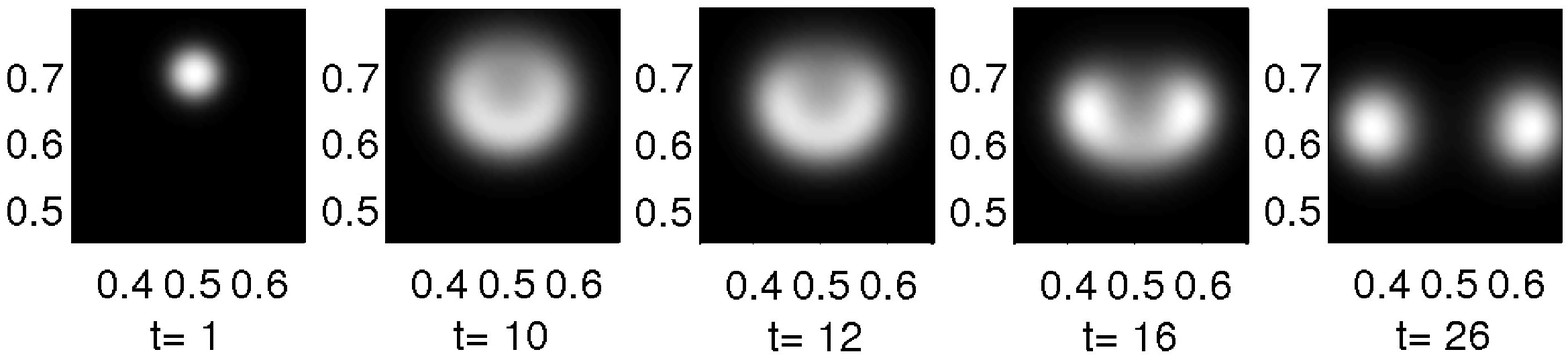}\vspace*{-0.4cm}}
\subfigure{\includegraphics[width=6in]{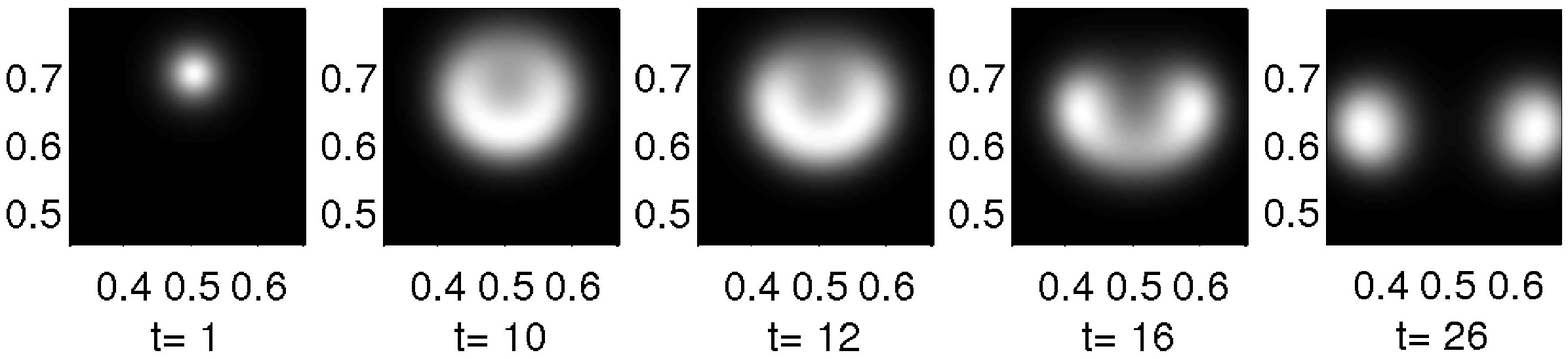}\vspace*{-0.4cm}}
\caption{\em One-spot pattern in the unit square $[0, 1] \times [0,
1]$.  Let $D=1.0$, $\epsilon=0.02$ and $\mathbf{x}_1 = (0.5, 0.7)$. We
set $\mathcal{A} = 9.9213$ (top row), $ \mathcal{A} = 12.643$
(middle row), and $\mathcal{A} = 13.285$ (bottom row), so
that $S_1 \approx 4.5>\Sigma_2$, $S_1 \approx 6.0>\Sigma_3$ and $S_1
\approx 6.4>\Sigma_4$, respectively. The numerical solutions for $v$
from \eqref{1:GS_2D} are computed using VLUGR
(cf.~\cite{vlugr_Blom:1966}), and are plotted at different instants in
time in the zoomed spatial region $[0.325, 0.675] \times [0.45, 0.8]$.}
\label{fig:1spt_smallS_zoom}
\end{figure}

For illustration, we consider a one-spot pattern with $D=1$ and
$\eps=0.02$ in the unit square $[0, 1] \times [0, 1]$ with a spot
centered at $\mathbf{x}_1 = (0.5, 0.7)$.  From \eqref{3:AS1spot}, we
then calculate $\mathcal{A}$ from $\mathcal{A} = S_1 + 2 \pi \nu
R_{1,1}\, S_1 + \nu \chi(S_1)$, where $R_{1,1}$ is the regular part of
the reduced-wave Green's function that can be calculated from
\eqref{3:squaregreen} of Appendix A.  We numerically study the
spot-splitting process by using VLUGR (cf.~\cite{vlugr_Blom:1966}) to
compute solutions to \eqref{1:GS_2D} for $\mathcal{A}=9.9213$,
$\mathcal{A} = 12.643$, and $\mathcal{A} = 13.285$, corresponding to
$S_1=4.5>\Sigma_2$, $S_1 = 6.0>\Sigma_3$, and $S_1 = 6.4>\Sigma_4$,
respectively.  The results are shown in
Fig.~\ref{fig:1spt_smallS_zoom} in the sub-region $[0.325, 0.675]
\times [0.45, 0.8]$. These computations show that spot
self-replication, leading to the creation of two distinct spots, is a
robust phenomena for \eqref{1:GS_2D} whenever $S_1>\Sigma_2$. Although
for $S_1=6.0$ and $S_1=6.4$ the initial instability leads to a
crescent pattern for the volcano profile for $V$, eventually two spots
are created from this instability. Therefore, these numerical results
support the conjecture that the unstable mode $m=2$ dominates any of
the other unstable modes with $m>2$ in the weakly nonlinear regime,
and eventually leads to the creation of two spots from a single spot
when $S_1>\Sigma_2$.

\subsection{Radially Symmetric Local Perturbations: Competition and Oscillatory Instabilities}\label{sec:eig_rad}

In \S \ref{sec:eig_nrad} the stability of the spot profile
to locally non-radially symmetric perturbations was studied
numerically. In this subsection, we examine the stability of the spot
profile to locally radially symmetric perturbations of the form
$N_j=N_j(\rho)$ and $\Phi_j=\Phi_{j}(\rho)$, with $\rho=|\mathbf{y}|$,
which characterize instabilities in the amplitudes of the spots. With
the assumption that $\tau\lam \ll {\mathbf O}(\eps^{-2}D)$,
(\ref{3:innereigen}) reduces to the following radially symmetric
eigenvalue problem on $0<\rho<\infty$: 
\bsub \label{4:innereig}
\begin{gather}
 \Phi_j^{\p\p} + \frac{\,1}{\,\rho} \Phi_j^{\p} - \Phi_j
 + 2 U_j V_j \Phi_j + V_j^2 N_j = \lam \Phi_j \,,\qquad
 N_j^{\p\p} + \frac{\,1}{\,\rho} N_j^{\p}  -  V_j^2 N_j  - 2 U_j V_j \Phi_j 
 = 0 \,, \label{4:innereig:ab} \\
\label{4:eigenbdc} \Phi_j^{\p}(0)=N_j^{\p}(0) =0\,;\qquad \Phi_j(\rho)
 \to 0\,, \quad  N_j(\rho) \to  C_j \ln \rho + B_j + o(1) \,, \;\;\; \mbox{as}
\;\;\;\rho \to \infty.
\end{gather}
\esub Here $U_j$ and $V_j$ is the radially symmetric solution of the
core problem \eqref{3:2Dcore_sol} for the $j^{\mbox{th}}$ spot.  From
the divergence theorem, the constant $C_j$ in \eqref{4:eigenbdc} is
given by $C_j \equiv \int_0^{\infty} (2 U_j V_j \Phi_j + V_j^2
N_j)\,\rho \,d\rho$.  We emphasize that the operator in
(\ref{4:innereig:ab}) for $N_j$ reduces to $N_{j}^{\p\p}+ \rho^{-1}
N_{j}^{\p}\approx 0$ for $\rho\gg 1$, and so we cannot impose that
$N_j\to 0$ as $\rho\to \infty$. Instead, we must allow for the
possibility of a logarithmic growth at infinity for $N_j$, as written
in (\ref{4:eigenbdc}).  This growth condition, which will lead to a
global coupling of the $k$ local eigenvalue problems, is in contrast
to the decay condition as $\rho\to\infty$ used in \S
\ref{sec:eig_nrad} for the stability analysis with respect to locally
non-radially symmetric perturbations.

To formulate our eigenvalue problem we must match the far-field
logarithmic growth of $N_j$ with a global outer solution for
$\eta$. This matching globally couples the local eigenvalue problems
near each spot. To determine the problem for the outer solution for
$\eta$, we use the fact that $v$ is localized near $\mathbf{x}_j$ for
$j=1,\ldots,k$. Then, from (\ref{3:2dinnvar}) and
(\ref{3:inn_stab_var}), we represent the last two terms in the $\eta$
equation of \eqref{3:eigen} in the sense of distributions to obtain
that $2 u v \phi + v^2 \eta \sim 2 \pi \eps \sqrt{D} A^{-1}
\sum_{j=1}^k C_j \, \delta(\mathbf{x} - \mathbf{x}_j)$, where $C_j
\equiv \int_0^{\infty} (2 U_j V_j \Phi_j + V_j^2 N_j) \,\rho \,
d\rho$ and $\delta(\mathbf{x}-\mathbf{x_j})$ is the Dirac delta
function.  Therefore, in the outer region, we obtain from
\eqref{3:eigen} that $\eta$ satisfies
\begin{equation}
  \Delta \eta - \frac{\, \left(1+\tau\lambda\right)}{D} \eta = 
  \frac{\,2\pi \eps }{\,A \sqrt{D}}\,\sum_{j=1}^k C_j 
 \delta(\mathbf{x} - \mathbf{x}_j)  \,, \quad \mathbf{x}\in \Omega \,; 
 \qquad \partial_n \eta=0 \,, \quad \mathbf{x}\in \partial\Omega \,.
\end{equation}
This outer solution can be represented in terms of a $\lambda$-dependent
Green's function as
\begin{equation}
\label{4:outsol} \eta =  -\frac{\, 2 \pi\eps}{\,A
\sqrt{D}}\,\sum_{j=1}^k  C_j G_{\lambda}(\mathbf{x}; \mathbf{x}_j) \,,
\end{equation}
where $G_{\lambda}(\mathbf{x};\mathbf{x}_j)$ satisfies
\bsub \label{4:Greenlamall}
\begin{gather}
 \Delta G_{\lambda} -\frac{\left(1+\tau\lambda\right)}{D}\,
G_{\lambda} = - \delta(\mathbf{x}-\mathbf{x}_j) \,, \quad \mathbf{x} \in
\Omega \,; \qquad \partial_n G_{\lambda} = 0 \,, \quad
 \mathbf{x}\in \partial\Omega \,, \label{4:Greenlam}\\
 G_{\lambda}(\mathbf{x}; \mathbf{x}_j) \sim - \frac{\,1}{\,2 \pi}
\ln| \mathbf{x} - \mathbf{x}_j| + R_{\lambda\,j,j} + o(1) 
\quad \mbox{as}\;\;\mathbf{x}  \to  \mathbf{x}_j \,. \label{4:Greenlamsing}
\end{gather}
\esub We remark that the regular part $R_{\lam j,j}$ of $G_{\lam}$
depends on $\mathbf{x}_j$, $D$, and $\tau \lambda$.

The matching condition between the outer solution \eqref{4:outsol} for
$\eta$ as $\mathbf{x} \to \mathbf{x}_j$ and the far-field behavior
\eqref{4:eigenbdc} as $\rho \to \infty$ of the inner solution $N_j$
near the $j^{\mbox{th}}$ spot, defined in terms of $\eta$ by
(\ref{3:inn_stab_var}), yields that
\begin{equation}
\label{4:matching} -\frac{\, 2 \pi\eps }{\,A\sqrt{D}} \, \left[
  C_j \left( -\frac{\, 1}{2\pi} \ln|\mathbf{x}-\mathbf{x}_j| +
  R_{\lam\, j,j} \right) \right] - \frac{\, 2 \pi\eps }{\,A\sqrt{D}} 
\sum_{i\neq j}^{k} C_i G_{\lam \, i, j} 
   \sim \frac{\, \eps}{A\sqrt{D}} \left[ C_j \ln|\mathbf{x}-\mathbf{x}_j|
  + \frac{C_j}{\nu} + B_j \right] \,,
\end{equation}
where $G_{\lam \, i,j}\equiv G_{\lam}(\mathbf{x}_i;\mathbf{x_j})$ and
$\nu={-1/\ln\eps}$. This matching condition provides the $k$ equations
\begin{equation}
\label{4:CBorig} C_j \left(1+ 2 \pi \nu R_{\lambda\, j,j} \right) + \nu B_j +
 2\pi \nu \sum_{i\neq j}^{k} C_i G_{\lambda \, i,j} = 0 \,, \quad
   j=1, \ldots, k \,.
\end{equation}
We remark that the constants $\tau$ and $D$ appear in the operator of
the $\lam$-dependent Green's function defined by \eqref{4:Greenlamall}.

\begin{figure}[htbp]
\begin{center}
\subfigure[$\mbox{Re}(\lambda_0)$ vs.~$S_j$] 
{\label{fig:eigC0:a} \includegraphics[height=1.8in, width=2.2in]{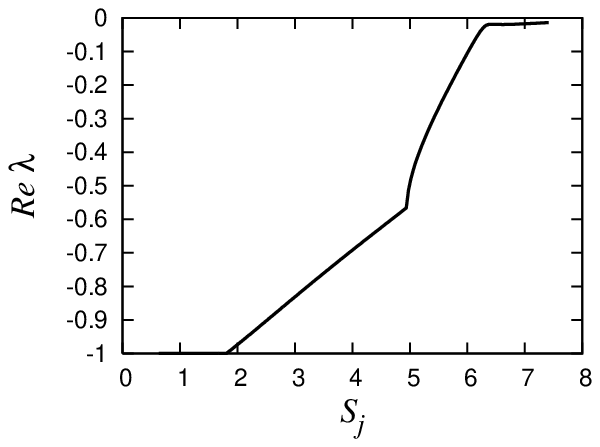}}
\subfigure[$\Phi_j$ vs.~$\rho$] 
{\label{fig:eigC0:b} \includegraphics[height=1.8in, width=2.2in]{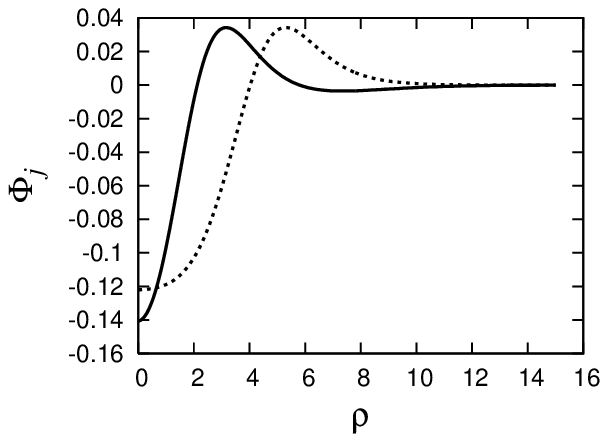}}
\subfigure[$N_j$ vs.~$\rho$] 
{\label{fig:eigC0:c} \includegraphics[height=1.8in, width=2.2in]{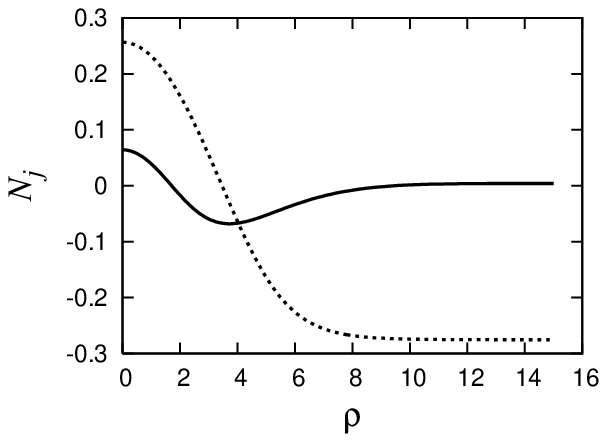}} 
\caption{{\em Left figure: The real part of the largest eigenvalue
$\mbox{Re}(\lambda_0)$ of \eqref{4:innereig} vs.~$S_j$ subject to the
condition that $N_j$ is bounded as $\rho\to\infty$. Middle figure:
$\Phi_j$ vs.~$\rho$ for $S_j=3.0$ (solid curve) and $S_j=6.0$ (dashed
curve). Right figure: $N_j$ vs.~$\rho$ for $S_j=3.0$ (solid curve)
and $S_j=6.0$ (dashed curve). }}
\label{fig:eigC0:bc}
\end{center}
\end{figure}

For $D=\mathcal{O}(1)$ and $\tau=\mathcal{O}(1)$, we consider the
leading-order theory in $\nu$ based on assuming that $\nu \equiv
{-1/\ln\eps}\ll 1$. Then, to leading-order in $\nu$, \eqref{4:CBorig}
yields that $C_j = 0$ for $j=1, \cdots, k$. This implies that $N_j$ in
\eqref{4:innereig} is bounded as $\rho\to \infty$, and so we can
impose that $N_{j}^{\p}(\rho) \to 0$ as $\rho \to \infty$. Therefore,
when $\tau=\mathcal{O}(1)$ and $D=\mathcal{O}(1)$, then, to leading
order in $\nu$, the eigenvalue problems \eqref{4:innereig} for
$j=1,\ldots,k$ are coupled together only through the determination of
the source strengths $S_1,\ldots,S_k$ from the nonlinear algebraic
system \eqref{3:ASsmallD}. For this leading-order theory, we compute
the real part of the principal eigenvalue $\lam_0$ of
\eqref{4:innereig} as a function of $S_j$, subject to the condition
that $N_{j}^{\p}(\rho) \to 0$ as $\rho\to\infty$. This computation is
done by discretizing \eqref{4:innereig} by finite differences and then
calculating the spectrum of the resulting matrix eigenvalue problem
using LAPACK (cf.~\cite{lapack_Anderson:1999}). The plot of
$\mbox{Re}(\lam_0)$ versus $S_j$ is shown in Fig.~\ref{fig:eigC0:a}
for the range $S_j<7.5$, which includes the value $S_j=\Sigma_2\approx
4.31$ corresponding to the spot self-replication threshold of \S
\ref{sec:eig_nrad}.  In Fig.~\ref{fig:eigC0:b} and
Fig.~\ref{fig:eigC0:c}, respectively, we plot the eigenfunctions
$\Phi_j(\rho)$ and $N_j(\rho)$ for two different values of $S_j$.  For
this leading-order-in-$\nu$ {\em local} eigenvalue problem, our
computations show that $\mbox{Re}(\lam_0)<0$ for $S_j<7.5$. Therefore,
we conclude that an instability can only be generated through the {\em
global} coupling of the local eigenvalue problems. This coupling
occurs when we do not make the $\nu\ll 1$ approximation in
\eqref{4:CBorig}.

Since $\nu={-1/\ln\eps}$ decreases only very slowly as $\eps$
decreases, the leading-order approximation $C_j=0$ for $j=1,\ldots,k$
to \eqref{4:CBorig}, which does not lead to any instabilities, is not
expected to be very accurate. Consequently, we must examine the effect
of the coupling in \eqref{4:CBorig}.  Since \eqref{4:innereig} is a
linear homogeneous problem, we introduce $\hat{B}_j$ by $B_j =
\hat{B}_j C_j$, and we define the new variables $\hat{\Phi}_j$ and
$\hat{N}_j$ by $\Phi_j = C_j \hat{\Phi}_j$ and $N_j = C_j
\hat{N}_j$. Then, \eqref{4:innereig} on $0<\rho<\infty$ becomes
\begin{subequations}
\label{4:innereig1}
\begin{gather}
 \hat{\Phi}_j^{\p\p} + \frac{\,1}{\,\rho} \hat{\Phi}_j^{\p} -\hat{\Phi}_j
 + 2 U_j V_j \hat{\Phi}_j + V_j^2 \hat{N}_j = \lam \hat{\Phi}_j  \,,
 \qquad
 \hat{N}_j^{\p\p} + \frac{\,1}{\,\rho} \hat{N}_j^{\p}  -  V_j^2 \hat{N}_j  - 
2 U_j V_j \hat{\Phi}_j = 0 \,,\label{4:innereig1:ab}\\
\label{4:eigenbdc1} \hat{\Phi}_j^{\p}(0)=\hat{N}_j^{\p}(0) =0 \,; \qquad
  \hat{\Phi}_j(\rho)
\to 0\,, \;\;\; \hat{N}_j(\rho) \to  \ln \rho + \hat{B}_j \,, \quad \mbox{as}
\quad \rho \to \infty \,.
\end{gather}
\end{subequations}
The constant $\hat{B}_j$ in (\ref{4:eigenbdc1}) is a function of the, as yet
unknown, complex-valued eigenvalue parameter $\lambda$. It also depends
on the source strength $S_j$ through the solution $U_j$ and $V_j$ to the
core problem \eqref{3:2Dcore_sol}.

In terms of $\hat{B}_j$, \eqref{4:CBorig} transforms to the
homogeneous linear system for $C_1,\ldots,C_k$ given by
\begin{equation}
\label{4:CB}
 C_j (1+ 2 \pi \nu R_{\lambda\, j,j} + \nu \hat{B}_j) + 2\pi\nu 
 \sum_{i\neq j}^k C_i G_{\lambda \, j, i} = 0 \,, \qquad
  j=1, \ldots, k \,.
\end{equation}
It is convenient to express \eqref{4:CB} in matrix form as
\bsub \label{4:Glammatrix}
\begin{equation}
\mathcal{M}\, \mathbf{c} = \mathbf{0} \,, \qquad
\mathcal{M} \equiv I + \nu \mathcal{B} + 2 \pi \nu
\mathcal{G}_{\lambda} \,, \label{4:CBsys} 
\end{equation}
where $I$ is the $k \times k$ identity matrix and $\mathbf{c} \equiv
(C_1,\ldots,C_k)^T$. In \eqref{4:CBsys}, $\mathcal{B}$ is a
diagonal matrix and $\mathcal{G}_{\lam}$ is the $\lambda$-dependent
symmetric Green's matrix defined by
\begin{equation}
 \mathcal{G}_{\lambda} \equiv \left(
\begin{array}{ccccc} R_{\lambda\, 1, 1} & G_{\lambda\, 1, 2} &\cdots
&G_{\lambda\, 1, k} \\ G_{\lambda\, 2, 1}  & R_{\lambda\, 2, 2}
&\cdots &G_{\lambda\, 2, k} \\ \vdots &\vdots &\vdots &\vdots\\
G_{\lambda\, k, 1}  & G_{\lambda\, k, 2} &\cdots &G_{\lambda\, k, k}
\end{array} \right)\,, \qquad \mathcal{B} \equiv \left( \begin{array}{ccccc}
\hat{B}_1 &0 &\cdots &0 \\
0 &\hat{B}_2, &\cdots &0 \\ \vdots &\vdots &\vdots &\vdots\\ 0 &0
&\cdots & \hat{B}_k \end{array} \right) \,. \label{4:GLmat}
\end{equation}
\esub 
Since $\bar{G}_{\lam ij}\neq G_{\lam j, i}$ when $\lambda$ is
complex-valued, where the overbar denotes complex conjugate,
$\mathcal{G}_\lam$ is not Hermitian.

The eigenvalue $\lambda$ is determined from the condition that
$\mbox{det}(\mathcal{M}) = 0$, so that there is a nontrivial solution
$\mathbf{c}\neq \mathbf{0}$ to \eqref{4:CBsys}. This condition leads,
effectively, to a transcendental equation for $\lambda$. The roots of
this equation determine the discrete eigenvalues governing the linear
stability of the $k$-spot quasi-equilibria to locally radially
symmetric perturbations near each spot. We summarize our formulation
of the global eigenvalue problem as follows:

\vspace*{0.1cm}\noindent{\bf \underline{Principal Result 4.2}:}\; {\em
Consider a $k-$spot quasi-equilibrium solution to the GS model
\eqref{1:GS_2D}. For $\eps \to 0$, with $D={\mathcal O}(1)$,
$A={\mathcal O}(-\eps\ln\eps)$, and $\tau\lam \ll {\mathcal
O}(\eps^{-2})$, the stability of this pattern to locally radially
symmetric perturbations near each spot is determined by the condition
$\mbox{det}(\mathcal{M})=0$, where $\mathcal{M}$ is defined in
\eqref{4:Glammatrix}. The diagonal matrix $\mathcal{B}$ in
\eqref{4:Glammatrix} is determined in terms of $S_j$ and $\lambda$ by
the local problems \eqref{4:innereig1} for $j=1,\ldots,k$.  If the
principal eigenvalue $\lambda_0$ of this global eigenvalue problem is
such that $\mbox{Re}(\lambda_0)<0$, then the $k-$spot
quasi-equilibrium solution is linearly stable to locally radially
symmetric perturbations near each spot, and it is linearly unstable if
$\mbox{Re}(\lambda_0)>0$.}

\vspace*{0.1cm}

We remark that \eqref{4:Glammatrix} couples the local spot solutions
in two distinct ways. First, the $\lambda$-dependent terms
$R_{\lambda\, j,j}$ and $G_{\lambda \, i, j}$ in the Green's matrix
$\mathcal{G}_\lambda$ in \eqref{4:Glammatrix} depend on $D$,
on $\tau \lambda$, and on the spatial configuration
$\mathbf{x}_1,\ldots,\mathbf{x}_k$ of spots, as well as the
geometry of $\Omega$. Secondly, the constant $\hat{B}_j$ in the matrix
$\mathcal{B}$ depends on $\lambda$ and on $S_j$. Recall that the
source strengths $S_1,\ldots,S_k$ are coupled through the nonlinear
algebraic system \eqref{3:ASsmallD}, which involves $\ac$, $D$, the
reduced-wave Green's function, and the spatial configuration
of spot locations.

From our numerical study of this global eigenvalue problem in \S
\ref{sec:12inf}--\ref{sec:sym}, there are two mechanisms through which
a $k$-spot quasi-equilibrium pattern can lose stability.  Firstly, for
$k\geq 1$ there can be a complex conjugate pair of eigenvalues that
crosses into the unstable half-plane $\mbox{Re}(\lam_0)>0$. This
instability as a result of a Hopf bifurcation initiates an oscillatory
profile instability, whereby the spot amplitudes undergo temporal
oscillations. Such an instability typically occurs if $\tau$ is
sufficiently large. Alternatively, for $k \geq 2$, the principal
eigenvalue $\lambda_0$ can be real and enter the unstable right
half-plane $\mbox{Re}(\lam_0)>0$ along the real axis
$\mbox{Im}(\lam_0)=0$. This instability, due to the creation of a
positive real eigenvalue, gives rise to an unstable sign-fluctuating
perturbation of the spot amplitudes and it initiates a competition
instability, leading to spot annihilation events. This instability can
be triggered if $D$ is sufficiently large or, equivalently, if the
inter-spot separation is too small.

The global eigenvalue problem leading to the stability formulation in
Principal Result 4.2 is a new result, and essentially can be viewed as
an extended NLEP theory that accounts for all terms in powers of
$\nu$. In Appendix B, we summarize the main results for the
leading-order-in-$\nu$ NLEP stability theory of \cite{2Dmulti_Wei:2003}
based on the parameter range $D={\mathcal O}(\nu^{-1})$ and
$A={\mathcal O}(\eps[-\ln\eps]^{1/2})$, and we show how our global
eigenvalue problem can be asymptotically reduced to the leading order
NLEP problem of \cite{2Dmulti_Wei:2003} in this parameter regime for
$D$ and $A$.

\subsection{Symmetric Spot Patterns and a Circulant Matrix}\label{sec:eig_circ}

The global eigenvalue problem is rather challenging to investigate in
full generality owing to the complexity of the two different coupling
mechanisms in \eqref{4:Glammatrix}. However, for the special case
where the spot configuration $\mathbf{x}_1,\ldots,\mathbf{x}_k$ is
such that $\mathcal{G}$, and consequently $\mathcal{G}_\lam$, are
circulant matrices, then the complexity of this eigenvalue problem
reduces considerably.  For this special arrangement of spot locations,
the spots have a common source strength $S_c=S_j$ for $j=1,\ldots,k$,
where $S_c$ satisfies the nonlinear algebraic equation
\eqref{3:circscalar}.  Hence, the inner problem \eqref{4:innereig1} is
the same for each spot, which enforces that $\hat{B}_j \equiv
\hat{B}_c$ for $j=1,\ldots,k$, where
$\hat{B}_c=\hat{B}_{c}(\lam,S_c)$. Therefore, we can write
$\mathcal{B} = \hat{B}_c I$ in \eqref{4:Glammatrix}. Moreover, let
$\mathbf{v}$ be an eigenvector of the $\lambda$-dependent Green's
matrix $\mathcal{G}_\lam$ with eigenvalue
$\omega_\lam=\omega_\lam(\tau\lam)$, i.e.~$\mathcal{G}_\lam \mathbf{v}
= \omega_\lam \mathbf{v}$. We note that the matrix $\mathcal{G}_\lam$
is also circulant when $\mathcal{G}$ is circulant. Then, the condition
that $\mathcal{M}$ in \eqref{4:Glammatrix} is a singular matrix
reduces to $k$ transcendental equations in $\lambda$. We summarize the
result as follows:

\vspace*{0.1cm}\noindent{\bf \underline{Principal Result 4.3:}} {\em 
Under the conditions of Principal Result 4.2, suppose that the spot
configuration $\mathbf{x}_1,\ldots,\mathbf{x}_k$ is such that 
$\mathcal{G}$, and consequently $\mathcal{G}_\lam$, are circulant
matrices. Then, the eigenvalue condition $\mbox{det}(\mathcal{M})=0$ for
\eqref{4:Glammatrix} reduces to the $k$ transcendental equations for
$\lambda$ given by}
\begin{equation}
\label{4:CBreduced} f_j \equiv 1 + \nu \hat{B}_c + 2 \pi \nu 
 \omega_{\lambda j}(\tau\lam)  = 0 \,,
\end{equation}
{\em where $\omega_{\lam j}(\tau\lam)$ for $j=1,\ldots,k$ is any
eigenvalue of the matrix $\mathcal{G}_\lam$. The $k$-distinct
eigenvectors $\mathbf{v}$ of ${\mathcal G}_\lam$ determine the choices
for $\mathbf{c}=(C_1,\ldots,C_k)^T$. By equating real and imaginary
parts, (\ref{4:CBreduced}) can be reduced to}
\begin{equation}
\mbox{Re} (f_j) \equiv 1/\nu + \mbox{Re} (\hat{B}_c) + 2 \pi \mbox{Re} (
\omega_{\lam j} (\tau \lambda)) = 0 \,,\qquad
\mbox{Im} (f_j) \equiv \mbox{Im}
(\hat{B}_c) + 2 \pi \mbox{Im} ( \omega_{\lam j} (\tau \lambda)) = 0 \,.
 \label{4:CBreduced_1} 
\end{equation}
\vspace*{0.1cm}

When $\mathcal{G}_\lam$ is a circulant and symmetric matrix, then its
spectrum, as needed in Principal Result 4.3, can be determined
analytically. To do so, let the row vector $\mathbf{a} = (a_1, \ldots,
a_k)$ denote the first row of $\mathcal{G}$.  Since $\mathcal{G}_\lam$
is circulant it follows that all of the other rows of $\mathcal{G}$
can be obtained by cycling the components of the vector
$\mathbf{a}$. In addition, since $\mathcal{G}_\lam$ is also
necessarily a symmetric matrix it follows that $a_2 = a_k$, $a_3 =
a_{k-1}$, $\ldots$, and $a_j = a_{k+2-j}$ for $j=2,\ldots, \lceil
{k/2} \rceil$, where the ceiling function $\lceil x \rceil$ is the
smallest integer not less than $x$.  We recall that if a $k \times k$
matrix is circulant, its eigenvectors $\mathbf{v}_j$ and eigenvalues
$\omega_{\lam\, j}$, which consist of the $k^{\mbox{th}}$ roots of
unity, are given by
\begin{equation}
\omega_{\lam j}= \sum_{m=0}^{k-1} a_{m+1} e^{2 \pi (j-1)m/k} \,,
 \qquad \mathbf{v}_j = (1, e^{2 \pi (j-1)/k} \,, \ldots, e^{2
 \pi(j-1)(k-1)/k})^T \,, \qquad j=1,\ldots, k\,.
\end{equation} 
Here $a_m$ is the $m^{\mbox{th}}$ component of the row vector
$\mathbf{a}$.

For illustration, let $k=3$. Then, the eigenpairs are $\mathbf{v}_1 =
(1,1,1)^{T}$ with $\omega_{\lam 1} = a_1 + a_2 + a_3$, $\mathbf{v}_2 = (1,
e^{2 \pi i /3}, e^{4 \pi i/3})^{T}$ with $\omega_{\lam 2} = a_1 + a_2 e^{2
\pi i /3} + a_3 e^{4 \pi i /3}$, and $\mathbf{v}_3 = (1, e^{4 \pi
i/3}, e^{2 \pi i /3})^{T}$ with $\omega_{\lam 3} = a_1 + a_2 e^{4 \pi i
/3} + a_3 e^{2 \pi i /3}$. Then, since the matrix is also symmetric,
we have $a_2 = a_3$, so that $\omega_{\lam 2} = \omega_{\lam 3} = a_1
- a_2$, which yields one eigenvalue of multiplicity two. Then, since any
linear combination of $\mathbf{v}_2$ and $\mathbf{v}_3$ is also an
eigenvector, we take the real part of $\mathbf{v}_2$ as one such
vector, and the imaginary part of $\mathbf{v}_2$ as the other. In
summary, we can take $\mathbf{v}_1 = (1, 1, 1)^T$, $\mathbf{v}_2 = (1,
-0.5, -0.5)^T$ and $\mathbf{v}_3 = (0, \sqrt{3}/2, -\sqrt{3}/2)^T$ as
the three eigenvectors of $\mathcal{G}_\lam$ for the three-spot
pattern.

In general, the symmetry of $\mathcal{G}_\lam$ implies that $a_j =
a_{k+2-j}$ and $\mathbf{v}_j = \mathbf{v}_{k+2-j}$, and $v_{jm} = v_{j
(k+2-m)}$ for $m = 2, \ldots, \lceil {k/2} \rceil$ and $j=2,\ldots,
\lceil {k/2} \rceil$. Here $v_{jm}$ denotes the $m^{\mbox{th}}$
component of the column vector $\mathbf{v}_j$.  Since the eigenvalue
is $\omega_{\lam j} = \mathbf{v}_j \cdot \mathbf{a} $, we have
$\omega_{\lam j} = \omega_{\lam (k+2-j)}$.  This generates $\lceil
{k/2} \rceil-1$ eigenvalues with multiplicity two, whose eigenvectors
can be obtained by taking any linear combination of two complex
conjugate eigenvectors. For instance, for the $j^{\mbox{th}}$
eigenvalue $\omega_{\lam j}$ the corresponding two eigenvectors can be
taken as the real and imaginary parts of $\mathbf{v}_j$,
respectively. In summary, we know that all of the eigenvectors can be
chosen to be real, but that the eigenvalues in general will be complex when
$\lambda$ is complex. This leads to the following result for
the spectrum of the $k\times k$ symmetric and circulant Green's matrix 
${\mathcal G}_\lam$ whose first row vector is $\mathbf{a}=(a_1,\ldots,a_m)$:
\begin{equation}
\label{4:kappa}
\begin{cases}
\omega_{\lam 1} &= \sum_{m=1}^k a_m,\quad v^{T}_1 = (1, \ldots, 1), \\
\omega_{\lam j} &= \sum_{m=0}^{k-1} \cos\left( \frac{2 \pi
(j-1)\,m}{k}\right)
a_{m+1}, \quad \mbox{eigenvalues with multiplicity 2},\\
 \mathbf{v}^T_j &= \left(1, \cos \left(\frac{2 \pi (j-1)}{k}\right), \ldots,
 \cos \left( \frac{2 \pi (j-1)(k-1)}{k} \right)\,\right) \,,\\
 \mathbf{v}^T_{k+2-j} &= \left(0, \sin \left( \frac{2 \pi (j-1)}{k}\right),
 \ldots, \sin \left( \frac{2 \pi (j-1)(k-1)}{k}\right)\,\right) \,,\quad
 j=2,\ldots, \lceil {k/2} \rceil +1 \,.
\end{cases}
\end{equation}
Note that if $k$ is even, then $\omega_{\lam ( \lceil {k/2}\rceil +1)}
= \sum_{m=1}^k (-1)^{m-1} a_m$ is a simple eigenvalue with eigenvector
$(1, -1, \cdots,1, -1)^T$.

For a symmetric spot pattern, we can simply substitute \eqref{4:kappa}
into \eqref{4:CBreduced_1} of Principal Result 4.3 to derive the 
transcendental equation associated with the $j^{\mbox{th}}$ eigenvector
$\mathbf{v}_j$ of $\mathcal{G}_\lam$. This yields,
\begin{equation}
 \nu^{-1}+ \mbox{Re}(\hat{B}_c) + 2 \pi \
\sum_{m=1}^k \mbox{Re}(a_m) v_{jm}   = 0 \,,  \qquad
  \mbox{Im}(\hat{B}_c) + 2 \pi \sum_{m=1}^k 
\mbox{Im}(a_m) v_{jm}  = 0 \,, \quad j=1, \ldots , \lceil {k/2} \rceil +1 \,.
 \label{4:eig}
\end{equation}
Here the row vector $\mathbf{a}$ has components $a_1 = R_{\lam
\, 1,1}$, and $a_m = G_{\lam\, 1,m}$ for $m=2, \ldots, k$, where
$G_{\lam\, 1,m}$ and $R_{\lam \, 1,1}$ are determined from the
$\lambda$-dependent Green's function and its regular part satisfying
\eqref{4:Greenlamall}. In (\ref{4:eig}), $v_{jm}$ denotes the $m^{\mbox{th}}$
component of the column vector $\mathbf{v}_j$. We remark that when $\lambda$
is real, (\ref{4:eig}) can be reduced to only one equation.

At the onset of a competition instability, where an unstable
eigenvalue first enters the unstable right-half plane
$\mbox{Re}(\lam)>0$ along the real axis $\mbox{Im}(\lam)=0$, the
result (\ref{4:eig}) can be simplified further.  Upon differentiating
the problem for $U_j$ and $V_j$ in (\ref{3:2Dcore_sol}) with respect
to $S_j$, and then comparing the resulting system with the eigenvalue
problem in (\ref{4:innereig1}), it readily follows that
\begin{equation}
   \hat{\Phi}_j = \frac{\partial}{\partial S_j} V_j \,, \qquad
   \hat{N}_j = \frac{\partial}{\partial S_j} U_j \,, \qquad \mbox{when}
   \quad \lam=0 \,. \label{3:newphi}
\end{equation}
Therefore, for the special case where $S_j=S_c$ for $j=1,\ldots,k$,
the constant $\hat{B}_c$ in (\ref{4:eig}) when $\lam=0$ can be
calculated in terms of the derivative of $\chi$ by
\begin{equation}
     \hat{B}_c = \chi^{\p}(S_c) \,, \quad \mbox{when} \quad \lam = 0 \,.
  \label{3:newbc}
\end{equation}
A plot of $\chi^{\p}(S_c)$ is shown below in Fig.~\ref{fig:core:c}.
Thus, for a configuration of spots for which the Green's matrix is
circulant symmetric, the threshold condition for a competition
instability, corresponding to setting $\lam=0$ in \eqref{4:eig} and
recalling \eqref{3:circscalar}, is to solve the coupled system
\begin{equation}
    \chi^{\p}(S_c) + 2\pi \omega_{\lam j} = -\nu^{-1} \,, \qquad
   {\mathcal A} = S_c \left( 1 + 2\pi \nu \theta \right) + \nu \chi(S_c)\,;
   \qquad {\mathcal A} = \eps^{-1} \nu  A \sqrt{D} \,, \qquad
   \nu = \frac{-1}{\ln\eps} \,. \label{3:newres}
\end{equation}
Since ${\mathcal G}$ and ${\mathcal G}_\lam$ coincide at
$\lam=0$, then $\theta$ in (\ref{3:newres}) is the eigenvalue of
${\mathcal G}$ with eigenvector $\mathbf{e}=(1,\ldots,1)^T$,
i.e. ${\mathcal G}\mathbf{e}=\theta \mathbf{e}$, and $\omega_{\lam j}$
for $j=1,\ldots,\lceil {k/2}\rceil + 1$ is any of the other
eigenvalues of $\mathcal{G}$ as given in \eqref{4:kappa} when
$\lam=0$. We remark that when $k$ is even, our computational results
in \S \ref{sec:12inf} -- \ref{sec:sym} below will show that the most
unstable mode for a competition instability is the sign-fluctuating
mode $(1, -1, \cdots,1, -1)^T$, which corresponds to setting $j=\lceil
{k/2} \rceil + 1$ in (\ref{3:newres}) and using $\omega_{\lam ( \lceil
{k/2}\rceil +1)} = \sum_{m=1}^k (-1)^{m-1} a_m$.  Here $a_1=R_{1,1}$,
and $a_m=G_{1, m}$ for $m=2,\ldots,k$, is the first row of ${\mathcal G}$.

Owing to the considerable reduction in complexity of the global
eigenvalue problem when ${\mathcal G}$ and ${\mathcal G}_\lam$ are
circulant matrices, in \S \ref{sec:12inf}--\ref{sec:sym} we will only
compute competition and oscillatory stabilities thresholds of
quasi-equilibrium spot patterns that lead to this special matrix
structure. The numerical approach used to compute these thresholds
associated with the stability formulation of Principal Result 4.3 is
outlined in Appendix C.

\begin{figure}[htbp]
\subfigure[$\chi^{\p}$ vs.~$S$] { \label{fig:core:c}
\includegraphics[width=3.0in, height=1.8in]{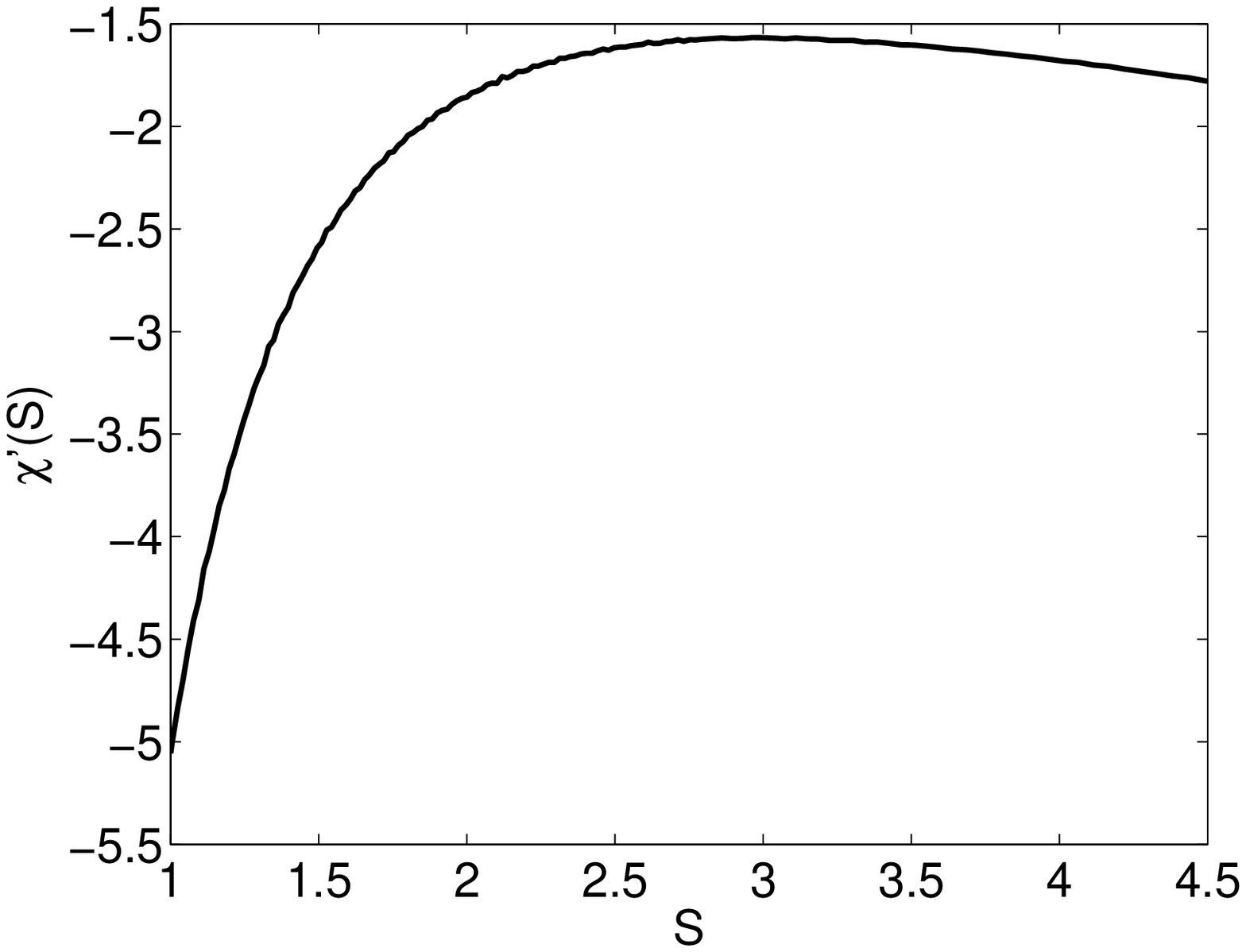}}
\subfigure[$\gamma$ vs.~$S$]{ \label{fig:AS_inf:f}
\includegraphics[width=3.0in,height=1.85in,clip]{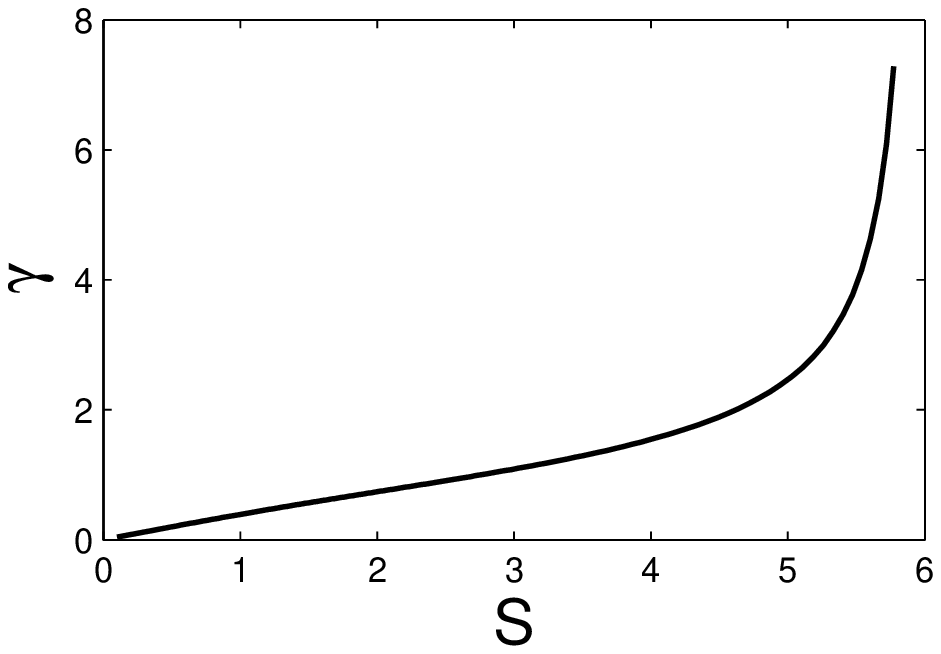}}
\caption{{\em Left figure: Plot of numerically computed
$\chi^{\p}(S)$, as defined in \eqref{3:2Dcore_sol}. Right figure: Plot
of the numerically computed $\gamma(S)$ as defined in \eqref{3:dyn} of
Principal Result 3.1.}}
\end{figure}

Finally, we remark that the numerical implementation of our asymptotic
theory in \S \ref{sec:dyn} and \S \ref{sec:eig} for spot dynamics and
stability relies on the computation of the three functions $\chi(S)$,
$\chi^{\p}(S)$, and $\gamma(S)$, defined in terms of the core problem
by \eqref{3:2Dcore_sol} and \eqref{3:dyn}. The result for $\chi(S)$
was given previously in Fig.~\ref{fig:core:b}, while $\chi^{\p}(S)$
and $\gamma(S)$ are plotted in Fig.~\ref{fig:core:c} and
Fig.~\ref{fig:AS_inf:f}, respectively.

\setcounter{equation}{0}
\setcounter{section}{4}
\section{One- and Two-Spot Patterns on the Infinite Plane and in Large Domains} \label{sec:12inf}

In this section we apply the asymptotic theory of \S \ref{sec:quasi}
-- \ref{sec:eig} to the special case of either a one- or a two-spot
pattern on the infinite plane $\mathbb{R}^2$. For this problem we can
set $D=1$ in (\ref{1:GS_2D}) without loss of generality. When
$\Omega=\mathbb{R}^2$, the reduced-wave Green's function and its
regular part satisfying \eqref{3:Green} with $D=1$ is simply
\begin{equation}
  G(\mathbf{x}; \mathbf{x}_j) = \frac{\,1}{\,2 \pi} \, K_0 
 \left( \left|\mathbf{x}-\mathbf{x}_j \right|\right) \,, \qquad
  R_{jj} = \frac{1}{\;2 \pi} (\ln 2 - \gamma_e) \,. \label{4:green_inf}
\end{equation}
Here $\gamma_e$ is Euler's constant, and $K_{0}(r)$ is the modified Bessel
function of the second kind of order zero.

A $k$-spot quasi-equilibrium solution with spots at $\mathbf{x}_j\in
\mathbb{R}^2$ for $j=1,\ldots,k$ is given in \eqref{3:quasi} of
Principal Result 2.1. From \eqref{3:ASsmallD} and \eqref{4:green_inf},
the source strengths $S_1,\ldots,S_k$ satisfy the nonlinear algebraic
system
\begin{equation}
   \ac = S_j \Big( 1 +
 \nu (\ln 2 - \gamma_e) \Big) + \nu \sum_{i\neq j}^{k}
  S_i K_0 \left|\mathbf{x}_i-\mathbf{x}_j \right| + \nu \chi(S_j) \,, 
\quad j=1,\ldots,k \,; \qquad
  \ac = \eps^{-1} \nu A \,, \quad \nu = -\frac{1}{\ln\eps} \,. \label{4:ASinf} 
\end{equation}

To determine the stability of the $k$-spot quasi-equilibrium solution
to locally radially symmetric perturbations, we must find the
eigenvalues of the global eigenvalue problem of \S \ref{sec:eig_rad}
consisting of \eqref{4:innereig1} coupled to \eqref{4:CB}. The
$\lambda$-dependent Green's function and its regular part, satisfying
\eqref{4:Greenlamall} with $D=1$ in $\mathbb{R}^2$, are
\begin{equation}
  G_\lam(\mathbf{x}; \mathbf{x}_j) = \frac{\,1}{\,2 \pi} \, 
  K_0 \left(\sqrt{1 + \tau \lambda}\, |\mathbf{x}-\mathbf{x}_j|
\right)\,, \qquad R_{\lambda \,j j} = \frac{1}{\;2 \pi}
\Big(\ln 2 - \gamma_e - \log\sqrt{1 + \tau \lambda} \Big) \,. 
\label{4:greenlam_inf}
\end{equation}
Since $\lambda$ can be complex, we must take $\log{z}$ to be the
principal branch of the logarithm function, and choose for
$z=\sqrt{1+\tau \lam}$ the principal branch of the square root, in
order that $K_{0}(z)$ decays when $z\to \infty$ along the positive
real axis.  In terms of this Green's function, the homogeneous
linear system \eqref{4:CB} becomes
\begin{equation}
 C_j \Big( 1+ \nu (\ln 2 - \gamma_e - \log
\sqrt{1 + \tau \lambda} + \hat{B}_j) \Big)   
  + \nu \sum_{i\neq j}^k  C_i K_0 \left( \sqrt{1 + \tau \lambda}\,
    |\mathbf{x}_i-\mathbf{x}_j| \right) = 0 \,, \qquad j=1,\ldots, k \,.
\label{4:CBinf}
\end{equation}

Below, we consider the specific cases of one- and two-spot solutions
in $\mathbb{R}^2$. The numerical procedure used below to compute the
competition and oscillatory instability thresholds is summarized in
Appendix C.

\subsection{A One-Spot Solution in $\mathbb{R}^2$}\label{sec:1inf}

For a one-spot solution with spot at the origin, \eqref{4:ASinf} reduces
to the scalar nonlinear algebraic equation
\begin{equation}
\label{4:AS1} \ac = S_1 + \nu S_1 (\ln 2 - \gamma_e) + \nu \chi(S_1 ) \,,
 \qquad \ac = \eps^{-1} \nu A  \,, \qquad \nu= -\frac{1}{\ln\eps}  \,.
\end{equation}
In order that $C_1\neq 0$ in \eqref{4:CBinf}, we require that $\lambda$
satisfy the nonlinear transcendental equation
\begin{equation}
\label{4:CB1} \hat{B}_1 + \frac{\,1}{\,\nu} + \Big(\ln 2 -
\gamma_e - \log \sqrt{1 + \tau \lambda} \Big) = 0 \,.
\end{equation}
Here $\hat{B}_1=\hat{B}_{1}(\lambda,S_1)$ is to be computed from
\eqref{4:innereig1} in terms of the solution $U_1$ and $V_1$ to the
core problem (\ref{3:2Dcore_sol}). Our computational results lead to a
phase diagram in the $A$ versus $\tau$ parameter-plane characterizing
the stability of a one-spot solution. In the numerical results below
we have fixed $\eps=0.02$.

\begin{figure}[htbp]
\centering \subfigure[$A$ vs.~$\tau$] { \label{fig:k1Hopf:a}
  \includegraphics[width=3.0in,height=1.8in]{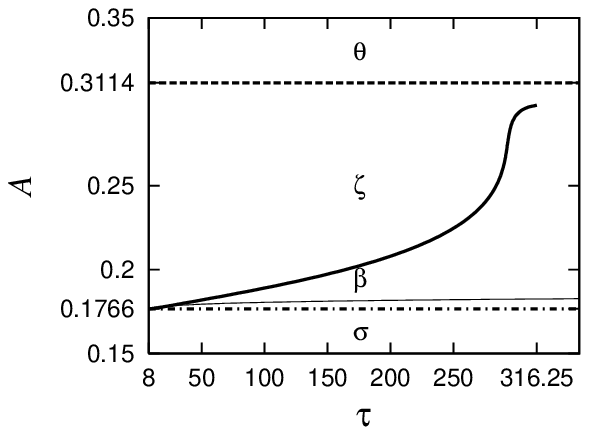}}
\subfigure[$\mbox{Im}(\lambda)$ vs.~$\mbox{Re}(\lambda)$] 
{\label{fig:k1Hopf:b}\includegraphics[width=3.0in, height=1.8in
]{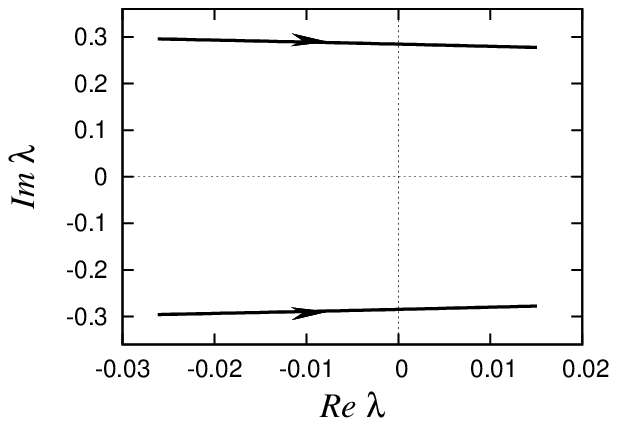}}
\caption[One-spot solution in the infinite plane: phase diagram and
spectrum]{{\em One-spot solution in $\mathbb{R}^2$ with $\eps=0.02$. (a)
Phase-diagram in the parameter-plane $A$ vs.~$\tau$. The solid curve
shows the Hopf bifurcation threshold $\tau_H$, the lower horizontal
thin solid line plots our existence threshold $A_f \approx 0.1766$,
the dot-dashed horizontal line is the existence threshold
$A_{fw}\approx 0.1758$ from NLEP theory, and the upper heavy dotted
horizontal line indicates the spot-splitting threshold $A_s= 0.311$.
Regime $\sigma$: No quasi-equilibrium solution exists;\; Regime
$\beta$: Oscillations in the spot amplitude; \; Regime $\zeta$: Stable
one-spot solution;\; Regime $\theta$: spot self-replication regime. \;
(b) Fix $A=0.18$, the spectrum in the complex $\lambda$ plane is shown
for $\tau \in [30, 40]$. At $\tau=30$, $\lambda = -0.026 \pm
0.296\,i$. At $\tau = 36.05$, $\lambda = -0.000049 \pm 0.285\, i$. At
$\tau=40$, $\lambda = 0.015 \pm 0.278\, i$. As $\tau$ increases a
complex conjugate pair of eigenvalues enters the right half-plane. }}
\label{fig:k1Hopf}
\end{figure}

From \eqref{4:AS1} we compute that there is no quasi-equilibrium
one-spot solution when $A\geq A_f=0.1766$. This existence threshold
for $A$, which was obtained by varying $S_1$ in \eqref{4:AS1} on
$S_1\in [0.22, 7.41]$, is shown by the lower thin solid line in
Fig.~\ref{fig:k1Hopf:a}. From \S \ref{sec:eig_nrad}, the spot
self-replication threshold $A_{s}\approx 0.311$ for $A$ is obtained by
setting $S_1=\Sigma_2 \approx 4.31$ and $\chi(\Sigma_2)\approx -1.783$
in \eqref{4:AS1}. It is shown by the upper horizontal dotted line in
Fig.~\ref{fig:k1Hopf:a}. Moreover, in Fig.~\ref{fig:k1Hopf:a} the
solid curve is the Hopf bifurcation threshold $A$ versus $\tau_H$ as
computed from (\ref{4:CB1}).  At this Hopf bifurcation value, a
complex complex conjugate pair of eigenvalues first enters the right
half-plane. The resulting parameter-plane as shown in
Fig.~\ref{fig:k1Hopf:a} is divided into four regions with the
following solution behavior. In Regime $\sigma$ with $A < A_f$, the
quasi-equilibrium solution does not exist; in Regime $\beta$, enclosed
by the thin solid horizontal line $A_f= 0.1766$ and the heavy solid
Hopf bifurcation curve, the quasi-equilibrium solution is unstable to
an oscillatory profile instability; in Regime $\zeta$ it is stable;
and in Regime $\theta$, it is unstable to spot self-replication.  In
Fig.~\ref{fig:k1Hopf:b}, we fix $A=0.18$ and we show that the real
part $\mbox{Re}(\lam)$ increases with $\tau$ on the range $\tau \in
[30, 40]$. Note that for $A=0.18$, the Hopf bifurcation threshold
$\tau_H = 36.06$ is where the complex conjugate pair of eigenvalues
intersect the imaginary axis.

\begin{figure}[htbp]
\centering \subfigure[$\varepsilon=0.02$] {
\label{fig:k1NLEP:a}
  \includegraphics[width=2.2in,height=1.8in]{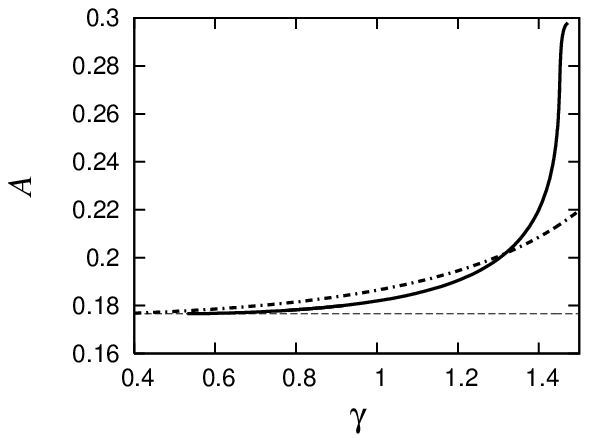}}
\subfigure[$\varepsilon=0.01$] 
{\label{fig:k1NLEP:b}
  \includegraphics[width=2.2in,height=1.8in]{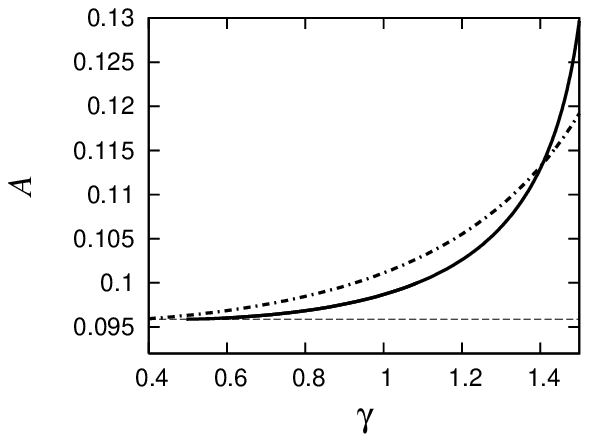}}
\subfigure[$\varepsilon=0.005$] { \label{fig:k1NLEP:c}
  \includegraphics[width=2.2in,height=1.8in]{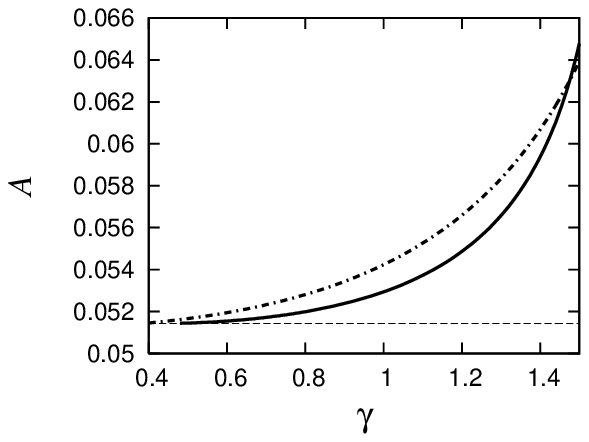}}
\caption{{\em One-spot solution in $\mathbb{R}^2$. In the $A$
  vs.~$\gamma\equiv {-\ln\tau/\ln\eps}$ plane, we plot our
 existence threshold $A_f$ by the thin dashed horizontal line, our Hopf
 bifurcation threshold $A$ by the heavy solid curve, and the stability
 threshold $A_{sw}$ from the NLEP theory by the dot-dashed curve
 (a) $\eps=0.02$, $A_f = 0.17659$; $A_{fw} = 0.17575$; (b)
 $\eps=0.01$, $A_f = 0.095875$, $A_{fw} = 0.095342$; (c)
 $\eps=0.005$, $A_f = 0.051432$, $A_{fw} = 0.051133$. Here
 $A_{fw}$ is the existence threshold from NLEP theory given in
 \eqref{4:1spot_NLEP}.}}
\label{fig:k1NLEP}
\end{figure}

For three values of $\eps$, in Fig.~\ref{fig:k1NLEP} we compare the
stability threshold obtained from (\ref{4:CB1}) with that obtained
from the NLEP analysis of \cite{2D_Wei:2001}, as summarized in
Appendix B. From \eqref{4:AfWei} and \eqref{4:tauWei} of Appendix B,
there are two threshold values $A_{fw}$ and $A_{sw}$ predicted from
the leading-order-in-$\nu$ NLEP theory of \cite{2D_Wei:2001}. For
$A<A_{fw}$ a one-spot quasi-equilibrium solution does not exist. For
$A<A_{sw}$ this solution is unstable to an oscillatory instability.
These threshold values, with $A_{sw}$ depending on $\tau$ and the
ground-state solution $w$ of (\ref{4:groundstate}), are
\begin{equation}
 A_{fw} \equiv 2 \eps \sqrt{\frac{\,b_0}{\,\nu}} \,, \quad
 b_0 \equiv \int_0^{\infty} w^2 \rho\, d\rho \approx
  4.9347 \,; \qquad
 A_{sw} \equiv A_{fw} \left[1 - 
 \left(\frac{\gamma}{4-\gamma}\right)^2 \right]^{-1/2} \,, \qquad
  \gamma \equiv -\frac{\ln\tau}{\ln\eps} \,. \label{4:1spot_NLEP}
\end{equation}
For $\eps=0.02$, we get $A_{fw} \approx 0.1757$, which agrees very
well with our existence threshold $A_f \approx 0.1766$. In
Fig.~\ref{fig:k1NLEP} we compare the NLEP stability threshold $A_{sw}$
versus $\gamma={-\ln\tau/\ln\eps}$ with our Hopf bifurcation threshold
$A$ versus $\gamma={-\ln\tau_H/\ln\eps}$.  We recall from Appendix B
that the NLEP theory predicts instability when $A<A_{sw}$ for $0\leq
\tau\leq \eps^{-2}$, but has no conclusion regarding stability or
instability if $A>A_{sw}$. Our Hopf bifurcation threshold predicts
that stability changes when we cross the heavy solid curves in
Fig.~\ref{fig:k1NLEP}.

Next, we study the possibility of real-valued unstable eigenvalues
that have crossed along the real-axis $\mbox{Im}(\lam)=0$ into the
unstable half-plane $\mbox{Re}(\lam)>0$. The threshold condition for
this is to set $\lam=0$ and $\hat{B}_1=\chi^{\prime}(S_1)$ in
(\ref{4:CB1}) to obtain that $S_1$ satisfies
$\chi_{1}^{\p}(S_1)+\nu^{-1}+\ln{2}=\gamma_e$. By differentiating
(\ref{4:AS1}) for $\ac$ with respect to $S_1$, we obtain that this
threshold coincides with the minimum value $S_{1m}$ of $S_1$ for the
curve $\ac$ versus $S_1$ as defined by (\ref{4:AS1}). For $S>S_{1m}$,
we compute numerically that $\mbox{Re}(\lam)<0$, while 
$\mbox{Re}(\lam)>0$ for $S<S_{1m}$. The critical value $S_{1m}$, for which
$\lam=0$, corresponds to the fold point in the bifurcation diagram of
$u(0)$ versus $\ac$ (such as shown in Fig.~\ref{fig:AS:b}.

\subsection{A Two-Spot Solution in $\mathbb{R}^2$}\label{sec:2inf}

Next, we consider the case of two spots centered at
$\mathbf{x}_1=(\alpha, 0)$ and $\mathbf{x}_2=(-\alpha, 0)$, where
$\alpha>0$ and $\alpha \gg {\mathcal O}(\eps)$. We look for a
symmetric solution where $S_1=S_2=S_c$. Then, from (\ref{4:ASinf}),
$S_c$ satisfies the nonlinear algebraic equation
 \begin{equation}
 \label{3:ASinf}
 \ac = S_c \left[ 1 + \nu (\ln{2} - \gamma_e) + 
   \nu K_0 \left( 2\alp\right) \right]  + \nu \, \chi(S_c) \,, \qquad
  \ac = \eps^{-1} \nu A \,, \qquad   \nu=-{1/\ln\eps} \,.
 \end{equation}

\begin{figure}[htpb]
\centering \subfigure[$\ac$ vs.~$S_c$]{
\includegraphics[width=2.2in,height=1.6in,clip]{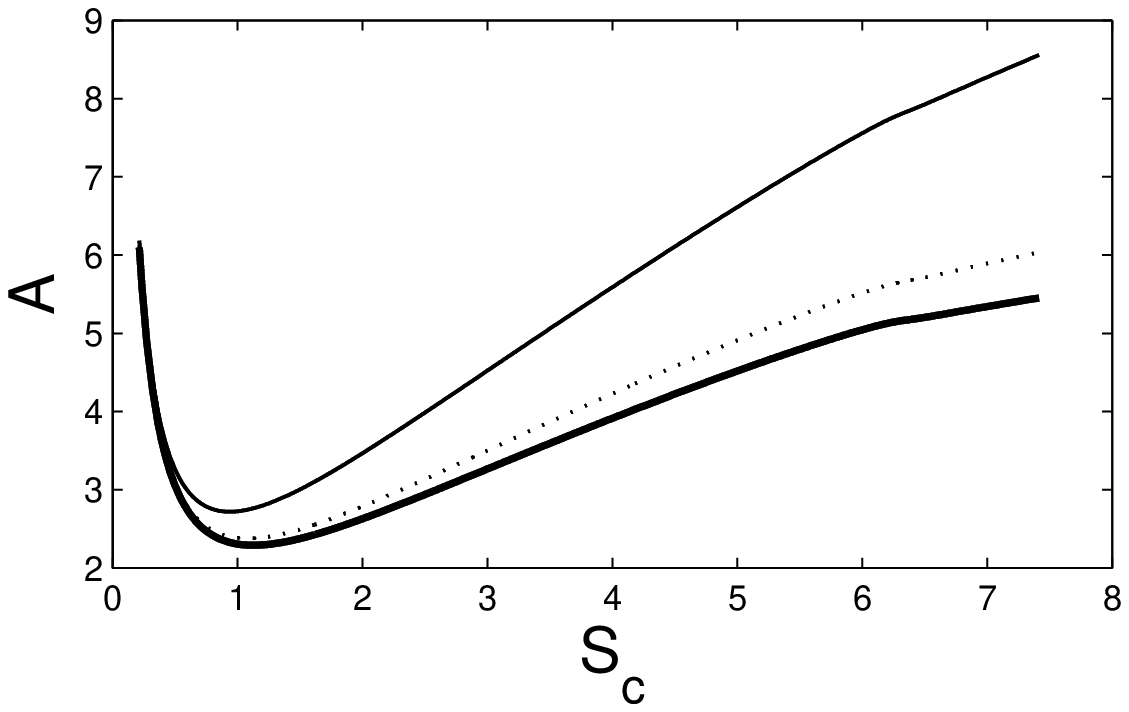}
\label{fig:AS_inf:a}} \subfigure[$\ac_f$ vs.~$\alpha$]{
\includegraphics[width=2.2in,height=1.6in,clip]{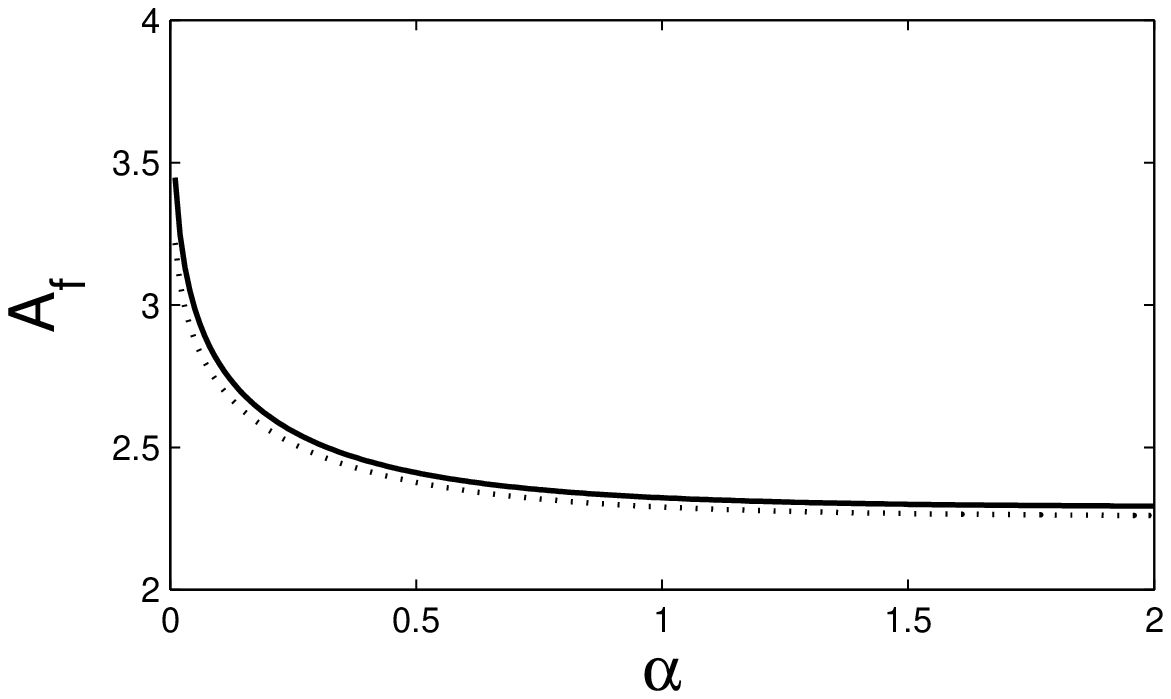}
\label{fig:AS_inf:b}} \subfigure[$S_c$ vs.~$\alpha$]{
\includegraphics[width=2.2in,height=1.6in,clip]{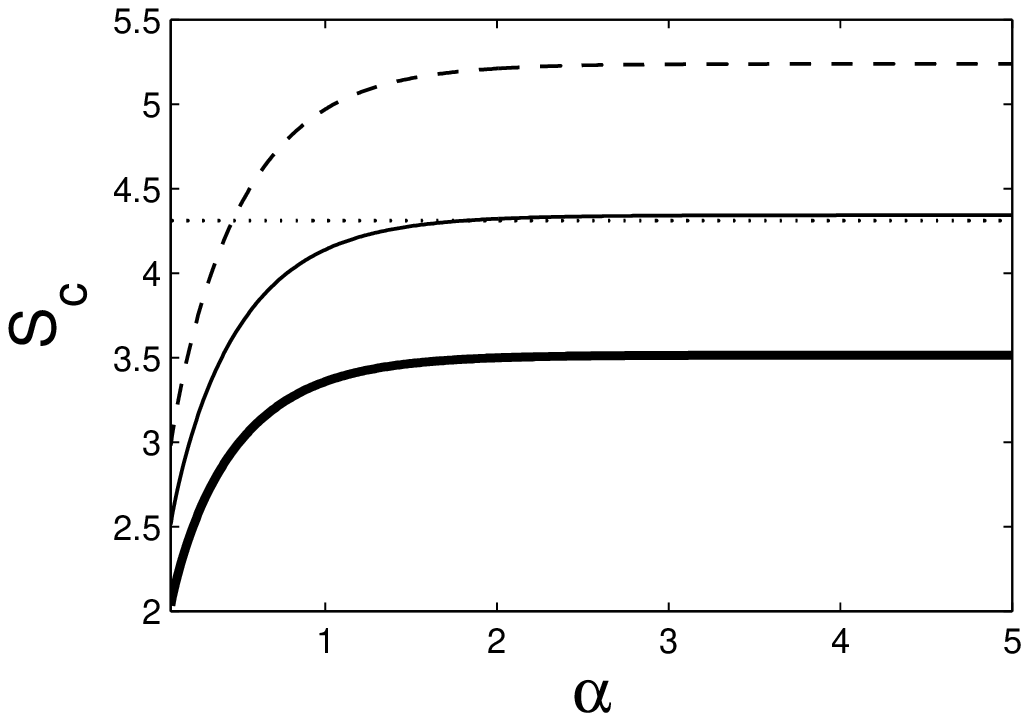}
\label{fig:AS_inf:c}} 
\caption[Two spots in the infinite plane]
{{\em Two spots at $(\pm\alpha, 0)$ in $\mathbb{R}^2$ 
with $\eps = 0.02$. (a) $\ac$ vs.~$S_c$ for $\alpha=0.1$ (solid
curve), $\alpha=0.5$ (dashed curve) and $\alpha=1.0$ (heavy solid
curve). (b) The fold point $\ac_f$ vs.~$\alpha$; two-term asymptotic
result (solid curve), and numerical result from \eqref{3:ASinf}
(dotted curve). (c) $S_c$ vs.~$\alpha$ for $\ac=3.5$ (heavy solid curve),
$\ac=4.0$ (solid curve) and $\ac=4.5$ (dashed curve).}}
\label{fig:AS_inf}
\end{figure}

For $\eps=0.02$, we first discuss the existence of the
quasi-equilibrium two-spot solution.  In Fig.~\ref{fig:AS_inf:a} we
plot $\ac$ versus $S_c$ for three different values of $\alpha$,
showing the existence of a fold point $\ac_f$ in the graph of $S_c$
versus $\ac$. This fold point $\ac_f(\alpha)$ is shown in
Fig.~\ref{fig:AS_inf:b} by the dotted curve. The solid curve in
Fig.~\ref{fig:AS_inf:b} is the corresponding two-term asymptotic
result for $\ac_{f}$ obtained by expanding \eqref{3:ASinf} in powers
of $\nu$ as in \eqref{3:circscalar_p}. This result shows that a
quasi-equilibrium two-spot pattern exists for a spot separation
distance of $2\alpha$ only when $\ac>\ac_f(\alp)$.

\begin{figure}[htbp]
\centering \subfigure[$A$ vs.~$\alpha$] {
\label{fig:k2Hopf:a}
  \includegraphics[width=3.0in, height=1.8in]{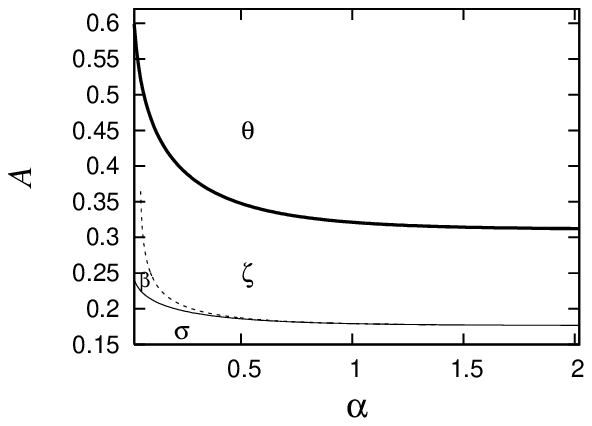}}
\subfigure[$\tau$ vs.~$\alpha$] 
{\label{fig:k2Hopf:b}
    \includegraphics[width=3.0in, height=1.8in]{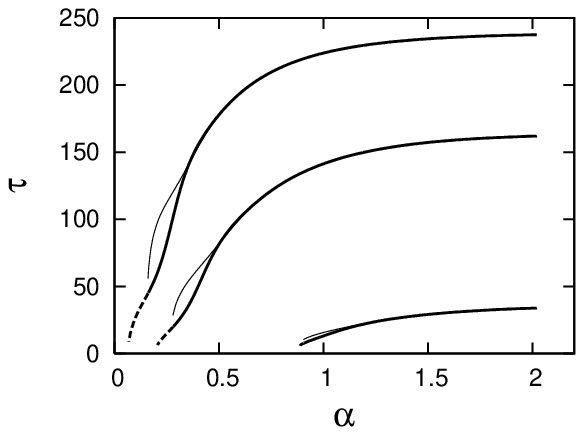}
}\\
\subfigure[$\lambda_i$ vs.~$\alpha$] { \label{fig:k2Hopf:c}
  \includegraphics[width=3.0in, height=1.8in]{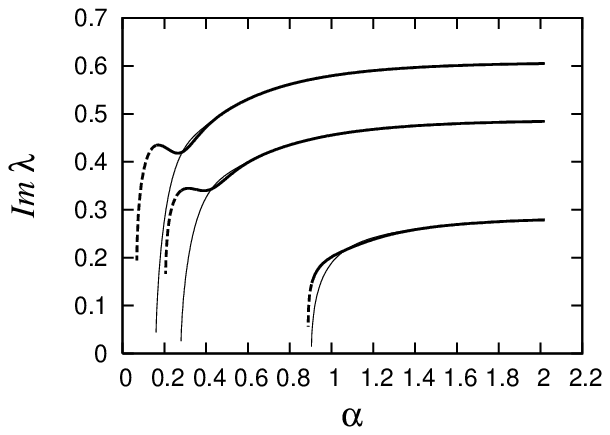}
}
\subfigure[$\mbox{Im}(\lambda)$ vs.~$\mbox{Re}(\lambda)$] 
{
    \label{fig:k2Hopf:d}
    \includegraphics[width=3.0in, height=1.8in]{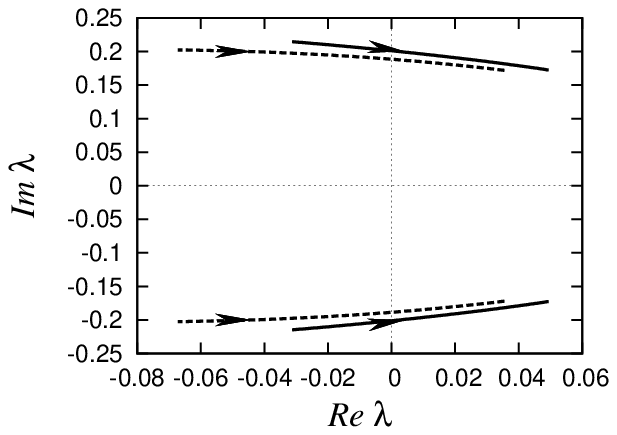} }
\caption[Two-spot solution in the infinite plane: Phase diagram]{{\em
Two-spot solution in $\mathbb{R}^2$. (a) We plot the existence
threshold $A_f$ vs.~$\alpha$ (lower curve), the spot-splitting
threshold $A_{s}$ vs.~$\alpha$ (top curve), and the competition
instability threshold (dotted curve). (b) Hopf bifurcation threshold
$\tau_{H1}$ for in-phase oscillations (bottom of each pair of curves)
and $\tau_{H2}$ for out-of-phase oscillations (top of each pair of
curves), are plotted versus $\alpha$ for $A=0.18$ (bottom pair),
$A=0.2$ (middle pair), and $A=0.22$ (top pair). The dotted portions on
$\tau_{H1}$ correspond to the parameter regime $\beta$ in subfigure a)
where competition instabilities occur for any $\tau$. (c) Imaginary
part $\lambda_i$ of $\lambda$ vs.~$\alpha$ at the Hopf bifurcations
thresholds for $A=0.18$ (bottom), $A=0.2$ (middle), and $A=0.22$
(top). The heavy solid curves are the eigenvalues for in-phase
oscillations, and the thin solid curves are the eigenvalues for
out-of-phase oscillations. The dotted portions lie on $\tau_{H1}$
where a competition instability occurs. (d) Eigenvalue path in the
complex plane for $A=0.18$ and $\alpha=1$ on the range $\tau \in [10,
20]$, with the arrow indicating the direction as $\tau$ increases. The
eigenvalues associated with in-phase (solid curves) and out-of-phase
(dashed curves) oscillations enter the right half plane at $\tau_{H1}$
and $\tau_{H2}$, respectively.}}
\label{fig:k2Hopf}
\end{figure}

Next, we discuss the possibility of self-replicating instabilities for
an inter-spot separation of $2\alpha$. In Fig.~\ref{fig:AS_inf:c}, we
plot $S_c$ versus $\alpha$ for $\ac=3.5$ (solid curve), $\ac=4.0$
(dashed curve) and $\ac=4.5$ (heavy solid curve). Notice that for a
fixed $\ac > \ac_f$, there are two solutions for $S_c$ for a given
$\alpha$. Since the solution branch with the smaller value for $S_c$
is unstable, we only plot the large solution branch for $S_c$ in this
figure. Fig.~\ref{fig:AS_inf:c} shows that for a fixed $\alpha$, $S_c$
is an increasing function of $\ac$, whereas for a fixed $\ac$, then
$S_c$ increases as $\alpha$ increases.  For the solid curve in
Fig.~\ref{fig:AS_inf:c} with $\ac=3.5$, $S_c<\Sigma_2$ for all
$\alpha$, so that the spots never split for any inter-separation
distance.  Alternatively, the dashed curve for $\ac=4.0$ and the heavy
solid curve for $\ac=4.5$ in Fig.~\ref{fig:AS_inf:c} intersect the
spot-splitting threshold $S_c = \Sigma_2$ at $\alpha \approx 1.81$ and
$\alpha \approx 0.46$, respectively. This threshold initiates a
spot-replication event. The self-replication threshold $\ac_{s}$
versus $\alp$ is obtained by setting $S_c=4.31$ in \eqref{3:ASinf}.
Moreover, from (\ref{3:ASinf}), we readily observe that $\ac_{s}$ is a
decreasing function of $\alpha$ and asymptotes to its minimum value
$\ac_{sm} = 3.98$ as $\alpha \to \infty$. This minimum value is
obtained by setting $S_c = \Sigma_2$ and $K_0(2 \alpha) \to 0$ in
\eqref{3:ASinf}. We conclude that for any $\ac > \ac_{sm}$, a two-spot
pattern on $\mathbb{R}^2$ will be linearly unstable to a
peanut-splitting instability if the inter-spot separation distance
$2\alpha$ is sufficiently large. Thus, for $\ac>\ac_{sm}$, a spot
self-replication instability is, essentially, {\em under-crowding
instability}, that is triggered only when the two spots are too
isolated from each other.  In terms of the original feed-rate
parameter $A$, the spot-splitting threshold $A_{s}\equiv {\eps
\ac_{s}/\nu}$ is given by the heavy solid curve in the phase
diagram of Fig.~\ref{fig:k2Hopf:a}.

Next, we compute the stability thresholds for either competition and
oscillatory instabilities.  The $\lambda$-dependent Green's function
is given in \eqref{4:greenlam_inf}. Since the $\lambda$-dependent Green's
matrix $\mathcal{G}_{\lam}$ is circulant, then Principal Result 4.3
applies. The eigenvectors $\mathbf{v}_j$ and eigenvalues
$\omega_{\lam\, j}$ of $\mathcal{G}_\lam$ for $i=j,2$ are
\begin{align*}
\omega_{\lambda \, 1} &= \omega_{\lambda\, 1}(\tau\lam) = G_{\lambda
 12} + R_{\lambda 11} = \frac{1}{2\pi} \left[ (\ln 2 - \gamma_e -
 \log\sqrt{1 + \tau \lambda}) + K_0(2 \alpha \sqrt{1 +\tau
 \lambda})\right] \,, \qquad \mathbf{v}_1 \equiv (1,1)^T \,, \\
 \omega_{\lambda \, 2} &= \omega_{\lambda\, 2}(\tau\lam) = R_{\lambda
 11} - G_{\lambda 12} =\frac{1}{2\pi}\left[ (\ln 2 - \gamma_e -
  \log \sqrt{1 + \tau \lambda}) - K_0( 2 \alpha \sqrt{1 +
 \tau \lambda}) \right] \,, \qquad \mathbf{v}_2 \equiv (1,-1)^T \,.
\end{align*}
Then, from \eqref{4:CBreduced} of Principal Result 4.3, we must determine
the roots $\lam$ of the two transcendental equations
\begin{equation}
  \nu^{-1} + \hat{B}_c + 2 \pi \omega_{\lambda\, j}(\tau\lam)  =
  0 \,, \quad j=1,2 \,, \label{4:CB2}
\end{equation}
where $\hat{B}_c(\lam,S_c)$ is to the common value for $\hat{B}_j$ for
$j=1,2$ in \eqref{4:innereig1}.

Our numerical results below show that in some region of the $A$ versus
$\alpha$ parameter-plane, the two-spot solution becomes unstable when
$\tau$ is increased due to the creation of an unstable eigenvalue
$\lam$ from $\omega_{\lam \, 1}$ as a result of a Hopf
bifurcation. Since this oscillatory instability is associated with the
eigenvector $\mathbf{v}_1=(1,1)^{t}=(C_1,C_2)^T$, it leads to the
initiation of a simultaneous in-phase oscillation in the amplitudes of
the two spots. In another region of the $A$ versus $\alpha$ parameter
plane we will show numerically that an unstable real positive
eigenvalue $\lam$ for $\omega_{\lam\, 2}$ can occur, even when
$\tau=0$, if the inter-spot distance $2\alp$ is below some
threshold. Since this instability is associated with the
eigenvector $\mathbf{v}_2=(1,-1)^{t}=(C_1,C_2)^T$, it leads to the
initiation of a sign-fluctuating instability in the amplitudes of the
two spots, which leads ultimately to the annihilation of one of the
two spots. Since this instability is triggered when $2\alpha$ is
sufficiently small, it is also referred to as a competition or an
overcrowding instability. The threshold for a competition instability
associated with the dominant sign-fluctuating mode
$\mathbf{v}_2=(1,-1)^T$ is obtained from the formulation
(\ref{3:newres}). The threshold condition for this instability is
reduced to finding the curve $A=A(\alp)$, by eliminating the source
strength $S_c$ from the following nonlinear algebraic system:
\begin{equation}
   \chi^{\p}(S_c) + \ln{2} - \gamma_e - K_{0}(2\alp) = -\nu^{-1}
 \,, \qquad A = \frac{\eps}{\nu} \left( S_c \left[1 + \nu \left(
 \ln{2} - \gamma_e + K_{0}(2\alp)\right)\right] + \nu \chi(S_c) \right) \,.
 \label{4:new2spike}
\end{equation}

We fix $\eps=0.02$ in the computations below.  In
Fig.~\ref{fig:k2Hopf:a} the two-spot existence threshold $A_f$ as a
function of $\alpha\in [0.02, 2.02]$ is shown by the lower solid
curve, while the spot self-replication threshold $A_{s}$ versus
$\alpha$ discussed above is shown by the top heavy solid curve. To
obtain $A_f$ we determined the minimum value of $A$ as $S_c$ is varied
in \eqref{3:ASinf}.  In Fig.~\ref{fig:k2Hopf:a} the middle dotted
curve is the threshold for the onset of a competition instability,
obtained from (\ref{4:new2spike}). These three curves separate the $A$
versus $\alpha$ parameter-plane of Fig.~\ref{fig:k2Hopf:a} into four
distinct regions.  In Regime $\sigma$, the quasi-equilibrium two-spot
solution does not exist. In Regime $\beta$, the two-spot
quasi-equilibrium solution undergoes a competition instability for any
$\tau\geq 0$. This instability ultimately leads to the annihilation of
one of the spots. We note that the competition threshold approaches the
existence threshold when $\alpha$ becomes large. This is because when 
$\alpha\gg 1$ the spots are uncoupled, and there is no competition
instability for a one-spot solution in $\mathbb{R}^2$. In Regime
$\zeta$, the two-spot pattern is stable to a competition instability,
but becomes unstable to an oscillatory profile instability when
$\tau$ is large enough. As $\tau$ increases above a certain threshold
$\tau_H$, a Hopf bifurcation occurs and the two-spot solution is
unstable to the dominant in-phase, or synchronous, spot amplitude
oscillation. In Regime $\theta$ the two-spot quasi-equilibrium
solution is unstable to spot self-replication for any $\tau\geq 0$.

For three values of $A$, in Fig.~\ref{fig:k2Hopf:b} we plot the Hopf
bifurcation thresholds $\tau_{H1}$ and $\tau_{H2}$ associated with the
eigenvectors $\mathbf{v}_1 =(1,1)^T$ and $\mathbf{v}_2 =(1,-1)^T$,
respectively. From this figure we observe that in the limit 
$\alpha\to \infty$, corresponding to two isolated spots, the
thresholds $\tau_{H1}$ and $\tau_{H2}$ approximate the Hopf
bifurcation threshold $\tau_H$ of an isolated one-spot solution.

From Fig.~\ref{fig:k2Hopf:b}, the Hopf bifurcation curves for
$\tau_{H1}$, which each have a dotted portion, end at the values
$\alpha_f = 0.888, 0.206, 0.068$ for the curves $A=0.18, 0.2, 0.22$,
respectively. These critical values of $\alpha$ correspond to the
existence threshold $A = A_f(\alpha)$ for the quasi-equilibrium solution.
Moreover, the Hopf bifurcation threshold $\tau_{H2}$
terminates at another set of critical values $\alpha_c = 0.904, 0.28,
0.16$ for the curves $A=0.18, 0.2, 0.22$, respectively. The reason for
this disappearance is seen in Fig.~\ref{fig:k2Hopf:c}, where we plot
the imaginary part of the eigenvalues at the Hopf bifurcation
thresholds for $\mathbf{v}_1=(1,1)^T$ (curves with dotted portions)
and $\mathbf{v}_2=(1,-1)^T$ (curves without dotted portions). From
this figure it is clear that as $\alpha\to\alpha_c$ from above, the
complex conjugate pair of eigenvalues associated with $\tau_{H2}$
merge onto the real axis at the origin, which is precisely the
crossing point for the onset of a competition instability.

 When the spots are too close in the sense that $\alpha_f < \alpha <
\alpha_c$, then there is a positive real (unstable) eigenvalue for any
$\tau\geq 0$, which initiates a competition instability.  If in
addition, $\tau<\tau_{H1}$, it is the only unstable eigenvalue. In
Fig.~\ref{fig:k2Hopf:b} and Fig.~\ref{fig:k2Hopf:c}, the dotted
segments of the curves correspond to the range of $\alpha$ where this
competition instability occurs. Finally, we observe from
Fig.~\ref{fig:k2Hopf:b} that $\tau_{H1} < \tau_{H2}$. Therefore, for
each $\alp>\alp_c$, the in-phase synchronous oscillatory instability
of the spot amplitudes is always the dominant instability as $\tau$ is
increased.

For $A=0.18$, $\alpha =1.0$, and $10\leq \tau \leq 20$, in
Fig.~\ref{fig:k2Hopf:d} we plot the path of the eigenvalues in the
complex plane, corresponding to both $\mathbf{v}_1$ (solid curves) and
$\mathbf{v}_2$ (dotted curves), as $\tau$ is increased.  For each
case, $\mbox{Re}(\lam)$ increases as $\tau$ increases. The two Hopf
bifurcation values are, respectively, $\tau_{H1} = 13.1$ and
$\tau_{H2} = 15.6$.

\begin{figure}[htbp]
 \includegraphics[height=1.8in, width=3.0in]{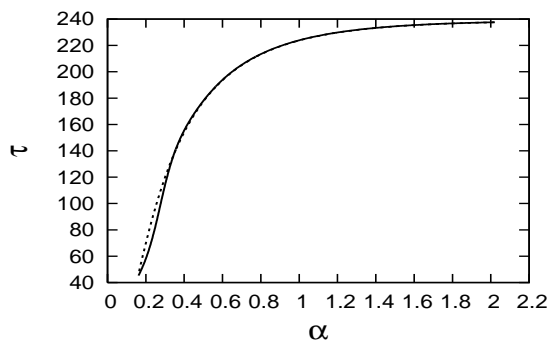}
 \caption{{\em Two-spot solution in $\R^2$. Fix $A=0.22$ and
$\eps=0.02$. The Hopf bifurcation threshold $\tau_{H1}$ vs.~$\alp$
for a synchronous oscillatory instability obtained from the full
stability problem \eqref{4:CB2} (solid curve) is compared with that
obtained from the approximate system (\ref{4:largetau}) resulting from the
large-$\tau$ approximation (dotted curve).}}
\label{fig:largetau}
\end{figure}

Next, we derive an approximation for the Hopf bifurcation threshold
valid for $\tau\gg 1$. For $\tau\gg 1$ and $\alp$ not too small, we
exploit the exponential decay of the modified Bessel-function for
large argument to obtain that the $\lambda$-dependent Green's matrix
${\mathcal G}_\lam$ is essentially a diagonal matrix.  Therefore, in
this large-$\tau$ limit we can substitute $\omega_{\lam j} \approx
R_{\lam 11}$ for $j=1,2$ into \eqref{4:CB2}. In this way, and by using
\eqref{4:greenlam_inf} for $R_{\lam 11}$, we obtain 
that the limiting Hopf bifurcation threshold $\tau_H$ and the critical
eigenvalue $\lam=i\lam_i$ are the solutions to the approximate system
\begin{equation}
    \mbox{Im}\left(\hat{B}_c\right) =\frac{\pi}{4}\,, \qquad
  \mbox{Re}\left(\hat{B}_c\right) + \ln{2} -\gamma_e
  - \frac{1}{2}\ln(\tau \lam_i)  = -\nu^{-1} \,. \label{4:largetau}
\end{equation}
For $\eps=0.02$ and $A=0.22$, in Fig.~\ref{fig:largetau} we compare
the threshold obtained from (\ref{4:largetau}) with that obtained from
the full transcendental equation \eqref{4:CB2} (as shown initially in
Fig.~\ref{fig:k2Hopf:b}). This comparison, on the range $\alp\ge 0.16$
for which no competition instabilities occur, shows that
(\ref{4:largetau}) gives a good approximation provided that $\alp$ is
not too small.

Finally, we derive the ODE system that determines the slow dynamics of the
two-spot pattern when $S_c < \Sigma_2$. Provided that no other profile
instabilities are present, and that $\ac>\ac_f(\alp)$, we can explicitly
evaluate the terms in (\ref{3:dyn}) of Principal Result 3.1 to obtain
the DAE ODE system
\begin{equation}
 \frac{\,d \alpha }{d\xi} =  - \gamma(S_c) S_c
K^{\p}_0 \left( 2\alpha \right) \,, \qquad 
 \gamma(S_c)\equiv \frac{-2} {\int_{0}^{\infty} \phi^*
V^{\p}_0\rho\, d\rho, } \,, \quad \xi = \eps^2 t \,,\label{3:2ode}
\end{equation}
with $S_c$ determined in terms of $\alp$ by (\ref{3:ASinf}).  Since
$K_0'(r)<0$ for any $r> 0$ and $|K_0'(r)|$ is a decreasing function of
$r$, (\ref{3:2ode}) shows that the two spots are repelling and that
their common speed $\alp_c \equiv |\frac{\, d\alpha}{d\xi}|$ is a
decreasing function of $\alp$.  Since $K_{0}(2\alp)$ is an
exponentially small when $\alp\gg 1$, the two-spot dynamics
\eqref{3:2ode} is metastable in this limit, and is similar to the
metastable repulsive dynamics studied in \cite{metastable_Ei:2002} for
a two-spot solution to the GM model in $\R^2$.

Next, we discuss the possibility of dynamically-triggered
instabilities induced by the slow evolution of the two-spot pattern.
With regards to spot self-replication, we observe that since the two
spots are dynamically repelling, and $S_c$ is a monotonically
increasing function of $\alpha$ (see Fig.~\ref{fig:AS_inf:c}), it
follows that a dynamically-triggered spot self-replication instability
will occur when $\ac> \ac_{sm}\approx 3.98$. To illustrate this let
$\ac=4.0$. Then, the dashed curve of $S_c$ versus $\alpha$ in
Fig.~\ref{fig:AS_inf:c} shows that $S_c>\Sigma_2$ only if $\alpha \ge
1.81$. Now consider an initial two-spot quasi-equilibrium solution
with $\alpha=1$ at $t=0$. For this initial value of $\alp$, then
$\ac>\ac_f$ (see Fig.~\ref{fig:AS_inf:b}). Then, since
$\alp^{\prime}>0$ and $\alpha\to \infty$ as $t\to \infty$ from the ODE
\eqref{3:2ode}, it follows that $\alpha=1.81$ at some long time
$t={\mathcal O}(\eps^{-2})$, which triggers the spot self-replication
instability. From Principal Result 3.2 of \S \ref{sec:dyn}, the spots
will split in a direction perpendicular to the $x$-axis.

Thus, for the GS model \eqref{1:GS_2D}, we conclude that there is a
wide parameter range for which two spots in $\mathbb{R}^2$ will
undergo a dynamically-triggered spot self-replication event that is
induced by their slow repulsive dynamics. In contrast, we remark that
since the Schnakenburg model of \cite{Schnaken_KWW:2008} involves the
Neumann Green's function rather than the reduced-wave Green's
function, the infinite-plane problem for this related RD model is
ill-posed.

Finally, we remark that the monotone increasing behavior of the Hopf
bifurcation threshold $\tau_{H1}$ versus $\alp$ (see
Fig.~\ref{fig:k2Hopf:b}), coupled to the repulsive spot dynamics of
the two-spot pattern, precludes the existence of any
dynamically-triggered oscillatory instability for a two-spot solution
in $\mathbb{R}^2$.  Namely, consider an initial two-spot
quasi-equilibrium with initial inter-separation distance $2\alp_0$,
and with a fixed $\tau$ with $\tau < \tau_{H1}(\alp_0)$, where
$\alp_0>\alp_c$. Then, since $\alp$ increases as $t$ increases, and
$\tau_{H1}$ is monotone increasing in $\alp$, it follows that
$\tau<\tau_{H1}(\alp)$ for any $t>0$. In a similar way, we can show
that there are no dynamically-triggered competition instabilities in
$\mathbb{R}^2$.

\subsection{Two-Spot Patterns in a Large Square Domain}\label{sec:2inf_large}

Next, we consider a two-spot pattern in the square domain $[0, L]
\times [0, L]$ with spots located symmetrically about the midline of
the square at $\mathbf{x}_{1,2}=(L/2 \pm \alpha, L/2)$, where $\alpha
> 0$. For this arrangement, ${\mathcal G}$ in \eqref{3:ASmatrix} is
circulant symmetric, with matrix entries that can be numerically
calculated from \eqref{3:squaregreen} of Appendix A.  We fix the
parameter values $D=1$, $\ac=4.2$, and $\eps=0.02$. Since the
limiting solution for $L \to \infty$ reduces to the infinite-plane
problem, a large square domain can be used to approximate a two-spot
evolution for the infinite-plane problem of \S \ref{sec:2inf}. Our
goal here is to examine the asymptotic prediction of spot
self-replication.

\begin{figure}[htbp]
\centering
\includegraphics[width=3.5in, height=1.8in]{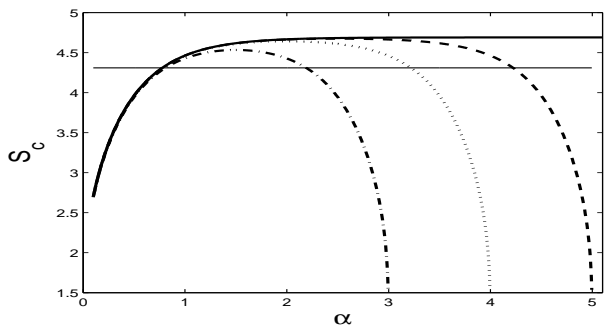}\
\caption{{\em Comparison of two-spot patterns in a square domain and
in an infinite domain. Fixing $\eps=0.02$, $D=1$ and
$\mathcal{A} =4.2$, we plot $S_c$ vs.~$\alpha$ for an
infinite domain (heavy solid curve), and for square domains
with side-lengths $L=10$ (dashed curve), $L=8$ (dotted curve),
and $L=6$ (dot-dashed curve). The thin solid line is the
peanut-splitting threshold $\Sigma_2\approx 4.31$, which intersects the
four curves at $\alpha_s \approx 0.775$ for the infinite domain,
$\alpha_{s1} \approx 0.775$ for $L=10$, $\alpha_{s2} \approx 0.780$
for $L=8$, and $\alpha_{s3} \approx 0.805$ for $L=6$.}}
\label{fig:2spt_cmp_A4d2}
\end{figure}

In Fig.~\ref{fig:2spt_cmp_A4d2} we plot $S_c$ versus $\alpha$ for a
square of side-lengths $L=6,8,10$. By symmetry, the equilibrium spot
locations are at $\alp_{e}={L/4}$. In this figure, we also plot the
infinite-plane result for $S_c$ versus $\alp$, obtained from
(\ref{3:ASinf}).  The resulting four curves of $S_c$ versus $\alp$ are
essentially indistinguishable on the range $\alp\ll L$ where the
finite-boundary effects are insignificant. From this figure, we
observe that the spot-splitting threshold $S_c=\Sigma_2\approx 4.31$,
represented by the thin solid line, intersects each of the four curves
at some critical values of $\alp$, which initiates a spot
self-replication event. We observe from Fig.~\ref{fig:2spt_cmp_A4d2}
that $S_c<\Sigma_2$ when $\alp$ becomes close to ${L/2}$. Therefore,
for a spot near an edge of the square, the Neumann boundary condition
effectively introduces an image spot outside of the square in close
proximity to the interior spot, which then eliminates the possibility
of spot self-replication.

\begin{figure}[htbp]
\begin{center}
\includegraphics[width=7.0in]{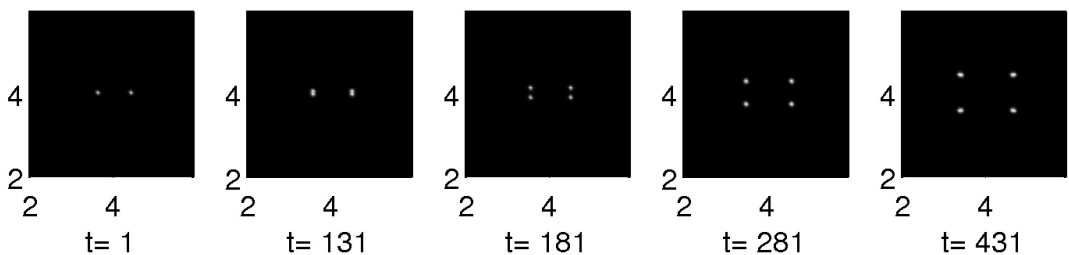}
\caption{{\em A two-spot pattern in the square $[0, 8] \times [0, 8]$.
Let $\mathcal{A} = 4.2$, $D=1.0$, $\eps=0.02$ with initial spot
locations $\mathbf{x}_{1,2} = (4 \pm 0.4, 4)$ at time $t=0$. Then,
$S_c \approx 3.77$ at $t=0$, but $S_c>\Sigma_2\approx 4.31$ at a later
time. The numerical solution to \eqref{1:GS_2D}, represented as a
gray-scale image of $v$, is plotted in the zoomed region $[2, 6]
\times [2, 6]$ at times $t=1, 41, 131, 181$, $231, 281, 331, 431$. The
spots undergo splitting in a direction perpendicular to their motion
consistent with Principal Result 3.2.}}
\label{fig:2spt_L8}
\end{center}
\end{figure}

For $L=8$, we now compare our asymptotic prediction for spot
self-replication with corresponding full numerical results computed
from \eqref{1:GS_2D} using VLUGR (cf.~\cite{vlugr_Blom:1966}). We take
an initial two-spot pattern with $\alpha=0.4<\alp_s$ at time $t=0$. At
this initial time, we calculate from \eqref{3:circscalar} that $S_c
\approx 3.77$, so that the initial two-spot pattern is stable to spot
self-replication. Then, since the equilibrium state is at $\alp_e=2.0$
and the two-spot dynamics is repulsive, it follows that $\alpha$
increases as $t$ increases and will eventually exceed the
spot-splitting threshold $\alpha_{s3}\approx 0.780$. This prediction
of a dynamically-triggered instability, which very closely
approximates that for the infinite-plane problem of \S
\ref{sec:2inf}, is confirmed by the full numerical results shown
in Fig.~\ref{fig:2spt_L8}.

\setcounter{equation}{0}
\setcounter{section}{5}
\section{Symmetric Spot Patterns in the Square and Disk} \label{sec:sym}

In this section, we implement the asymptotic theory of \S
\ref{sec:quasi}--\ref{sec:eig} to study competition, oscillatory, and
spot self-replication instabilities for various spot patterns for
which the Green's matrix ${\cal G}$ in (\ref{3:Gmatrix}) has a
circulant matrix structure. The thresholds for these instabilities are
favorably compared with results from full numerical simulations
obtained by using the PDE solver VLUGR (cf.~\cite{vlugr_Blom:1966})
for a square domain and the finite element code of
\cite{stripe_KSWW:2006} for the disk.

In all of the numerical simulations below and in \S \ref{sec:asy} we
have set $\eps=0.02$ so that $\nu=-{1/\ln\eps} =0.2556$. For given GS
parameters $\ac$ and $D$, and for an initial configuration of spot
locations $\mathbf{x}_1,\ldots,\mathbf{x}_k$, we compute the source
strengths $S_1,\ldots,S_k$ from the nonlinear algebraic system
\eqref{3:ASsmallD}. The initial condition for the full numerical
simulations is the quasi-equilibrium solution
\eqref{3:quasi} with the values for $V_j(0)$ as plotted in
Fig.~\ref{fig:core:b}, corresponding to the computed values of
$S_j$. Since this initial condition provides a decent, but not
sufficiently precise, initial $k-$spot pattern, we only begin to track
the spot locations from the full numerical simulations after the
completion of a short transient period. To numerically identify the
locations of the spots at any time, we determine all local maxima of
the computed solution $v$ by identifying the maximal grid values. This
simple procedure is done since it is expensive to interpolate the grid
values to obtain more accurate spot locations at every time step.  In
this way, both the spatial trajectories and the amplitudes 
of the spots are obtained from the full numerical results. In
many of the numerical experiments below, we show a gray-scale contour plot
of the numerical solution for $v$. The bright (white) regions
correspond to the spot regions where $v$ has a large amplitude, while
the dark region is where $v$ is exponentially small.

\subsection{A Four-Spot Pattern in the Unit Square} \label{sec:sym_sq}

We first consider a four-spot pattern in the unit square $\Omega =
[0,1]\times[0,1]$ with $\eps=0.02$ fixed.  The four spots are
centered at the corners of a square symmetrically placed inside
$\Omega$ as shown in Fig.~\ref{fig:k4_squ_drift}. For this
configuration, the Green's matrix ${\cal G}$ is circulant symmetric,
so that there is a solution to \eqref{3:ASmatrix} for which the spots
have a common source strength $S_c$. From \eqref{3:circscalar}, this
common source strength $S_c$ satisfies the nonlinear algebraic
equation
\begin{equation}
  \ac = S_c + 2\pi \nu \theta S_c + \nu \chi(S_c) \,, \qquad A =
   {\mathcal{A} \eps/(\nu \sqrt{D})} \,, \qquad \nu\equiv
   {-1/\ln\eps} \,, \qquad \theta \equiv
   R_{1,1} + G_{1,2} + G_{1,3} + G_{1,4} \,. \label{3:circsquare}
\end{equation}
Numerical values for $R_{11}$ and $G_{1j}$ for $j=2,\ldots,4$ are
obtained from the reduced-wave Green's function for the unit square as
given in \eqref{3:squaregreen} of Appendix A.

\begin{figure}[htbp]
\begin{center}
\includegraphics[ width=3.0in, height=1.8in]{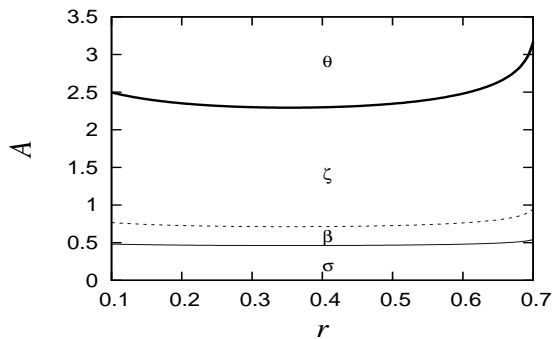}
\caption{{\em For $\eps=0.02$, $k=4$, $D=1$, we plot the phase
diagram $A$ vs.~$r$ for a four-spot quasi-equilibrium pattern in a
square with spots centered along the diagonals of the unit square at a
distance $r$ from the midpoint of the square.}}
\label{fig:k4_squ_phase}
\end{center}
\end{figure}

We let $r$ denote the distance from each spot to the center
$(0.5,0.5)$ of the square, so that $r \in (0, 1/\sqrt{2})$.  For
$D=1$, in Fig.~\ref{fig:k4_squ_phase} we plot the phase diagram of $A$
versus $r$ for $0.1 \leq r \leq 0.7$. In this phase diagram, the spot
self-replication threshold is obtained by setting
$\chi\left(\Sigma_2\right)\approx -1.79$ and $\Sigma_2\approx 4.31$ in
(\ref{3:circsquare}). The existence threshold for the four-spot
quasi-equilibrium pattern is obtained by determining, for each fixed
$r$, the minimum point of the curve $A$ versus $S_c$ from
(\ref{3:circsquare}). Finally, as similar to the case of a two-spot
pattern on the infinite plane as studied in \S \ref{sec:2inf}, the
competition instability threshold in Fig.~\ref{fig:k4_squ_phase} is
set by the sign-fluctuating eigenvector. Therefore, from
(\ref{3:newres}) this threshold of $A$ versus $r$ is obtained by
numerically solving
\begin{equation}
    \chi^{\p}(S_c) + 2\pi \sum_{m=1}^{4} (-1)^{m-1} a_m = -\nu^{-1} \,;
 \qquad \quad a_1=R_{11} \,; \quad a_j= G_{1j} \,, \quad j=2,\ldots,4 \,,
  \label{3:newresk_sq}
\end{equation}
together with (\ref{3:circsquare}).  Recall that a plot of
$\chi^{\p}(S_c)$, as needed in \eqref{3:newresk_sq}, was shown in
Fig.~\ref{fig:core:c}.

The phase diagram of Fig.~\ref{fig:k4_squ_phase} consists of four
distinct parameter regimes. In Regime $\sigma$ the four-spot
quasi-equilibrium solution does not exist. In Regime $\beta$ the
quasi-equilibrium solution exists but is unstable to a competition
instability. In Regime $\zeta$ the solution is unstable to an
oscillatory profile instability when $\tau$ exceeds a Hopf bifurcation
threshold $\tau_H$. In Regime $\theta$ the solution is unstable to
spot self-replication.

\begin{figure}[htbp]
\centering
\subfigure[$\tau_H$ vs.~$r$]
{\label{fig:k4_squ_D1_Hopf:a} 
\includegraphics[ width=3.0in, height=1.8in]{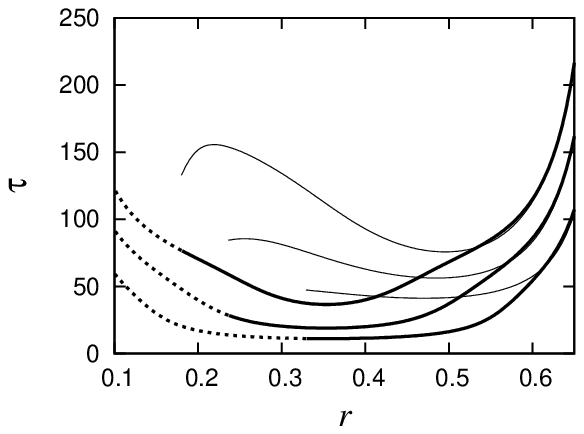}}
\subfigure[$\mbox{Im}\lambda$ vs.~$r$]
{\label{fig:k4_squ_D1_Hopf:b} 
\includegraphics[ width=3.0in, height=1.8in]{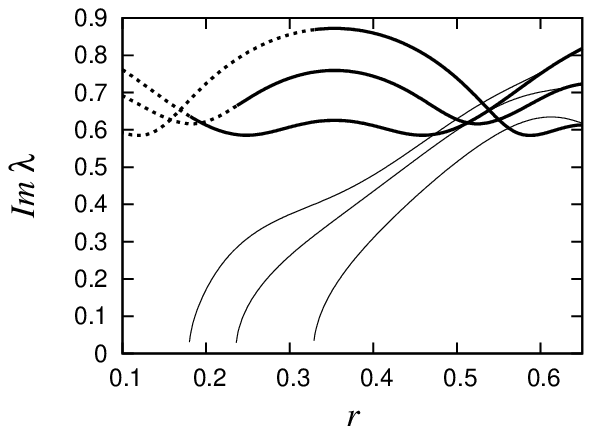}}
\caption{{\em Fix $\eps=0.02$, $k=4$, and $D=1$. For a few values
of $A$ we plot the Hopf bifurcation threshold $\tau_H$ (left figure)
and pure imaginary eigenvalue $\lambda$ (right figure) for a four-spot
pattern in the unit square. The in-phase oscillation eigenvector
$(1,1,1,1)^{T}$ corresponds to the heavy solid curves, while the
out-of-phase oscillation eigenvector $(1,-1,1,-1)$ corresponds to the
thin solid curves.  From top to bottom, the three heavy solid curves
at $r=0.65$ correspond to $A=1.0, 0.9, 0.8$.}}
\label{fig:K4_squ_D1_Hopf}
\end{figure}

\begin{figure}[htbp]
\begin{center}
\includegraphics[width=7.0in]{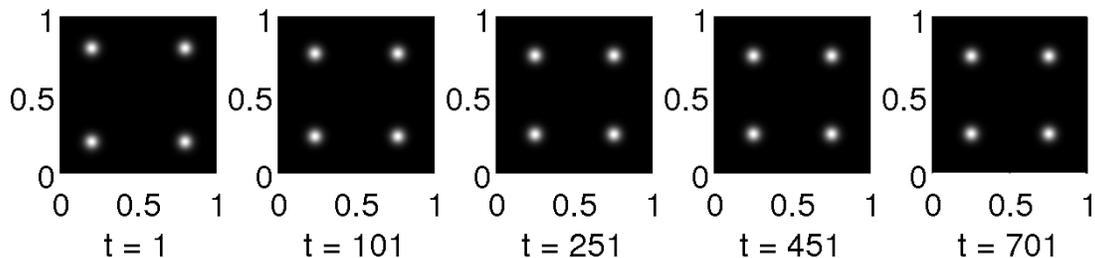}
\caption{{\em Experiment 6.1: Fix $\eps=0.02$, $k=4$, $D=1$,
$A=1.0$, and $\tau=1$. Starting from $r(0)={3/(5\sqrt{2})}$, we
observe the slow drift of the four spots towards their equilibrium
locations within the unit square. At $t=701$, they are at
(approximately) $(0.75, 0.75)$, $(0.25,0.75)$, $(0.25,0.25)$ and
$(0.75, 0.25)$, namely $r=0.5/\sqrt{2}$. }}
\label{fig:k4_squ_drift} 
\end{center}
\end{figure}

Next, we calculate the Hopf bifurcation threshold $\tau_H$ and the
pure imaginary eigenvalue $\mbox{Im}\lambda$ corresponding to an
oscillatory profile instability. For $D=1$, and for a few values of
$A$, in Fig.~\ref{fig:k4_squ_D1_Hopf:a} we plot $\tau_H$ vs.~$r$
associated with the in-phase oscillation eigenvector $(1,1,1,1)^{T}$
(heavy solid curves) and with the out-of-phase oscillation eigenvector
$(1,-1,1,-1)$ ( thin solid curves). The dotted portions of the heavy
solid curves correspond to where a competition instability occurs.
From Fig.~\ref{fig:k4_squ_D1_Hopf:a}, we observe that $\tau_H$ is set
by the synchronous oscillatory instability and that, at each fixed
$r$, $\tau_H$ increases with $A$. In Fig.~\ref{fig:k4_squ_D1_Hopf:b},
we plot $\mbox{Im}\lambda$ vs.~$r$ at the Hopf threshold. From this
figure, we observe that as $r$ decreases, the pure imaginary
eigenvalue for the out-of-phase oscillation eigenvector
$(1,-1,1,-1)^{T}$ (thin solid curve) approaches zero, which implies
the onset of a competition instability with eigenvalue $\lambda = 0$.

We now perform a few numerical experiments to test the predictions of the
asymptotic theory. In Experiments 6.1--6.3 we fix $D=1$ and $\eps=0.02$, 
and we vary $A$, $\tau$, and the initial distance $r(0)$ of the spots
from the midpoint of the square. In these experiments we will focus on
competition and oscillatory instabilities of the four-spot pattern.

\vspace*{0.1cm}\noindent{\em {\underline{Experiment 6.1:}}\; (Slow
Drift of the Spots):} We fix $r(0)={3/(5\sqrt{2})}$ and $A=1$,
corresponding to Regime $\zeta$ of the phase diagram in
Fig.~\ref{fig:k4_squ_phase}. We set $\tau=1$, which is below the
numerically computed Hopf bifurcation threshold of $\tau_H= 44.8$ at
$r(0)$. Thus, the initial four-spot pattern is predicted to be stable
to all three instabilities. From the dynamics in Principal Result 3.1
we can verify that the four spots drift slowly towards the midpoint of
the square, and reach an equilibrium state with $r\approx
{1/(2\sqrt{2})}$ for $t\gg 1$. For $\tau=1$, there is no
dynamically-triggered oscillatory profile instability.  The full
numerical results in Fig.~\ref{fig:k4_squ_drift} confirm this prediction
of the asymptotic theory.

\begin{figure}[htbp]
\begin{center}
\includegraphics[width=7.0in]{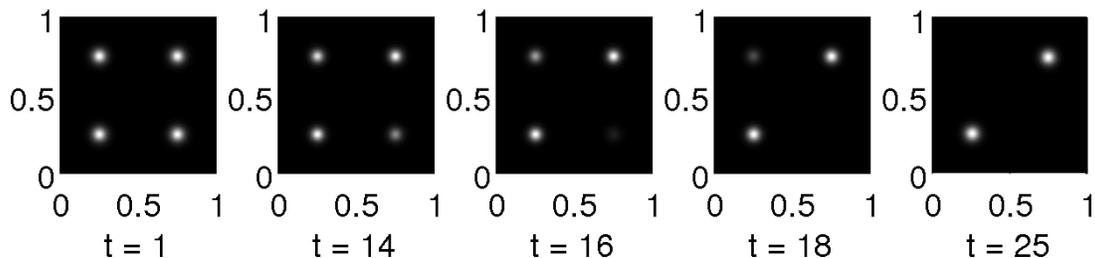}
\caption{{\em Experiment 6.2: Fix $\eps=0.02$, $D=1$,
$A=0.6$, and $\tau=1$. This plot shows a competition instability for the
four-spot pattern with spots initially located at a distance
$r={1/(2\sqrt{2})}$ from the midpoint of the square.}}
\label{fig:k4_squ_compet:a} 
\end{center}
\end{figure}

\begin{figure}[htbp]
\centering
\includegraphics[ width=3.0in,height=1.8in]{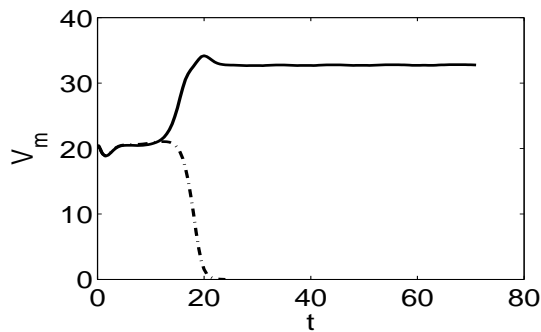}
\caption{{\em Experiment 6.2: Fix $\eps=0.02$,
$r(0)={1/(2\sqrt{2})}$, $D=1$, $A=0.6$, and $\tau=1$. We plot the
amplitude of the two spots at $(0.25,0.25)$ (heavy solid curve) and
$(0.25,0.75)$ (dashed curve).}} \label{fig:k4_squ_compet:b} 
\end{figure}

\vspace*{0.1cm}\noindent{\em {\underline{Experiment 6.2:}}\; ( A
Competition Instability):} Let $r={1/(2\sqrt{2})}$, corresponding
(roughly) to the equilibrium distance from the midpoint for the
four-spot pattern. We choose $A=0.6$ corresponding to Regime $\beta$
in the phase diagram of Fig.~\ref{fig:k4_squ_phase}. In addition, we
take $\tau=1$ for which $\tau<\tau_H$. Therefore, we predict that the
initial four-spot pattern is unstable to a competition instability
This is confirmed by the full numerical results computed from the GS
model (\ref{1:GS_2D}) shown in Fig.~\ref {fig:k4_squ_compet:a}. In
this figure, we observe that the two spots at $(0.25, 0.75)$ and
$(0.75, 0.25)$ decay very fast, and that after $t=20$, these two spots
are annihilated.  The spot amplitudes as a function of $t$ are shown
in Fig.~\ref{fig:k4_squ_compet:b}.

\begin{figure}[htbp]
\centering
\subfigure[$\tau=10$]
{\includegraphics[ width=2.2in,height=1.8in]{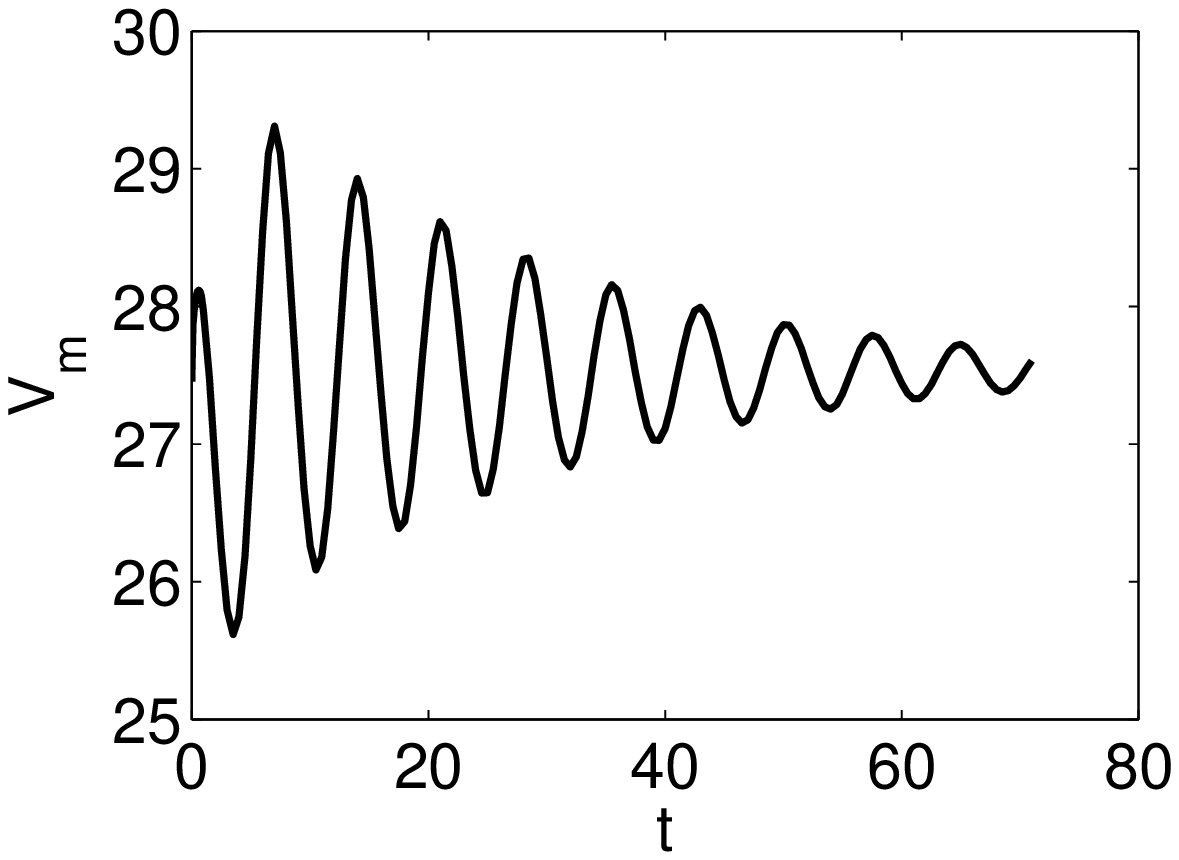}}
\subfigure[$\tau=11$]
{\includegraphics[ width=2.2in,height=1.8in]{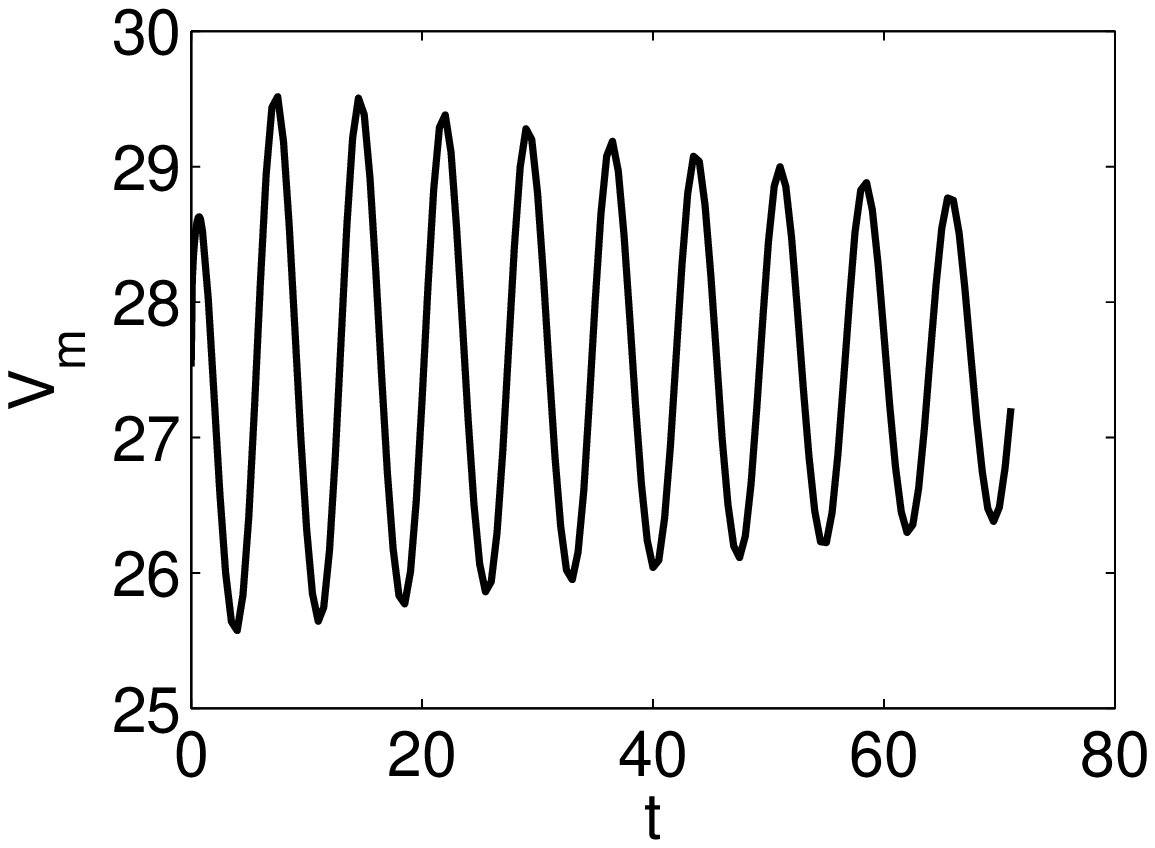}}
\subfigure[$\tau=12$]
{\includegraphics[ width=2.2in,height=1.8in]{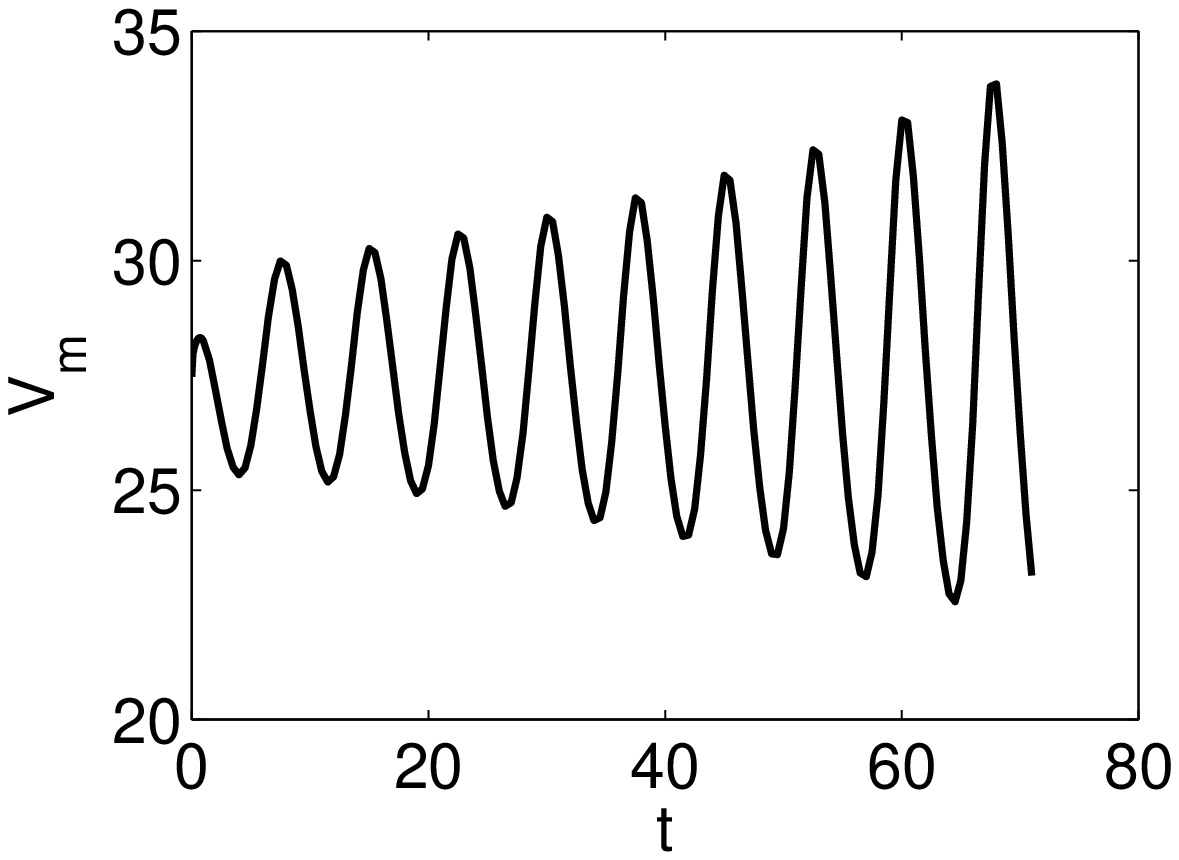}}
\caption{{\em Experiment 6.3: Fix $\eps=0.02$, $r(0)={1/(2\sqrt{2})}$,
$D=1$, and $A=0.8$. For the initial four-spot pattern in a square, we
plot the numerically computed spot amplitude $v_m=v_m(t)$ for the spot
at $(0.75, 0.75)$ for $\tau = 10, 11, 12$. When $\tau=12 > \tau_H$,
the oscillation is unstable.}}
\label{fig:k4_squ_prof} 
\end{figure}

\vspace*{0.1cm}\noindent{\em {\underline{Experiment 6.3:}}\; (An
Oscillatory Instability):} We fix $r={1/(2\sqrt{2})}$ at $t=0$ and
choose $A=0.8$. This choice corresponds to Regime $\zeta$ in
Fig.~\ref{fig:k4_squ_phase}. With these parameter values, the
numerically computed Hopf bifurcation threshold predicted by the
asymptotic theory is $\tau_H = 11.0$ when $t=0$. To confirm this
prediction, we compute full numerical solutions to the GS model
(\ref{1:GS_2D}) for $\tau=10, 11, 12$. For these three choices of
$\tau$, in Fig.~\ref{fig:k4_squ_prof} we plot the numerically computed
amplitude $v_m=v_m(t)$ of the spot at $(0.75, 0.75)$. Our full
numerical computations show that all the spot amplitudes coincide with
the one in Fig.~\ref{fig:k4_squ_prof}. This synchronous oscillatory
instability was predicted by the asymptotic theory. Our numerical
results show that an unstable oscillation develops on the range
$11<\tau<12$, which was closely predicted by our asymptotic theory.
Our numerical results show that this unstable oscillation 
leads to the annihilation of the spot (not shown). As a result, we
conjecture that the Hopf bifurcation is subcritical.

\subsection{A One-Spot Pattern in the Unit Disk} \label{sec:sym_cir}

Let $\Omega$ be the unit disk, and consider a one-spot
quasi-equilibrium solution centered at $\mathbf{x}_1\in \Omega$, with
$r\equiv |\mathbf{x}_1|$. From \eqref{3:circscalar} and
\eqref{3:pval}, the source strength $S$ for this spot is determined in
terms of $A$ by
\begin{equation}
   \ac = S + 2\pi \nu R_{1,1} S + \nu \chi(S) \,, \qquad
   A = {\mathcal{A} \eps/(\nu \sqrt{D})} \,, \qquad 
  \nu\equiv {-1/\ln\eps} \,. \label{5:onespot}
\end{equation}
Here the regular part $R_{1,1}\equiv R(\mathbf{x}_1;\mathbf{x}_1)$ of
$G$, which depends on $r=|\mathbf{x}_1|$, is calculated from
\eqref{3:simpleG} of Appendix A. For this case, the spot dynamics from
Principal Result 3.1 shows that the spot will drift with speed
${\mathcal O}(\eps^2)$ along a ray towards the center of the disk.

\begin{figure}[htbp]
 \includegraphics[height=1.8in, width=3.0in]{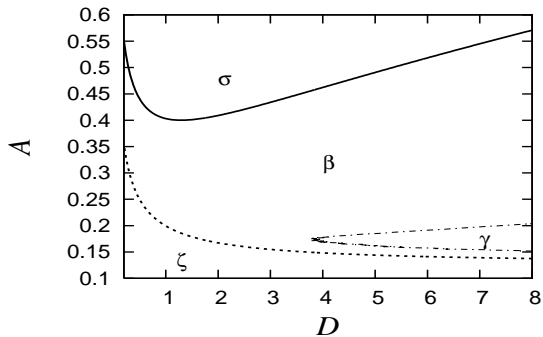}
 \caption{{\em  Plot of the phase diagram $A$ vs.~$D$ for $\eps=0.02$
 for a one-spot solution centered at the origin in the unit disk. It is
 only in Regime $\sigma$ that spot self-replication can occur. The
 solid curve is the spot self-replication threshold $A_{s}(D)$ of
 \eqref{3:onespot_disk}, while the lower dotted curve is the existence
 threshold $A_{f}(D)$. In Regime $\zeta$ a one-spot solution
 does not exist. In Regimes $\beta$ and $\gamma$ an oscillatory
 instability for a spot at the origin occurs only if
 $\tau>\tau_H(0)$. In Regime $\gamma$, $\tau_{H}^{\p\p}(0)>0$, while in
 Regime $\beta$, $\tau_H^{\p\p}(0)<0$, where $\tau_H=\tau_H(r)$. Thus,
 only in Regime $\gamma$ can we find a value of $\tau$ and an initial
 spot location for which a dynamical oscillatory instability is triggered
 for a spot that slowly drifts to the center of the unit disk.}}
\label{fig:disk_phase}
\end{figure}

For a one-spot solution centered at the origin, the spot
self-replication threshold $A=A_{s}(D)$ is given by
\begin{equation}
   A_{s}(D) = \frac{\eps}{\nu \sqrt{D}} \left[ \Sigma_2 \left(1 + 2\pi \nu
  R_{1,1} \right) + \nu \chi\left(\Sigma_2\right) \right] \,, 
  \label{3:onespot_disk}
\end{equation}
where $\chi\left(\Sigma_2\right)\approx -1.79$ and $\Sigma_2\approx
4.31$. Here $R_{1,1}\equiv R_{1,1}(\mathbf{0};\mathbf{0})$ is given
explicitly in (\ref{3:disk_r}).  Moreover, the existence threshold for
a one-spot solution centered at the origin is obtained by replacing
$\Sigma_2$ in (\ref{3:onespot_disk}) with $S$, and then computing the
minimum value, denoted by $A_f=A_f(D)$, of the resulting expression
with respect to $S$. The self-replication and existence thresholds for
a one-spot solution centered at the origin are shown in the phase
diagram of Fig.~\ref{fig:disk_phase}.

\begin{figure}[htbp]
\centering \subfigure[$\tau_H$ vs.~$r$ for $A=0.16$]{\label{fig:k1cirHopf:a}
  \includegraphics[width=3.0in,height=1.8in]{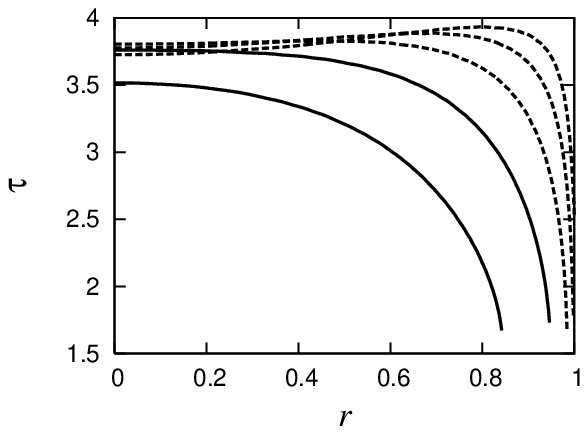} }
  \subfigure[$\tau_H$ vs.~$r$ for $A=0.18$] {\label{fig:k1cirHopf:b}
    \includegraphics[width=3.0in,height=1.8in]{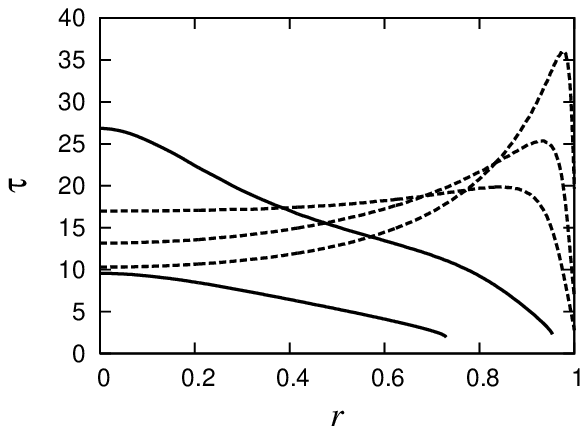} }
\caption[One-spot solution in the unit disk: dynamical profile
instability]{{\em A One-spot solution in the unit disk at a distance
$r=|\mathbf{x}_1|$ from the origin. Plots of $\tau_H$ vs.~$r$ for
different $A$ and $D$. (a) Fix $A=0.16$. The solid curves are for
$D=4, 5$, arranged from lower to upper $y-$intercepts,
respectively. The dashed curves are for $D=6, 7, 8$, arranged from
upper to lower $y$-intercepts, respectively. (b) Fix $A=0.18$.  The
solid curves are for $D=2,3$ arranged from lower to upper
$y-$intercepts, respectively. The dashed curves are for $D=4.3, 5, 6$,
arranged from upper to lower $y-$intercepts, respectively. }}
\label{fig:k1cirHopf}
\end{figure}

Next, we study oscillatory instabilities of the spot profile.  For
each $r=|\mathbf{x}_1|$ in $0<r<1$, we suppose that $A>A_{f}(D,r)$,
where $A_f=A_{f}(D,r)$ is the existence threshold for a one-spot
quasi-equilibrium solution centered at $\mathbf{x}_1$. In
Fig.~\ref{fig:k1cirHopf:a}, we plot the numerically computed Hopf
bifurcation threshold $\tau_H$ versus $r$ for fixed $A=0.16$, but for
different values of $D$. The computations were done using the global
eigenvalue problem of Principal Result 4.3. For $D= 4$ and $D=5$ we
observe from Fig.~\ref{fig:k1cirHopf:a} that the maximum of
$\tau_H=\tau_H(r)$ is obtained at the equilibrium location $r=0$, and
that $\tau_H$ decreases monotonically as $r$ increases. In contrast,
for $D=6, 7, 8$ Fig.~\ref{fig:k1cirHopf:a} shows that the curves of
$\tau_H=\tau_H(r)$ are convex near $r=0$, so that $\tau_{H}$ has a
local minimum at $r=0$. Therefore, when $D$ is sufficiently large, we
conclude that we can obtain a dynamically-triggered oscillatory
instability in the spot amplitude. To illustrate this, suppose that
$D=8$. Then, we calculate $\tau_H(0) \approx 3.73$ and $\tau_H(0.612)
\approx 3.88$.  Suppose that we take $\tau=3.8$ with the initial spot
location at time $t=0$ at $r=0.612$. Then, since
$\tau<\tau_{H}(0.612)$, the spot is stable at $t=0$. However, since
the motion of the spot is towards the origin and $\tau>\tau_{H}(0)$,
it follows that a dynamically-triggered oscillatory profile
instability will occur before the spot reaches the center of the disk.
A qualitatively similar scenario occurs for other values of $A$. In
particular, for $A=0.18$, in Fig.~\ref{fig:k1cirHopf:b} we plot
$\tau_H$ versus $r$ for various fixed values of $D$. For the values
$D=4.3, 5, 6$, we observe that $\tau_H(r)$ has a local minimum at the
equilibrium location $r=0$. Thus, for these values of $D$, we can
choose a value of $\tau$ and an initial spot location that will lead
to a dynamically-triggered oscillatory instability.

\begin{figure}[htbp]
 \centering 
\subfigure[$A=0.16$]{\includegraphics[width=2.3in, height
 = 1.3in]{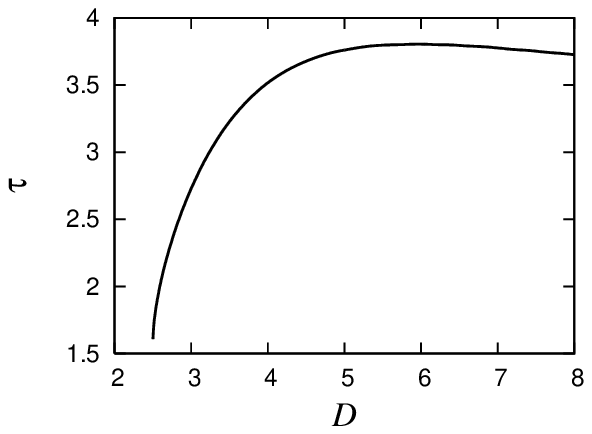}}
\subfigure[$A=0.17$]{\includegraphics[width=2.3in, height =
 1.3in]{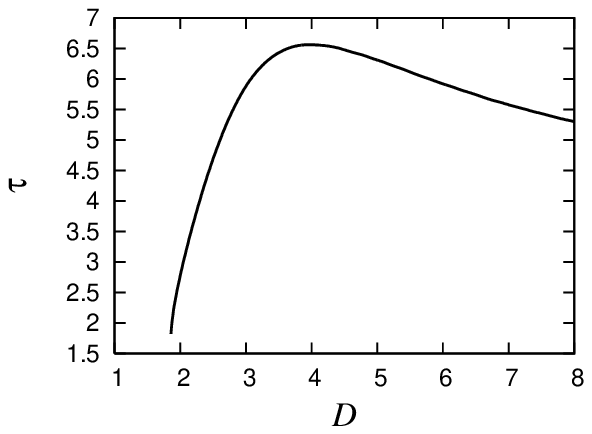}}
 \subfigure[$A=0.18$]{\includegraphics[width=2.3in, height =
 1.3in]{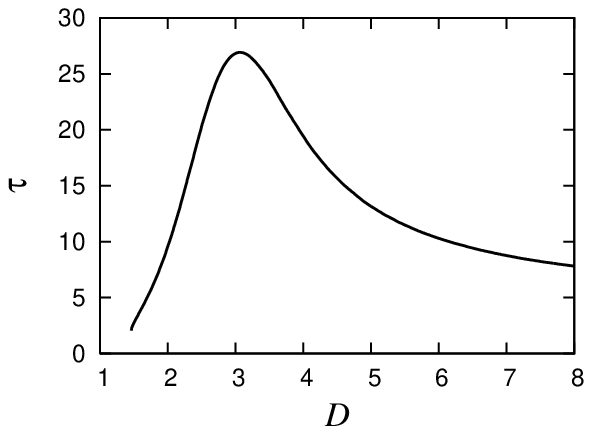}}\\
 \subfigure[$A=0.19$]{\includegraphics[width=2.3in, height =
 1.3in]{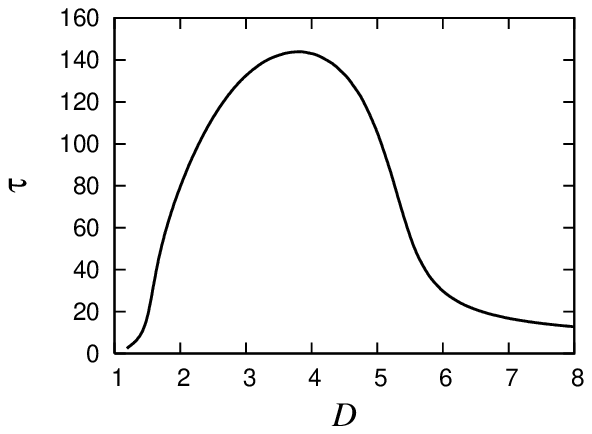}}
 \subfigure[$A=0.2$]{\includegraphics[width=2.3in, height =
 1.3in]{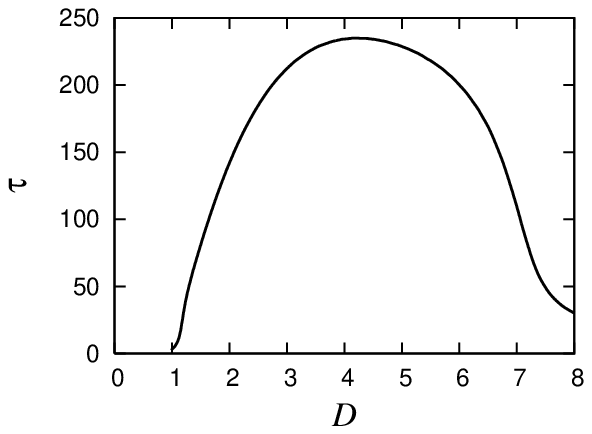}}
 \subfigure[$A=0.21$]{\includegraphics[width=2.3in, height =
 1.3in]{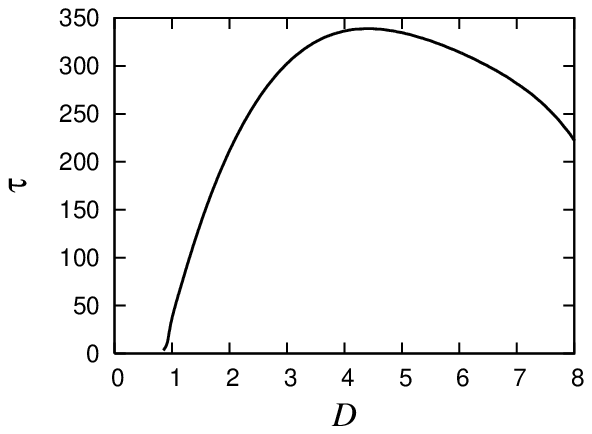}}
 \caption{{\em One-spot solution centered at the origin
 in the unit disk.  The Hopf bifurcation threshold $\tau_{H}$ vs.~$D$
 is plotted for $A=0.16,\ldots,0.21$. This shows that
 $\tau_{H}(D)$ has a maximum at some critical value of $D$.}}
 \label{fig:DynProf_Dtau}
\end{figure}

For a spot centered at the origin of the unit disk, and with
$\eps=0.02$, in Fig.~\ref{fig:DynProf_Dtau} we plot the Hopf
bifurcation threshold $\tau_{H}$ vs.~$D$ for $A=0.16,\ldots,0.21$.  We
observe that each curve is not monotone in $D$, and has a local
maximum at some value of $D$, with $\tau_H$ decreasing as $D$
increases.

In Fig.~\ref{fig:disk_phase} we plot a phase diagram in the $A$ versus
$D$ parameter plane for $\eps=0.02$ for a one-spot solution centered
at the origin in the unit disk. From this figure, there are four
distinct regions with different solution behavior. In Regime $\sigma$
the spot at the origin will undergo self-replication. In Regime
$\zeta$ there is no one-spot solution centered at the origin.  In
Regime $\beta$ the one-spot solution undergoes a Hopf bifurcation if
$\tau$ is large enough and the Hopf bifurcation threshold
$\tau_H=\tau_H(r)$, shown in Fig.~\ref{fig:k1cirHopf}, satisfies
$\tau_{H}^{\p\p}(0)<0$, and so has a local maximum at $r=0$. Finally,
in Regime $\gamma$, the one-spot solution has a Hopf bifurcation if
$\tau$ is large enough, but now the Hopf bifurcation threshold
satisfies $\tau_{H}^{\p\p}(0)>0$, and so has a local minimum at
$r=0$. Therefore, in Regime $\gamma$ one can choose an initial spot
location inside the unit disk, together with a value of $\tau$, for
which a dynamic oscillatory instability of a one-spot
quasi-equilibrium solution will be triggered as the spot drifts
towards its equilibrium location at the center of the unit disk.

We now argue that the solid curve $A=A_s(D)$ in
Fig.~\ref{fig:disk_phase} not only corresponds to the threshold
condition for the self-replication of a spot centered at the origin,
but also corresponds to the threshold for a dynamically-triggered spot
self-replication event for a one-spot pattern in the unit disk
starting from the initial spot location at radius $r=|\mathbf{x}_1|$
with $0<r<1$. To show this, we differentiate \eqref{5:onespot} with
respect to $r$ to obtain that ${dS/dr}=-2\pi\nu S \left( {d
R_{1,1}/dr} \right) + {\mathcal O}(\nu^2)$. It is readily verified
numerically by using \eqref{3:simpleG} of Appendix A that $R_{1,1}$
has a global minimum at $r=0$ with $R_{1,1}$ an increasing function of
$r$ on $0<r<1$, and $R_{1,1}\to +\infty$ as $r\to 1^{-}$. Therefore,
we have ${dS/dr}<0$ on $0<r<1$, so that $S$ is a
monotone decreasing function of $r$ for any fixed values of $A$ and
$D$. Moreover, by differentiating \eqref{5:onespot}, we readily
observe that $S$ is a monotone increasing function of $A$ at each
fixed $r$ and $D$. Therefore, it follows that the threshold curve for
a dynamically-triggered spot self-replication event is obtained by
setting $S = \Sigma_2$ and $r=0$ in \eqref{5:onespot} to obtain
$A_{s}(D)$ as given in \eqref{3:onespot_disk}.
Therefore, in Regime $\sigma$ of Fig.~\ref{fig:disk_phase} it follows
that we can choose an initial spot location and a value of $A$ for
which a dynamic spot self-replication instability will occur for a
spot which slowly drifts towards the center of the unit disk.

\begin{figure}[htbp]
\centering \subfigure[$\tau=3.2$] { \label{fig:k1cirNum1:a}
  \includegraphics[width=2.2in,height=1.3in]{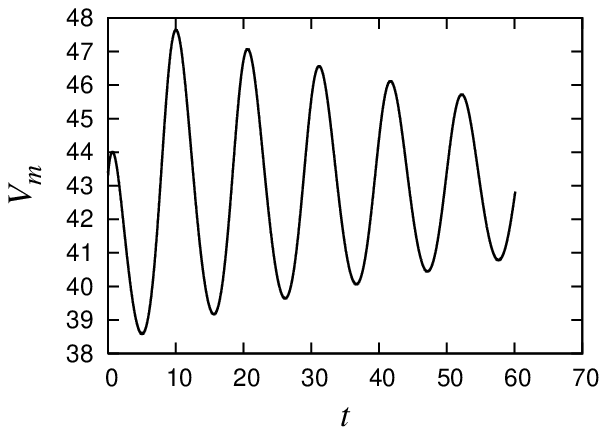}}
\subfigure[$\tau=3.3$] { \label{fig:k1cirNum1:b}
    \includegraphics[width=2.2in,height=1.3in]{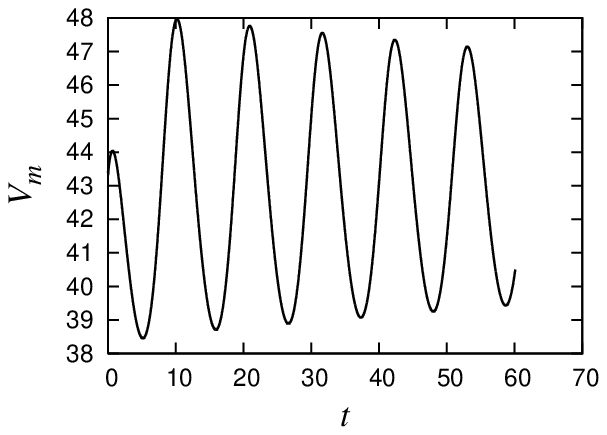}}
\subfigure[$\tau=3.4$] { \label{fig:k1cirNum1:c}
    \includegraphics[width=2.2in,height=1.3in]{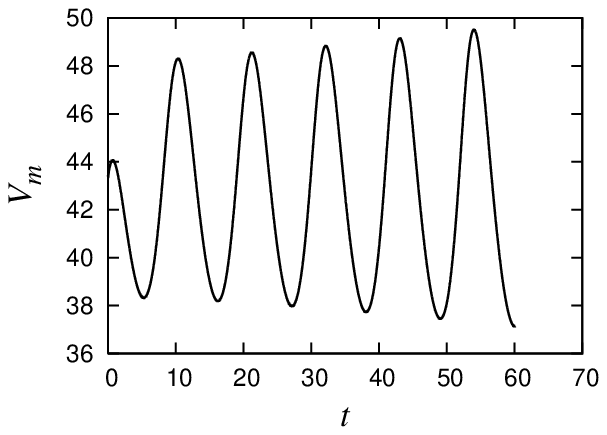}} 
\subfigure[$\tau=9.0$] { \label{fig:k1cirNum2:a}
  \includegraphics[width=2.2in,height=1.3in]{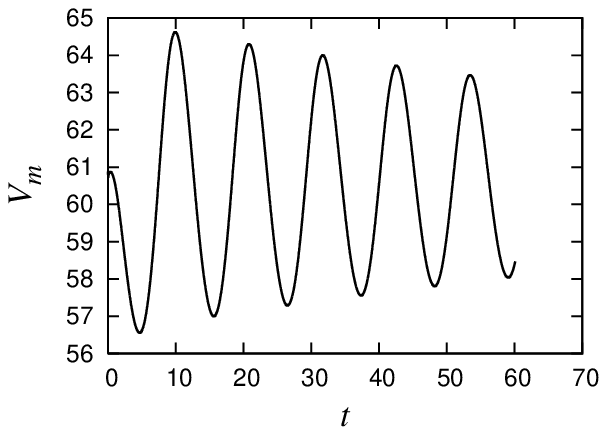}}
\subfigure[$\tau=9.5$] {\label{fig:k1cirNum2:b}
    \includegraphics[width=2.2in,height=1.3in]{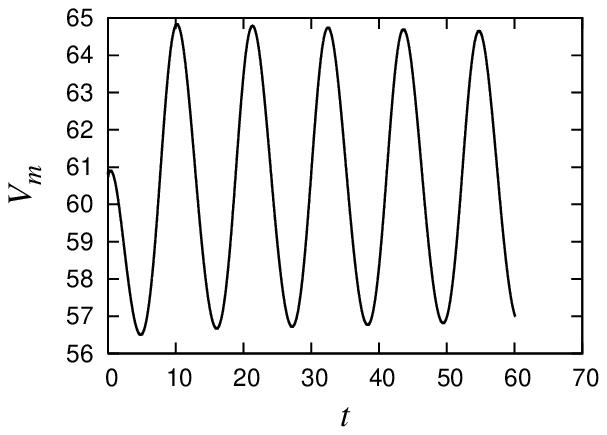}}
\subfigure[$\tau=10.0$] {\label{fig:k1cirNum2:c}
    \includegraphics[width=2.2in,height=1.3in]{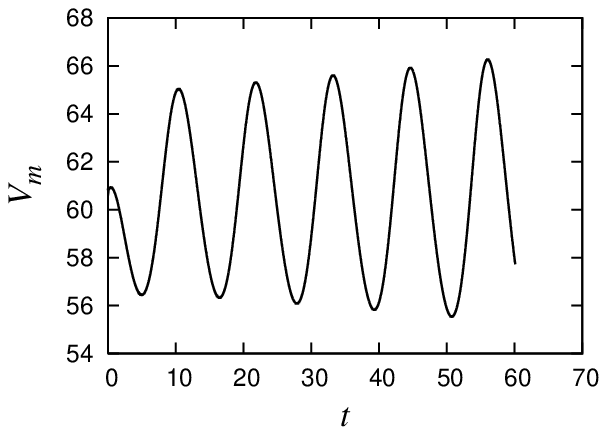}}
\caption{{\em Experiment 6.4: One-spot solution with spot at the center of the
unit disk. We fix $\eps=0.02$. Top Row: For $A=0.16$ and $D=4$, the
spot amplitude $v_m$ vs.~$t$ is plotted for (a) $\tau=3.2$; (b) $\tau=3.3$; 
(c) $\tau=3.4$. Bottom Row: For $A=0.18$ and $D=6$, the spot amplitude 
$v_m$ vs.~$t$ is plotted for (d) $\tau=9.0$; (e) $\tau=9.5$; 
(f) $\tau=10.0$. }}
\label{fig:k1cirNum12}
\end{figure}

\vspace*{0.1cm}\noindent{\em {\underline{Experiment 6.4:}}\; (One-spot
solution in the unit disk: Comparison with full numerical
simulations):} For a one-spot solution centered at the origin of the
unit disk we now compare our numerically computed Hopf bifurcation
threshold, as obtained from our global eigenvalue problem, with full
numerical results as computed from the full GS model
\eqref{1:GS_2D}. For three values of $\tau$, in the top row of
Fig.~\ref{fig:k1cirNum12}, we plot $v_m\equiv v(0,t)$ versus $t$ for
the GS parameters $\eps=0.02$, $A=0.16$, and $D=4$. This figure shows
that an unstable oscillation occurs near $\tau=3.4$, which agrees
rather closely with the Hopf bifurcation threshold $\tau_H \approx
3.5$ computed from our global eigenvalue formulation of Principal
Result 4.3 (see Fig.~\ref{fig:k1cirHopf:a}). We remark that if we
changed the parameters to $A=0.18$ and $D=6$, then $\tau_H \approx
10.3$ (see Fig.~\ref{fig:k1cirHopf:b}), as predicted by our global
eigenvalue problem. This compares rather closely with the result
$\tau\approx 10$ computed from the full GS model \eqref{1:GS_2D}
observed in the bottom row of Fig.~\ref{fig:k1cirNum12}.

\subsection{A One-Ring Pattern of $K$-Spots in the Unit Disk} 
\label{sec:sym_ring}

Next, we consider the special case when $k$ spots are equally-spaced
on a ring of radius $r$ centered at the origin of the unit
disk. Representing points as complex numbers in the unit disk, the
centers of the spots are at
\begin{equation}
 \mathbf{x}_j = r e^{2 \pi i j / k},\quad j=1,\ldots,k\,, \qquad
i\equiv\sqrt{-1} \,. \label{6:rpatt}
\end{equation}
For this symmetric arrangement of spots, the matrix ${\mathcal G}$ in
\eqref{3:ASmatrix} is circulant symmetric and is a function of the
ring radius $r$. Hence, $\mathcal{G} \mathbf{e} = \theta \mathbf{e}$
with $\mathbf{e}=(1,\ldots,1)^T$, where $\theta={p_k(r)/k}$ with
$p_{k}(r) \equiv\sum_{i=1}^{k}\sum_{j=1}^{k} {\mathcal G}_{i,j}$.  For
$D\gg 1$, we can analytically determine a two-term expansion for
$p_{k}(r)$ by using \eqref{3:G2GN}, the simple explicit form
\eqref{3:Neumannxy} for the Neumann Green's function, and Proposition
4.3 of \cite{trap_KTW:2005}. In this way, for $D\gg 1$, we get
\begin{equation}
  p_{k}(r) \equiv\sum_{i=1}^{k}\sum_{j=1}^{k} {\mathcal G}_{i,j} \sim
  \frac{k^2 D}{|\Omega|} + 
 \sum_{i=1}^{k}\sum_{j=1}^{k} {\mathcal G}_{i,j}^{(N)}
  = \frac{\, k^2 D}{\pi} - \frac{k}{\,2 \pi}
\left[\ln(k r^{k-1}) + \ln(1- r^{2k}) - r^2 k + \frac{3k}{4} 
 \right] + {\mathcal O}\left( D^{-1}\right) \,. 
\label{3:pk_dlarge}
\end{equation}

Since $\mathcal{G}$ is circulant, \eqref{3:ASmatrix} has a solution
for which the source strengths $S_j$ for $j=1,\ldots,k$ have a common
value $S_c$. From \eqref{3:circscalar} of \S \ref{sec:quasi}, it
follows that $S_c$ satisfies the scalar nonlinear algebraic equation
\begin{equation}
  \ac =  S_c \left( 1+\frac{\,2 \pi \nu }{\,k} p_k(r) \right) +\nu
\chi(S_c) \,, \qquad A = \frac{\eps}{\nu\sqrt{D}} {\cal A} \,, \quad
  \nu = \frac{-1}{\ln\eps} \,. \label{3:ring_sc}
\end{equation}
To determine the dynamics of the $k$ spots, we calculate
$\mathbf{f}_j$ in \eqref{3:dynbc} as
$\mathbf{f}_j = \pi k^{-1} S_c p_k^{\p}(r) \,e^{2 \pi i j / k}$.
Then, the dynamics \eqref{3:dyn} of Principal Result 3.1 shows that all the
spots remain on a ring of radius $r(t)$, which satisfies the ODE
\begin{equation}
\label{3:dynsym} \frac{\,d r}{d t} \sim  - \frac{\pi \eps^2}{\, k}
    \gamma(S_c) S_c \, p_k^{\p}(r) \,.
\end{equation}
This ODE is coupled to the nonlinear algebraic equation
\eqref{3:ring_sc} for $S_c$ in terms of $r$.

The equilibrium ring radius $r_e$ for \eqref{3:dynsym} satisfies
$p_{k}^{\p}(r)=0$, and is independent of $\ac$. For a
two-spot pattern, where $k=2$, then $r_e$ is given in Table \ref{tab:equil}
for various values of $D$. There is no simple explicit formula for
$p_{k}(r)$ when $D={\mathcal O}(1)$. However, for $D\gg 1$, $p_{k}(r)$
is given asymptotically in (\ref{3:pk_dlarge}).  By differentiating
this asymptotic result, it readily follows that $r_e$, for $D\gg 1$,
is the unique root on $0<r_e<1$ of ${[k-1]/(2k)} -
r^2={r^{2k}/(1-r^{2k})}$ for any $k\geq 2$. It is also readily shown
from \eqref{3:pk_dlarge} that $p^{\p\p}_k(r_e) > 0$, with
$p_{k}^{\p}(r)<0$ for $r<r_e$ and $p_{k}^{\p}(r)>0$ for
$r>r_e$. Hence, for $D\gg 1$, $p_k(r)$ attains its global minimum
value at the equilibrium radius $r_e$.  Therefore, since
$\gamma(S_c)>0$, then $r_e$ is a stable equilibrium point for the ODE
\eqref{3:dynsym} when $D\gg 1$ and, moreover, $r\to r_e$ for any
initial point $r(0)$, with $0<r(0)<1$, as $t\to \infty$. By
differentiating \eqref{3:ring_sc} for the source strength, this
condition on $p_{k}(r)$ also implies that $S_{c}(r)$ has a maximum
value at $r=r_e$ when $D\gg 1$.

\begin{table}
\centering
\begin{tabular}{c|c||c|c}
\hline $D$ &$r_e$  &$D$ &$r_e$ \\ \hline
0.8  &0.45779 &2.0  &0.45540\\
1.0  &0.45703 &3.0  &0.45483 \\
1.3  &0.45630 &4.0  &0.45454\\
1.5  &0.45596 &5.0  &0.45436\\ \hline
\end{tabular}
\caption[Equilibrium ring radius for a two-spot symmetric pattern in the 
unit disk]
{{\em Equilibrium ring radius $r_e$ for a two-spot pattern of the form
\eqref{6:rpatt} in the unit disk for different values of $D$.}}
\label{tab:equil}
\end{table}

\begin{figure}[htbp]
\subfigure[$k=3$ spots] { \label{fig:acrit:a}
  \includegraphics[width=2.2in, height=1.8in]{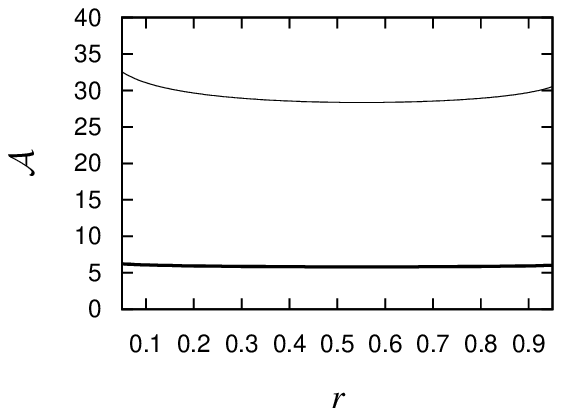}}
\subfigure[$k=4$ spots] { \label{fig:acrit:b}
  \includegraphics[width=2.2in, height=1.8in]{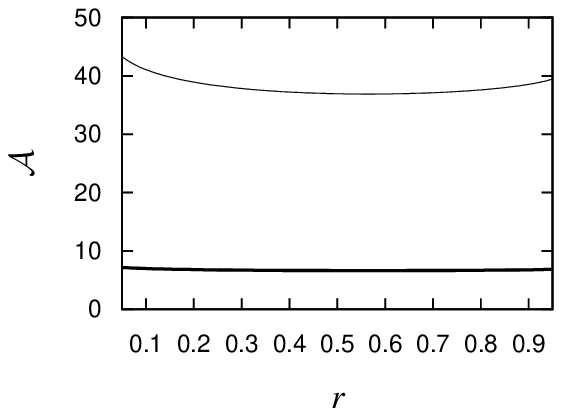}}
\subfigure[$k=5$ spots] { \label{fig:acrit:c}
  \includegraphics[width=2.2in, height=1.8in]{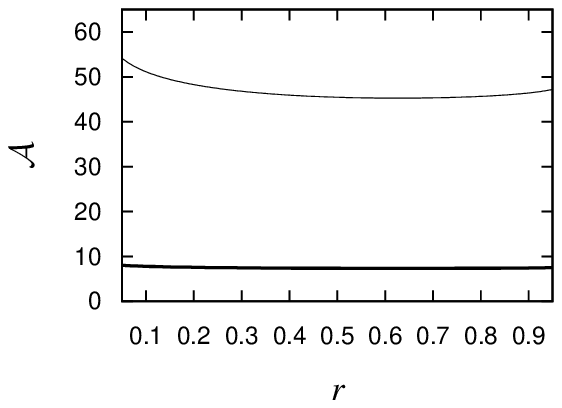}}
\caption{ {\em For $\eps=0.02$ and $D=3.912$ we plot the existence threshold 
${\mathcal A}_f$ (heavy solid curves) and the spot-splitting threshold 
${\mathcal A}_s$ (solid curves) as a function of the ring radius $r$
for either $k=3$ (left figure), $k=4$ (middle figure), or $k=5$ 
(right figure) equally spaced spots on a ring.}}
\label{fig:acrit}
\end{figure}

\vspace*{0.2cm}\noindent {\underline{{\bf Spot Self-Replication
Instabilities:}}} We first discuss spot self-replication instabilities
for equally spaced spots on a ring.  For $\eps=0.02$ and $D=3.912$, in
Fig.~\ref{fig:acrit} we plot the minimum value ${\mathcal
A}_{f}={\mathcal A}_{f}(r)$ of $\ac$ for which a quasi-equilibrium
ring pattern exists at each fixed $r$ for $k=3,4,5$ equally spaced
spots on a ring. These plots are obtained by determining the minimum
point of the ${\mathcal A}$ versus $S_c$ curve in
\eqref{3:ring_sc}. In these plots we also show the spot-splitting
threshold ${\mathcal A}_{s}={\mathcal A}_s(r)$ obtained by setting
$S_c=\Sigma_2\approx 4.31$ and $\chi(\Sigma_2)\approx -1.79$ in
\eqref{3:ring_sc}.

\begin{figure}[htpb]
\centering
\subfigure[$S_c$ vs.~$r$] {
\includegraphics[width=3.0in,height=1.8in]{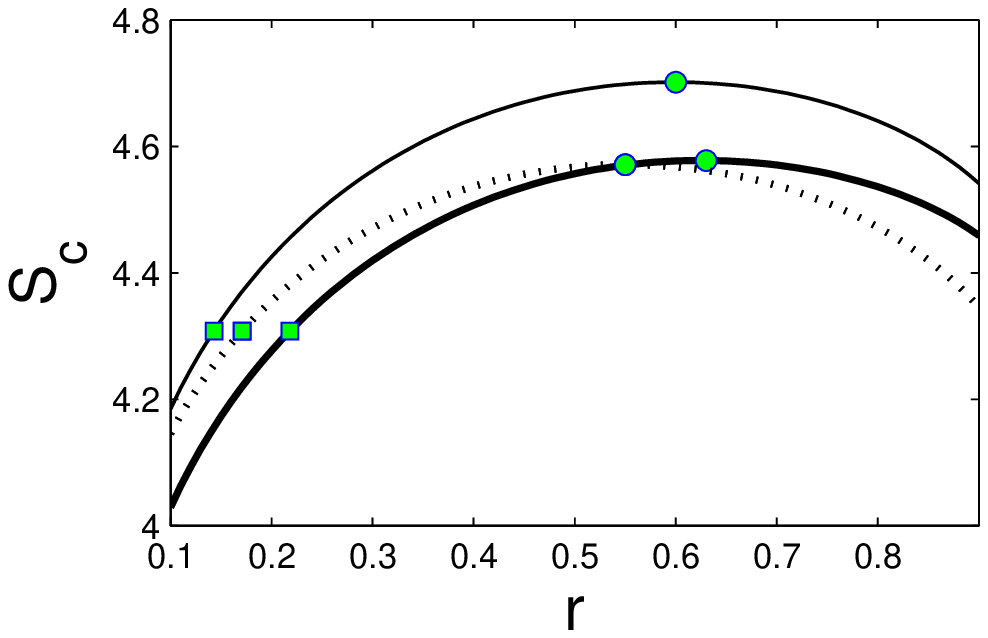}
\label{fig:1ring:a}} \subfigure[ $S_c$ vs.~$r$]{
\includegraphics[width=3.0in,height=1.8in]{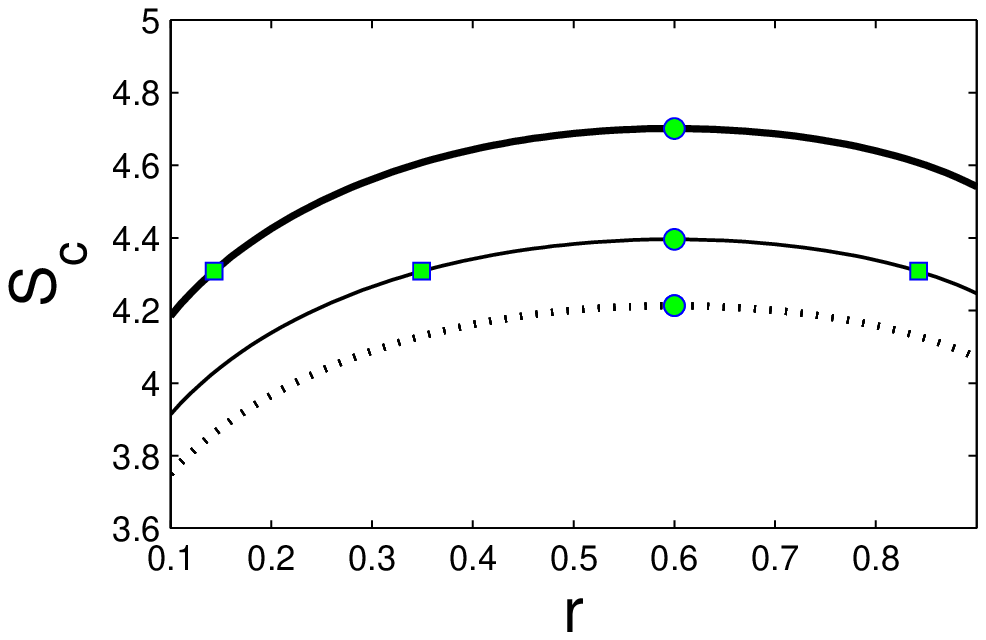}
\label{fig:1ring:b}} \caption[Symmetric pattern: all $k$ spots on a
ring in a unit circle]{{\em Fix $D=3.912$. (a) $S_c$ vs.~$r$ with
$k=3,\, \ac = 30$ (dotted curve), $k=4,\, \ac = 40$ (solid curve) and
$k=5,\, \ac = 48$ (heavy solid curve). (b) Fix $k=4$, we plot $S_c$
vs.~$r$ with $\ac=36$ (dotted curve), $\ac=37.5$ (solid curve), and
$\ac=40$ (heavy solid curve). The square dots indicate spot-splitting
thresholds $S_c=\Sigma_2\approx 4.31$, and the circular dots, where
$S_c$ achieves its maxima, denote equilibrium ring radii.}}
\label{fig:1ring}
\end{figure}

In Fig.~\ref{fig:1ring:a}, we fix $D = 3.912$ and plot $S_c$ versus
$r$ for $k=3,\, \ac =30$, for $k=4,\, \ac = 40$, and for $k=5,\, \ac =
48$. For these three patterns, the equilibrium states are,
respectively, $(r_{e1}, S_{ce1}) = (0.55, 4.57)$, $(r_{e2}, S_{ce2}) =
(0.60, 4.70)$, and $(r_{e3}, S_{ce3}) = (0.63, 4.58)$. These points
are the circular dots in Fig.~\ref{fig:1ring:a}. The square dots in
this figure correspond to the spot-replicating threshold $\Sigma_2
\approx 4.31$, which occurs at $r_{s1} = 0.17$, $r_{s2} = 0.14$, and
$r_{s3} =0.22$. Any portion of these curves above the square dots
correspond to ring radii where simultaneous spot-splitting will occur.

In Fig.~\ref{fig:1ring:b}, we fix $k=4$ and $D=3.912$ and plot $S_c$
versus $r$ for $\ac=36$, $\ac=37.5$, and $\ac=40$. The equilibria are
at $(r_{e4}, S_{ce4}) = (0.60, 4.21)$, $(r_{e5}, S_{ce5}) = (0.60,
4.40)$, and $(r_{e2}, S_{ce2}) = (0.60, 4.70)$, respectively. These
points correspond to maxima of $S_{c}(r)$. In this figure the
spot-splitting thresholds are the square dots.  For $\ac=36$ and
$k=4$, we observe that since $S_{ce4} = 4.21<\Sigma_2\approx 4.31$
when $r=r_e$, then $S_{c}<\Sigma_2$ for $r\neq r_e$, and
hence this four-spot pattern is stable to spot self-replication for
all values of the ring radius.  In contrast, consider the solid curve
in Fig.~\ref{fig:1ring:b} for $\ac=37.5$ and $k=4$. Then, if the
initial ring radius $r(0)$ is in the interval between the two square
dots in this figure, i.e.~$0.35 < r(0) < 0.84$, we predict that all
four spots will begin to split simultaneously starting at $t=0$.
Similarly, from the heavy solid curve for $\ac=40$ and $k=4$ in
Fig.~\ref{fig:1ring:b}, we predict that the four spots will split
simultaneously when $r(0) > r_{s2} = 0.14$.  This prediction is
confirmed in the full numerical results below.

The plots of ${\mathcal A}_s={\mathcal A}_s(r)$ and $S_c=S_c(r)$ in
Fig.~\ref{fig:acrit} and Fig.~\ref{fig:1ring}, respectively, also
clearly show the possibility of a dynamically-triggered spot-splitting
instability. For instance, consider the solid curve in
Fig.~\ref{fig:1ring:b} for $\ac=37.5$ and $k=4$ and an initial ring
radius $r(0)=0.2$. This initial point is below the spot-splitting
threshold shown in Fig.~\ref{fig:acrit:b}, and so the initial pattern
is stable to spot-splitting. However, eventually as $t$ increases the
ring radius will cross the threshold value $r\approx 0.35$ for
spot-splitting, and a dynamically-triggered simultaneous
spot-splitting event will occur.

This possibility of a dynamically-triggered spot-splitting instability
for equally spaced spots on a ring for the GS model is in distinct
contrast to the behavior found in \cite{Schnaken_KWW:2008} for the
Schnakenburg model. For this related RD model, the common source
strength for a pattern of $k$ equally-spaced spots on a ring is
independent of the ring radius $r$. This precludes the existence of a
dynamically-triggered spot-splitting instability for the Schakenburg
model.

\begin{figure}[htpb]
\begin{center}
{\includegraphics[width=3.3cm,clip]{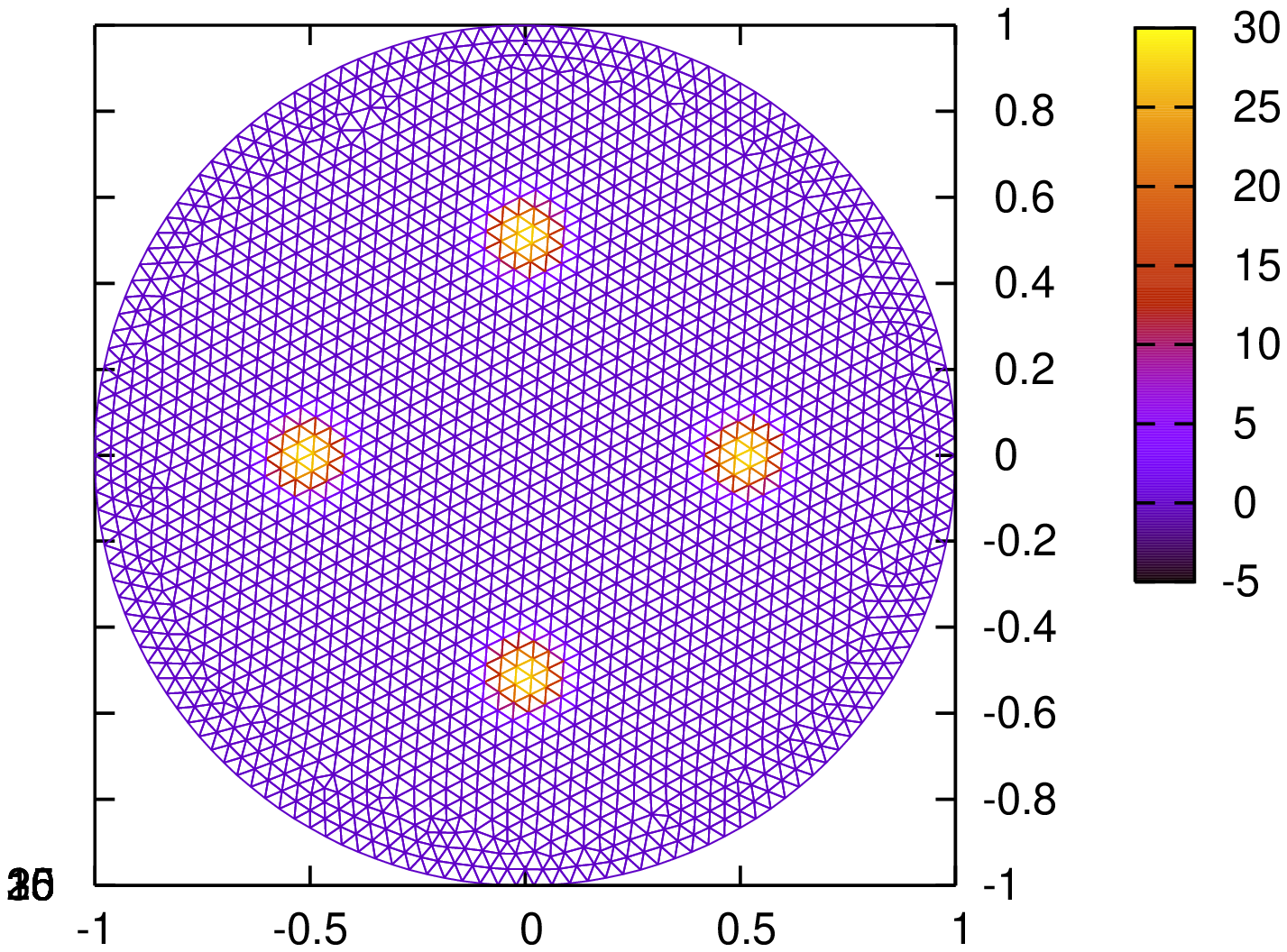}}
{\includegraphics[width=3.3cm,clip]{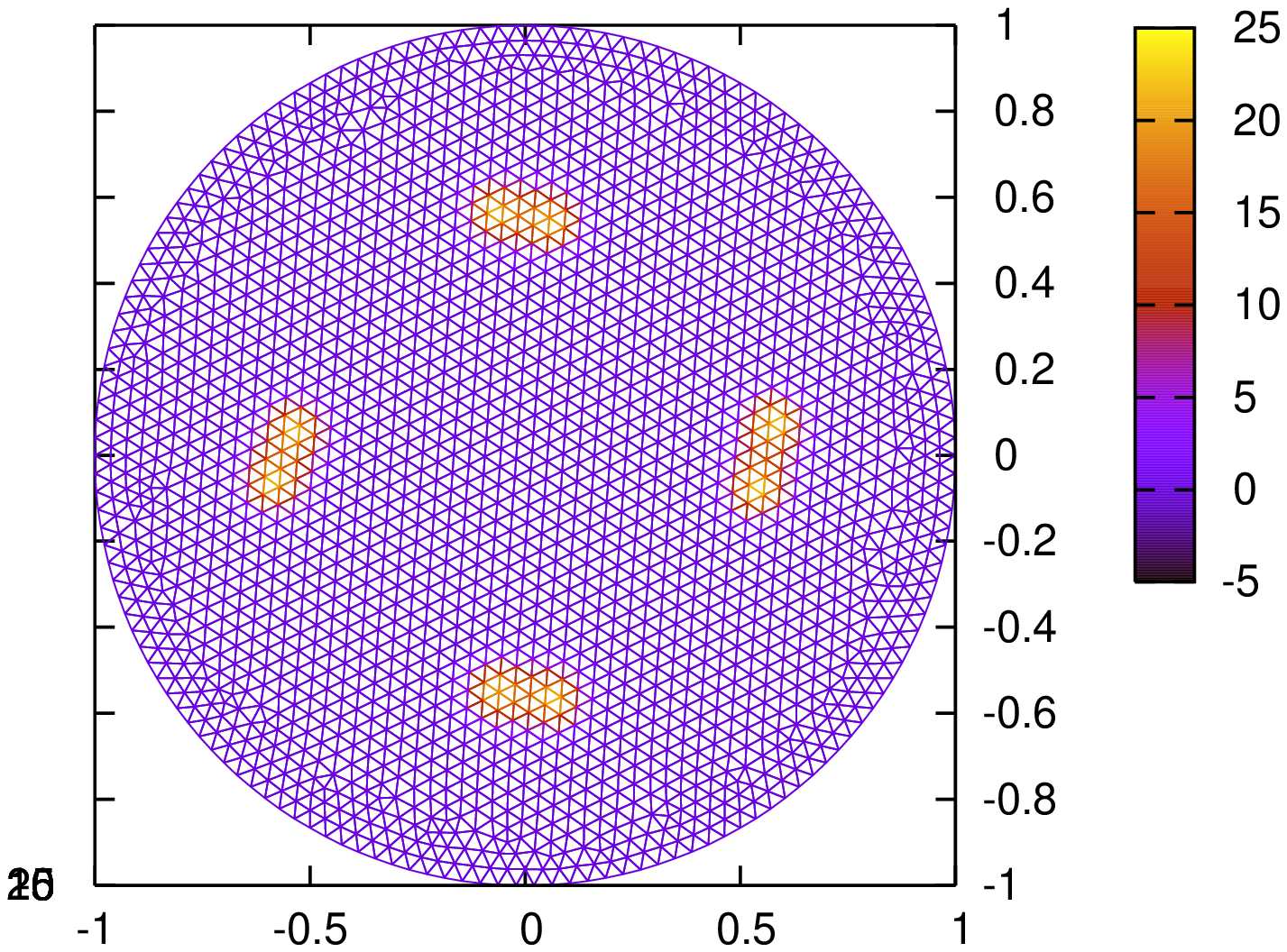}}
{\includegraphics[width=3.3cm,clip]{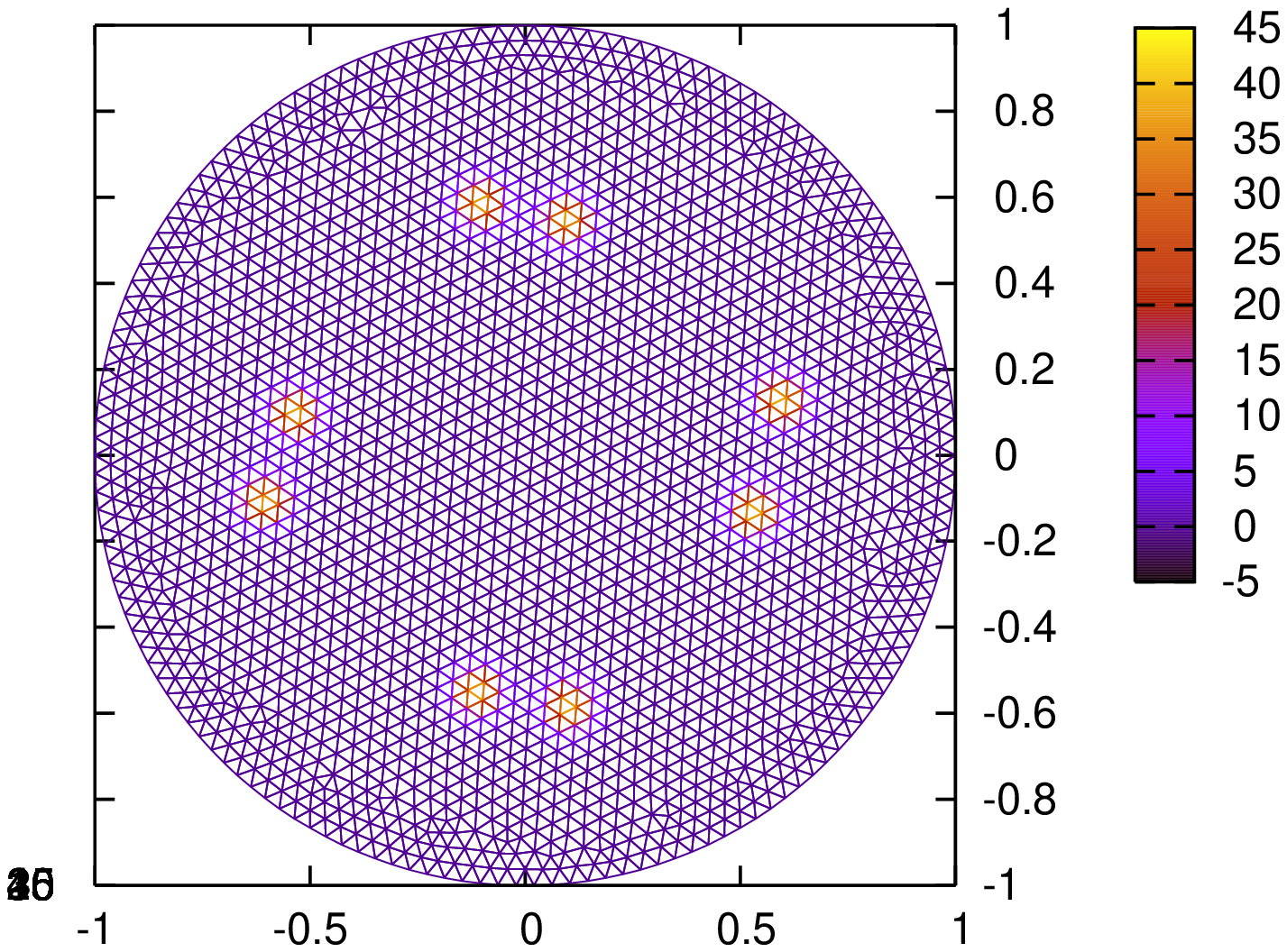}}
{\includegraphics[width=3.3cm,clip]{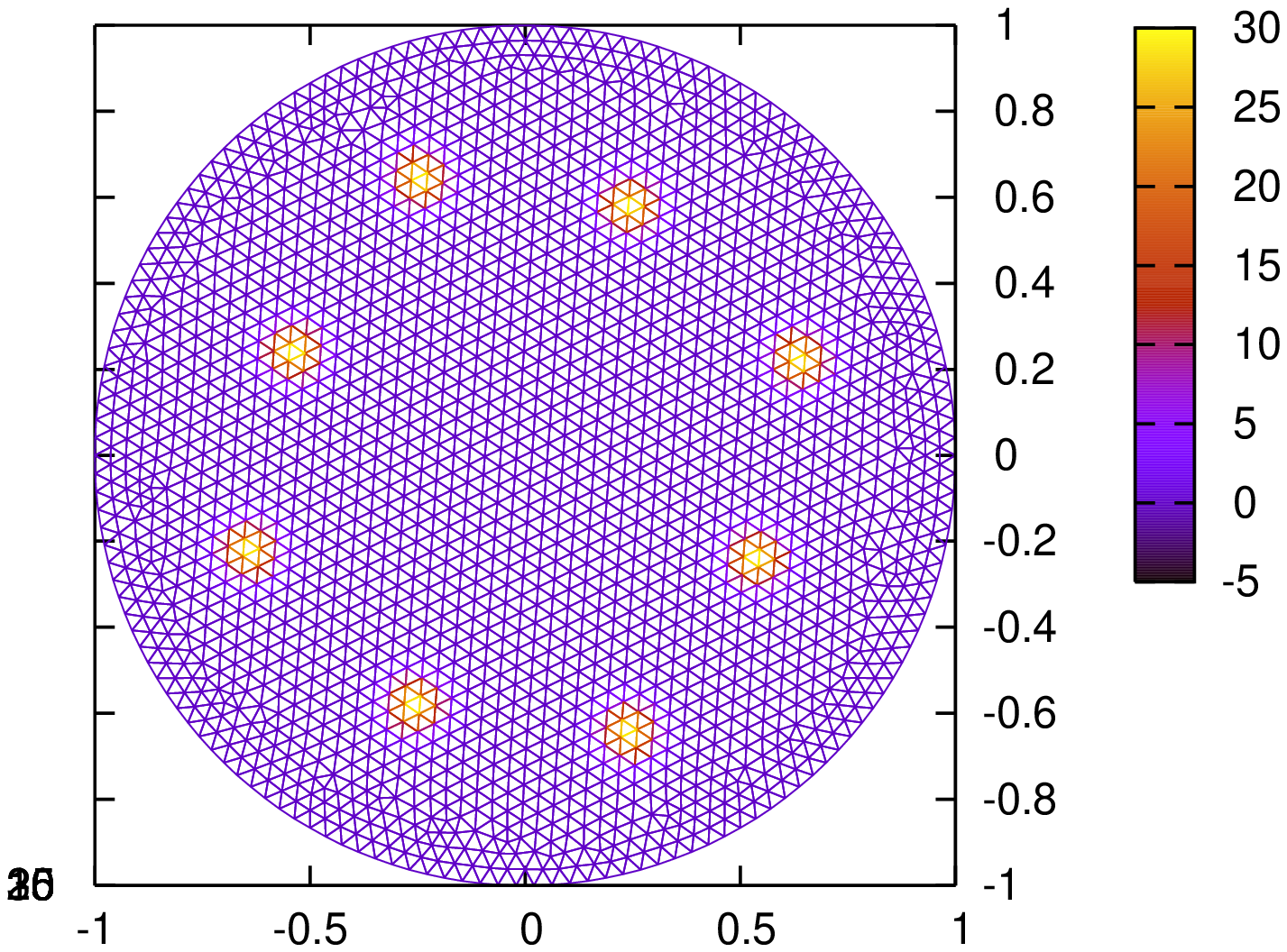}}
\caption[Four spots on a ring: spot-replication]{{\em Experiment 6.5:
Fix $D=3.912$, $\ac = 40$, $\tau=1$, and $\eps=0.02$. Consider an
initial four-spot pattern with equally-spaced spots on a ring of
radius $r=0.5$. The initial common source strength is $S_c \approx
4.69$. From left to right we plot $v$ at times $t=4.6,70,93,381$. All
spots undergo splitting, and the dynamics leads to an equilibrium
eight-spot pattern on a ring.}}
\label{fig:exp8}
\end{center}
\end{figure}

\vspace*{0.2cm}\noindent{\em {\underline{Experiment 6.5:}}\; (One-ring
pattern and spot-splitting when $D \gg {\mathcal O}(1)$):} For $\ac =
40$ and $D = 3.912$, we consider an initial four-spot pattern with
equally-spaced spots on an initial ring of radius $r = 0.5 > r_{s2}$
at time $t=0$. Since $S_c \approx 4.69$ at $t=0$, as computed from
\eqref{3:ring_sc}, our asymptotic theory predicts that all four spots
will begin to split simultaneously at $t=0$. The full numerical
results for $v$ computed from the GS model \eqref{1:GS_2D}, as shown
in Fig.~\ref{fig:exp8}, confirm this prediction, and also show that
the direction of splitting is perpendicular to the direction of spot
motion, as predicted by Principal Result 3.2. The splitting process
generates an equilibrium pattern of eight equally-spaced spots on a
ring.

\noindent {\underline{{\bf Phase Diagrams for Existence, 
Self-Replication, and Competition:}}} Next, we use our global eigenvalue
problem of Principal Result 4.3 of \S \ref{sec:eig_rad} to compute the
competition instability threshold as a function of the ring radius $r$
for various values of the parameters. Our results are shown in terms
of phase diagrams of $A$ versus $r$.

\begin{figure}[htbp]
\centering \subfigure[$D=0.2$] { \label{fig:k2cirD:a}
  \includegraphics[width=2.2in,height=1.8in]{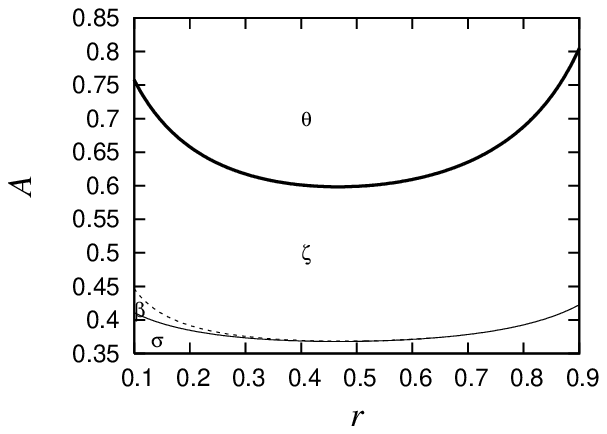}}
 \subfigure[$D=1.0$] { \label{fig:k2cirD:b}
  \includegraphics[width=2.2in,height=1.8in]{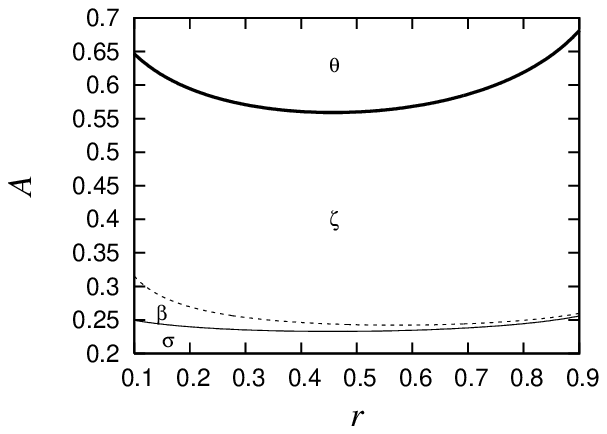}}
\subfigure[$D=5.0$] {\label{fig:k2cirD:c}
 \includegraphics[width=2.2in,height=1.8in]{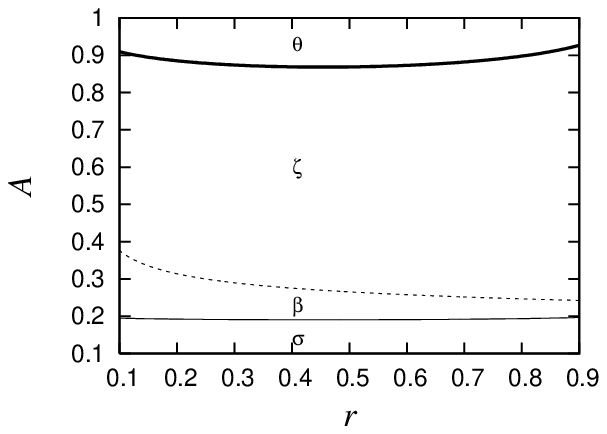} }
\caption[Two spots on a ring: the phase diagram]{{\em The phase diagram
$A$ vs.~$r$ for a two-spot quasi-equilibrium solution on a
ring of radius $r$ in the unit disk when $\eps=0.02$. The solid curve
is the existence threshold $A_f$, the dotted curve is the competition
instability threshold; the heavy solid curve is the spot
self-replication threshold $A_s$.  Regime $\sigma$: no two-spot
solution; Regime $\beta$: the two-spot solution is unstable to
competition; Regime $\zeta$ the two-spot solution is unstable to a
oscillation if $\tau>\tau_H(r)$; Regime $\theta$: the two-spot
solution is unstable to spot self-replication.  Subfigures from left
to right:: (a) $D=0.2$; (b) $D=1$; (c) $D=5$.}}
\label{fig:k2cirD}
\end{figure}

For $\eps=0.02$, in Fig.~\ref{fig:k2cirD} we plot the phase diagram
$A$ versus $r$ for $D=0.2$, $D=1.0$, and $D=5.0$, for a two-spot
quasi-equilibrium solution, showing the thresholds for the existence
of a quasi-equilibrium pattern, for a competition instability, and for
a spot self-replication instability.  For each fixed $r$, the
existence thresholds in these figures were computed by determining the
minimum of the curve $\ac$ versus $S_c$ in \eqref{3:ring_sc}, while
the spot self-replication thresholds were obtained by substituting
$S_c=\Sigma_2\approx 4.31$ and $\chi(\Sigma_2)\approx -1.79$ in
\eqref{3:ring_sc}. In addition, the competition instability
thresholds were computed from (\ref{3:newresk}), as we discuss below.

\begin{figure}[htbp]
\centering 
  \subfigure[$k=4$] { \label{fig:cirD0d2:b}
  \includegraphics[width=2.2in,height=1.8in]{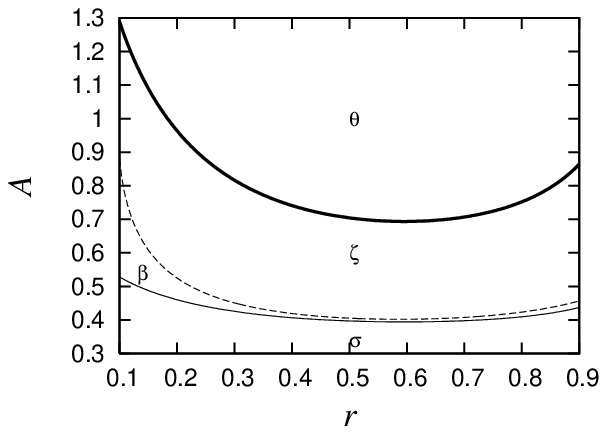}}
  \subfigure[$k=8$] {\label{fig:cirD0d2:c}
  \includegraphics[width=2.2in,height=1.8in]{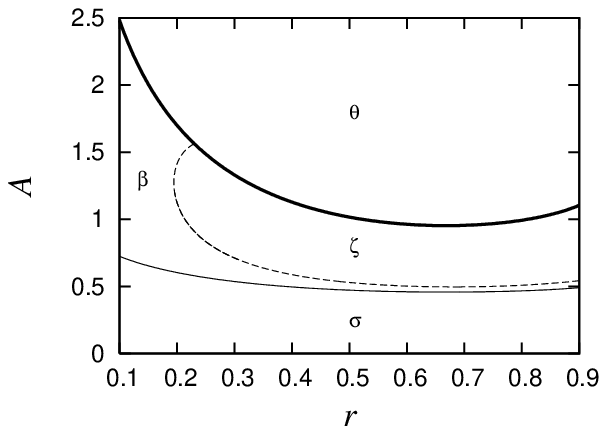}}
  \subfigure[$k=16$] {\label{fig:cirD0d2:d}
  \includegraphics[width=2.2in,height=1.8in]{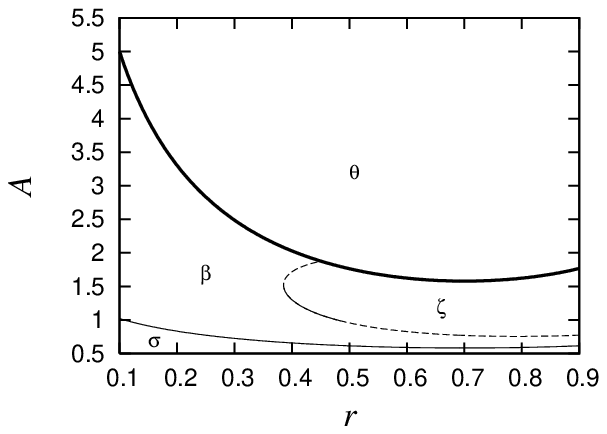}}
\caption[$k$ spots with $k>2$ on a ring: the phase diagram]{{\em 
 The phase diagram $A$ vs.~$r$ for a $k$-spot
quasi-equilibrium solution on a ring of radius $r$ in the unit
disk with $\eps=0.02$ and $D=0.2$. The thin solid curves are the existence
threshold $A_f$ for the quasi-equilibrium solution, the dotted curves are
the critical values of $A$ for a competition instability; the heavy solid 
curves are the spot self-replication threshold $A_s$. In Regime $\sigma$ the 
quasi-equilibrium solution does not exist. In Regime $\beta$ the 
quasi-equilibrium solution exists but is unstable to competition. In 
Regime $\zeta$ the solution is unstable to an oscillatory instability if 
$\tau$ exceeds a Hopf bifurcation threshold $\tau_H(r)$. In Regime 
$\theta$ the solution is unstable to spot self-replication. Subfigures from 
left to right: (a) $k=4$; (b) $k=8$; (c) $k=16$.}}
\label{fig:cirD0d2}
\end{figure}

The solution behavior in the four distinct parameter regimes of
Fig.~\ref{fig:k2cirD} is described in the caption of
Fig.~\ref{fig:k2cirD}. As expected, the subfigure for $D=0.2$ on the
range $0\leq r \leq 0.5$ is qualitatively similar to the phase diagram
shown in Fig.~\ref{fig:k2Hopf:a} for a two-spot solution on the
infinite plane.  In Fig.~\ref{fig:cirD0d2} we show similar phase
diagrams for $D=0.2$ and $\eps=0.02$, but for $k=4$, $k=8$, and
$k=16$, equally-spaced spots on a ring of radius $r$.

From \S \ref{sec:eig_rad} and the result (\ref{4:kappa}) for the
spectrum of the $\lambda$-dependent circulant Green's matrix
${\mathcal G}_{\lam}$, it follows for $k=4$ that the eigenvectors of
${\mathcal G}_\lam$ are $\mathbf{v}_1 = (1, 1, 1, 1)^T$, $\mathbf{v}_2
=(1, 0, -1, 0)^T$, $\mathbf{v}_3 = (1, -1, 1, -1)^T$, and
$\mathbf{v}_4 = (0, 1, 0, -1)^T$. Note that the last three of these
vectors have components with different signs. Our computational
results show that the competition instability threshold is set by the
sign-fluctuating instability $\mathbf{v}_3$, which has a larger
competition instability regime in the phase diagram of $A$ versus $r$
than does either $\mathbf{v}_2$ or $\mathbf{v}_4$. There is no
competition instability threshold associated with $\mathbf{v}_1$.
Similarly for $k=8$ and $k=16$, the instability associated with the
eigenvector of the form $\mathbf{v} = (1, -1, 1, -1, \ldots, 1, -1)^T$
is always the dominant competition mechanism.  In contrast, as we
discuss below, our numerical results show that the Hopf bifurcation
threshold $\tau_H$ for an oscillatory profile instability is set by
the eigenvector $\mathbf{v}_1$, which corresponds to a synchronous
oscillatory instability.

Since the sign-fluctuating eigenvector determines the competition
instability threshold in Fig.~\ref{fig:k2cirD} and
Fig.~\ref{fig:cirD0d2}, we get from (\ref{3:newres}) that, for $k$
even, this threshold is obtained by numerically solving the simple
coupled system
\bsub \label{3:newresk}
\begin{gather}
    \chi^{\p}(S_c) + 2\pi \sum_{m=1}^{k} (-1)^{m-1} a_m = -\nu^{-1} \,, 
  \qquad A = \frac{\eps}{\nu \sqrt{D} } \left[ S_c \left( 1 + 2\pi \nu
    \theta \right) + \nu \chi(S_c)\right]\,, \qquad \nu \equiv -\frac{1}
  {\ln\eps} \,, \label{3:newresk_1} \\
   \theta = \sum_{m=1}^{k} a_m \,, \quad a_1=R_{11} \,; \quad a_j=
 G_{1j} \,, \quad j=2,\ldots,k \,. \label{3:newresk_2}
\end{gather}
\esub Here $\theta$ is the eigenvalue of $\mathcal{G}$ with
eigenvector $\mathbf{e}=(1,\ldots,1)^T$. Numerical values for $R_{11}$
and $G_{1j}$, for $j=2,\ldots,k$, are obtained from the reduced-wave
Green's function for the unit disk given in \eqref{3:simpleG} of
Appendix A.

\vspace*{0.2cm}
\noindent {\underline{{\bf Oscillatory and Competition
Instabilities:}}} Next, we use our global eigenvalue problem to
compute Hopf bifurcation thresholds $\tau_H$ for an oscillatory
instability of $k$ equally-spaced spots on a ring. Such an instability
is the only instability mechanism in each Regime $\zeta$ of
Fig.~\ref{fig:cirD0d2} and Fig.~\ref{fig:k2cirD}. Since this
instability is not readily depicted in a phase diagram of the type
shown in Fig.~\ref{fig:cirD0d2} and Fig.~\ref{fig:k2cirD}, we now
discuss it some detail.

To focus the discussion, we will only consider an equally-spaced two-spot
pattern on a ring of radius $r$ in the unit disk. We show that both
dynamically-triggered oscillatory and competition instabilities are
possible for different ranges of the GS parameters, and we study the
spectrum of the linearization in some detail.

\begin{figure}[htbp]
\centering 
\subfigure[$\tau_H$ vs.~$r$] { \label{fig:k2cirHopf:b}
  \includegraphics[width=3.0in,height=1.8in]{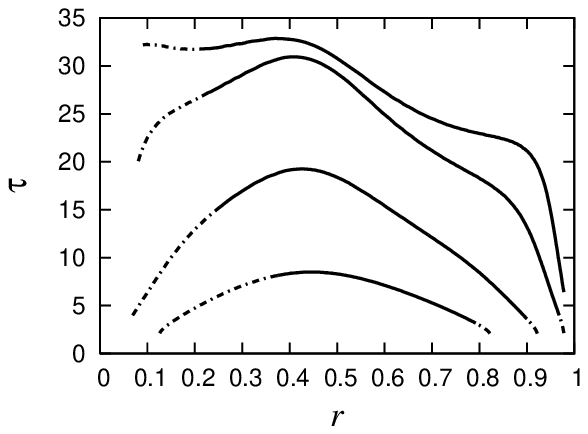}}
\subfigure[$\tau_H$ vs.~$r$] { \label{fig:k2cirHopf:c}
    \includegraphics[width=3.0in,height=1.8in]{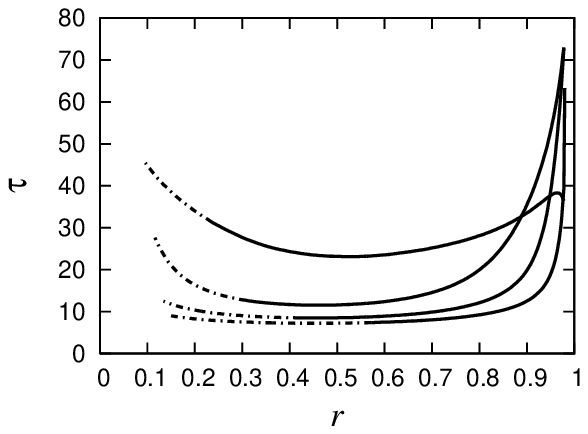} }
\caption[Two spots on a ring: a dynamical competition instability]{{\em
Two equally-spaced spots on a ring of radius $r$ in the unit disk for
$\eps=0.02$: (a) For $A=0.26$, the Hopf bifurcation threshold $\tau_H$
vs.~$r$ is plotted for $D=0.8, 1.0, 1.3, 1.5$, with the lower curves
corresponding to smaller values of $D$. The solid curves correspond to
regions in $r$ where an oscillatory profile instability occurs, and
the dotted portions of these curves correspond to where a competition
instability occurs for any $\tau>0$. (b) $\tau_H$ vs.~$r$ for the
larger values of $D$ given by $D = 2, 3, 4, 5$. The lower curves at
$r=0.6$ correspond to larger values of $D$.}}
\label{fig:k2cirHopf}
\end{figure}

For $A=0.26$, $k=2$, and $\eps=0.02$, the Hopf bifurcation threshold
$\tau_H$ versus the ring radius $r$, corresponding to
$\mathbf{v}_1=(1,1)^T$, is shown by the solid curves in
Fig.~\ref{fig:k2cirHopf}. The dotted segments on these curves is where
a competition instability occurs. In Fig.~\ref{fig:k2cirHopf:b} we
plot $\tau_H(r)$ for $D=0.8, 1.0, 1.3, 1.5$, while in
Fig.~\ref{fig:k2cirHopf:c} we plot $\tau_H(r)$ for $D=2, 3, 4,
5$. From the convexity of $\tau_H$ versus $r$ in
Fig.~\ref{fig:k2cirHopf:c}, we conclude that a dynamically-triggered
oscillatory profile instability can occur for a two-spot pattern when
$A=0.26$ and $D\geq 2$. We also observe that when $D=5$, a competition
instability occurs when $r \approx 0.568$. Therefore, since the
equilibrium ring radius is $r_e\approx 0.455$ for $D=5$ (see
Table~\ref{tab:equil}), we conclude that there can be a
dynamically-triggered competition instability for this value of $D$.
To illustrate this, let $D=5$ and $\tau=1$ and suppose that we have an
initial two-spot pattern with initial value $r=0.7$ at $t=0$. Then, as
the two spots move slowly towards each other, a dynamically-triggered
competition instability will be initiated before the ring reaches its
equilibrium radius at $r_e \approx 0.455$.

\begin{figure}[htbp]
\centering \subfigure[$\tau_H$ vs.~$r$] { \label{fig:k2cirSpec:a}
  \includegraphics[width=3.0in,height=1.8in]{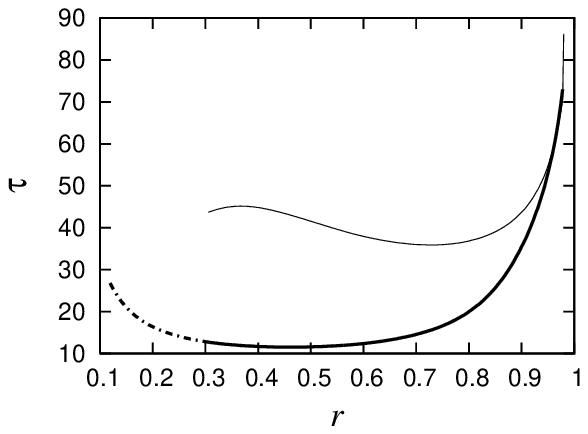}}
  \subfigure[$\mbox{Im} (\lambda)$ vs.~$\mbox{Re} (\lambda)$]
  {\label{fig:k2cirSpec:b}
  \includegraphics[width=3.0in,height=1.8in]{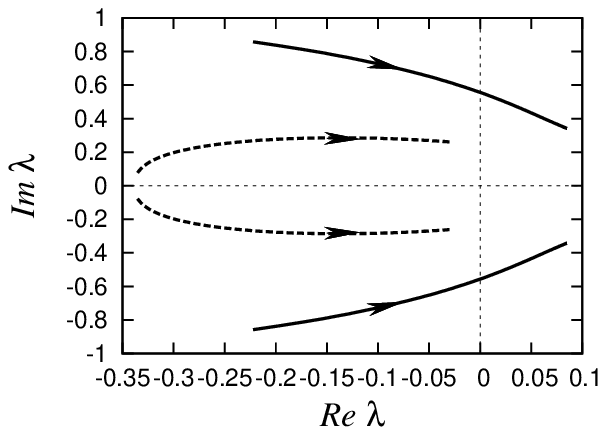} }
  \caption[Two spots on a ring: in-phase oscillation threshold]{{\em
  Two equally-spaced spots on a ring of radius $r$ in the unit
  disk. Fix $A=0.26$, $\eps=0.02$, and $D=3$. (a) $\tau_H$ vs.~$r$ for
  in-phase spot amplitude oscillation $\mathbf{v}_1=(1,1)^T$ (lower
  heavy solid curve), and $\tau_{H2}$ for the out-of-phase oscillation
  $\mathbf{v}_2=(1,-1)^T$ (upper solid curve). The dotted portions on
  $\tau_{H1}$ are where a competition instability occurs for any
  $\tau\geq 0$.  (b) For the equilibrium ring radius $r_e=0.45483$, we
  plot $\lambda$ in the complex plane on the range $\tau \in [3.7,
  36]$. The solid curve is for $\mathbf{v}_1=(1, 1)^T$, while the
  dotted curve is for $\mathbf{v}_2=(1, -1)^T$. As $\tau$ increases
  (direction of the arrows), $\mbox{Re}(\lam)$ increases.}}
  \label{fig:k2cirSpec}
\end{figure}

Next we fix $A=0.26$ and $D=3$. In Fig.~\ref{fig:k2cirSpec:a} we plot
the Hopf bifurcation thresholds $\tau_{H1}$ and $\tau_{H2}$ for
in-phase and out-of-phase spot amplitude oscillations,
respectively. Since $\tau_{H1}<\tau_{H2}$, the oscillatory
instability triggered as $\tau$ increases is a synchronous spot
amplitude oscillation. The dotted portion on $\tau_{H1}$, representing
the continuation of the heavy solid curve in
Fig.~\ref{fig:k2cirSpec:a}, shows where a competition instability
occurs for any $\tau\geq 0$.  In Fig.~\ref{fig:k2cirSpec:b}, we fix
two spots on an equilibrium ring of radius $r_e=0.45483$, and we plot
the path of the complex conjugate pair of eigenvalues in the complex
plane for the range $\tau\in [3.7, 36]$. We note that as $\tau$
increases, the real parts of both pairs of eigenvalues increase. The
eigenvalue for in-phase oscillations $\mathbf{v}_1=(1, 1)^T$, which is
given by the solid curve in Fig.~\ref{fig:k2cirSpec:b}, enters the
right half-plane at $\tau \approx 11.5$. Note that the imaginary part
of the eigenvalue for the out-of-phase oscillation $\mathbf{v}_2=(1,
-1)^T$ becomes very close to the negative real axis when $\tau < 3.7$.

Next, for $\eps=0.02$ and $A=0.26$, we treat $D$ as the bifurcation
parameter and we relate our results with those obtained from the NLEP
theory of \cite{2Dmulti_Wei:2003}. In Theorem 2.3 of
\cite{2Dmulti_Wei:2003}, as summarized in Appendix B, the
leading-order-in-$\nu$ NLEP analysis proves that if $D = {\mathcal
O}(\nu^{-1})$ and $L_0 < \frac{\eta_0}{(2 \eta_0 + k)^2}$, then the
small solution $u^-, v^-$ of \eqref{1:GS_2D} is stable for any $\tau$
sufficiently small. In addition, the NLEP theory proves that if $L_0 >
\frac{\eta_0}{(2\eta_0 + k)^2}$, then this solution is unstable for
any $\tau>0$. From \eqref{4:ldef}, we recall that $L_0 = \lim_{\eps
\to 0} \frac{2 \eps^2 \pi b_0}{A^2 |\Omega|}$ and $\eta_0 = \lim_{\eps
\to 0}\frac{|\Omega|}{2 \pi D \nu}$, where $b_0=\int_{0}^{\infty} w^2
\rho \, d\rho \approx 4.9347$ and $w(\rho)$ satisfies
\eqref{4:groundstate}. With $|\Omega| = \pi$ and $\nu={-1/\ln\eps}$,
and for the other GS parameter values as given, this stability bound
is $1.158 < D < 3.303$. For this range of $D$, the
leading-order-in-$\nu$ NLEP theory predicts that this solution is
stable for $\tau$ sufficiently small. Outside of this bound for $D$,
the NLEP theory of \cite{2Dmulti_Wei:2003} predicts that a competition
instability occurs for any $\tau>0$. This conclusion from NLEP theory
is independent of the ring radius $r$.

\begin{figure}[htbp]
\centering \subfigure[$D_c$ vs.~$r$] { \label{fig:k2cirComp:a}
  \includegraphics[width=3.0in,height=1.8in]{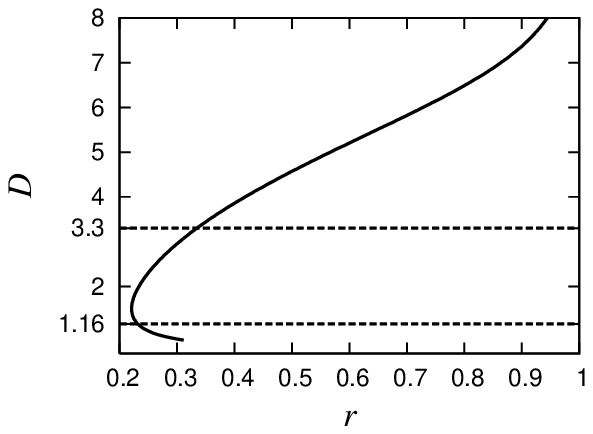}
} \subfigure[$\mbox{Im} (\lambda)$ vs.~$\mbox{Re} (\lambda)$] {
\label{fig:k2cirComp:b}
  \includegraphics[width=3.0in,height=1.8in]{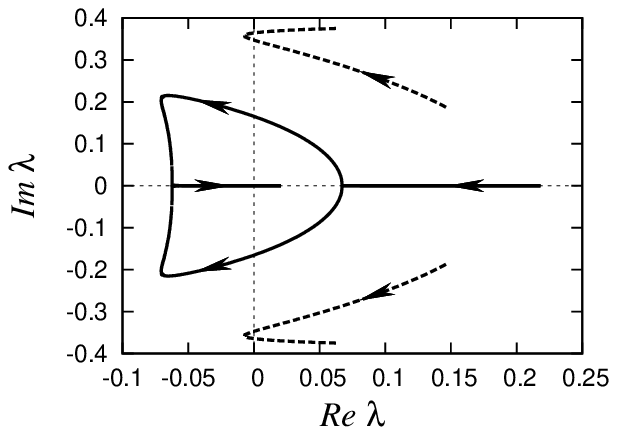}
} \caption[Two spots on a ring: a competition instability]{{\em Two
equally-spaced spots on a ring of radius $r$ in the unit disk. Fix
$A=0.26$ and $\eps=0.02$. (a) Plot of $D_c$ vs.~$r$ (solid curve),
where $D_c$ is the critical value of $D$ at which a competition
instability is initiated. The region $1.1583 < D < 3.3030$ between the
two horizontal dotted lines is where the leading-order-in-$\nu$ NLEP theory
predicts that no competition instability occurs.  (b) For $\tau=30$
and $r=0.3$, a plot of the spectrum in the complex plane is shown when
$D$ varies on the range $[0.8, 3.1]$, which corresponds to taking a
vertical slice in Fig.~\ref{fig:k2cirComp:a} at $r=0.3$. The solid
path in the spectrum corresponds to  $\mathbf{v}_2=(1,-1)^T$, while the
the dotted path is for $\mathbf{v}_1=(1,1)^T$. The arrows show the path 
as $D$ increases.}}
\label{fig:k2cirComp}
\end{figure}

\begin{figure}[htbp]
\begin{center}
\subfigure[ $0.80<D<.845$]
{\label{fig:spec_real_a} 
\includegraphics[width=3.0in, height=1.8in]{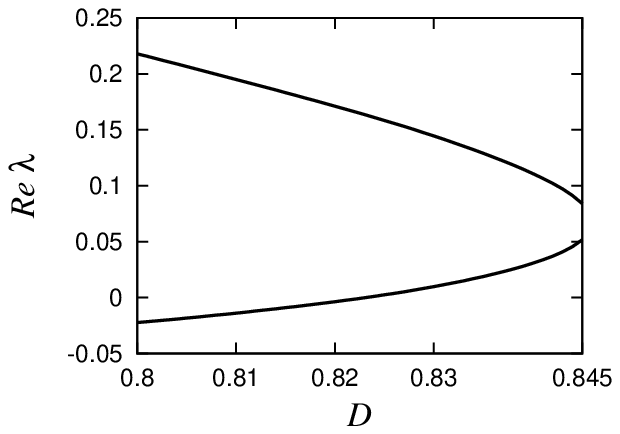}}
\subfigure[ $2.95<D<3.1$]
{\label{fig:spec_real_b} 
\includegraphics[ width=3.0in, height=1.8in]{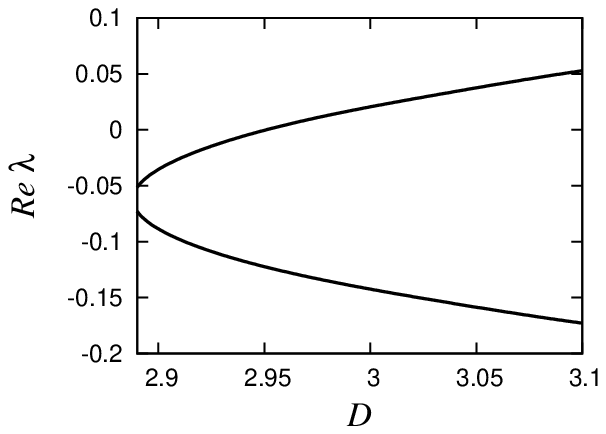}}
\caption{\em { Two equally-spaced spots on a ring of radius $r=0.3$ in the
unit disk. Fix $A=0.26$, $\eps=0.02$, and $\tau=30$, and consider the
sign-fluctuating eigenvector $\mathbf{v}_2=(1,-1)^T$. We plot the
eigenvalues of the global eigenvalue problem for two ranges of $D$ where they
are real. Left figure: $D \in [0.8, 0.845]$. Right figure: $D \in 
[2.95, 3.10]$.}}
\label{fig:spec_real}
\end{center}
\end{figure}

For $\eps=0.02$ and $A=0.26$, in Fig.~\ref{fig:k2cirComp:a} we plot
$D_c$ versus $r$ by the solid curve, where $D_c$ denotes the critical
value of $D$ where a competition instability is initiated, as computed
from (\ref{3:newresk}). To the left of this curve, oriented with
respect to the direction of increasing $D$, we predict that a
competition instability occurs for any $\tau > 0$, while to the right
of this curve we predict that the solution is stable when $\tau$ is
small enough. The two horizontal dotted lines with $D_1=1.158$ and
$D_2 = 3.303$ in this figure bound a region of stability as predicted
by the leading-order-in-$\nu$ NLEP theory of
\cite{2Dmulti_Wei:2003}. This prediction from NLEP theory does not
capture the dependence of the stability threshold on the ring radius
$r$. Our stability formulation, which accounts for all orders in
$\nu={-1/\ln\eps}$, shows that $D_c=D_{c}(r)$. For $\eps=0.02$, $\nu$
is certainly not very small, and this dependence of the threshold on
$r$ is very significant.

In Fig.~\ref{fig:k2cirComp:b}, for $A=0.26$, $\tau=30$, and $r=0.3$,
we plot the spectrum in the complex plane as $D$ varies over the range
$[0.8, 3.0]$. This range of $D$ corresponds to taking a vertical slice
at $r=0.3$ in Fig.~\ref{fig:k2cirComp:a} that cuts across the
stability boundary $D_{c}(r)$ at two values of $D$.  The arrows in
Fig.~\ref{fig:k2cirComp:b} indicate the direction of the path of
eigenvalues as $D$ is increased from $D=0.8$. The loop of spectra,
depicted by the solid curves in Fig.~\ref{fig:k2cirComp:b}, shows the
eigenvalues associated with the eigenvector $\mathbf{v}_2=(1,
-1)^T$. From Fig.~\ref{fig:k2cirComp:a} we predict instability when
$D$ is near $D=0.8$ or when $D$ is near $D=3.0$. This behavior is
shown by the closed loop of spectra in the complex $\lambda$ plane of
Fig.~\ref{fig:k2cirComp:b}, and in Fig.~\ref{fig:spec_real} where the
eigenvalues are plotted for the range of $D$ where they are real.

More specifically, with regards to the loop of spectra (solid curves)
in Fig.~\ref{fig:k2cirComp:b} associated with the
$\mathbf{v}_2=(1,-1)^T$ eigenvector, our results show that for $D=0.8$
there are two real eigenvalues at $\lambda \approx 0.218$ and $\lambda
\approx -0.0224$ (the negative eigenvalue is not shown in
Fig.~\ref{fig:k2cirComp:b}). From Fig.~\ref{fig:spec_real_a} we
observe that as $D$ is increased above $D=0.8$ the positive real
eigenvalues move along the horizontal axis in different directions,
and collide at $D\approx 0.845$, producing a complex conjugate pair of
unstable eigenvalues at this value of $D$. Then, for $D \approx 1.0$,
the complex conjugate pair of eigenvalues enters the stable left
half-plane $\mbox{Re}(\lam)<0$. For $D=2.89$ the eigenvalues merge
onto the negative real axis, as shown in
Fig.~\ref{fig:spec_real_b}. One eigenvalue then enters the right
half-plane at $D \approx 2.95$, while the other slides along the
negative real axis. Therefore, for $D>2.95$, there is a positive real
eigenvalue associated with the eigenvector $\mathbf{v}_2=(1,-1)^T$,
which generates a competition instability.

Alternatively, the dotted curves in Fig.~\ref{fig:k2cirComp:b}
correspond to the path of eigenvalues associated with the
$\mathbf{v}_1=(1,1)^T$ eigenvector, representing synchronous
oscillatory instabilities. This path depends sensitively on the value
chosen for $\tau$. For $\tau=30$, our computational results show that
as $D$ is increased above $D=0.8$, the real part of these eigenvalues
first starts to decrease and then crosses into the stable left
half-plane $\mbox{Re}(\lam)<0$ at $D \approx 1.33$. The real part of
these eigenvalues starts increasing when $D \approx 1.59$ and the path
then re-enters the unstable right half-plane at $D=1.89$ where
stability is lost at a Hopf bifurcation corresponding to synchronous
oscillations of the two spot amplitudes. This path of eigenvalues
remain in the unstable right half-plane $\mbox{Re}(\lam)>0$ for $D>1.89$.

For $\tau=30$, $A=0.26$, and $\eps=0.02$, we conclude that if $1.33 <
D < 1.89$, then the two-spot pattern is stable to both competition and
synchronous oscillatory instabilities. When $D > 2.95$ there is an
unstable real eigenvalue in the right half-plane associated with a
competition instability as well as a synchronous oscillatory
instability associated with the complex conjugate pair of
eigenvalues. On the range $1.89<D<2.95$ there is only a synchronous
oscillatory instability. We remark that if $\tau$ is taken to be
much smaller than our chosen value $\tau=30$, then for
$D>2.95$ there would be only a competition instability from a positive
real eigenvalue, and no oscillatory instability would occur.

\begin{figure}[htbp]
\centering \subfigure[$v_m$ vs.~$t$] 
{ \includegraphics[width=3.0in, height=1.8in]{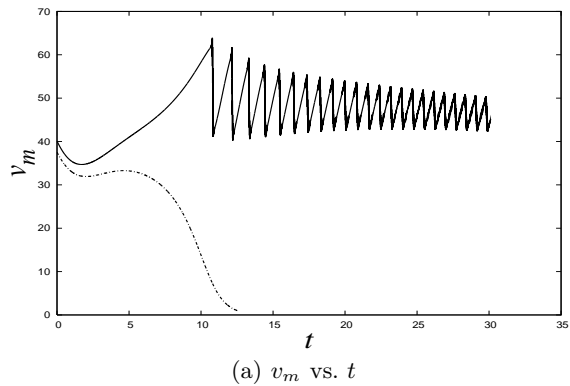} } 
\caption[Two spots on a ring: the spot amplitudes]{{\em Experiment 6.6:
  Two equally-spaced spots on a ring of radius $r=0.3$ in the unit
  disk. Fix $\eps=0.02, A=0.26$, $D=3$, and $\tau=30$.  The dot-dashed
  curve plots the amplitude $v_{m1}$  vs.~$t$ of one the spots, while
  the solid curve shows the amplitude $v_{m2}$ vs.~$t$ of the other spot.
  Both a competition and an oscillatory instability occur for the initial 
  pattern.}} \label{fig:compet_k2_amp}
\end{figure}

\vspace*{0.2cm}\noindent{\em {\underline{Experiment 6.6:}}\; (One-ring
pattern with two equally-spaced spots on a ring: Simultaneously
occurring instabilities):} For an equally-spaced two-spot pattern on a
ring of radius $r=0.3$, we now validate the theoretical prediction,
based on Fig.~\ref{fig:k2cirComp}, of having both a competition and
oscillatory instability occurring simultaneously for the GS parameter
values $A=0.26$, $D=3$, $\tau = 30$, and $\eps=0.02$. For these
values, Fig.~\ref{fig:k2cirComp:b} shows that there is an unstable
real eigenvalue together with an unstable complex conjugate pair of
eigenvalues. In Fig.~\ref{fig:compet_k2_amp} we plot the maxima of $v$
versus $t$ of the two spots, as computed from a full numerical
solution of \eqref{1:GS_2D}. This figure shows that one of the two
spots is annihilated before $t=13$, consistent with a competition
instability. In contrast, the amplitude of the other spot begins to
oscillate, but the oscillation ceases after the first spot is
annihilated. This occurs since after one spot has been destroyed, the
value $\tau=30$ is below the Hopf bifurcation threshold for a one-spot
solution in the unit disk.

\setcounter{equation}{0}
\setcounter{section}{6}
\section{Some Asymmetric Spot Patterns in the Unit Square and Disk}
\label{sec:asy}

In this section we study spot self-replication instabilities for some
spot patterns for which the Green's matrix is not circulant.  The GS
parameters are chosen so that competition and oscillatory
instabilities do not occur. The asymptotic results for spot
self-replication and spot dynamics are favorably compared with full
numerical results.

\subsection{The Unit Square} \label{sec:asy_square}
 We will consider three examples for the unit square $[0,1]\times
[0,1]$ for $\eps=0.02$ and $\tau=1.0$.

\begin{figure}[htpb]
\hspace*{-1cm}
\includegraphics[width=5.4in]{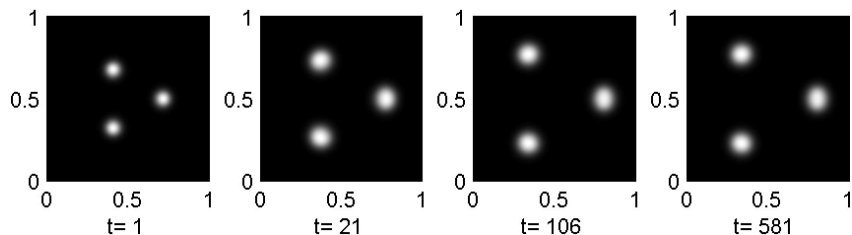}
\caption[A three-spot pattern: slow drifting spots]{{\em Experiment
  7.1: Fix $\ac=20$, $D=1$, $\tau=1$, and $\eps=0.02$. Consider a
  three-spot initial pattern with spots equally-spaced on a ring of
  radius $r=0.2$ centered at $(0.5, 0.5)$ at time $t=0$ in the unit
  square. The initial common source strength is $S_c \approx 3.71$ (see
  Table~\ref{tab:exp4}). The numerical solution $v$ is shown
  at various times.}}
\label{fig:exp4vlg}
\end{figure}

\vspace*{0.2cm} \noindent{\em {\underline{Experiment 7.1:}}\; (A
three-spot pattern: Slowly drifting spots):} Let $\ac=20$ and $D=1$,
and consider an initial three-spot pattern with spots equally-spaced
on a ring of radius $r = 0.2$ centered at the midpoint of the unit
square. The initial spot coordinates at $t=0$ are given in
Table~\ref{tab:exp4}. For this pattern, the matrix $\mathcal{G}$ in
\eqref{3:ASmatrix} is not circulant. However, by evaluating the
entries of ${\mathcal G}$ by using \eqref{3:squaregreen} of Appendix
A, we compute that the three spots still have an initial common source
strength $S_c \approx 3.71$.  Since $S_c < \Sigma_2$, we predict that
there is no spot self-replication initiated at $t=0$. From the full
numerical results shown in Fig.~\ref{fig:exp4vlg}, we observe that all
three spots drift slowly outwards, and approach equilibrium locations
inside $\Omega$ when $t$ is large. During this evolution, the source
strengths never exceed the threshold $\Sigma_2$, and so there is no
dynamically-triggered spot self-replication instability.

\begin{figure}[htpb]
\centering \subfigure[$x_j$ vs.~$t$] {
\includegraphics[width=2.4in,height=1.6in]{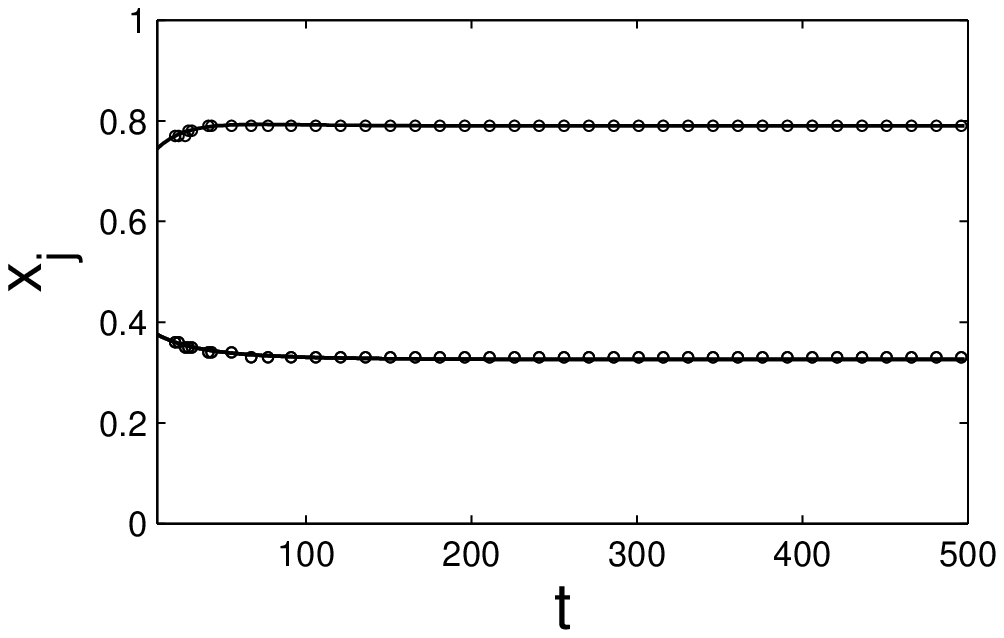}
\label{fig:exp4dyn:a}} \subfigure[$y_j$ vs.~$t$] {
\includegraphics[width=2.4in,height=1.6in]{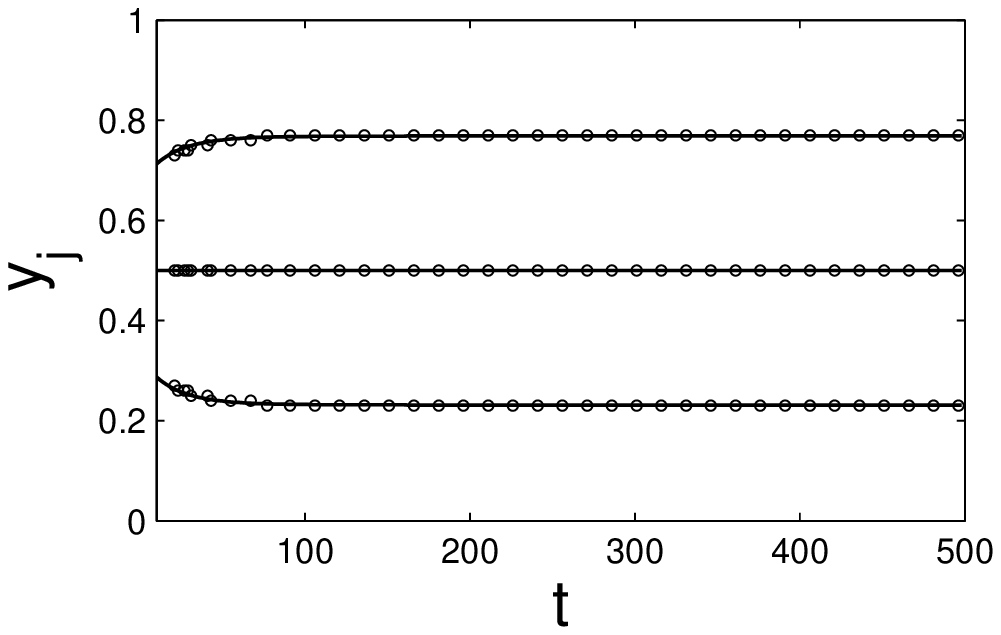}
\label{fig:exp4dyn:b}} \caption[Comparison of asymptotic and
numerical results for the dynamics of a three-spot pattern]{{\em
Experiment 7.1: Fix $\ac=20$, $D=1$, $\tau=1$, and
$\eps=0.02$. Comparison of the asymptotic dynamics of a $3-$spot
pattern (solid curves) with full numerical results (discrete markers)
starting from $t=10$. (a) $x_j$ vs.~t; (b) $y_j$ vs.~t.}}
\label{fig:exp4dyn}
\end{figure}

In Fig.~\ref{fig:exp4dyn} we show a very favorable comparison between
the spot trajectories, as obtained from the dynamics (\ref{3:dyn}) of
Principal Result 3.1, and the corresponding numerical results computed
from (\ref{1:GS_2D}). Since the initial locations of the
$1^{\mbox{st}}$ and $2^{\mbox{nd}}$ spots are exactly symmetric, it
follows that $G_{1,3} = G_{2,3}$, $R_{1,1} = R_{2,2}$, and $G_{1,2} =
G_{2,1}$ in the Green's matrix ${\mathcal G}$. Thus, from
Principal Result 3.1, these two spots move at the same speed and their
trajectories in the $x-$direction exactly overlap, as shown in
Fig.~\ref{fig:exp4dyn:a}.  Moreover, from the spot equilibrium
condition \eqref{3:finaleq1}, we can calculate the equilibrium
locations $x_{je}, y_{je}$, and source strengths $S_{je}$ for
$j=1,\ldots,3$. These values are given in Table~\ref{tab:exp4}, where
we observe that our equilibrium result closely predicts the final
equilibrium state $x_{jn}$, $y_{jn}$, $S_{jn}$ computed from
full numerical solutions of the GS model \eqref{1:GS_2D}.

\begin{table}
\centering
\begin{tabular}{c c c c c || c c c || c c c}
\hline\hline
 $j$ & $x_j$ & $ y_j$ & $S_j$ & $V_j(0)$ & $x_{jn}$ & $y_{jn}$ &$S_{jn}$
  & $x_{je}$ & $ y_{je}$  & $S_{je}$ \\ \hline
 1 &0.40  &0.67  &3.71  &34.2  &0.33 &0.77 &3.68  &0.33 &0.77 &3.68\\
 2 &0.40  &0.33  &3.71  &34.2  &0.33 &0.23 &3.68  &0.33 &0.23 &3.68\\
 3 &0.70  &0.50  &3.71  &34.2  &0.79 &0.50 &3.95  &0.79 &0.50 &3.95\\ \hline
\end{tabular}
\caption[Data for Experiment 7.1:]{{\em Experiment 7.1: The columns of
 $x_j$, $y_j$, $S_j$ and $V_j(0)$ correspond to the initial
 condition. The remaining data is with regards to the numerical and
 asymptotic results, respectively, for the equilibrium state.}}
 \label{tab:exp4}
\end{table}

\begin{figure}[htpb]
\hspace*{-1cm}
\includegraphics[width=5.3in]{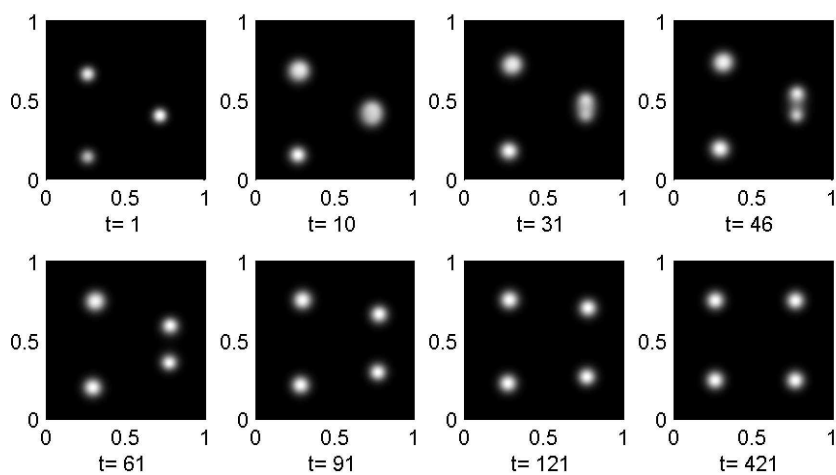}
\caption[A three-spot pattern: self-replication]{{\em Experiment 7.2:
Fix $\ac=20$, $D=1$, $\tau=1$, and $\eps=0.02$. Consider a three-spot
initial pattern with spots equally-spaced on a ring of radius $r =0.3$
centered at $(0.4, 0.4)$ in the unit square.  The initial source
strengths $S_j$ for $j=1,\ldots,3$ are given in Table~\ref{tab:exp5},
for which $S_3 > \Sigma_2$.  The full numerical solution for $v$ is
plotted at different instants in time.  The $3^{\mbox{rd}}$ spot
undergoes self-replication. Subsequently, all four spots slowly drift
towards a symmetric equilibrium $4-$spot pattern.}}
\label{fig:exp5vlg}
\end{figure}

\vspace*{0.2cm} 
\noindent{\em {\underline{Experiment 7.2:}}\; (A three-spot pattern:
Spot-splitting and dynamics after splitting):} Next, we fix $\ac=20$
and $D=1$ and consider an initial three-spot pattern on a ring of
radius $r = 0.3$ centered at $(0.4, 0.4)$ in the unit square. These GS
parameter values are the same as in Experiment 7.1, except that here
the initial spot configuration is different. The initial data are
given in the left side of Table~\ref{tab:exp5}. Since $S_3>\Sigma_2$,
the asymptotic theory predicts that the $3^{\mbox{rd}}$ spot will
undergo splitting beginning at $t=0$. In Fig.~\ref{fig:exp5vlg}, we
plot the numerical solution $v$ at different instants in time. From
this figure we observe that the $3^{\mbox{rd}}$ spot (the brightest
one at $t=1$) deforms into a peanut shape at $t=31$, and then splits
into two spots at $t=46$ in a direction perpendicular to its
motion. Subsequently, the four spots drift slowly to a symmetric
four-spot equilibrium state.

\begin{figure}[htpb]
\hspace{-0.7cm}\centering \subfigure[$x_j$ vs.~$t$] {
\includegraphics[width=2.4in,height=1.6in]{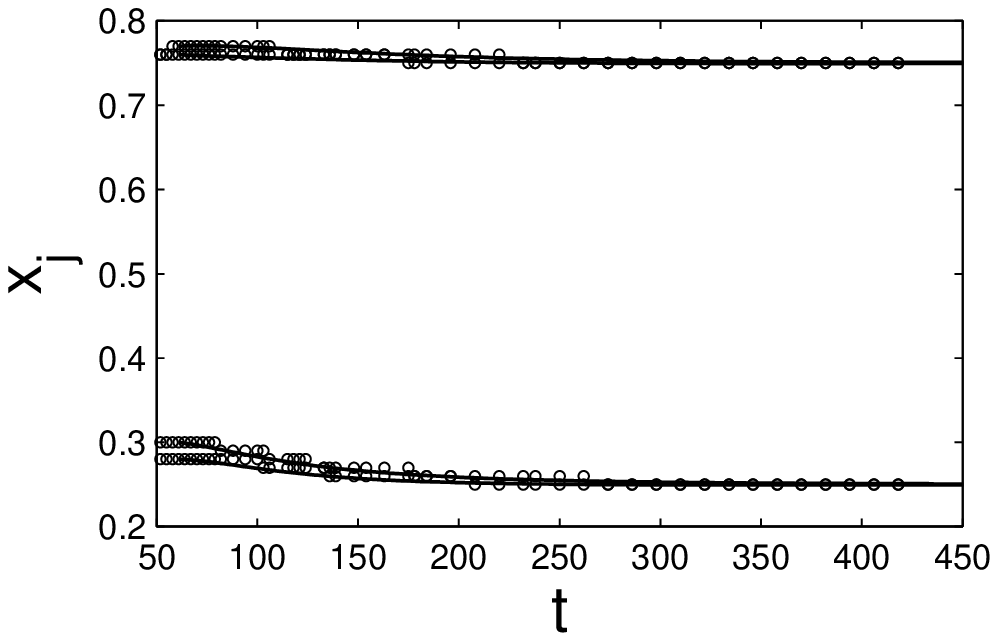}
\label{fig:exp5dyn:a}} \subfigure[$y_j$ vs.~$t$] {
\includegraphics[width=2.4in,height=1.6in]{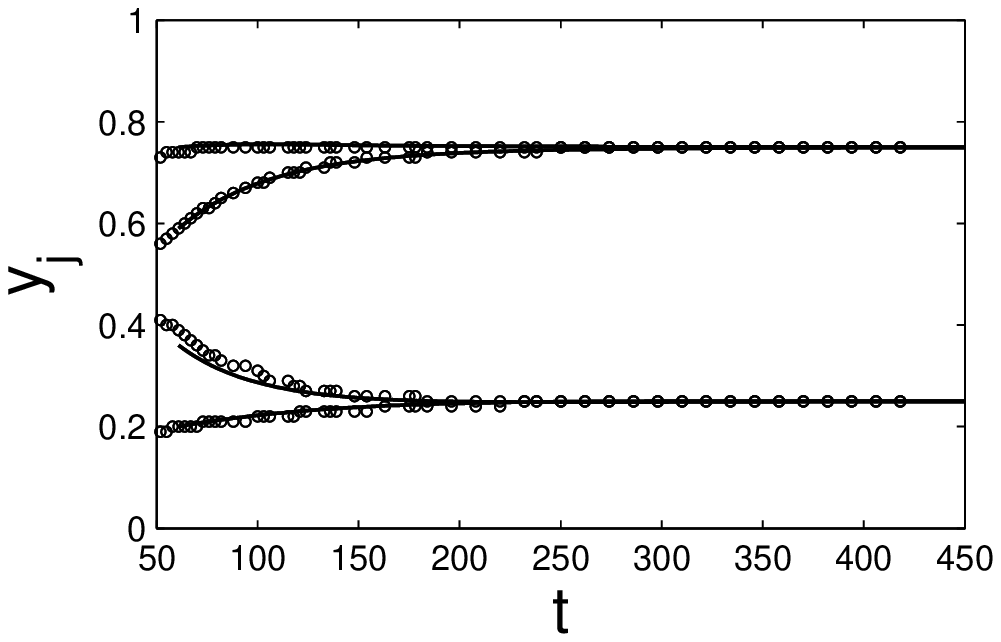}
\label{fig:exp5dyn:b}} 
\caption[Comparison of asymptotic and numerical results for the
dynamics of a four-spot pattern after a spot-splitting event] {{\em
Experiment 7.2: Fix $\ac=20$, $D=1$, $\tau=1$, and
$\eps=0.02$. Starting from $t=61$ after the spot-splitting event,
and using the initial values as in Table~\ref{tab:exp5}, we compare
the dynamics of the $4-$spot pattern from the asymptotic analysis
(solid curves) with the full numerical results (discrete markers). (a)
$x_j$ vs.~t; (b) $y_j$ vs.~t.}}
\label{fig:exp5dyn}
\end{figure}

After the splitting event, we use the new $4-$spot pattern $x_{jt}$, $
y_{jt}$ for $j=1,\ldots, 4$ at $t=61$, as given in
Table~\ref{tab:exp5}, as the initial conditions for the asymptotic DAE
system of Principal Result 3.1. At $t=61$, the source strengths
$S_{1t}, \ldots, S_{4t}$, as given in Table~\ref{tab:exp5}, are all
below the spot-splitting threshold. In Fig.~\ref{fig:exp5dyn}, we show
a very favorable comparison between the asymptotic trajectories of the
spots for $t\ge 61$ and the full numerical results computed from
(\ref{1:GS_2D}). Therefore, Principal Result 3.1 closely predicts the
motion of a collection of spots after a spot self-replication event.
We remark that in Fig.~\ref{fig:exp5dyn:a}, the $x-$coordinates
of the $1^{\mbox{st}}$ and $2^{\mbox{nd}}$ spots, as well as the
$3^{\mbox{rd}}$ and $4^{\mbox{th}}$ spots, essentially overlap 
as a result of their almost symmetric locations. 

From the spot equilibrium condition \eqref{3:finaleq1}, we calculate
the asymptotic result for the symmetric equilibrium locations
$x_{je}$, $y_{je}$ and source strengths $S_{je}$, for
$j=1,\ldots,4$. These results are given in Table~\ref{tab:exp5}, and
compare almost exactly with the results from the full numerical
simulations of the GS model (\ref{1:GS_2D}) at $t=421$ (not shown).

\begin{table}
\centering
\begin{tabular}{c c c c c || c c c || c c c}
\hline\hline
 $j$ & $x_j$ & $ y_j$ &$S_j$ & $V_j(0)$ & $x_{jt}$ & $ y_{jt}$ &$S_{jt}$
 & $x_{je}$    & $ y_{je}$    &$S_{je}$\\ \hline
 1  & 0.25 &0.66 &4.05 & 32.54 &0.30 &0.75 &3.44 &0.25 &0.75 &2.87 \\ 
 2  & 0.25 &0.14 &2.37 & 35.48 &0.28 &0.20 &3.01 &0.25 &0.25 &2.87 \\ 
 3  & 0.70 &0.40 &4.79 & 27.65 &0.76 &0.36 &2.40 &0.75 &0.25 &2.87\\
 4  &    - & -   & -   & -     &0.77 &0.59 &2.55 &0.75 &0.75 &2.87\\ \hline
\end{tabular}
\caption[Data for Experiment 3.4]{{\em Experiment 7.2:
The columns of $x_j$, $y_j$, $S_j$ and $V_j(0)$ correspond to the initial 
condition. The data $x_{jt}$, $y_{jt}$, and $S_{jt}$ correspond to
the initial conditions used for the asymptotic dynamics at $t=61$ after the
spot-splitting event. The final columns are the equilibrium results for the
$4$-spot pattern obtained from the asymptotic theory.}} \label{tab:exp5}
\end{table}

\begin{figure}[htpb]
\hspace*{-1cm}
\includegraphics[width=5.3in]{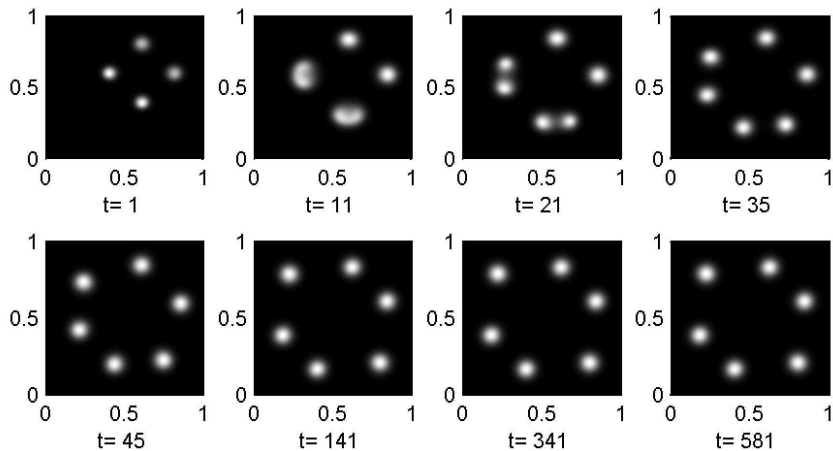}
\caption[An asymmetric four-spot pattern: two self-replication events]
{{\em Experiment 7.3: Fix $\ac=30$, $D=1$, $\tau=1$, and
$\eps=0.02$. Consider a four-spot initial pattern with spots
equally spaced on a ring of radius $r =0.2$ centered at $(0.6, 0.6)$
in the unit square. The initial source strengths $S_j$ for
$j=1,\ldots,4$ are given in Table~\ref{tab:exp6}, for which $S_3 >
\Sigma_2$ and $S_4>\Sigma_2$.  The full numerical solution for $v$ is
shown at different instants in time. The $3^{\mbox{rd}}$ and
$4^{\mbox{th}}$ spots undergo self-replication.  Subsequently, all six
spots move towards a symmetric $6-$spot equilibrium pattern.}}
\label{fig:exp6vig}
\end{figure}

\vspace*{0.2cm} 
\noindent{\em {\underline{Experiment 7.3:}}\; (An asymmetric four-spot
pattern):} We fix $\ac=30$ and $D=1$, and consider an initial $4-$spot
pattern with equally-spaced spots on a ring of radius $0.2$ centered
at $(0.6, 0.6)$ in the unit square. The initial spot locations and
source strengths are given on the left side of
Table~\ref{tab:exp6}. For this case, we find that $S_3> \Sigma_2$ and
$S_4 > \Sigma_2$, so that our asymptotic theory predicts that these
two specific spots undergo self-replication events starting at $t=0$.
The full numerical results for this pattern are given in
Fig.~\ref{fig:exp6vig}, where we observe that the $3^{\mbox{rd}}$ and
$4^{\mbox{th}}$ spots split in a direction perpendicular to their
motion. The resulting $6-$spot pattern, with initial locations
$x_{jt}$, $ y_{jt}$ for $j=1,\ldots,6$ at $t=21$, as given in
Table~\ref{tab:exp6}, are used as initial conditions for the
asymptotic DAE system in Principal Result 3.1. These six spots evolve
outwards as $t$ increases and eventually approach their equilibrium
states $x_{je}$, $y_{je}$, and $S_{je}$, for $j=1,\ldots,6$, as given
in Table~\ref{tab:exp6}, when $t=581$. The asymptotic result obtained
from the spot equilibrium condition \eqref{3:finaleq1} for the
$6-$spot equilibrium state is found to compare very favorably with the
full numerical results at $t=581$.

\begin{figure}[thbp]
\centering \subfigure[$x_j$ vs.~$t$] {
\includegraphics[width=2.4in,height=1.6in]{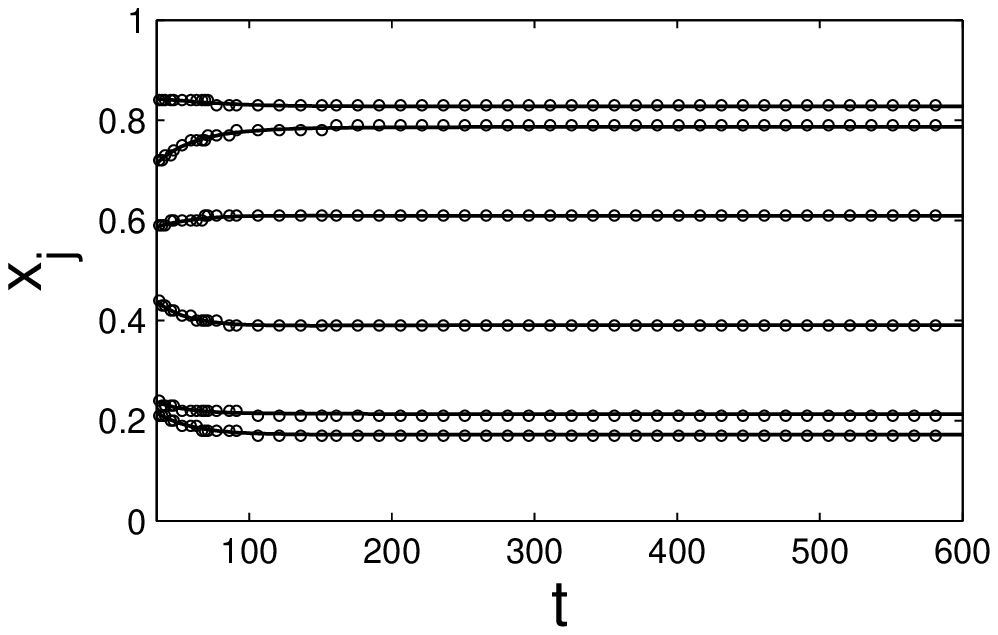}
\label{fig:exp6dyn:a}} \subfigure[$y_j$ vs.~$t$] {
\includegraphics[width=2.4in,height=1.6in]{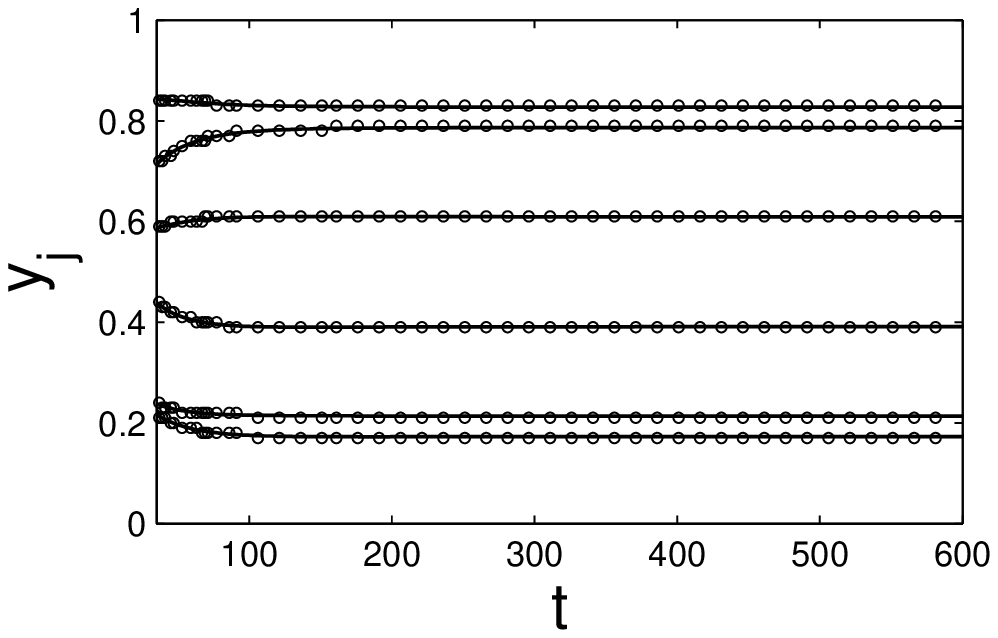}
\label{fig:exp6dyn:b}} 
\caption[Comparison of asymptotic and numerical results for the
dynamics of a four-spot pattern after two spot-splitting events] {{\em
Experiment 7.3: Fix $\ac=30$, $D=1$, $\tau=1$, and
$\eps=0.02$. For $t\geq 21$, after the two spot-splitting
events, we compare the asymptotic dynamics of the $6$ spots (solid
curves) with full numerical results (discrete markers).  (a) $x_j$
vs.~t; (b) $y_j$ vs.~t.}}
\label{fig:exp6dyn}
\end{figure}

A very favorable comparison of analytical (solid curves) and full
numerical results (discrete markers) for the spot trajectories after
$t=21$ is shown in Fig.~\ref{fig:exp6dyn}. This agreement further
validates our spot equilibrium condition \eqref{3:finaleq1}.

\begin{table}
\centering
\begin{tabular}{c c c c c|| c c c ||c c c}
\hline
 $j$   &$x_j$  & $ y_j$  &$S_j$   & $V_j$  & $x_{jt}$    & $ y_{jt}$
 &$S_{jt}$ & $x_{je}$    & $ y_{je}$    &$S_{je}$\\ \hline
 1  &0.80 & 0.60 & 2.82 & 36.21  &0.84 &0.58  &3.13 &0.83 &0.61  &2.91\\
 2  &0.60 &0.80  & 2.82 & 36.21  &0.58 &0.84  &3.13 &0.61 &0.83  &2.91\\
 3  &0.40 & 0.60 & 5.69 & 19.07  &0.25 &0.49  &2.93 &0.17 &0.39  &2.91\\
 4  &0.60 & 0.40 & 5.69 & 19.07  &0.26 &0.66  &2.73 &0.21 &0.79  &3.07\\
 5  &-    &-     &-     &-       &0.49 &0.25  &2.93 &0.39 &0.17  &2.91\\
 6  &-    &-     &-     &-       &0.66 &0.26  &2.73 &0.79 &0.21  &3.07\\\hline
\end{tabular}
\caption[Data for Experiment 5.4]{{\em Experiment 7.3: The
columns of $x_j$, $y_j$, $S_j$ and $V_j(0)$ correspond to the initial
condition. The data $x_{jt}$, $y_{jt}$, and $S_{jt}$ correspond to the
initial conditions used for the asymptotic dynamics at $t=21$ after
the two spot-splitting events. The final columns are the equilibrium
results for the $6$-spot pattern.}} \label{tab:exp6}
\end{table}

\subsection{The Unit Disk: A Ring Pattern of Spots with a Center Spot} 
\label{sec:asy_disk}
 
In this subsection, we let $\eps=0.02$ and $\tau=1$, and we construct a
spot pattern for which $k-1$ spots are equally spaced on a ring of
radius $r$, with $0<r<1$, and with an additional spot located at the
center of the unit disk. The $k-1$ spots on the ring have a common
source strength $S_c$, while the $k^{\mbox{th}}$ spot at the origin
has a source strength denoted by $S_k$. For this pattern, the centers
of the spots, written as complex numbers inside the unit
disk, are at
\[ \mathbf{x}_j = r e^{2 \pi i j /(k-1)} \,, \quad j=1,\ldots,k-1 \,;
\qquad \mathbf{x}_k = 0 \,, \qquad k\geq 3\,.\] We will analyze the
dynamics of the spots and the possibility of spot-splitting for this
pattern for the limiting case $D\gg 1$, for which the Green's function
$G$ can be approximated by the Neumann Green's function
$G^{(N)}$. Recall, that $G^{(N)}$ has a very simple analytical formula
in the unit disk (see Appendix A.2). As a result, the dynamics of the
spots and the conditions for spot self-replication can be obtained in
a very explicit form.

For this pattern, Principal Result 3.1 shows that the $k^{\mbox{th}}$
spot at the center of the disk is stationary. In addition, by
differentiating \eqref{3:pk_dlarge}, it readily follows that the
dynamics of the remaining $k-1$ spots on the ring satisfy
\begin{equation}
\mathbf{x}^{\p}_j \sim - 2 \pi \eps^2 \gamma(S_c) \left[
\left(\frac{\,S_c}{\,2}\right) \frac{p_{k-1}^{\p}(r)}{k-1} 
 e^{2 \pi i j/(k-1)} + S_k G_{r}(r;0) \, e^{2 \pi i j/(k-1)} \right]\,,
  \qquad j=1,\ldots,k-1 \,.  \label{3:cs_xj_1}
\end{equation}
Here $G(r;0)$ is the radially symmetric reduced-wave Green's function
with singularity at the origin, and $p_{k-1}(r)$ is defined in
(\ref{3:pk_dlarge}), with limiting asymptotics for $D\gg 1$ also given in
(\ref{3:pk_dlarge}).  Since $\mathbf{x}_j=r \, e^{2\pi i j/(k-1)}$ for
$j=1,\ldots,k-1$, then (\ref{3:cs_xj_1}) reduces to an ODE for the
ring radius given by \bsub
\label{3:cs_all}
\begin{equation}
  \frac{d r}{dt} \sim - 2 \pi \eps^2 \gamma(S_c) \left[
\left(\frac{\,S_c}{\,2}\right) \frac{p_{k-1}^{\p}(r)}{k-1} 
   + S_k G_{r}(r;0) \right]\,, \label{3:cs_xj}
\end{equation}
in terms of $S_c=S_c(r)$ and $S_k=S_k(r)$. From \eqref{3:ASsmallD}, we
obtain that $S_c=S_c(r)$ and $S_k=S_k(r)$ are the solutions to the
coupled two-component nonlinear algebraic system
\begin{equation}
 \ac = S_c \left( 1+ \frac{2\pi\nu}{k-1} p_{k-1}(r) \right) +\nu
\chi(S_c)+ 2\pi\nu G(r;0) S_k  \,, \qquad
 \ac = 2\pi \nu (k-1)  G(r;0) S_c   +\nu \chi(S_k)+ S_k\left(1 + 2\pi\nu
 R_{0,0} \right) \,, \label{3:1ring1hole}
\end{equation}
where $R_{0,0}$ is the regular part of $G(r;0)$ at the origin.
\esub
The problem \eqref{3:cs_all} is a DAE system for the ring radius, consisting
of the ODE (\ref{3:cs_xj}) coupled to the two nonlinear constraints
(\ref{3:1ring1hole}). The core problem for an individual spot enters only
through the determination of $\chi(S)$ and $\gamma(S)$.

Since $p_{k-1}(r)$ is only known explicitly when $D\gg 1$, we use the
large $D$ asymptotics \eqref{3:pk_dlarge}, together with
\begin{equation}
  G(r;0) \sim D + G^{(N)}(r;0) = D -\frac{1}{2\pi} \ln r + \frac{r^2}{4 \pi} -
\frac{3}{8\pi}, \qquad R_{0,0} \sim D + R^{(N)}_{0,0} = D -\frac{3}{8\pi}
 \,, \qquad D\gg 1 \,, \label{3:grbig}
\end{equation}
to calculate the equilibrium ring radius. By setting $r^{\p}=0$ in
(\ref{3:cs_xj}), we obtain the following transcendental equation for the
equilibrium ring radius $r_e$ when $D\gg 1$:
\begin{equation}
\frac{S_k}{S_c} \left( 1 - r^2 \right) + \frac{(k-2)}{2} -
 r^2 (k-1) - (k-1)\,\frac{r^{2k-2}}{1 - r^{2k-2}} = 0 \,.
\label{3:1ring1hole:a}
\end{equation}
By solving the three nonlinear algebraic equations in
(\ref{3:1ring1hole}) and (\ref{3:1ring1hole:a}) we obtain equilibrium
values $r_e$, $S_{ce}$, and $S_{ke}$.

\begin{figure}[htpb]
\begin{center}
\subfigure[$S_c$ vs.~$r$]
{\includegraphics[width=3.0in, height=1.8in,clip]{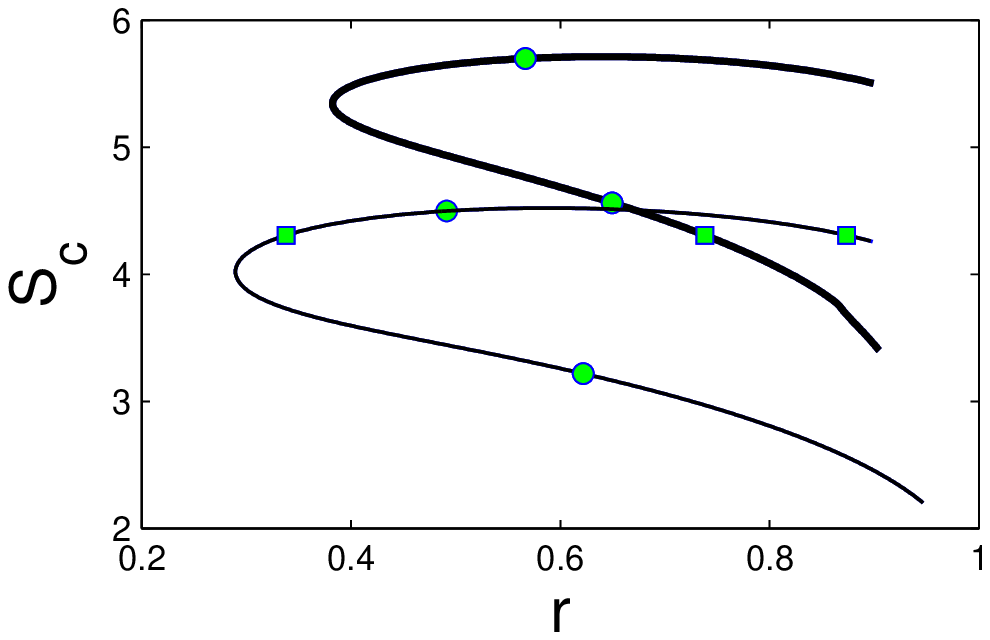}
\label{fig:1ring1hole:ac} }
\subfigure[$S_k$ vs.~ $r$]
{\includegraphics[width=3.0in, height=1.8in,clip]{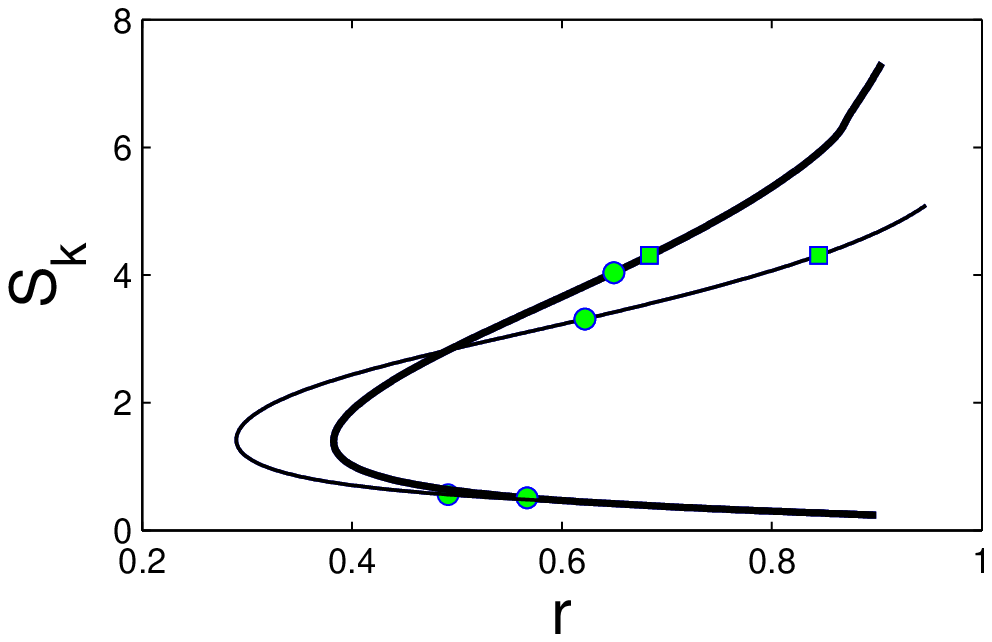}
\label{fig:1ring1hole:ak} }\\
\subfigure[$S_c$ vs.~$r$]
{\includegraphics[width=3.0in, height=1.8in,clip]{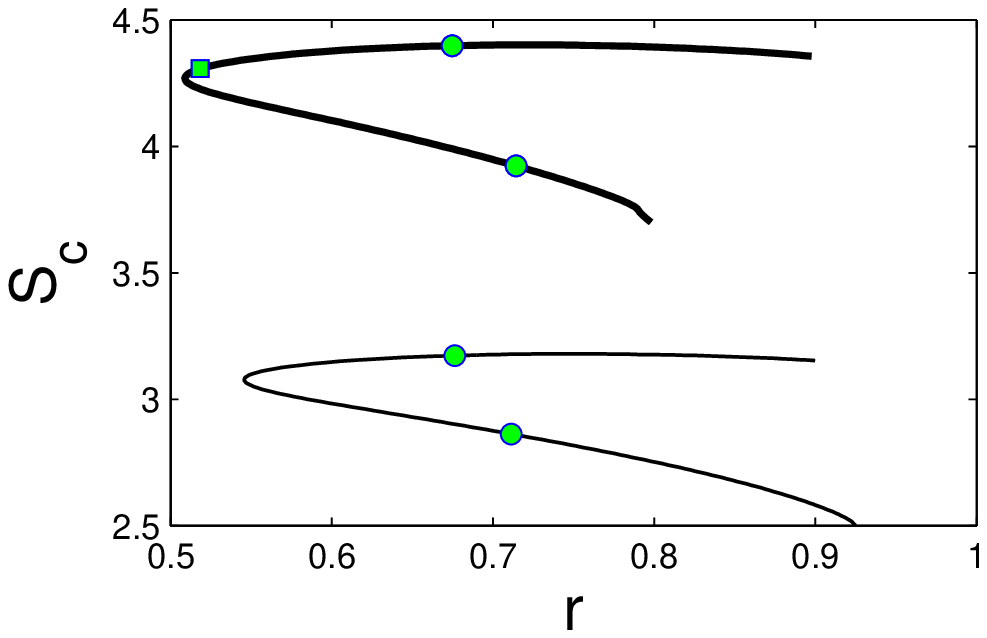}
\label{fig:1ring1hole:bc} }
\subfigure[$S_k$ vs.~$r$]
{\includegraphics[width=3.0in,height=1.8in,clip]{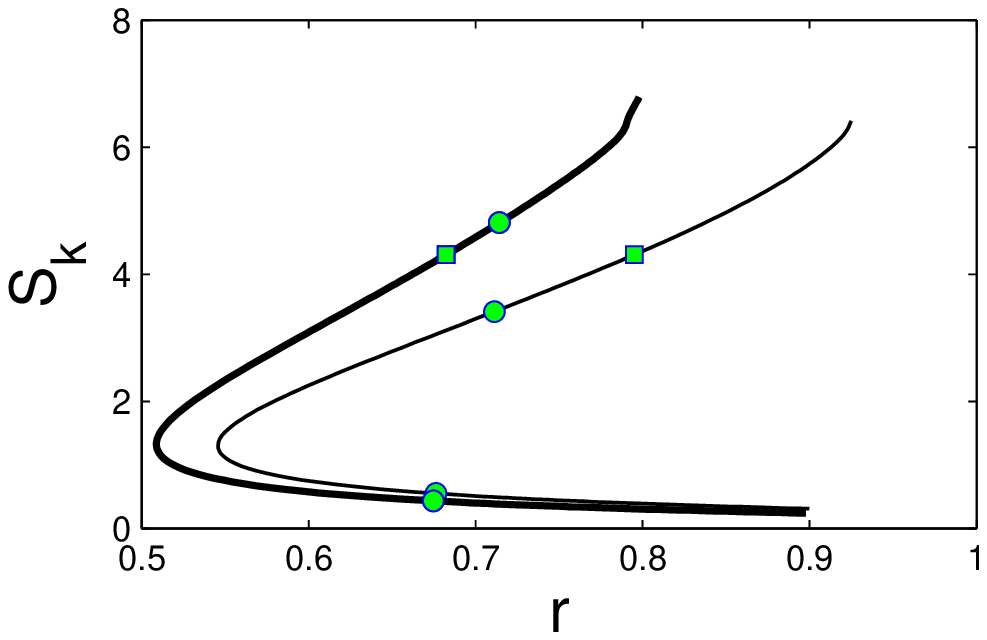} 
\label{fig:1ring1hole:bk} }
\caption[One center-spot and all other spots on a ring]{{\em Fix
$D=3.912$, $\eps=0.02$. (a) The solid curves correspond to $\ac=22,\,
k=3$ and the heavy solid curves correspond to $\ac=38,\, k=4$. The
open dots correspond to spot-splitting thresholds, and the solid dots
correspond to equilibrium ring radii. (b) The heavy solid curves
correspond to $\ac=82,\, k=10$ and the solid curves correspond to
$\ac=60,\, k=10$. Left: $S_c$ vs.~$r$; Right: $S_k$ vs.~$r$. Each
upper (lower) branch for $S_c(r)$ corresponds to the lower (upper)
branch for $S_k(r)$.}} \label{fig:1ring1hole}
\end{center}
\end{figure}

In Fig.~\ref{fig:1ring1hole:ac} and Fig.~\ref{fig:1ring1hole:ak}, we
fix $D = 3.912$ and we plot the numerically computed source strengths
$S_c$ and $S_k$, respectively, versus the ring radius $r$ for $k=3,
\ac = 22$, and for $k=4, \ac = 38$, as computed from
(\ref{3:1ring1hole}). In this computation we used the large $D$
asymptotics of (\ref{3:pk_dlarge}) and (\ref{3:grbig}) in
(\ref{3:1ring1hole}). In Fig.~\ref{fig:1ring1hole:ac} the upper
(lower) branch of the $S_c=S_c(r)$ curve corresponds to the lower
(upper) branch of the $S_k=S_k(r)$ curve shown in
Fig.~\ref{fig:1ring1hole:ak}.  For each pattern, there are two
equilibrium ring radii, which are marked by the solid dots. The
spot-splitting threshold $\Sigma_2$ is also marked on the curves in
Fig.~\ref{fig:1ring1hole:ac} and Fig.~\ref{fig:1ring1hole:ak} by the
open dots. We now show how these figures can be readily
used to predict spot-splitting behavior.

\begin{figure}[htpb]
\begin{center}
{\includegraphics[width=3.3cm,clip]{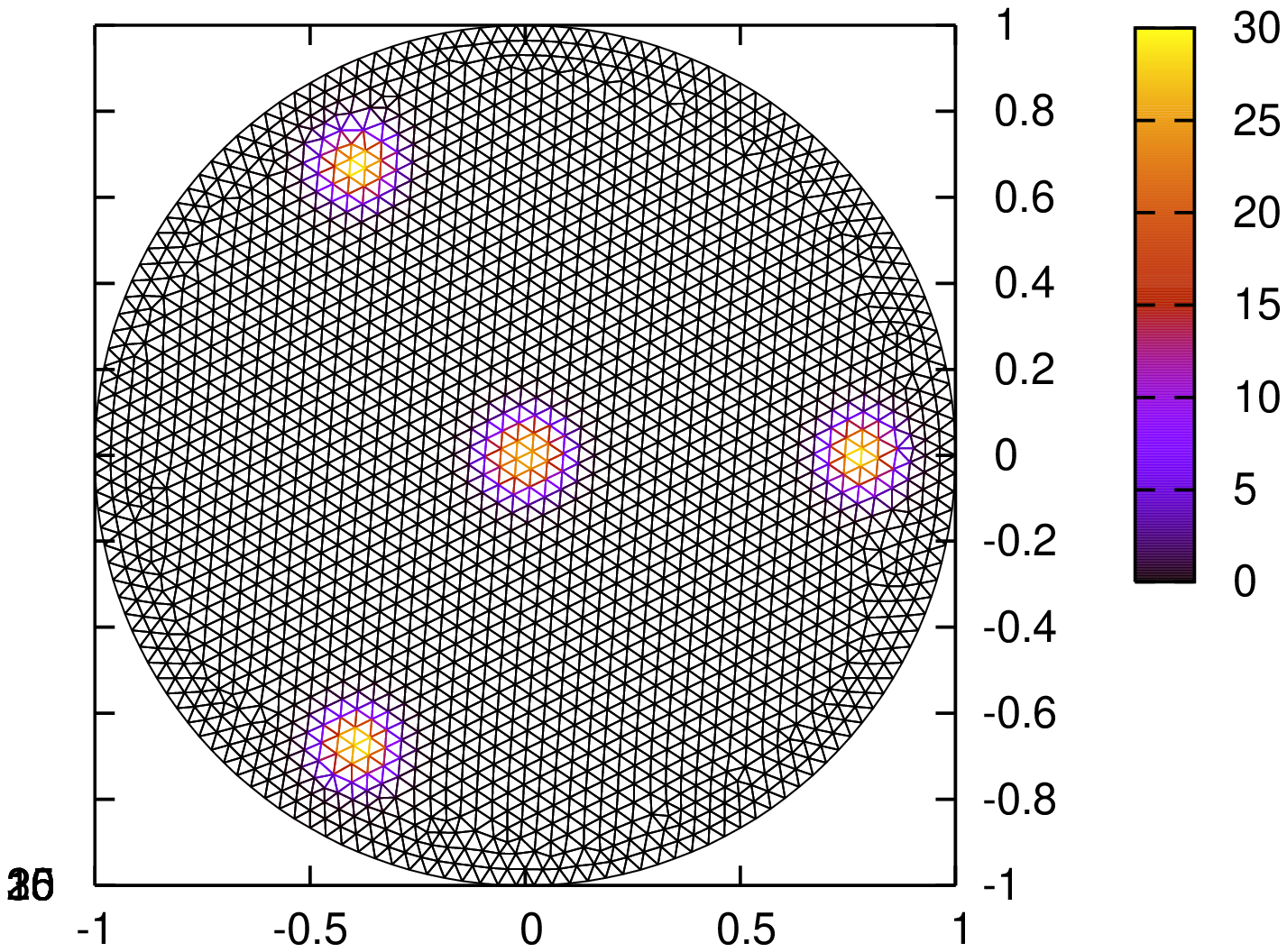}}
{\includegraphics[width=3.3cm,clip]{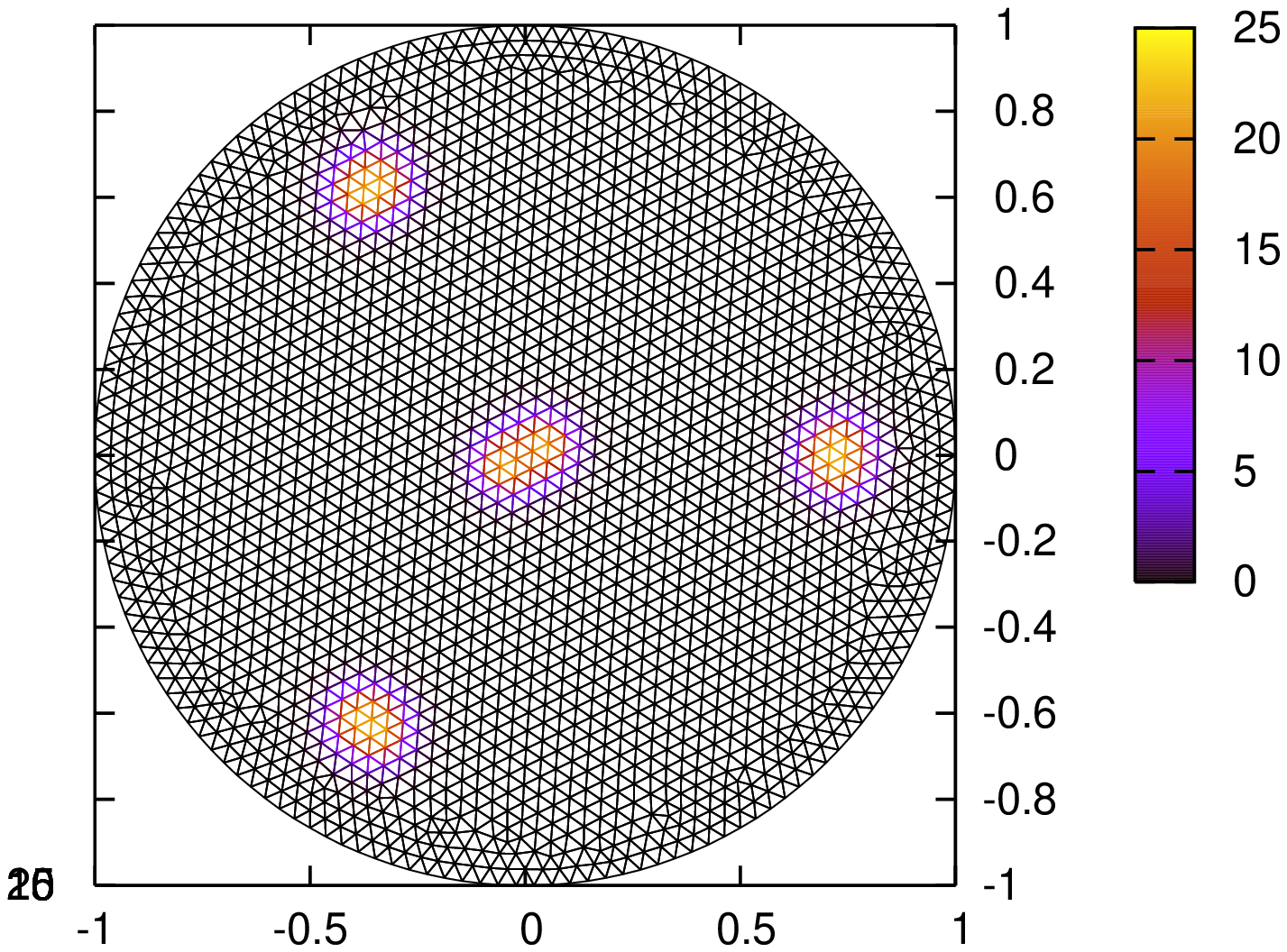}}
{\includegraphics[width=3.3cm,clip]{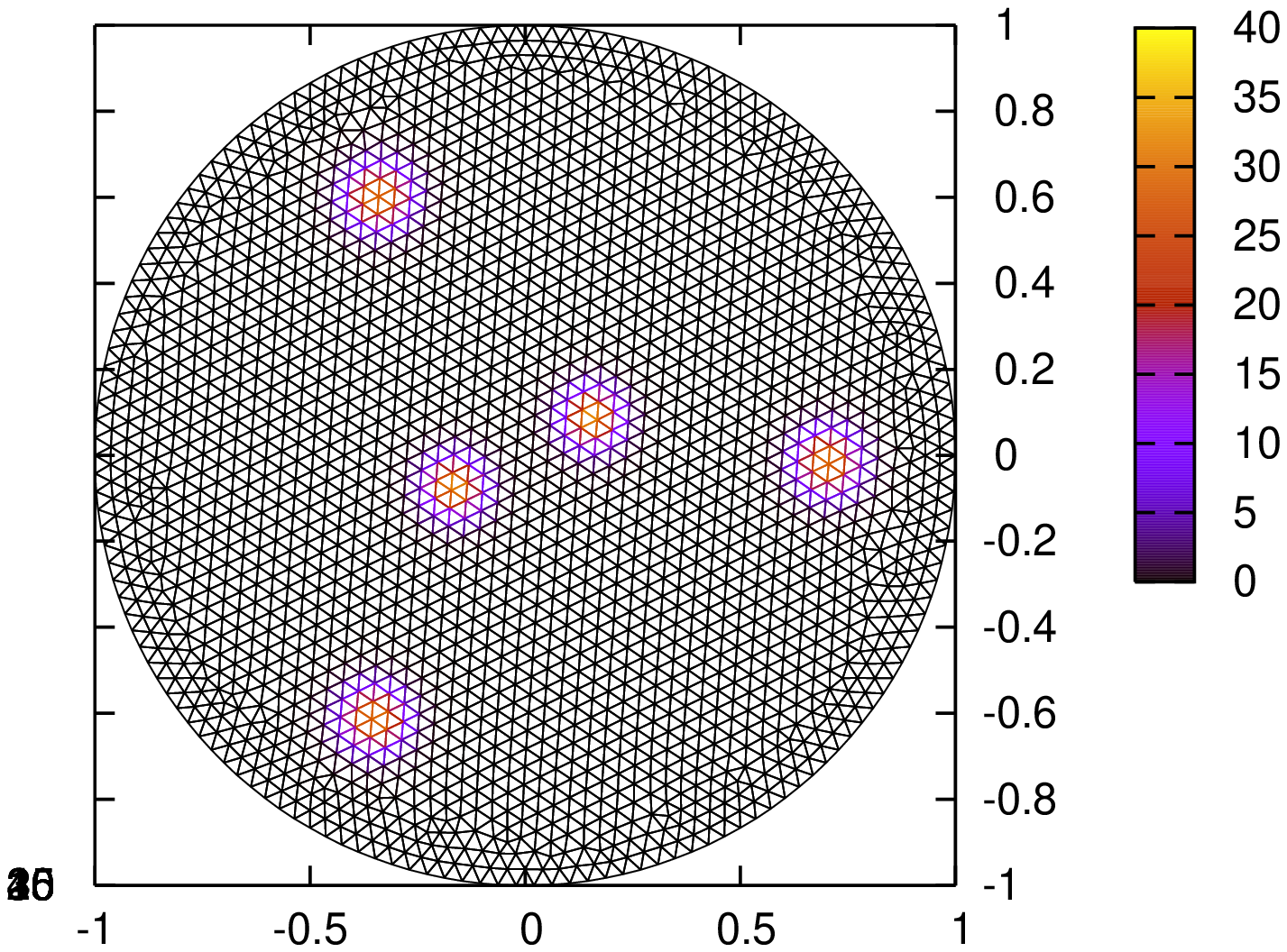}}
{\includegraphics[width=3.3cm,clip]{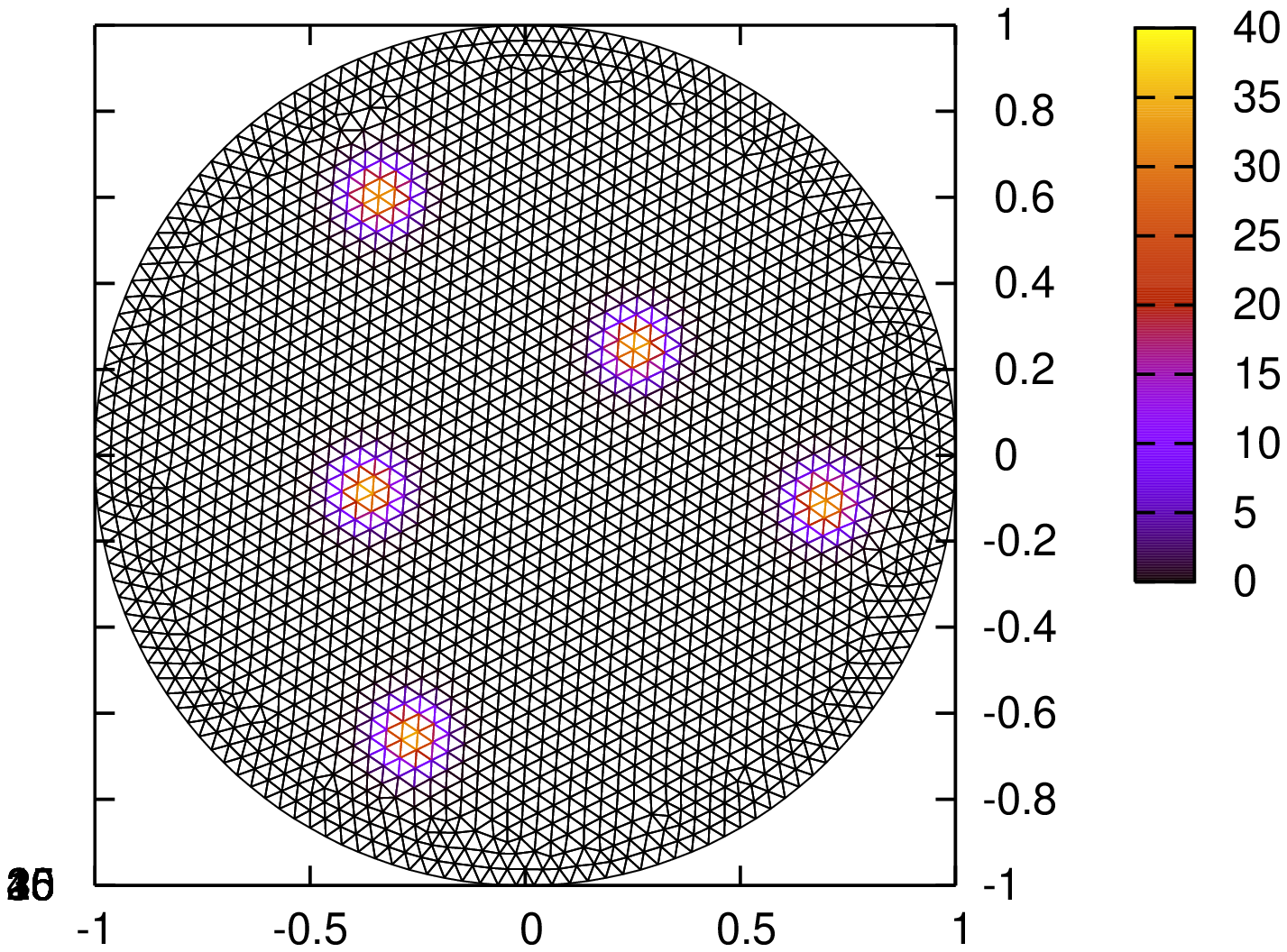}}
\end{center}
\begin{center}
{\includegraphics[width=3.3cm,clip]{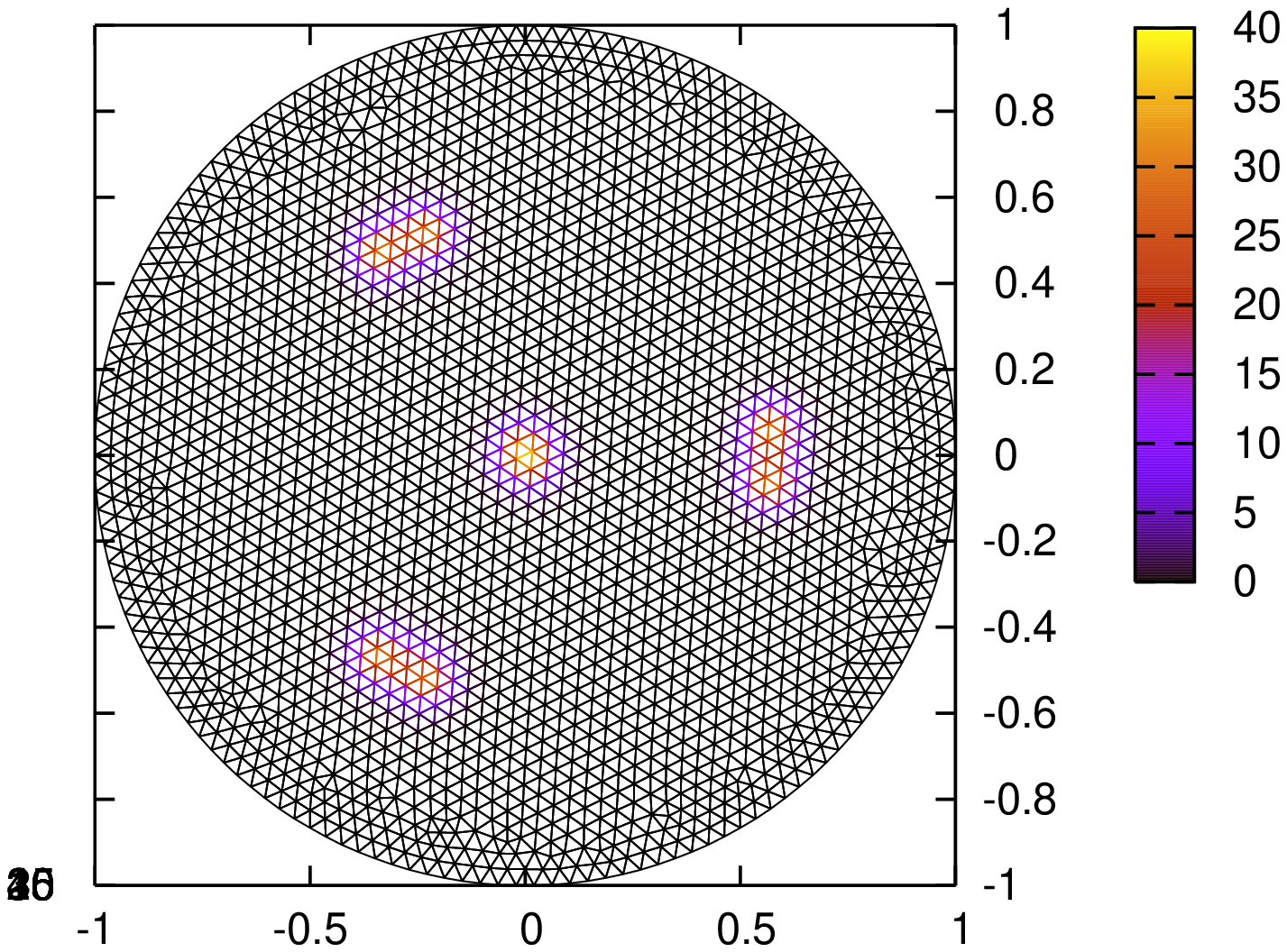}} 
{\includegraphics[width=3.3cm,clip]{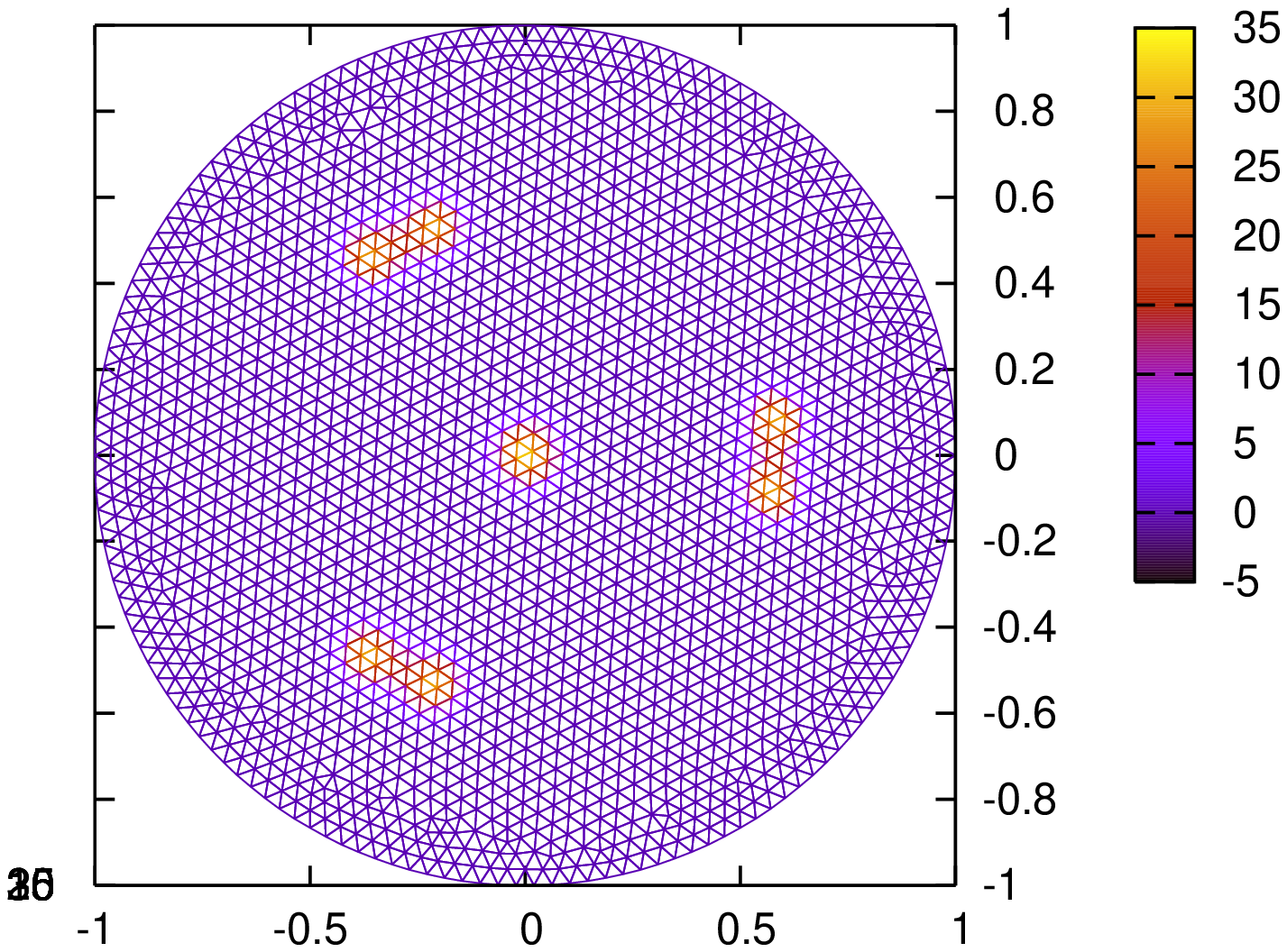}}
{\includegraphics[width=3.3cm,clip]{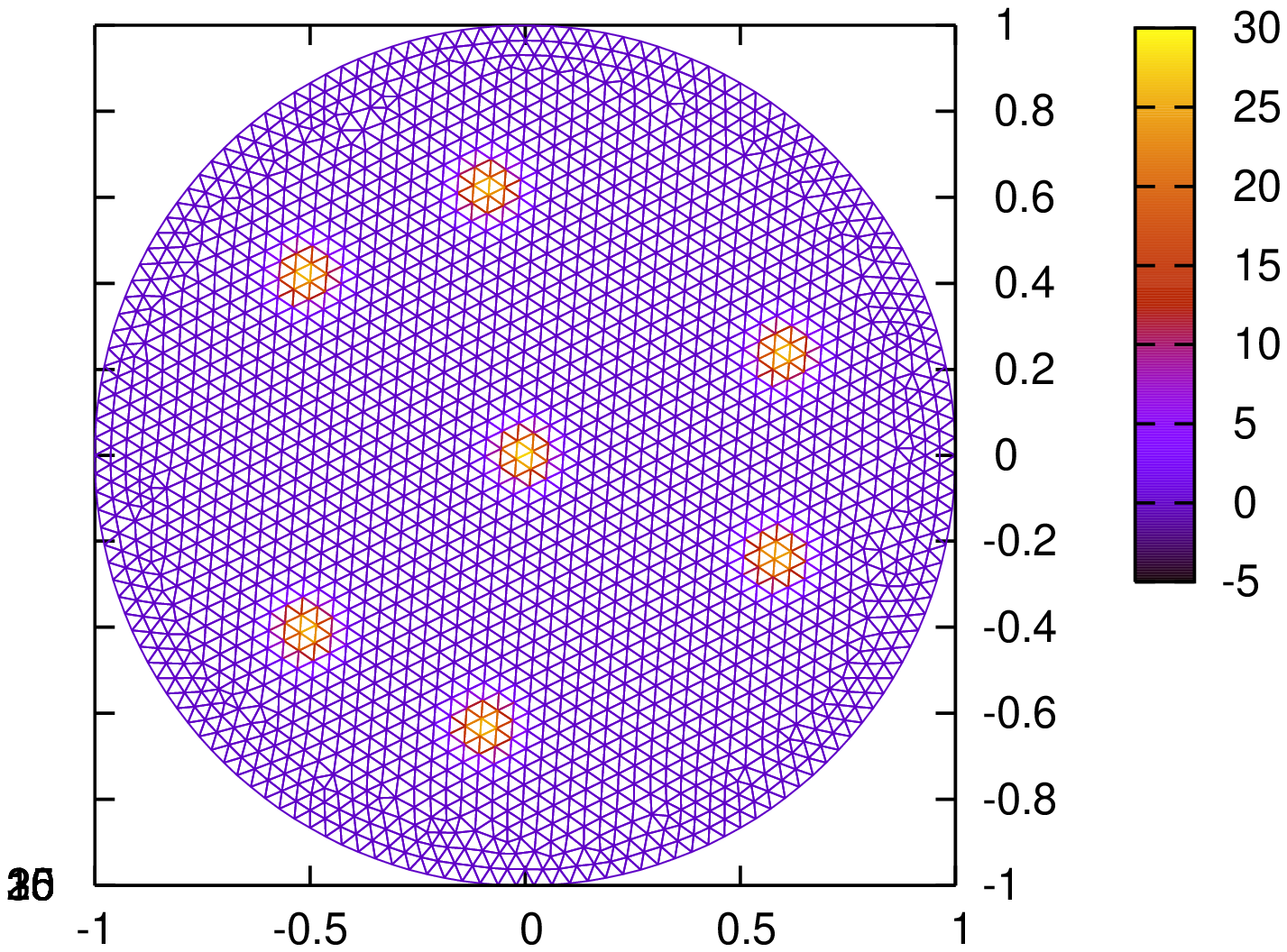}}
{\includegraphics[width=3.3cm,clip]{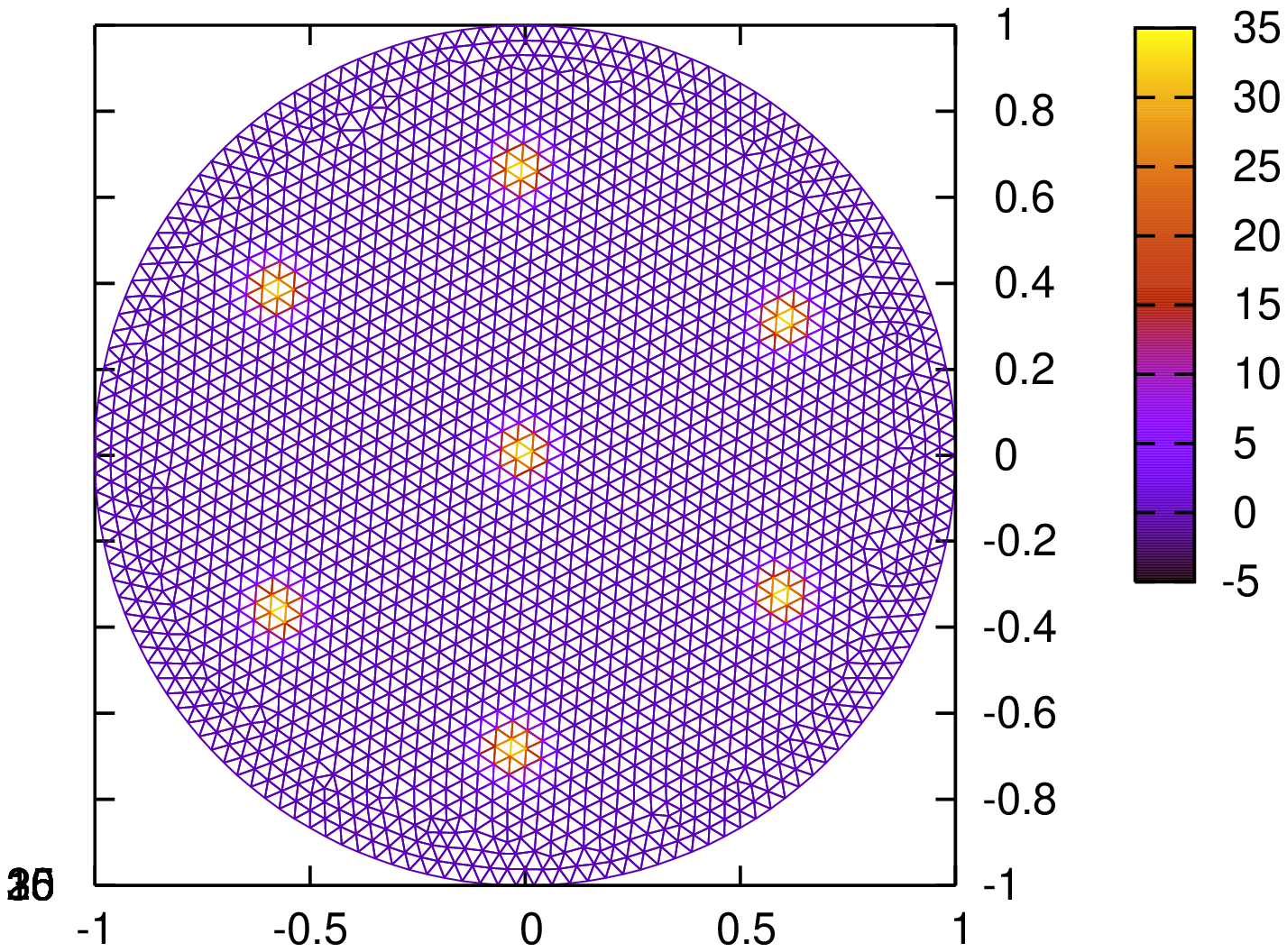}}
\end{center}
\begin{center}
\caption[One center-spot and three spots on a ring: different
splitting patterns]{{\em Experiment 7.4: Fix $\ac=38$, $D=3.912$, and
$\eps=0.02$. The numerical results for $v$ computed from the GS model
\eqref{1:GS_2D} at different instants in time are shown. The initial
condition has three equally-spaced spots on a ring together with a
center-spot. Top row: initial ring radius $r=0.8$: Plots of $v$ from
left to right at times $t=4.6,46,102,298$. Bottom row: initial ring
radius $r=0.5$: Plots of $v$ from left to right at times
$t=4.6,56,140,298$.}} \label{fig:exp9}
\end{center}
\end{figure}

\vspace*{0.2cm}\noindent{\em {\underline{Experiment 7.4:}}\; (A
one-ring pattern with a center-spot for $D\gg {\mathcal O}(1)$:
Different types of spot-splitting):} We fix $\ac=38$ and $D=3.912$ and
consider an initial pattern of one center-spot and three
equally-spaced spots on a ring of initial radius $r=0.8$ at time
$t=0$.  We then predict the dynamical behavior of this initial spot
pattern from the heavy solid curves in Fig.~\ref{fig:1ring1hole:ac}
and Fig.~\ref{fig:1ring1hole:ak}.  From the data used to plot
Fig.~\ref{fig:1ring1hole:ak}, we obtain from the upper branch of the
$S_k$ versus $r$ curve that $S_k=5.38>\Sigma_2$ at time
$t=0$. Correspondingly, from the lower branch of the $S_c$ versus $r$
curve in Fig.~\ref{fig:1ring1hole:ac}, we obtain that $S_c<\Sigma_2$
at $t=0$. Therefore, we predict that the center-spot will undergo
spot-splitting starting at $t=0$. The other three spots remain on a
ring whose radius decreases slowly in time. For this parameter set,
the full numerical results computed from the GS model \eqref{1:GS_2D},
shown in the top row of Fig.~\ref{fig:exp9}, confirm the prediction
based on the asymptotic theory.

In contrast, suppose that the initial ring radius is decreased from
$r=0.8$ to $r=0.5$. Then, from the heavy solid curve in
Fig.~\ref{fig:1ring1hole:ac}, we now calculate that $S_c\approx
4.92>\Sigma_2$ from the lower branch of the $S_c$ versus $r$ curve.
Correspondingly, from the upper branch of the $S_k$ versus $r$ curve
in Fig.~\ref{fig:1ring1hole:ak} we get $S_k < \Sigma_2$ at
$t=0$. Thus, for this initial ring radius, we predict that the three
spots on the ring undergo simultaneous spot-splitting starting at
$t=0$, while the center-spot does not split. The full numerical
results shown in the second row of Fig.~\ref{fig:exp9}, as computed
from the GS model \eqref{1:GS_2D}, fully support this prediction based
on the asymptotic theory.

We remark that based on the relative locations of the equilibrium ring
radius and the spot-splitting threshold in
Fig.~\ref{fig:1ring1hole:ac} and Fig.~\ref{fig:1ring1hole:ak}, we
conclude that it is not possible to obtain a dynamically-triggered
spot-self replication instability for these parameter values. This
more exotic instability is shown in our final example.

\vspace*{0.2cm}\noindent{\em {\underline{Experiment 7.5:}}\; (A
one-ring pattern with a center-spot for $D\gg {\mathcal O}(1)$: A
dynamically-triggered instability):} In Fig.~\ref{fig:1ring1hole:bc}
and Fig.~\ref{fig:1ring1hole:bk}, we fix $k=10$ and $D=3.912$ and plot
$S_c$ and $S_k$ versus $r$ for $\ac=82$ and $\ac=60$. In these figures,
each solid dot and empty circle denotes an equilibrium ring radius and
the spot-splitting threshold, respectively.

Consider first the solid curve in Fig.~\ref{fig:1ring1hole:bc} and
Fig.~\ref{fig:1ring1hole:bk} corresponding to $\ac=60$. For this value
of $\ac$ it follows that only the center-spot can split and the
minimum ring radius that leads to this spot-splitting is
$r=0.80$. Based on the relative locations of the equilibrium state and
the spot-splitting threshold on the solid curve for $S_k=S_k(r)$ in
Fig.~\ref{fig:1ring1hole:bk}, we conclude that a dynamically-triggered
spot self-replication instability is not possible.

\begin{figure}[htpb]
\begin{center}
{\includegraphics[width=3.3cm,clip]{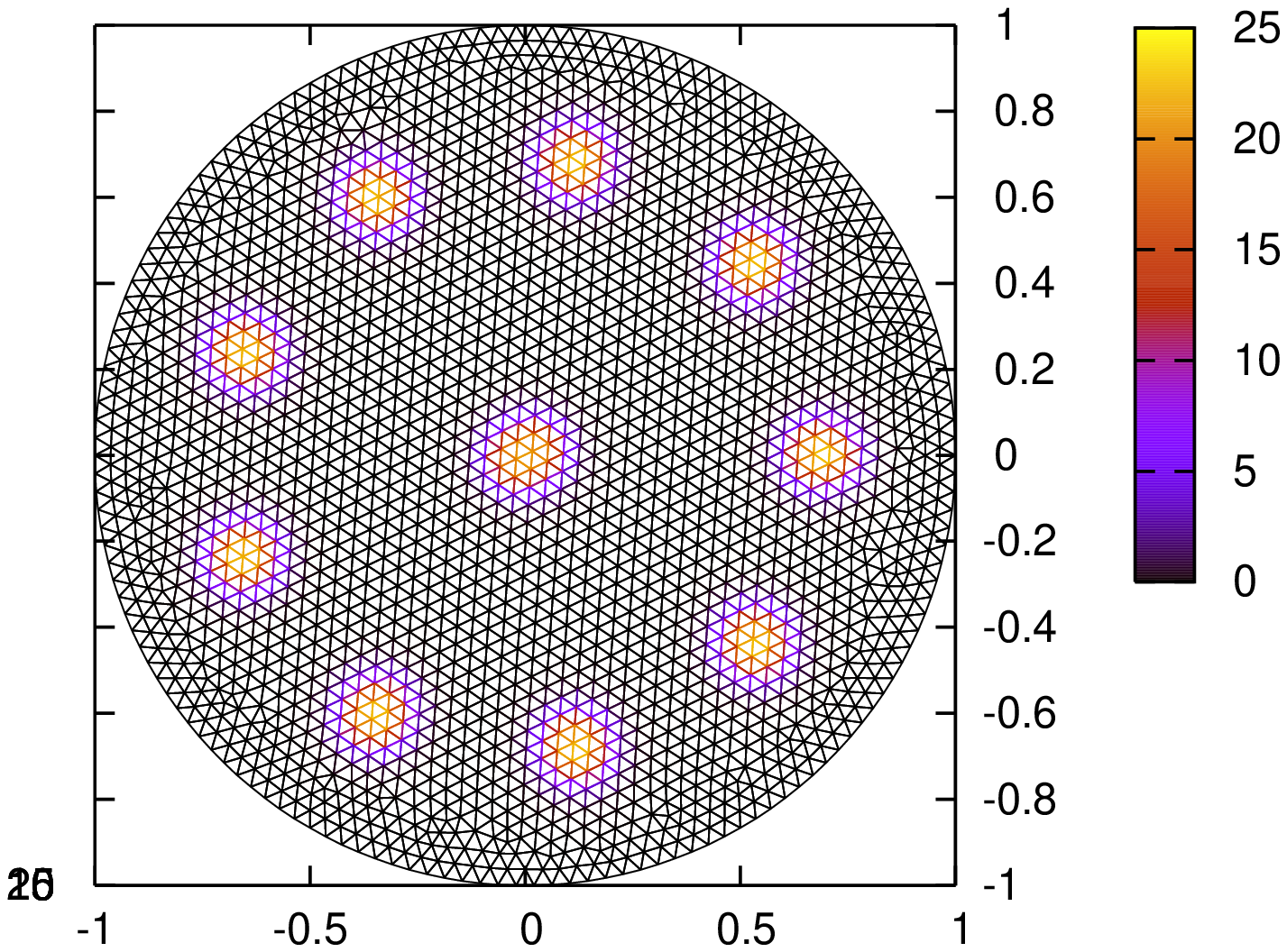}}
{\includegraphics[width=3.3cm,clip]{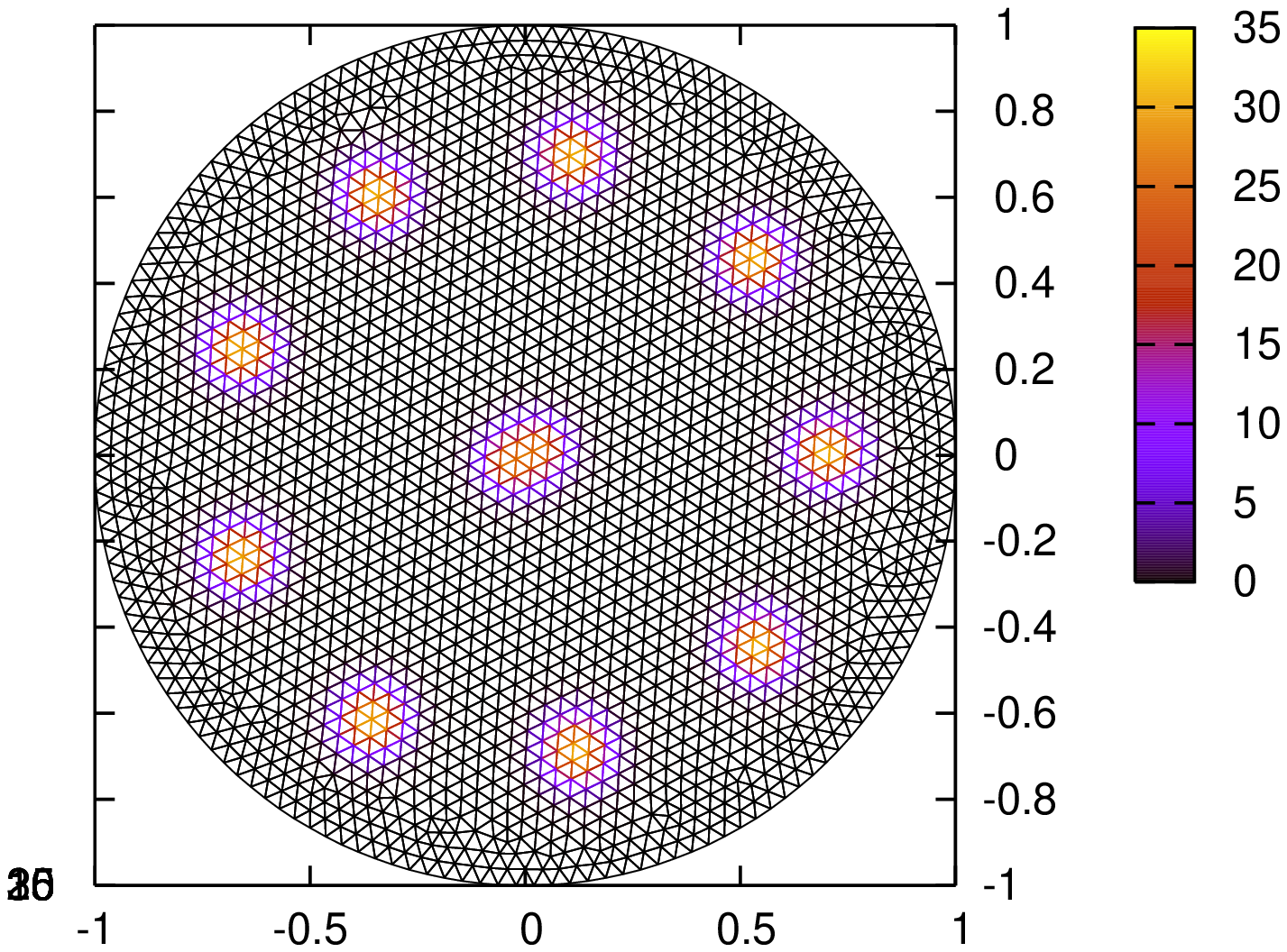}}
{\includegraphics[width=3.3cm,clip]{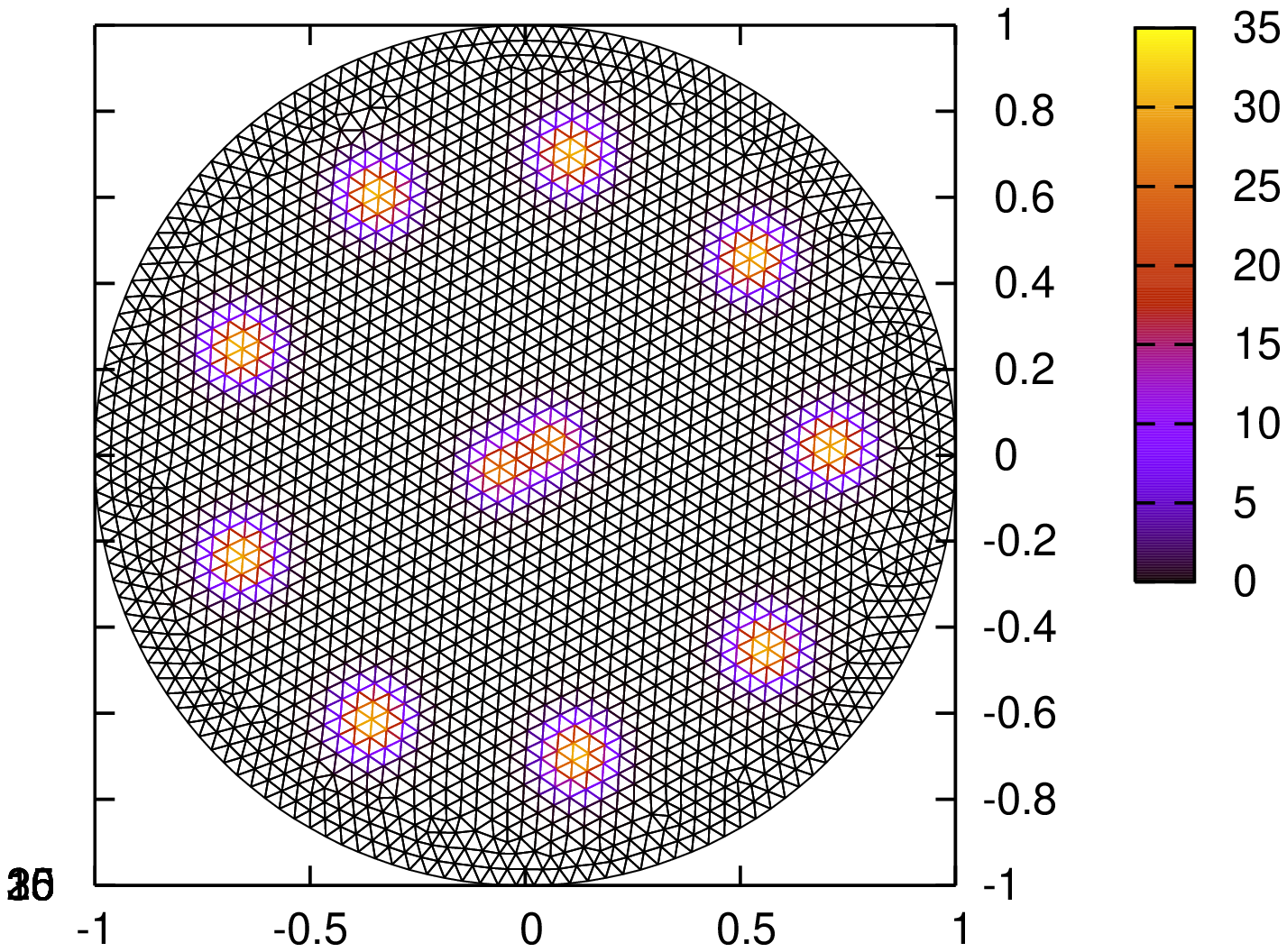}}
{\includegraphics[width=3.3cm,clip]{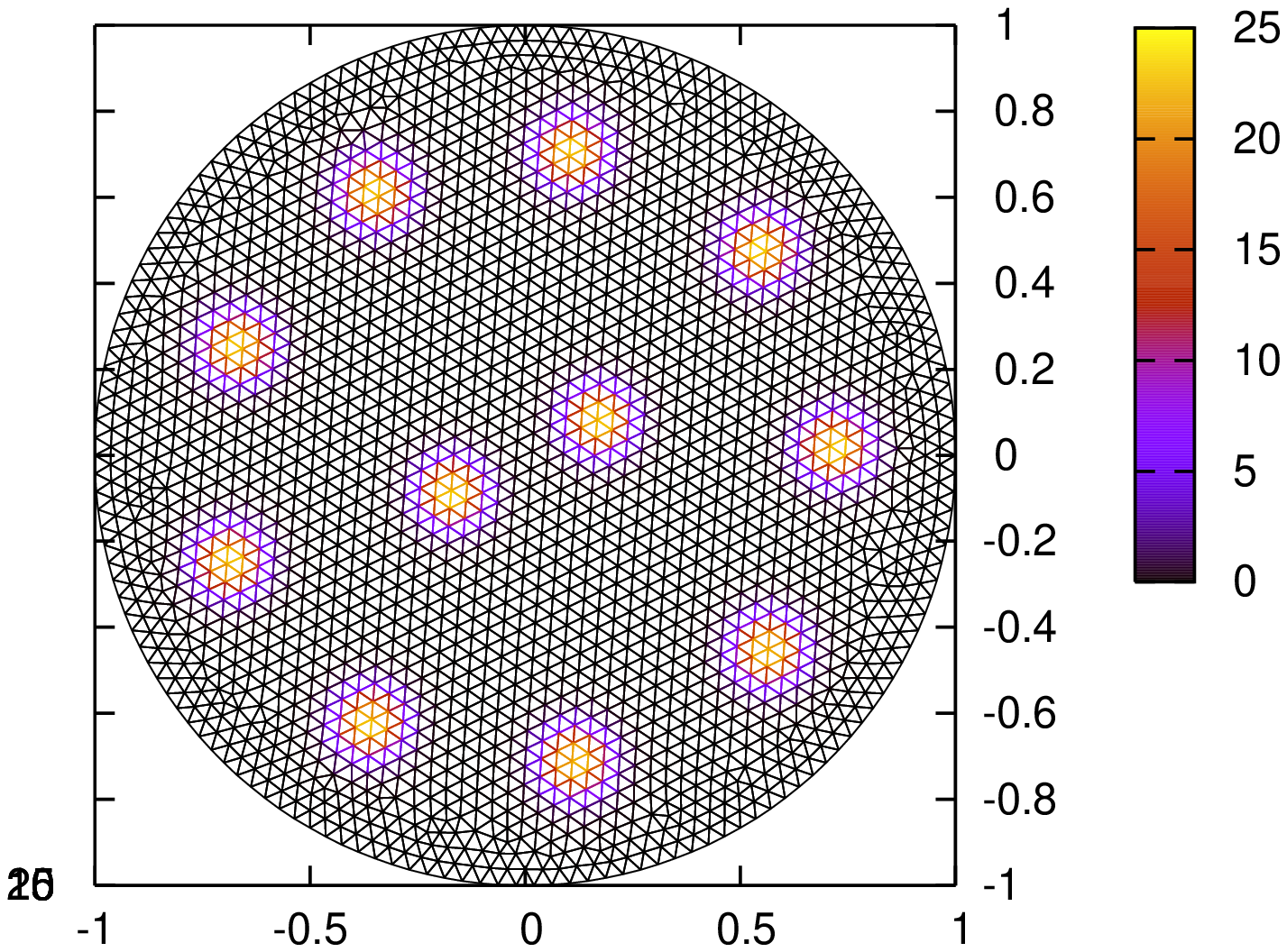}}
\caption[One center-spot and nine spots on a ring: a dynamically-triggered
 instability]{{\em Experiment 7.5: Fix $\ac=82$, $D=3.912$, and
 $\eps=0.02$. The full numerical results computed from the GS model
 \eqref{1:GS_2D} are shown at different instants in time. The initial
 pattern has nine spots equally spaced on a ring of initial radius
 $r=0.62$ together with a center-spot. From left to right the plots of
 $v$ are shown at times $t=60,93,144,214$.}} \label{fig:exp10}
\end{center}
\end{figure}

Next, suppose that $\ac$ is increased to $\ac=82$, and consider an
initial ring radius of $r=0.62$ at time $t=0$. Then, from the heavy
solid curve in Fig.~\ref{fig:1ring1hole:bc} we calculate the initial
source strengths as $S_c \approx 4.03$ on the lower branch of the
$S_c$ versus $r$ curve, and correspondingly $S_k = 3.38$ at $t=0$ on
the upper branch of the $S_k$ versus $r$ curve in
Fig.~\ref{fig:1ring1hole:bk}. Therefore, we predict that the initial
pattern is stable to spot-splitting. However, for $t>0$, we predict
from Fig.~\ref{fig:1ring1hole:bk} that the center-spot will eventually
split at the ring radius $r=0.68$ when $S_k$ crosses above
$\Sigma_2$. This occurs at a time before the ring radius has
approached its equilibrium state at $r_{e1} = 0.71$. Therefore, our
asymptotic theory predicts that a dynamical-triggered spot-splitting
instability will occur for this parameter set. This behavior is
confirmed by the full numerical results shown in Fig.~\ref{fig:exp10}
computed from the GS model \eqref{1:GS_2D}.

\setcounter{equation}{0}
\setcounter{section}{7}
\section{Discussion} \label{sec:discussion}

For the GS model \eqref{1:GS_2D} in a two-dimensional domain in the
semi-strong interaction regime $D={\mathcal O}(1)$, and for
$A={\mathcal O}(-\eps\ln\eps)$, the slow dynamics and instability
mechanisms of quasi-equilibrium multi-spot patterns was studied by a
hybrid asymptotic-numerical method.  By using a singular perturbation
approach that accounts for all terms in powers of $\nu={-1/\ln\eps}$,
a multi-spot quasi-equilibrium pattern was constructed by
asymptotically matching a local approximation of the solution near
each spot to a global representation of the solution defined in terms
of the reduced-wave Green's function. The local problem near each
spot, referred to as the core problem, consisted of a
radially-symmetric BVP system that must be solved numerically.  In the
global, or outer, representation of the solution, each spot at a given
instant in time is asymptotically approximated as a Coulomb
singularity for $u$ of strength $S_j$ at location
$\mathbf{x}_j\in\Omega$ for $j=1,\ldots,k$. In Principal Result 3.1 a
DAE system was derived for the slow time evolution of the coordinates
$\mathbf{x}_j$ and $S_j$ for $j=1,\ldots,k$, which characterizes the
slow dynamics of the $k$-spot quasi-equilibrium pattern.  The method
used to construct quasi-equilibria is an extension of that developed
in \cite{sumlog_ward} (see also \cite{trap_KTW:2005}, \cite{coombs},
\cite{pillay}) to sum logarithmic asymptotic expansions for 
linear PDE models in perforated 2-D domains.

The stability of the multi-spot quasi-equilibrium solution to three
distinct types of ${\mathcal O}(1)$ time-scale instabilities was 
studied. From a numerical study of a local eigenvalue problem near each
spot, an explicit criterion for the initiation of a peanut-splitting
instability was obtained, and it was verified numerically
that this instability triggers a nonlinear spot self-replication
event. In addition, a singular perturbation method was used to formulate
a global eigenvalue problem characterizing the onset of either oscillatory
or competition instabilities for the multi-spot pattern. This global
eigenvalue problem was studied for special multi-spot patterns for which
a certain Green's matrix is circulant.

Our hybrid asymptotic-numerical framework for spot dynamics and
stability was implemented numerically for some spot patterns in $\R^2$
and in the unit disk and square. The results from our theory were
shown to very compare favorably with those obtained from full
numerical simulations of the GS model (\ref{1:GS_2D}). One key finding
is that dynamically-triggered instabilities can occur for the GS model
in a wide parameter range. The qualitative differences in multi-spot
solution behavior between the GS model \eqref{1:GS_2D} and the
Schnakenburg model of \cite{Schnaken_KWW:2008} were highlighted.

The numerics required to implement the hybrid asymptotic-numerical
theory is rather simple, provided that the reduced-wave Green's
function for the domain can be readily determined. For the domains
considered in this paper, this Green's function is available
analytically, whereas for other domains it can be computed rapidly by
using fast multipole methods for linear PDE's (cf.~\cite{CHL}). When
this Green's function is available, the implementation of the hybrid
asymptotic-numerical approach only requires the numerical solution of
a coupled DAE system for the spot dynamics and a BVP to compute
competition and oscillatory instability thresholds. The DAE system for
$\mathbf{x}_j$ and $S_j$ involves two functions (i.e.~$\chi(S_j)$ and
$\gamma(S_j)$) that are determined in terms of the solution to the
core problem. These functions are readily pre-computed by solving some
simple BVP's.  Similarly, from a BVP eigenvalue problem, the condition
for the onset of self-replication is readily pre-computed in terms of
a threshold value of the source strength $S_j$. 

In this way, a similar hybrid approach should be readily applicable to
study spot dynamics and spot stability for related two-component RD
models of the general form \bsub\label{d:class}
\begin{equation}
  v_t =\eps^2 \Delta v - v + f(u) v^p \,, \qquad
  \tau u_t = D \Delta u - (u-u_0) + \eps^{-2} g(u) v^m \,, \qquad
  \mathbf{x}\in \Omega \,,
\end{equation}
\esub for $p>0$, $m>0$, and for certain classes of $f(u)$ and
$g(u)$. The key conditions that are needed for our approach are that
there is a core problem near each spot for which $v\to 0$ and $u$ has
a logarithmic growth at infinity in terms of the stretched inner
variable, and that the the spots are coupled together via a
quasi-static linear elliptic concentration field for $u$. Such a
generalization of our overall approach for the GS model is undertaken
in \cite{CIRW} for the simple case of a two-spot evolution in $\R^2$.
The work in \cite{CIRW} is the two-dimensional counterpart of the
study of \cite{DK_1D} of two-spike solutions for a general class of RD
systems on the infinite line.

We conclude this paper by discussing a few open problems for the GS
model related to spot dynamics, equilibria, and stability.  The first
main problem is to develop a weakly nonlinear stability theory for the
three different types of instabilities. Our numerical results suggest
that both the peanut-splitting instability and the oscillatory
instability are subcritical, as they lead to spot self-replication and
spot-annihilation, respectively. In particular, with regards to
self-replication, it would be interesting to perform a weakly
nonlinear analysis of the time-dependent core problem
\eqref{2Dreduced} to show that the initial peanut-splitting
instability does not saturate in the weakly nonlinear regime, but
instead triggers a nonlinear spot self-replication event.

For the equilibrium problem, a main open problem is to use the spot
equilibrium condition \eqref{3:finaleq1} to calculate detailed
bifurcation diagrams in terms of $A$ and $D$ of steady-state $k$-spot
patterns in arbitrary domains. One goal would be to determine whether
asymmetric equilibria, characterized by equilibrium spot patterns with
different source strengths, are possible and to then determine the
multiplicity of equilibrium solutions.  For these asymmetric
multi-spot equilibria, it should be possible to calculate oscillatory
and competition instability thresholds from our global eigenvalue
problem.  The analysis in this paper has largely been restricted to
the case where the Green's matrix is circulant, which allows for spot
patterns with a common source strength.

Another open problem is to study weak translational instabilities
associated with multi-spot equilibrium patterns. Such small eigenvalue
instabilities, with growth rate $\lam={\mathcal O}(\eps^2)$, are
characterized by instabilities of stationary points of the DAE system
of Principal Result 3.1 governing spot dynamics. Based on the
numerical results in Fig.~15 of \cite{Schnaken_KWW:2008}, one simple
scenario where such a weak instability occurs is for a one-ring
pattern of spots inside a disk.  More specifically, we expect that an
equilibrium pattern consisting of $k$ equally-spaced spots on a ring
concentric within the unit disk will be weakly unstable with growth
rate $\lam={\mathcal O}(\eps^2)$ when $k$ exceeds some
threshold. Similar thresholds for weak instabilities have been studied
in \cite{gueron} for a ring of spots in the context of a simple
interacting particle system model, and in \cite{Boatto_1} and
\cite{Boatto_2} for a ring of Eulerian fluid point vortices on the
surface of a sphere.

Finally, it would be very interesting to exhibit a parameter regime
for the GS model where the quasi-equilibrium spot pattern undergoes
repeated episodes of either spot self-replication, when a spot is too
far from any of its neighbors, or spot-oscillation leading to annihilation,
which is triggered when two slowly-drifting spots become too closely
spaced. For $D$ small, but with $D>>{\mathcal O}(\eps^2)$, such a
replication-annihilation ``attractor'' should be realizable by
considering the instability thresholds for two-spot interactions in
$\R^2$. In this paper, a key first step in the construction of this
``attractor'' has been undertaken by showing that the GS model can
robustly support the existence of dynamically-triggered instabilities
for a collection of spots that are initially stable at $t=0$.

\section*{Acknowledgments} 
M.~J.~W. was supported by NSERC (Canada). We are grateful to
Prof.~P.~Zegeling for the use of his code VLUGR2 to calculate
full numerical solutions of the GS model. MJW is grateful to
Prof.~J.~Wei for helpful comments. We are grateful to the referees for their 
very helpful comments.

\appendix
\renewcommand{\theequation}{\Alph{section}.\arabic{equation}}
\newsection{The Reduced-Wave and Neumann Green's Function}
\label{appendix:green}

In this appendix we give detailed analytical results for both the
reduced-wave Green's function and the Neumann Green's function for the
unit disk and for a rectangle. These results are needed in \S
\ref{sec:sym} and \S \ref{sec:asy} in order to numerically implement
the analytical theory of \S \ref{sec:quasi}--\ref{sec:eig} for spot
dynamics and instabilities.

\subsection{Green's Function for a Unit Disk}

Let $\Omega =\{\mathbf{x}:|\mathbf{x}|\leq 1\}$.  Upon introducing
polar coordinates $(x_0, y_0) = (\rho_0 \cos \theta_0, \rho_0 \sin
\theta_0)$, the reduced-wave Green's function of (\ref{3:Green})
satisfies
\begin{equation}
\label{3:circlegreen} G_{\rho \rho} + \frac{\,1}{\,\rho}G_{\rho} +
\frac{\,1}{\,\rho^2}G_{\theta \theta} - \frac{1}{D} G = -
\frac{1}{\,\rho}\,\delta(\rho-\rho_0) \delta(\theta-\theta_0)\,, \quad
 0<\rho<1 \,, \,\,\,\, 0\leq \theta < 2\pi \,,
\end{equation}
with boundary conditions $G(\rho, \theta+ 2 \pi) = G(\rho,
\theta)$, $G_{\rho}(1, \theta) = 0$, and where $G$ is well-behaved as 
$\rho\to 0$.

To determine $G$, we first extract the $\theta$ dependence by introducing
the complex Fourier series 
\begin{equation}
\label{3:fourier} G(\rho,\theta;\rho_0,\theta_0) = \frac{\,1}{\,2
\pi} \sum^{\infty}_{n=-\infty} \tilde{G}_n(\rho;\rho_0,\theta_0)
e^{-i n \theta} \,, \qquad
 \tilde{G}_n(\rho;\rho_0,\theta_0) = \int^{2 \pi}_0 e^{i n \theta} 
G(\rho,\theta;\rho_0,\theta_0) \, d \theta \,.
\end{equation}
From \eqref{3:circlegreen} we obtain that
$\tilde{G}_{n \rho \rho} + \rho^{-1}\tilde{G}_{n \rho} -
 n^2 \rho^{-2}\tilde{G}_{n} - D^{-1} \tilde{G}_n = - 
\rho^{-1} \delta(\rho-\rho_0) e^{i n \theta_0}$ on $0<\rho<1$, with
boundary conditions $\tilde{G}_{n \rho}(1;\rho_0,\theta_0) = 0$
and $\tilde{G}_n(0;\rho_0,\theta_0)< \infty$. The solution to this 
problem has the form
\begin{equation}
\label{3:coefG} \tilde{G}_{n}(\rho;\rho_0,\theta_0) = \begin{cases} &A_1 I_n
\left(\frac{\,\rho}{\,\sqrt{D}} \right)\,, \qquad 0<\rho<\rho_0 \,,
\\ &A_2 \left[ I_n \left(\frac{\,\rho}{\,\sqrt{D}} \right) -
\frac{I^{\p}_n \left(\frac{\,1}{\,\sqrt{D}} \right)}{K^{\p}_n
\left(\frac{\,1}{\,\sqrt{D}} \right)} K_n
\left(\frac{\,\rho}{\,\sqrt{D}} \right)\right]\,, \qquad \rho_0<\rho<1 \,,
\end{cases}
\end{equation}
where $I_{n}(r)$ and $K_{n}(r)$ are modified Bessel functions of order
$n$. In \eqref{3:coefG}, the coefficients $A_1$ and $A_2$ are obtained
by making $\tilde{G}_{n}$ continuous at $\rho_0$ and by requiring that
the jump in the derivative of $\tilde{G}_n$ at $\rho_0$ is
$[\tilde{G}_{n\rho}]_{\rho_0}=-\frac{e^{i n \theta_0}}{\rho_0}$.
Then, upon using the Wronskian determinant for $K_n$ and $I_n$, we
readily determine $A_1$ and $A_2$, and obtain
\begin{equation}
\label{3:expcoefG} \tilde{G}_{n}(\rho;\rho_0,\theta_0) = 
\left[ K_n
\left(\frac{\,\rho_{>}}{\,\sqrt{D}} \right) - \frac{K^{\p}_n
\left(\frac{\,1}{\,\sqrt{D}} \right) }{I^{\p}_n
\left(\frac{\,1}{\,\sqrt{D}} \right) } I_n
\left(\frac{\,\rho_{>}}{\,\sqrt{D}} \right) \right] \,e^{i n \theta_0}
I_n \left(\frac{\,\rho_{<}}{\,\sqrt{D}} \right)\,, \qquad 0<\rho< 1\,,
\end{equation}
where we have define $\rho_{<}=\min(\rho_0,\rho)$ and
$\rho_{>}=\max(\rho_0,\rho)$.  Therefore, the Fourier expansion for
the reduced-wave Green's function satisfying \eqref{3:circlegreen} is
\begin{equation}
 G(\rho, \theta; \rho_0,\theta_0 ) =
\frac{\,1}{\,2 \pi} \sum^{\infty}_{n=-\infty}  e^{-i n
(\theta-\theta_0)} \left[ K_n \left(\frac{\,\rho_{>}}{\,\sqrt{D}}
\right) - \frac{K^{\p}_n \left(\frac{\,1}{\,\sqrt{D}} \right) }{I^{\p}_n
\left(\frac{\,1}{\,\sqrt{D}} \right) } I_n
\left(\frac{\,\rho_{>}}{\,\sqrt{D}} \right) \right]  I_n
\left(\frac{\,\rho_{<}}{\,\sqrt{D}} \right) \,, \qquad 0<\rho<1\,.
\label{3:expansionG}
\end{equation}

The infinite series representation (\ref{3:expansionG}) converges
very slowly. To improve the convergence properties, we must extract
from $G$ the free-space Green's function $G_{f}$, which has the
Fourier representation
\begin{equation}
\label{3:freegreen} G_f(\rho,\theta;\rho_0,\theta_0) = \frac{\, 1}{\, 2\pi}
 \left( K_0\left(\frac{ |\mathbf{x}-\mathbf{x}_0| } {\,\sqrt{D}}\right)\right)
 = \frac{\,1}{\,2 \pi} \sum^{\infty}_{n=-\infty}  e^{-i n
(\theta-\theta_0)}   K_n \left(\frac{\,\rho_{>}}{\,\sqrt{D}} \right)
 I_n \left(\frac{\,\rho_{<}}{\,\sqrt{D}} \right) \,, \qquad  0<\rho<1\,,
\end{equation}
where $|\mathbf{x}-\mathbf{x}_0|\equiv \sqrt{\rho^2 + \rho^2_0 - 2
\rho \rho_0 \cos(\theta-\theta_0)}$. Then, we can decompose $G$ in
\eqref{3:expansionG} as
\begin{equation}
\label{3:uniformG} G(\rho,\theta;\rho_0,\theta_0) = \frac{\,1}{\,2 \pi}
K_0\left(\frac{ |\mathbf{x}-\mathbf{x}_0|}{\,\sqrt{D}}\right) -
\frac{\,1}{\,2 \pi} \sum^{\infty}_{n=-\infty} e^{-i n
(\theta-\theta_0)} \frac{K^{\p}_n \left(\frac{\,1}{\,\sqrt{D}} \right)}
{I^{\p}_n \left(\frac{\,1}{\,\sqrt{D}} \right) } I_n
\left(\frac{\,\rho_0}{\,\sqrt{D}} \right) I_n
\left(\frac{\,\rho}{\,\sqrt{D}} \right) \,, \qquad 0<\rho<1 \,.
\end{equation}
In \eqref{3:uniformG}, the first term arises
from the source at $(\rho_0, \theta_0)$, while the infinite sum
represents the effects of the boundary conditions. Then, we use
$K_{-n}(z) =K_{n}(z)$ and $I_{-n}(z) = I_{n}(z)$ to further simplify
\eqref{3:uniformG} to
\begin{multline}
G(\rho,\theta;\rho_0,\theta_0) = \frac{\,1}{\,2 \pi}
K_0\left(\frac{\,R}{\,\sqrt{D}}\right) - \frac{\,1}{\,2 \pi}
\frac{K^{\p}_0 \left(\frac{\,1}{\,\sqrt{D}} \right) } {I^{\p}_0
\left(\frac{\,1}{\,\sqrt{D}} \right) } I_0
\left(\frac{\,\rho_0}{\,\sqrt{D}} \right) I_0
\left(\frac{\,\rho}{\,\sqrt{D}} \right) + \frac{1}{\pi} \delta_M \\ -
\frac{\,1}{\,\pi} \sum^{M}_{n=1} \cos (n (\theta-\theta_0)) \frac{K^{\p}_n
\left(\frac{\,1}{\,\sqrt{D}} \right) }{I^{\p}_n
\left(\frac{\,1}{\,\sqrt{D}} \right) } I_n
\left(\frac{\,\rho_0}{\,\sqrt{D}} \right) I_n
\left(\frac{\,\rho}{\,\sqrt{D}} \right) \,, \qquad 0<\rho<1 \,.
 \label{3:simpleG}
\end{multline}
Since $K^{\p}_n \left(\frac{\,1}{\,\sqrt{D}} \right) < 0$, and
$I^{\p}_n \left(\frac{\,1}{\,\sqrt{D}} \right) > 0$ for any integer $n
\geq 0$, then $\delta_M $ is a small error term bounded by
\[ 
 | \delta_M| \leq P_M \equiv -\sum^{\infty}_{n=M+1} \frac{K^{\p}_n
\left(\frac{\,1}{\,\sqrt{D}} \right) }{I^{\p}_n
\left(\frac{\,1}{\,\sqrt{D}} \right) } I_n
\left(\frac{\,\rho_0}{\,\sqrt{D}} \right) I_n
\left(\frac{\,\rho}{\,\sqrt{D}} \right) \,.\] 

In Table~\ref{tab:green}, we give the number $M$ of Fourier terms in
\eqref{3:simpleG} required to calculate the reduced-wave Green's
function within a tolerance of $10^{-8}$. The series for $P_M$ is
found to converge fairly fast, especially when the singular point is
not close to the boundary of the unit circle and/or $D$ is not too
large. Equation \eqref{3:simpleG} for $G$ with $P_M<10^{-8}$ is the
one used in the numerical simulations in \S
\ref{sec:sym}--\ref{sec:asy}. From \eqref{3:simpleG} we can readily
extract the self-interaction, or regular part, $R$ of the reduced-wave
Green's function, as written in \eqref{3:gloc}. The gradients of $R$
and $G$ that are needed in Principal Result 3.1 to determine the
motion of a collection of spots are determined numerically from
\eqref{3:simpleG}.

\begin{table}
\centering
\begin{tabular}{c  c  c  c }
 \hline
$\rho_0$ &$\rho$ &$M$  &$P_M$ \\
\hline\hline 0.8 &0.8 &31  &8.1737e-009\\ 
0.8 &0.5 &16
 &6.7405e-009 \\ 
 0.8 &0.2 &8
&9.5562e-009 \\ 
0.5 &0.5 &11  &4.3285e-009 \\
\hline
\end{tabular}
\caption{{\em The number of Fourier terms needed to determine $G$ to
within a tolerance $10^{-8}$ for $D=1$.}}
\label{tab:green}
\end{table}

\subsection{Neumann Green's Function for a Unit Disk}
For $D \gg {\mathcal O}(1)$, we recall from \eqref{3:G2GN} that the
reduced-wave Green's function and its regular part can be approximated
by the Neumann Green's function $G^{(N)}(\mathbf{x};\mathbf{x}_0)$ and
its regular part $R^{(N)}(\mathbf{x};\mathbf{x}_0)$ satisfying
\eqref{3:gneum}. For the unit disk, this Neumann Green's function and
its regular part are defined in cartesian coordinates by (see
eq.~(4.3) of \cite{trap_KTW:2005}), 
\bsub \label{3:Neumannxy}
\begin{align}
   G^{(N)}(\xb;\xib) &= \frac{1}{2\pi} \left( - \ln|\xb-\xib| -
   \ln\left| \xb|\xib|- \frac{\xib}{|\xib|} \right| + \frac{1}{2}
   (|\xb|^2 + |\xib|^2 ) - \frac{3}{4} \right) \,,
   \label{gr:neum_disk_g} \\ 
   R^{(N)}(\xib;\xib) & = \frac{1}{2\pi}
   \left( - \ln\left( 1 - |\xib|^2 \right) + |\xib|^2 - \frac{3}{4}
   \right) \,. \label{gr:neum_disk_r} 
\end{align}
\esub
The gradients of these quantities can be obtained by a simple calculation as
\begin{equation*}
\nabla G^{(N)}(\mathbf{x};\mathbf{x}_0) = - 
  \frac{\,1}{\,2 \pi} \left[ 
\frac{\left(\mathbf{x}-\mathbf{x}_0\right)}{\,|\mathbf{x}-\mathbf{x}_0|^2} +
\frac{|\mathbf{x}_0|^2}{  \bar{\mathbf{x}} |\mathbf{x}_0|^2 -
\bar{\mathbf{x}_0}} - \mathbf{x} \right] \,,  \qquad
\nabla R^{(N)}(\mathbf{x}_0;\mathbf{x}_0) = 
 \frac{\,1}{\,2 \pi} \left( \frac{\, 2 - |\mathbf{x}_0|^2
}{\, 1 - |\mathbf{x}_0|^2} \right) \mathbf{x}_0 \,,
\end{equation*}
where the overbar denotes complex conjugate.  In \S
\ref{sec:sym}--\ref{sec:asy} these results allow us to explicitly
determine the dynamics of a collection of spots inside a disk when $D$
is large, as characterized in Principal Result 3.1 of \S
\ref{sec:dyn}.

When $\Omega$ is the unit disk, and for two specific sets of source
and observation points in $\Omega$, in Table~\ref{tab:green2} we give
numerical results for the two-term approximation (see equation
(\ref{3:G2GN}))
\begin{equation}
 G \sim \tilde{G} \equiv \frac{D}{|\Omega|} + G^{(N)} \,, \qquad
 R \sim  \tilde{R} \equiv \frac{D}{|\Omega|} + R^{(N)} \,, \label{a:two-term}
\end{equation}
with $|\Omega|=\pi$, which relates the reduced-wave and Neumann
Green's functions when $D$ is large. The results for $G^{(N)}$,
$R^{(N)}$, $G$, and $R$, were computed from \eqref{3:Neumannxy} and
\eqref{3:simpleG}. The results in Table~\ref{tab:green2} are believed
to be correct to the number of digits shown, and confirm that
$G\approx \tilde{G}$ and $R\approx \tilde{R}$ when $D$ is large.

\begin{table}
\begin{center}
\begin{tabular}{c|cccc|cccc}
\hline
& \multicolumn{4}{c}{$(\rho,\rho_0,\theta-\theta_0)=(0.3,0.5,0.0)$} & 
 \multicolumn{4}{c}{$(\rho,\rho_0,\theta-\theta_0)=(0.8,0.2,{\pi/6})$} \\
\hline 
$D$ & $ G$ & $\tilde{G}$ & $R$ & $\tilde{R}$ &  $G$ & $\tilde{G}$ & $R$ & 
 $\tilde{R}$ \\ 
 \hline
0.1 &0.12182  &0.22154 &-.15651 &-.00196 & 0.02473 & 0.06198 &-.16303
 &-.07467 \\
1.0 &0.48316 &0.50802  &0.24622 &0.28452 &0.33490  &0.34846
&0.19617  &0.21181 \\
4.0 &1.45566  &1.46295 &1.22822 &1.23945 &1.29922 &1.30339 &1.16243
 &1.16674\\
10.0  &3.36978 &3.37281 &3.14465 &3.14931 &3.21150 &3.21325
 &3.07484  &3.07660\\
\hline
\end{tabular}
\end{center}
\caption{{\em Comparison of results for the reduced-wave Green's function
$G$ and its regular part $R$ for the unit disk with the two-term large
$D$ approximation $\tilde{G}$ and $\tilde{R}$, as given in
\eqref{a:two-term}. The comparison is made at two sets of source and
observation points in $\Omega$.  For $D$ large, we note that
$G\approx \tilde{G}$ and $R\approx \tilde{R}$, as expected.}}
\label{tab:green2}
\end{table}

\subsection{Green's Function for a Rectangle}
Next, we calculate the reduced-wave Green's function $G(x, y ; x_0,
y_0)$ in the rectangular domain $\Omega = [0,L] \times [0, d]$. Since
the free-space Green's function $G_f$ is given by
$G_f(\mathbf{x};\mathbf{x}_0)= (2\pi)^{-1} K_0  \left( 
{|\mathbf{x}-\mathbf{x}_0|/\sqrt{D}} \right)$, we can use the
method of images to write the solution to \eqref{3:Green} as
\bsub \label{3:squaregreen}
\begin{gather}
\label{3:squaregreen_a} G(\mathbf{x};\mathbf{x}_0) =\frac{1}{\,2 \pi}
\sum_{n=-\infty}^{\infty}\sum_{m=-\infty}^{\infty} \sum^4_{l=1} K_0
\left( \frac{|\mathbf{x}-\mathbf{x}^{(l)}_{mn}|}{\sqrt{D}} \right) \,,\\
\mathbf{x}^{(1)}_{mn} = (x_0 + 2 n L, y_0 + 2 m d)\,, 
 \quad \mathbf{x}^{(2)}_{mn} = (- x_0 + 2 n L, y_0 + 2 m d) \,, \\
 \mathbf{x}^{(3)}_{mn} = (x_0 + 2 n L, -y_0 + 2 m d)\,,
  \quad \mathbf{x}^{(4)}_{mn} = (- x_0 + 2 n L, - y_0 + 2 m d) \,.
\label{3:squaregreen_b}
\end{gather}
\esub Since $K_0(r)$ decays exponentially as $r\to\infty$, then the
infinite series in \eqref{3:squaregreen} converges fairly rapidly when
$D={\mathcal O}(1)$. The regular part of the reduced-wave Green's
function together with the gradients of $G$ and $R$, as required in
Principal Result 3.1, are readily evaluated numerically from
\eqref{3:squaregreen}.

\subsection{Neumann Green's Function for a Rectangle}
For $D \gg {\mathcal O}(1)$ is large, we can approximate $G$ and
its regular part $R$ by the corresponding Neumann Green's function
$G^{(N)}$ and its regular part $R^{(N)}$ by the two-term expansion in
\eqref{3:G2GN}. This latter Green's function, calculated by an
analytical Ewald-type summation method, was given analytically in
formula (4.13) of \cite{Schnaken_KWW:2008} as \bsub \label{3:squareNG}
\begin{equation}
 G^{(N)}(\mathbf{x};
\mathbf{x}_0) = - \frac{\,1}{\, 2 \pi} \ln |\mathbf{x}-\mathbf{x}_0|
+ R^{(N)}(\mathbf{x}; \mathbf{x}_0) \,,
\end{equation}
where $R^{(N)}(\mathbf{x}; \mathbf{x}_0)$ is given explicitly by
\begin{multline}
 R^{(N)}(\mathbf{x}; \mathbf{x}_0) = - \frac{\,1}{\,2 \pi}
\sum_{n=0}^{\infty} \ln(|1 - q^n  z_{+,+}| |1 - q^n z_{+,-}| |1 -
q^n z_{-,+}| |1 - q^n \zeta_{+,+}||1 - q^n \zeta_{+,-}| |1 - q^n
\zeta_{-,+}|
|1 - q^n \zeta_{-,-}|) \\ +
\frac{\,L}{\,d}\left[\frac{\,1}{\,3}- \frac{\max(x,x_0)}{\, L} +
\frac{\,1}{\,2}\left( \frac{\,x_0^2 + x^2}{\,L^2} \right) \right] -
\frac{1}{\, 2 \pi} \ln \left( \frac{\,|1 -
z_{-,-}|}{|r_{-,-}|}\right) - \frac{1}{\, 2 \pi} \sum_{n=1}^{\infty}
\ln|1 - q^n z_{-,-}|\,. \label{A:lasteq}
\end{multline}
Here points in the rectangle are written as complex numbers. In
\eqref{A:lasteq}, $q \equiv e^{- 2 L \pi / d}$, while the eight
complex-valued constants $z_{\pm, \pm}$ and $\zeta_{\pm, \pm}$ are
given by
\begin{gather}
z_{+, \pm} \equiv \exp(\mu (-|x+x_0| + i (y \pm y_0)) / 2 )\,, \qquad
z_{-, \pm} \equiv \exp(\mu (-|x-x_0| + i (y \pm y_0)) / 2 )\,, \\
\zeta_{+, \pm} \equiv \exp(\mu (|x+x_0| - 2L + i (y \pm y_0)) / 2)\,, \qquad
 \zeta_{-, \pm} \equiv \exp(\mu (|x-x_0| - 2L + i (y \pm
y_0)) / 2 ) \,.
\end{gather}
\esub
Here $\mu$ is defined by $\mu = {2 \pi /d}$. The self-interaction term
$R^{(N)}(\mathbf{x}_0;\mathbf{x}_0)$, and the gradients of $R^{(N)}$ and
$G^{(N)}$ can be calculated readily from \eqref{3:squareNG}. In particular,
for the unit square $[0,1]\times [0,1]$ with a singularity at the middle
$\mathbf{x}_0\equiv (0.5,0.5)$ the self-interaction term $R^{(N)}_{0,0}\equiv
R^{(N)}(\mathbf{x}_0;\mathbf{x}_0)$ calculated from \eqref{3:squareNG}
is simply
\begin{equation}
   R^{(N)}_{00}=-\frac{1}{\pi}\sum_{n=1}^{\infty} \ln\left( 1- q^n\right) +
  \frac{1}{12} - \frac{1}{2\pi}\ln(2\pi) \,, \qquad q\equiv e^{-2\pi}\,.
 \label{A:RN00}
\end{equation}

\appendix \setcounter{section}{1}
\renewcommand{\theequation}{\Alph{section}.\arabic{equation}}
\newsection{Survey of NLEP Theory: Comparison of Quasi-Equilibria
and Stability}
\label{appendix:nlep}

In \cite{2D_Wei:2001}, the existence and stability of a one-spot
pattern to the GS model on the infinite plane $\mathbb{R}^2$ was
studied. For this infinite-plane problem, we can set $D=1$ in
(\ref{1:GS_2D}).  In the inner region near the spot, the leading-order
analysis of \cite{2D_Wei:2001} in terms of $\nu\equiv {-1/\ln\eps}$
and for $A={\mathcal O}\left(\eps [-\ln\eps]^{1/2}\right)$, showed
that $u\sim u_0$, where $u_0$ is locally constant near the
spot. Moreover, $v\sim v_0 = {w(\rho)/(A u_0)}$, where $w(\rho)$ is
the radially symmetric ground-state solution of 
\begin{equation}
\label{4:groundstate} \Delta_{\rho} w -w + w^2 = 0 \,, \quad 0<\rho<\infty \,;
 \qquad w(0)>0 \,, \,\,\,  w^{\p}(0) = 0 \,; \qquad
 w\to 0 \,\,\, \mbox{as} \,\,\, \rho \to \infty \,,
\end{equation}
where $\Delta_\rho w \equiv w^{\p\p} + \rho^{-1} w^{\p}$. For $\eps\to 0$,
it was shown in \cite{2D_Wei:2001} that $u_0$ satisfies the quadratic equation
\begin{equation}
\label{4:Wei1spt} 1 - u_0 \sim \frac{\, L}{\,u_0}, \qquad
L \equiv \frac{\eps^2}{\nu \,A^2} \int_0^{\infty}
w^2 \rho\, d \rho \,, \qquad  \nu \equiv -\frac{1}{\ln \eps} \,.
\end{equation}
Therefore, the existence condition for a one-spot equilibrium solution is that
\begin{equation}
\label{4:AfWei} L \leq \frac{\,1}{\,4}\quad
\Rightarrow\;\;\;  A \geq A_{fw} \equiv 2
\eps \sqrt{\frac{\,b_0}{\,\nu}} \,, \qquad
b_0 \equiv \int_0^{\infty} w^2(\rho) \rho\, d\rho \,.
\end{equation}
Define $L_0$ and $\gamma$ by $L_0 = \lim_{\eps \to 0} L =
{\mathcal O}(1)$ and $\gamma \equiv \nu\,\ln \tau={\mathcal
O}(1)$, so that $\tau={\mathcal O}(\eps^{-\gamma})$, and
assume that $0\leq \gamma < 2$. Then, the following radially symmetric
nonlocal eigenvalue problem (NLEP) was derived in equation (5.5) of
\cite{2D_Wei:2001}:
\begin{equation}
\label{4:NLEP1spt}
 \Delta_{\mathbf{\rho}}\psi_0 - \psi_0  + 2 w \psi_0 - \frac{2(1-u_0)
 (2-\gamma)}{2 u_0 +(1-u_0)(2-\gamma)}\, w^2 \,
 \frac{\int_0^{\infty} w \psi_0 \, \rho \, d\rho}
 {\int_0^{\infty} w^2 \, \rho \, d\rho} = \lambda \psi_0 \,, \qquad
  0 < \rho < \infty \,,
\end{equation}
with $\psi_0\to 0$ as $\rho \to \infty$. An analysis of this NLEP
in \cite{2D_Wei:2001} led to Theorem 2.2 of \cite{2D_Wei:2001}, which we
summarize as follows:
\begin{enumerate}
\item There exists a saddle node bifurcation at $L_0 =
\frac{\,1}{\,4}$, such that there are two equilibrium solutions
$u_{0}^{\pm}$ given by $u_0^{\pm} = {\left( 1 \pm \sqrt{1 - 4 {L}_0}
\right)/2}$ for ${L}_0 < \frac{\,1}{\,4}$, and no equilibrium
solutions when $L_0 > \frac{\,1}{\,4}$.

\item Assume that $0 \leq \gamma < 2$ and $L_0 < \frac{\,1}{\,4}$. Then,
the solution branch $(u^+_0, v^+_0)$ is linearly unstable.

\item Assume that $0 \leq \gamma < 2$. Then, the other solution branch
$(u^-_0, v^-_0)$ is linearly unstable if
\begin{equation}
\label{4:tauWei} L_0 > \frac{\,1}{\,4}\left[1 -
\left(\frac{\gamma}{4 -\gamma}\right)^2 \right]\,,\,\,\,
\Rightarrow\;\;\; A < A_{sw} \equiv A_{fw} \left[1 - 
 \left(\frac{\gamma}{4 -\gamma}\right)^2 \right]^{-1/2}.
\end{equation}

\item Assume that $\gamma=0$ and $L_0 < \frac{\,1}{\,4}$. Then,
$(u^-_0, v^-_0)$ is linearly stable.

\item If $0 \leq \gamma < 2$, the stability of
$(u^-_0, v^-_0)$ is unknown for $A > A_{sw}$.

\end{enumerate}

We remark that the saddle-node bifurcation value of $A$, obtained from
\eqref{4:Wei1spt}, has the scaling $A=O\left(\eps
[-\ln\eps]^{1/2}\right)$ as $\eps\to 0$.  The stability
results listed above from \cite{2D_Wei:2001} also relate only to this
range in $A$. However, the quasi-equilibrium spot solution given in
Principal Result 2.1 of \S 2, and the spot self-replication threshold
of Principal Result 4.1, occur for the slightly larger range
 $A={\mathcal O}\left(\eps [-\ln\eps]\right)$. Consequently,
the occurrence of spot self-replication behavior for the GS model
\eqref{1:GS_2D} was not observed in the scaling regime for $A$
considered in \cite{2D_Wei:2001}. In \S \ref{sec:1inf} we compare the
Hopf Bifurcation threshold predicted by  \eqref{4:tauWei} with that
obtained from our global eigenvalue problem of \S \ref{sec:eig_rad}.

\subsection{NLEP Theory for $k$-Spot Patterns}

In \cite{2Dmulti_Wei:2003} a related theoretical approach, based on
the rigorous analysis of other NLEP's, was used to obtain a
leading-order asymptotic theory for the existence and stability of
$k$-spot patterns to the GS model \eqref{1:GS_2D} in a bounded
two-dimensional domain. In the inner region near the
$j^{\mbox{th}}$ spot centered at $\mathbf{x}_j$, the
leading-order-in-$\nu$ scaling in \cite{2Dmulti_Wei:2003} showed that
$u\sim u_j$, where $u_j$ is a constant, and that $v$ satisfies $v\sim
v_j \equiv \frac{1}{\,A u_j} w(\rho)$, where $w(\rho)$ is the radially
symmetric solution of \eqref{4:groundstate}.  Moreover, the $u_j$ for
$j=1,\ldots,k$ satisfy the nonlinear algebraic system
(cf.~\cite{2Dmulti_Wei:2003})
\begin{equation}
\label{4:quad} 1 - u_j - \frac{\eta_{\eps} L_{\eps}}{u_j} \sim
\sum_{i=1}^k \frac{L_{\eps}}{u_i} \,,
\end{equation}
where $L_{\eps}$ and $\eta_\eps$ are defined in terms of the area $|\Omega|$
of the domain by
\begin{equation}
 L_{\eps} \equiv \frac{2 \eps^2 \pi  b_0}{A^2 |\Omega|} \,, \qquad
\eta_{\eps} \equiv \frac{|\Omega|}{2 \pi D \nu} \,, \qquad b_0 \equiv
 \int_0^{\infty} w^2 \rho \, d \rho\,, \qquad \nu \equiv -\frac{1}{\ln\eps}
  \,. \label{4:ldef}
\end{equation}
For the case where $u_j=u_0$ for $j=1,\ldots,k$, then \eqref{4:quad}
is a quadratic equation for $u_j$, which leads to the following three
main parameter regimes of \cite{2Dmulti_Wei:2003}:
\begin{equation}
\label{4:Wexist}
\begin{cases} & 4 k L_0 < 1, \qquad \mbox{for}\;\;\eta_{\eps}
\to 0\;\;\Leftrightarrow \; D \gg  {\mathcal O}(\nu^{-1}), \\
& 4 \eta_{\eps} L_{\eps} < 1, \qquad \mbox{for}\;\;\eta_{\eps}
\to \infty \;\;\Leftrightarrow \; D =  {\mathcal O}(1), \\
& 4 (k+\eta_0) L_0 < 1, \qquad \mbox{for}\;\;\eta_{\eps} \to \eta_0
\;\;\Leftrightarrow \; D =  {\mathcal O}(\nu^{-1}) \,, 
\end{cases} \qquad  L_0 = \lim_{\eps \to 0} L_{\eps}={\mathcal O}(1) \,,
  \qquad  \eta_0 = \lim_{\eps \to 0} \eta_{\eps}={\mathcal O}(1) \,. 
\end{equation}

From \eqref{4:ldef}, the second line in \eqref{4:Wexist} is for
$A=O\left(\eps [-\ln\eps]^{1/2}\right)$ with $D={\mathcal O}(1)$,
while the third line in \eqref{4:Wexist} is for $A=O\left(\eps\right)$
with $D={\mathcal O}(\nu^{-1})$.  $\nu={-1/\ln\eps}$. Thus, the
$k$-spot quasi-equilibrium patterns of \cite{2Dmulti_Wei:2003} do not
correspond to the range $A=O\left(\eps[-\ln\eps]\right)$ and
$D={\mathcal O}(1)$, considered in \S \ref{sec:quasi}, where spot-splitting
can occur.

As shown in \cite{2Dmulti_Wei:2003}, the leading-order
quasi-equilibrium solution satisfies $u_j\sim u_0^{\pm}$ in the inner region.
The global representation for $v$ is $v\sim v_{0}^{\pm}$. Here
$u_{0}^{\pm}$ and $v_{0}^{\pm}$ satisfy
\begin{equation}
 u_0^{\pm} \sim \begin{cases} & \frac{\,1}{\,2}\left(1 \pm \sqrt{1 - 4
k L_0}\right), \qquad \mbox{for}\;\; D \gg {\mathcal O}(\nu^{-1})\,,
\\ & \frac{\,1}{\,2}\left(1 \pm \sqrt{1-4 \eta_{\eps}
L_{\eps}}\right)\,, \qquad \mbox{for}\;\; D = {\mathcal O}(1)\,, \\
&\frac{\,1}{\,2}\left(1 \pm \sqrt{1- 4 (k+\eta_0) L_0 }\right) \,,
\qquad \mbox{for}\;\; D = {\mathcal O}(\nu^{-1})\,,
\end{cases}  \qquad
v_{0}^{\pm} \sim \sum_{j=1}^k \frac{1}{A u_0^{\pm}}\,
w\left(\eps^{-1}|\mathbf{x} - \mathbf{x}_j|\right) \,. \label{4:Wequil}
\end{equation}
The stability analysis in \cite{2Dmulti_Wei:2003} was largely based on
the derivation and analysis of the radially symmetric NLEP
\bsub \label{4:NLEPcmp}
\begin{equation}
 \Delta_{\mathbf{\rho}}\psi - \psi  + 2 w \psi - \gamma\, w^2 \,
 \frac{\int_0^{\infty} w \psi \rho \, d\rho}{\int_0^{\infty} w^2 \rho \,
 d\rho} = \lambda \psi \,, \qquad 0<\rho<\infty \,; \qquad
  \psi\to 0 \,, \quad \rho\to \infty \,. \label{4:NLEPcmp_1}
\end{equation}
In terms of $u_0=u_{0}^{\pm}$ and $\tau={\mathcal O}(1)$, in
\eqref{4:NLEPcmp_1} there are two choices for $\gamma$ given by
\begin{equation}
\label{4:gam} \gamma = \frac{2 \eta_{\eps}  L_{\eps} (1 + \tau
\lambda) + k L_{\eps}}{(u_0^2 + L_{\eps} \eta_{\eps})(1 + \tau
\lambda) + k L_{\eps}} \,, \qquad \mbox{or}\qquad \gamma = \frac{2
\eta_{\eps}  L_{\eps}}{u_0^2 + L_{\eps} \eta_{\eps}} \,.
\end{equation}
\esub

Assuming that the third condition in \eqref{4:Wexist} holds, Theorem
2.3 of \cite{2Dmulti_Wei:2003} gives a stability result for $k$-spot
equilibrium solution based on an analysis of the NLEP
\eqref{4:NLEPcmp}, together with certain properties of
$\mathcal{F}(\mathbf{x}_1,\ldots,\mathbf{x}_k)$ defined by
\begin{equation}
    \mathcal{F}(\mathbf{x}_1,\ldots,\mathbf{x}_k) = -\sum_{i=1}^{k}
  \sum_{j=1}^{k} \left(\mathcal{G}\right)_{ij} \,. \label{4:fcal}
\end{equation}
Here $\mathcal{G}$ is the Green's matrix of \eqref{3:Gmatrix} defined
in terms of the reduced-wave Green's function. From
\cite{2Dmulti_Wei:2003}, an equilibrium spot pattern must be at a
critical point of $\mathcal{F}$.  We denote by ${\cal H}_0$ the
Hessian of $\mathcal{F}$ at this critical point.

Suppose that $4(\eta_\eps + k) L_{\eps}<{1/4}$. Then, Theorem 2.3 of
\cite{2Dmulti_Wei:2003} proves that the large solutions $(u^+, v^+)$
are all linearly unstable. For the small solutions $(u^-, v^-)$, this
theorem provides the following results for three ranges of $\eta_\eps$:

\begin{enumerate}
\item Assume that $\eta_{\eps} \to 0$, so that $D \gg {\mathcal
O}(\nu^{-1})$. Then,

\begin{itemize}
\item If $k=1$, and all the eigenvalues of the Hessian $\mathcal{H}_0$ are
negative, then there exists a unique $\tau_1 > 0$, with 
$\tau_1={\mathcal O}(1)$, such that for 
$\tau< \tau_1$, $(u^-, v^-)$ is linearly stable, while for $\tau > \tau_1$, 
it is linearly unstable.
\item If $k>1$, $(u^-, v^-)$ is linearly unstable for any $\tau \geq 0$.
\item If the Hessian $\mathcal{H}_0$ has a strictly positive
eigenvalue, then $(u^-, v^-)$ is linearly unstable for any $\tau \geq
0$.
\end{itemize}

\item Assume that $\eta_{\eps} \to \infty$, so that $D \ll {\mathcal
O}(\nu^{-1})$. Then,
\begin{itemize}
\item If all the eigenvalues of the Hessian $\mathcal{H}_0$ are negative,
then $(u^-, v^-)$ is linearly stable for any $\tau > 0$.
\item If the Hessian $\mathcal{H}_0$ has a strictly positive eigenvalue,
$(u^-, v^-)$ is linearly unstable for any $\tau \geq 0$.
\end{itemize}

\item Assume that $\eta_{\eps} \to \eta_0$, so that $D= {\mathcal
O}(\nu^{-1})$. The following results hold:
\begin{itemize}
\item If $L_0 <  \frac{\eta_0}{(2 \eta_0 + k)^2}$, and
all eigenvalues of $\mathcal{H}_0$ are negative, then
$(u^-, v^-)$ is linearly stable for $\tau={\mathcal O}(1)$ sufficiently
small or $\tau={\mathcal O}(1)$ sufficiently large.

\item If $k=1$, $L_0 > \frac{\eta_0}{(2 \eta_0 + 1)^2}$,
and all the eigenvalues of $\mathcal{H}_0$  are negative,
then there exists $\tau_2 > 0$, $\tau_3 > 0$, such that
for $\tau< \tau_2$, $(u^-, v^-)$ is linearly stable,
while for $\tau > \tau_3$, it is linearly unstable.

\item If $k > 1$ and $L_0 > \frac{\eta_0}{(2 \eta_0 + k)^2}$,
then $(u^-, v^-)$ is linearly unstable for any $\tau > 0$.

\item If the Hessian $\mathcal{H}_0$ has a strictly positive eigenvalue,
$(u^-, v^-)$ is linearly unstable for any $\tau \geq 0$.
\end{itemize}
\end{enumerate}

We now relate these results of \cite{2Dmulti_Wei:2003} with those
obtained in \S \ref{sec:quasi}--\ref{sec:eig}. Firstly, the condition
of \cite{2Dmulti_Wei:2003} that an equilibrium $k$-spot configuration
$\mathbf{x}_1,\ldots,\mathbf{x}_k$ must be at a critical point of
$\mathcal{F}$ in \eqref{4:fcal} is equivalent to the equilibrium
result in \eqref{3:finaleq1} of Principal Result 3.1 provided that we
replace the source strengths $S_{je}$ in \eqref{3:finaleq1} with their
leading-order-in-$\nu$ asymptotically common value $S_c\sim
\mathcal{A}$, as obtained from \eqref{3:s2t}.  The stability condition
of \cite{2Dmulti_Wei:2003} expressed in terms of the sign of the
eigenvalues of the Hessian $\mathcal{H}_0$ of $\mathcal{F}$ is
equivalent to the statement that the equilibrium spot configuration
$\mathbf{x}_1,\ldots,\mathbf{x}_k$ is stable with respect to the slow
motion ODE dynamics of Principal Result 3.1 of \S \ref{sec:dyn} when
we use the leading-order approximation $S_j\sim S_c$ for
$j=1,\ldots,k$.  Therefore, the condition on the Hessian
$\mathcal{H}_0$ in \cite{2Dmulti_Wei:2003} relates to the small
eigenvalues of order ${\mathcal O}(\eps^2)$ in the linearization.

In addition, the second result of Theorem 2.3 of
\cite{2Dmulti_Wei:2003} for $\eta_\eps\to \infty$, as written above,
includes the range $D={\mathcal O}(1)$. The condition $4(\eta_\eps +
k) L_{\eps}<{1/4}$, together with \eqref{4:ldef} relating $L_\eps$ to
$A$, shows that this second result of \cite{2Dmulti_Wei:2003} holds
when $A={\mathcal
O}\left(\eps\left[-\ln\eps\right]^{1/2}\right)$. Therefore, for this
range of $A$ and $D$, the leading order NLEP analysis in
\cite{2Dmulti_Wei:2003} predicts that there is an equilibrium solution
branch whose stability depends only on the eigenvalues of the Hessian
matrix $\mathcal{H}_0$. This result is very similar to that obtained
from our global eigenvalue problem in Principal Result 4.2 of \S
\ref{sec:eig_rad} (see Fig.~\ref{fig:eigC0:a}) for the nearby
parameter range $A={\mathcal O}(-\eps\ln\eps)$. There, we showed to
leading-order in $\nu$ that a $k$-spot quasi-equilibrium solution is
stable to competition or oscillatory instabilities when
$D={\mathcal O}(1)$ and $A={\mathcal O}\left(-\eps\ln\eps\right)$.

With the exception of the condition on the Hessian matrix of
$\mathcal{F}$, these stability results of \cite{2Dmulti_Wei:2003}
depend only on the number $k$ of spots, and not on their spatial
locations within $\Omega$. This is qualitatively very different to
that obtained in \S \ref{sec:12inf}--\S \ref{sec:sym} from our
globally coupled eigenvalue problem, which accounts for all terms in
powers of $\nu$.

\subsection{Comparison of the Quasi-Equilibrium Solutions}

In this subsection we show that our asymptotic construction of
quasi-equilibrium $k$-spot solutions, as given in Principal Result 2.1
of \S \ref{sec:quasi}, can be reduced to that given in
\cite{2Dmulti_Wei:2003} when $\ac={\mathcal O}(\nu^{1/2})$ in
\eqref{3:ASsmallD}.  Note that when $\ac={\mathcal O}(\nu^{1/2})$,
then from \eqref{3:pval},
$A=O\left(\eps\left[-\ln\eps\right]^{1/2}\right)$, which is the key
parameter regime of \cite{2Dmulti_Wei:2003}.

For $\ac={\mathcal O}(\nu^{1/2})$, \eqref{3:ASsmallD} together with
the core problem \eqref{3:2Dcore_sol}, suggest that we expand $U_j$,
$V_j$, $S_j$, and $\chi$, in \eqref{3:2Dcore_sol} as
\begin{equation}
\label{4:NLEPexpand}
\begin{aligned}
\chi &\sim \nu^{-1/2} (\chi_{0j} + \nu \chi_{1j} + \cdots) \,,
\quad U_j \sim \nu^{-1/2} (U_{0j} + \nu U_{1j} + \cdots) \,, \\
S_j &\sim \nu^{1/2} (S_{0j} + \nu S_{1j} + \cdots) \,, \quad V_j \sim
\nu^{1/2} (V_{0j} + \nu V_{1j} + \cdots) \,.
\end{aligned}
\end{equation}
Upon substituting \eqref{4:NLEPexpand} into \eqref{3:2Dcore_sol}, and
collecting powers of $\nu$, we obtain the leading-order problem
\begin{equation}
\Delta U_{0j}  = 0 \,, \qquad   U_{0j} \to \chi_{0j}\;\;\;
\mbox{as}\;\;\rho \to \infty \,; \qquad
\Delta V_{0j} - V_{0j} + U_{0j}V_{0j}^2 = 0\,, \qquad  V_{0j} \to
0\;\;\;\mbox{as}\;\;\rho \to \infty \,. \label{4:coreEX0}
\end{equation}
The solution to \eqref{4:coreEX0} is $U_{0j} = \chi_{0j}$, and
$V_{0j}={w/\chi_{0j}}$, where $w(\rho)$ satisfies (\ref{4:groundstate}). 
At next order, we get
\begin{subequations}
\label{4:coreEX1}
\begin{gather}
\Delta U_{1j}  = U_{0j}V_{0j}^2 \,, \qquad U_{1j} \to S_{0j} \ln \rho +
\chi_{1j}\;\;\;\mbox{as}\;\;\rho \to \infty \,,
\label{4:coreEX1:a} \\
\Delta V_{1j} - V_{1j} + 2U_{0j}V_{0j} V_{1j} = -U_{1j} V_{0j}^2 \,,
\qquad V_{1j} \to 0\;\;\;\mbox{as}\;\;\rho \to \infty \,.
\end{gather}
\end{subequations}
At one higher order, we find that $U_{2j}$ satisfies
\begin{equation}
\label{4:coreEX2} \Delta U_{2j}  = 2 U_{0j}V_{0j} V_{1j} + V_{0j}^2
U_{1j}\,, \qquad U_{2j} \to S_{1j} \ln \rho + \chi_{2j}\;\;\;
\mbox{as}\;\;\rho \to \infty \,.  \\
\end{equation}

Upon applying the divergence theorem to \eqref{4:coreEX1:a}, and using
$U_{0j} = \chi_{0j}$ and $V_{0j} ={w/\chi_{0j}}$, we obtain
\begin{equation}
\label{4:S0} S_{0j} = \int_0^{\infty} U_{0j}V_{0j}^2\, \rho \,d\rho =
\frac{\,b_0}{\,\chi_{0j}}\,, \qquad b_0 \equiv \int_0^{\infty} w^2 \,
\rho \, d\rho \,,
\end{equation}
which determines $S_{0j}$ in terms of $\chi_{0j}$.
We then decompose the solution to \eqref{4:coreEX1} in the form
\[   U_{1j}  = \frac{\,1}{\,\chi_{0j}} \Big(\chi_{0j} \chi_{1j}
+ \hat{U}_{1j} \Big) \,, \qquad
V_{1j}  = \frac{\,1}{\,\chi_{0j}^3} \Big( - \chi_{0j} \chi_{1j}\, w
+ \hat{V}_{1j}\Big)\,, \]
where $\hat{U}_{1j}$ and $\hat{V}_{1j}$ are the unique solutions on
$0<\rho<\infty$ of
\begin{equation}
\Delta_{\rho} \hat{U}_{1j}  = w^2\,, \qquad \Delta_{\rho} \hat{V}_{1j} - 
 \hat{V}_{1j} + 2 w \hat{V}_{1j} = -\hat{U}_{1j}\, w^2 \,;
 \qquad \hat{V}_{1j} \to 0\, \quad \hat{U}_{1j} - b_0 \ln \rho \to 0 
  \,, \quad \mbox{as} \,\,\, \rho\to \infty \,.
  \label{4:coreEX1hat}
\end{equation}
Similarly, by applying the divergence theorem to \eqref{4:coreEX2}, we
calculate $S_{1j}$ as
\begin{equation}
\label{4:S1} S_{1j} = - \frac{\,b_0 \chi_{1j}}{\,\chi_{0j}^2} +
\frac{\,b_1 }{\,\chi_{0j}^3} \,, \qquad b_1 \equiv \int_0^{\infty} (w^2
\hat{U}_{1j} + 2 w \hat{V}_{1j}) \, \rho \, d\rho \,.
\end{equation}
The BVP solver COLSYS (cf.~\cite{colsys_Ascher:1979}) is used to
numerically compute the ground-state solution $w(\rho)$ together with
the solutions $\hat{U}_{1j}$ and $\hat{V}_{1j}$ of
\eqref{4:coreEX1hat}. Then, from a simple numerical quadrature, we
estimate $b_0 \approx 4.9347$ and $b_1 \approx 0.8706$.

Finally, upon substituting \eqref{4:S0} and \eqref{4:S1} into the
nonlinear algebraic system \eqref{3:ASsmallD} of \S \ref{sec:quasi},
we obtain that 
\begin{equation}
 \ac =\nu^{1/2}\left(\chi_{0j} +
\frac{\,b_0}{\,\chi_{0j}} \right) +
\nu^{3/2}\Big[\frac{\,b_1}{\,\chi_{0j}^3} + \left(1 -
\frac{\,b_0}{\,\chi_{0j}^2} \right) \chi_{1j}  
  +2\pi \left(\frac{\,b_0}{\,\chi_{0j}} R_{j,j} + \sum_{i\neq j}^k
\frac{\,b_0}{\,\chi_{0i}} G_{ij} \right) \Big]\,. \label{4:NLEPAS}
\end{equation}
For a prescribed $\ac$ with $\ac=\ac_0\nu^{1/2} + \ac_1 \nu^{3/2} +
\cdots$, we could then calculate $\chi_{0j}$ and $\chi_{1j}$ from
\eqref{4:NLEPAS}.  Since $\ac={\mathcal O}(\nu^{1/2})$ corresponds to
$A=O\left(\eps\left[-\ln\eps\right]^{1/2}\right)$, we conclude that it is
in this parameter range that the coupled core problem
\eqref{3:2Dcore_sol} reduces to the scalar ground-state problem for
$v$, as was exploited in \cite{2Dmulti_Wei:2003}.

Finally, we show that \eqref{4:NLEPAS} reduces to the expression
\eqref{4:quad} used in \cite{2Dmulti_Wei:2003}. To see this, we note
from \eqref{3:2dinnvar} that $\ac = {A \sqrt{D} \nu/\eps}$ and $u_j =
{\eps U_j/(A\sqrt{D})} \approx {\eps \nu^{-1/2}
\chi_{0j}/(A\sqrt{D})}$. Moreover, for $D \gg {\mathcal O}(1)$, we use
$R_{j,j} = \frac{D}{|\Omega|} + {\mathcal O}(1)$, and $G_{ij} =
\frac{D}{|\Omega|} + {\mathcal O}(1)$ for $i\neq j$ (see equation
(\ref{3:G2GN}) of \S \ref{sec:quasi}). Then, from \eqref{4:NLEPAS}, we
obtain
\begin{equation}
\frac{A \sqrt{D} \nu}{ \eps} \sim \frac{A \sqrt{D} \nu}{\eps} u_j +
\frac{\,b_0 \eps}{A \sqrt{D}\,u_j}  + \frac{2\pi\nu D}{|\Omega|}
\left[\frac{\,b_0 \eps}{A \sqrt{D}\,u_j} + \sum_{i \neq j}^k
\frac{\,b_0 \eps}{A \sqrt{D}\,u_i} \right] \,. \label{4:alim}
\end{equation}
Upon using \eqref{4:ldef} for $L_{\eps}$ and $\eta_{\eps}$, \eqref{4:alim} 
reduces to the following result, in agreement with \eqref{4:quad} of
\cite{2Dmulti_Wei:2003}:
\begin{equation}
\label{4:NLEP0} 1 - u_j = \frac{\,b_0 \eps^2}{A^2 D \nu \,u_j}  +
\frac{ 2 \pi }{|\Omega|} \frac{\,b_0 \eps^2}{A^2 }  \sum_{i=1}^k
\frac{\,1}{\,u_i} = \frac{L_{\eps} \eta_{\eps}}{\,u_j} +\sum_{i=1}^k
\frac{ L_{\eps}}{\,u_i} \,.
\end{equation}

\subsection{Comparison of the Global Eigenvalue Problem with the NLEP of
\cite{2Dmulti_Wei:2003}}

Next, we show for $\ac={\mathcal O}(\nu^{1/2})$ and $D={\mathcal
O}(\nu^{-1})$ that our eigenvalue problem, consisting of
\eqref{4:innereig} coupled to \eqref{4:CBorig}, can be reduced to the
NLEP \eqref{4:NLEPcmp} derived in \cite{2Dmulti_Wei:2003}. To show
this, we expand $\Phi_j$, $N_j$, $B_j$, and $C_j$, in
\eqref{4:innereig} as
\begin{equation*}
B_j = B_{j0} + \nu B_{j1} + \cdots\,, \qquad C_j = \nu( C_{j0} + \nu
C_{j1} + \cdots ) \,, \qquad N_j = N_{j0} + \nu N_{j1} + \cdots \,, \qquad
\Phi_j = \nu \psi_{j} + \cdots\,.
\end{equation*}
Upon substituting these expansions, together with $U_{j}\sim
\nu^{-1/2} \chi_{0j}$ and $V_j \sim \nu^{1/2} {w/\chi_{0j}}$, into
\eqref{4:innereig} for $N_j$, we conclude that $N_{j0}=B_{j0}$, and that
$N_{j1}$ satisfies
\begin{equation}
\label{4:eigenN2} \Delta_{\rho} N_{j1}  = \frac{\,w^2}{\chi_{0j}^2} B_{j0}
 + 2 w \psi_j \,, \quad 0<\rho<\infty \,; \quad N_{j1} \to C_{j0} \ln
 \rho + B_{j1} \;\;\; \mbox{as}\;\;\rho \to \infty \,.
\end{equation}
Upon applying the divergence theorem to \eqref{4:eigenN2}, we calculate
$C_{j0}$ as
\begin{equation}
\label{4:divC0} C_{j0} = \frac{\,b_0}{\chi_{0j}^2} B_{j0} + 2
\int_0^{\infty} w \psi_j \rho \, d \rho \,, \qquad b_0 \equiv 
 \int_{0}^{\infty} w^2 \rho \, d\rho \,.
\end{equation}

Then, we let $D={D_{0}/\nu}\gg 1$ where $D_0={\mathcal O}(1)$, and we
impose that $\chi_{0j}=\chi_{0}$ for $j=1,\ldots,k$, so that to
leading-order each local spot solution is the same. For $D\gg 1$, the
$\lambda$-dependent Green's function $G_{\lam \, ij}$ and its regular
part, $R_{\lam\, j,j}$, satisfying \eqref{4:Greenlamall}, have the
leading-order behavior
\begin{equation}
    G_{\lam \, i j} \sim \frac{D}{|\Omega|(1+\tau \lam)} + {\mathcal
 O}(1) \,, \quad i\neq j \,; \qquad R_{\lam \, j, j} \sim
 \frac{D}{|\Omega|(1+\tau \lam)} + {\mathcal O}(1) \,.  \label{4:glead}
\end{equation}
Upon using (\ref{4:glead}), together with $B_{j}\sim B_{j0}$ and
$C_j\sim \nu C_{j0}$, we obtain that \eqref{4:CBorig}, with $D={D_0/\nu}$,
reduces to
\[C_{j0} + \frac{2 \pi D_0}{(1 + \tau \lambda) |\Omega|}
\sum_{i=1}^k C_{i0} + B_{j0} = 0 \,, \qquad j=1,\ldots,k \,. \] 
Upon substituting \eqref{4:divC0} for $C_{j0}$ into this equation, we obtain
that $B_{j0}$ for $j=1,\ldots,k$ satisfies
\begin{equation}
 \left(1+ \frac{b_0}{\chi_0^2} \right) B_{j0} + \frac{2 \pi D_0}
{(1 + \tau \lambda) |\Omega|}\frac{b_0}{\chi_0^2} \sum_{i=1}^k
B_{i0}
 = - 2 \int_0^{\infty} w \psi_j \, \rho \, d \rho - \frac{4 \pi
D_0}{(1 + \tau \lambda) |\Omega|} \sum_{i=1}^k \int_0^{\infty} w
\psi_i \, \rho \, d \rho \,. \label{4:bnlep}
\end{equation}

Next, we define the vectors $\mathbf{p}_0$ and $\Psi$ by
$\mathbf{p}_0 \equiv (B_{10}, \ldots, B_{k0})^{T}$ and $\Psi \equiv
\left( \int_0^{\infty} \omega \psi_1 \, \rho \, d\rho\,, \ldots \,,
 \int_0^{\infty} \omega \psi_k \, \rho \, d\rho\right)^{T}$. Then,
\eqref{4:bnlep} can be written in matrix form in terms
of $\mathbf{e}=(1,\ldots,1)^T$ as
\begin{equation*}
\left[ \left(1+\frac{b_0}{\chi_0^2} \right) I + \frac{2 \pi
D_0}{(1 + \tau \lambda) |\Omega|} \frac{b_0}{\chi_0^2}\, \mathbf{e}
  \mathbf{e}^T \right]\,\mathbf{p}_0 = - \left( 2\, I + \frac{4 \pi
D_0}{(1 + \tau \lambda) |\Omega|} \mathbf{e}\mathbf{e}^T \right)  \Psi \,.
\end{equation*}
The matrix multiplying $\mathbf{p}_0$ is an invertible rank-one
perturbation of the identity matrix, and hence
\begin{equation}
 \mathbf{p}_0 = \mathcal{D} \Psi \,, \qquad \mathcal{D} \equiv -
  \left[ \left(1+\frac{b_0}{\chi_0^2} \right)I + \frac{2 \pi D_0}{(1 +
  \tau \lambda) |\Omega|} \frac{b_0}{\chi_0^2}\, \mathbf{e}\mathbf{e}^T
  \right]^{-1} \left( 2 \, I + \frac{4 \pi D_0}{(1 + \tau \lambda)
  |\Omega|} \mathbf{e}\mathbf{e}^T \right) \,. \label{4:p0val} 
\end{equation}
We then re-write some quantities in the symmetric matrix
${\mathcal{D}}$ in terms of $L_{\eps}$ and $\eta_\eps$, as defined in
\eqref{4:ldef}, to get ${b_0/\chi_0^2} = {\eta_{\eps} L_{\eps}/u_0^2}$
and ${2 \pi D_0/{\left[(1 + \tau
      \lambda)|\Omega|\right]}}=\left[\eta_{\eps}(1 + \tau \lambda)
  \right]^{-1}$.  Since the eigenvalues of the matrix $\mathbf{e e}^T$
are either $k$ or $0$, the two distinct eigenvalues $r_1$ and $r_2$ of
$\mathcal{D}$ are
\begin{equation}
r_1 = - \frac{2\chi_0^2}{b_0} \frac{L_{\eps} \eta_{\eps} (1+\tau\lam) + k
L_{\eps}}{ (u_0^2 + L_{\eps} \eta_{\eps})(1+\tau\lam) 
+ k L_{\eps}} \,, \qquad
r_2 = - \frac{2\chi_0^2}{b_0}  \frac{L_{\eps} \eta_{\eps}}{u_0^2 + L_{\eps}
\eta_{\eps}} \,. \label{4:B0}
\end{equation}

Finally, to obtain an NLEP we substitute $U_j\sim \nu^{-1/2}\chi_{0}$,
$V_j\sim \nu^{1/2} {w/\chi_0}$, $\Phi_j\sim \nu \psi_j$, and $N_{j0}
\sim B_{j0}$, into \eqref{4:innereig} for $\Phi_j$, which leads to the
following radially symmetric vector NLEP for $\psi \equiv
(\psi_1,\ldots,\psi_k)^T$:
\begin{equation*}
 \Delta_{\mathbf{\rho}}\mathbf{\psi} - \mathbf{\psi}  + 2 w \mathbf{\psi}
 + \frac{w^2 b_0 }{\chi_0^2} 
  \frac{\int_{0}^{\infty} \rho w \mathcal{D} \mathbf{\psi} \, d\rho}{
  \int_{0}^{\infty} \rho w^2 \, d\rho} = \lam \mathbf{\psi} \,, \qquad
 0 < \rho < \infty \,; \qquad \mathbf{\psi}\to 0 \,, \quad \mbox{as} \,\,\,
  \rho \to \infty \,.
\end{equation*}
By diagonalizing this vector NLEP by using the eigenpairs of
$\mathcal{D}$, we obtain the scalar NLEP
\begin{equation}
 \Delta_{\rho}\psi_c - \psi_c  + 2 w \psi_c
 + w^2 \left( \frac{r_i b_0}{\chi_0^2} \right)
  \frac{\int_{0}^{\infty} \rho w \psi_c \, d\rho}{
  \int_{0}^{\infty} \rho w^2 \, d\rho} = \lam \psi_c \,, \qquad 
 0 < \rho < \infty \,; \qquad i=1,2 \,,
 \label{4:vector_nlep}
\end{equation}
where $r_1$ and $r_2$ are the eigenvalues of $\mathcal{D}$ given in
\eqref{4:B0}. This NLEP is identical to the NLEP \eqref{4:NLEPcmp}
studied rigorously in \cite{2Dmulti_Wei:2003}.  Therefore, we conclude
that our global eigenvalue problem in \S \ref{sec:eig_rad} reduces to
leading-order in $\nu$ to the NLEP problem of \cite{2Dmulti_Wei:2003}
when $A=O\left(\eps\left[-\ln\eps\right]^{1/2}\right)$ and
$D={\mathcal O}(\nu^{-1})$.

\appendix
\setcounter{section}{2}
\renewcommand{\theequation}{\Alph{section}.\arabic{equation}}

\newsection{Numerics for the Global Eigenvalue Problem: Competition
and Oscillatory Instabilities} \label{app:num}

In \S \ref{sec:12inf}--\S \ref{sec:sym} stability thresholds are
computed for spot configurations $\mathbf{x}_1,\ldots,\mathbf{x}_k$
for which the Green's matrix $\mathcal{G}$ and the $\lambda-$dependent
Green's matrix $\mathcal{G}_{\lambda}$ are circulant.  Therefore, we
must solve the coupled eigenvalue problem \eqref{4:CBreduced} and
\eqref{4:innereig1}, where the common spot profile and source strength
$S_c$ satisfy \eqref{3:2Dcore_sol} and \eqref{3:circscalar},
respectively. In solving \eqref{3:2Dcore_sol} and \eqref{4:innereig1},
we have chosen the finite interval $[0, L]$ with $L=15$ to adequately
approximate the infinite domain.

One numerical approach is to calculate the eigenvalue $\lambda$ by a
fixed-point iterative type method for a given set of parameters $A$,
$D$, $\tau$, and $\eps$, and for a given configuration of spots. The
outline of this method is as follows:

\begin{enumerate}

\item Given $\tau, \eps, A$, and $D$,  calculate 
$S_c$, $\chi(S_c)$, and the common spot profile $U_c, V_c$ from 
\eqref{3:2Dcore_sol} and (\ref{3:circscalar}).

\item For the $n^{\mbox{th}}$ iteration, starting from
a known eigenvalue $\lambda^{n}$, solve \eqref{4:innereig1} for the common
value $\hat{B}_c$.

\item Next, with $\hat{B}_c$ known, compute a new approximation
$\lambda^{n+1}$ to $\lambda$ by numerically solving
(\ref{4:CBreduced}) using Newton's method. This requires the
evaluation of the matrix eigenvalues of the $\lambda$-dependent
Green's matrix in \eqref{4:kappa}.

\item Repeat step $2$ using the updated value $\lambda^{n+1}$ until
the algorithm converges to some $\lambda$, with an error that is less
than a given tolerance. Then, $\lambda$ is the eigenvalue of the globally
coupled eigenvalue problem.

\end{enumerate}

This fixed-point iteration approach for $\lambda$ converges rather
slowly, and its success relies on the initial guess. In addition,
since we need to study the stability for a range of values of the
parameter $\tau$, the computational effort required with this simple
scheme is rather large. In \S \ref{sec:12inf}--\ref{sec:sym}, this
scheme is used to plot the path of certain eigenvalues in the complex
plane as a parameter is varied.

Our primary numerical method used in \S \ref{sec:12inf}--\ref{sec:sym}
is based on the assumption that an instability occurs at some critical
values of the parameters. For instance, to compute the Hopf
bifurcation threshold $\tau_H$, we assume that there is a complex
conjugate pair of pure imaginary eigenvalues when $\tau =
\tau_H$ for which $\lambda_r \equiv \mbox{Re}(\lambda) = 0$ and
$\lambda_i \equiv \mbox{Im}(\lambda) \neq 0$. Instead of solving for
$\lambda$ for each given $\tau$, we fix $\lambda_r = 0$, and solve for
the thresholds $\tau_H$ and $\lambda_i$, associated with a specific
eigenvalue $\omega_{\lambda\, j}$ and eigenvector $\mathbf{v}_j$ of
$\mathcal{G}_{\lambda}$. A rough outline of the algorithm that we use
is as follows:

\begin{enumerate}
\item Given $\eps$, $A$, and $D$,  calculate 
$S_c$, $\chi(S_c)$, and the common spot profile $U_c, V_c$ from 
\eqref{3:2Dcore_sol} and (\ref{3:circscalar}).

\item Fix $\lambda_r = 0$. In the $n^{\mbox{th}}$ iteration, starting
from the current approximation $\tau^{(n)}, \lambda_{i}^{(n)}$, 
solve the BVP \eqref{4:innereig1} with boundary conditions
$\hat{N}^{\p}_c(L) = 1/L $ and $\hat{\Phi}^{\p}_c(L) = 0$ using COLSYS
(cf.~\cite{colsys_Ascher:1979}). This yields $\hat{B}_c$ as $\hat{B}_c
= \hat{N}_c(L) - \ln L$, and its derivative
$\partial_{\lambda_i}\hat{B}_{c}$ is calculated numerically by varying
$\lambda_i$ using a centered difference scheme.

\item Calculate $\mathcal{G}_{\lambda}$ and evaluate its $j^{\mbox{th}}$
eigenvalue $\omega_j (\tau \lambda_i)$. The partial derivatives
$\partial_{\tau} \omega_j$ and $\partial_{\lambda_i} \omega_{j}$ are
computed numerically by a centered difference scheme.

\item Calculate the residuals $\mbox{Re}(f_j^{(n)})$ and
$\mbox{Im}(f_j^{(n)})$ in \eqref{4:CBreduced_1}, and then calculate
the Jacobian
\[ \mathcal{J} \equiv \left( \begin{array}{cc} \partial_{\tau} 
 \mbox{Re}(f_j^{(n)}) & \partial_{\lambda_i}  \mbox{Re}(f_j^{(n)}) 
\\ \partial_{\tau}  \mbox{Im}(f_j^{(n)}) & \partial_{\lambda_i} 
 \mbox{Im}(f_j^{(n)}) \end{array} \right) = \left( \begin{array}{cc}
 \partial_{\tau}  \mbox{Re}(\hat{B}_c)   & \partial_{\lambda_i} 
 \mbox{Re}(\hat{B}_c) + 2 \pi \partial_{\lambda_i} \mbox{Re}(\omega_j) \\
\partial_{\tau}  \mbox{Im}(\hat{B}_c)   & \partial_{\lambda_i} 
 \mbox{Im}(\hat{B}_c) + 2 \pi \partial_{\lambda_i} \mbox{Im}(\omega_j)
 \end{array} \right) \,.\]

\item Use Newton's method to update  $
\left( \begin{array}{c} \tau^{(n+1)} \\ \lambda_i^{(n+1)}  
\end{array} \right)= \left( \begin{array}{c} \tau^{(n)} \\ 
\lambda_i^{(n)}  \end{array} \right) - \mathcal{J}^{-1}
 \left( \begin{array}{c} \mbox{Re}(f_j^{(n)}) \\ \mbox{Im}(f_j^{(n)}) 
 \end{array} \right) \,.$  Then go to step $2$ and 
iterate further until reaching a specified tolerance.

\end{enumerate}

To compute competition instability thresholds, we can use either the
simpler formulation \eqref{3:newres}, or else employ a similar
approach to that outlined above for computing oscillatory instability
thresholds.  The competition instability threshold occurs at $\lambda
= 0$ when the largest eigenvalue first enters the right half plane
along the $\mbox{Im}(\lam)=0$ axis. For this case, the
$\lambda-$dependent matrix satisfies $\mathcal{G}_{\lambda = 0} =
\mathcal{G}$, and both $\hat{B}_c$ and $\omega_j$ are real-valued.  We
fix all the other parameters, and choose either $A$, $D$ or the spot
location as the bifurcation parameter. For instance, if we treat $A$
as the bifurcation parameter, then the algorithm to compute the
competition stability threshold is as follows:

\begin{enumerate}
\item Set $\lambda = 0$ and fix all of the parameters except for $A$.
Calculate $\mathcal{G}$ and its $j^{\mbox{th}}$ eigenvalue $\omega_j$.

\item In the $n^{\mbox{th}}$ iteration, starting from initial guess
$A^{(n)}$ for $A$, we calculate $S_c$ and $U_c, V_c$ from \eqref{3:circscalar} 
and \eqref{3:2Dcore_sol}.

\item Solve the BVP \eqref{4:innereig1} with $\lambda = 0$ and
  $\hat{N}^{\p}_c(L) = 1/L $ and $\hat{\Phi}^{\p}_c(L) = 0$. Compute
  $\hat{B}_c$ as $\hat{B}_c = \hat{N}_c(L) - \ln L$.
    
\item Substitute $\hat{B}_c$ and $\omega_j$ in \eqref{4:CBreduced}, and
calculate the real-valued residual $f_j^{(n)}$.
    
\item Compute $\frac{\,\partial
\hat{B}_c}{\,\partial S_c}$ numerically by a centered difference scheme.
Also compute $\frac{\,\partial f_j^{(n)}}{\,\partial
A} \equiv \frac{\,\partial \hat{B}_c}{\,\partial A} = \frac{\,\partial
\hat{B}_c}{\,\partial S_c}\, \frac{\,\partial S_c}{\, \partial A}$.
    
\item In \eqref{4:CBreduced}, use Newton's method to update $A^{(n+1)}
= A^{(n)} - \left( {\partial f_j^{(n)}/\partial A}\right)\, f_j^{(n)}$.
Then, go to step $2$ and iterate further until reaching a specified
tolerance.

\end{enumerate}

 In computing the thresholds for an oscillatory instability for the
infinite-domain problem in \S \ref{sec:12inf}, and for the unit square
and disk in \S \ref{sec:sym}, we must evaluate the $\lambda$-dependent
Green's matrix ${\cal G}_{\lam}$. This is done by replacing $D$ with
${D/(1+\tau\lam)}$ in the explicit formulae of Appendix A. However,
since $\lambda$ is in general complex, in order to use the results in
Appendix A we must be able to evaluate the required modified Bessel
functions $K_{n}(z)$ and $I_{n}(z)$ of a complex argument.  This was
done using the special function software of \cite{ZJ}.

 In \S \ref{sec:12inf}--\S \ref{sec:sym} the numerical approach
outlined in this appendix was used to compute oscillatory and
competition instability thresholds for various spot configurations in
the infinite plane as well as the unit square and disk.

\end{document}